\documentclass[11pt,a4paper,twoside]{report}
\usepackage[utf8]{inputenc} %Para las tildes en formato UTF-8. !Instalé latex-ucs
\usepackage[english]{babel}
\usepackage{graphicx}
\usepackage{amssymb,amsmath,wasysym}
\usepackage[toc,page]{appendix}
\usepackage{axodraw}
\usepackage{fancyhdr}
\usepackage{epstopdf}
\usepackage{cite}
\usepackage{cancel}
\usepackage{slashed}

\usepackage[bookmarks=false]{hyperref}
\newcommand{\mgut}{M_\text{GUT}}

\newcommand{\qa}{Q_\alpha}
\newcommand{\qb}{Q_\beta}
\newcommand{\qba}{\bar{Q}_{\dot{\alpha}}}
\newcommand{\qbb}{\bar{Q}_{\dot{\beta}}}
\newcommand{\da}{D_\alpha}
\newcommand{\dba}{\bar{D}_{\dot{\alpha}}}
\newcommand{\ta}{\theta^\alpha}
\newcommand{\tb}{\theta^\beta}
\newcommand{\tba}{\bar{\theta}^{\dot{\alpha}}}
\newcommand{\tbb}{\bar{\theta}^{\dot{\beta}}}
\newcommand{\xt}{\theta}
\newcommand{\bt}{\bar{\theta}}

\newcommand{\uu}{U(1)_Y}
\newcommand{\sud}{SU(2)_L}
\newcommand{\sut}{SU(3)_C}

\newcommand{\s}{\smallskip}

\renewcommand{\headrulewidth}{0.5pt}
\renewcommand{\footrulewidth}{0pt}
\fancypagestyle{plain}{
	\fancyhead{}
	\renewcommand{\headrulewidth}{0pt}
}

\begin{document}
%...Title Page
\begin{titlepage}

\vspace*{-2cm}
\includegraphics[width=50mm]{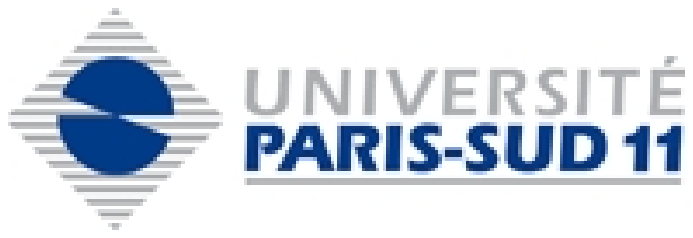}
\hfill
\includegraphics[width=20mm]{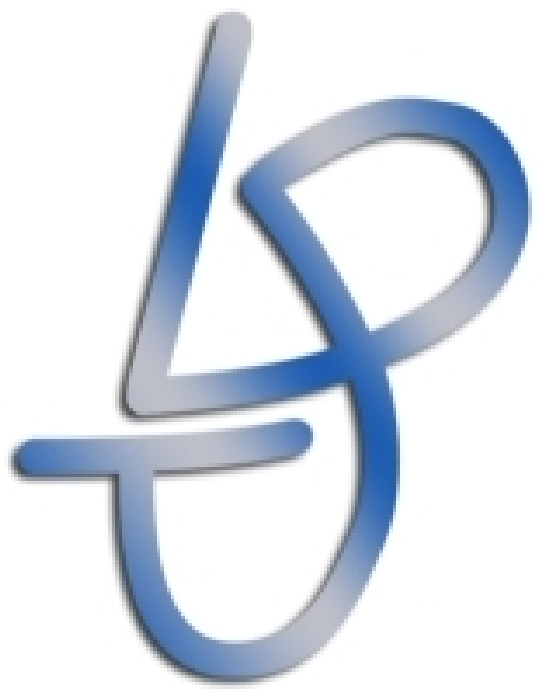}
\vspace{1cm}
\begin{center}
\fbox{{\bf UNIVERSITÉ PARIS-SUD 11}}

\vskip1cm

{\bf THÈSE DE DOCTORAT}
\vskip1cm

Spécialité: {\bf  PHYSIQUE THÉORIQUE}\\
\vskip0.3cm
Présentée\\
 pour obtenir le grade de
\vskip0.75cm
\large {\bf Docteur de l'Université Paris - Sud}

\vskip0.75cm

par

\vskip0.5cm

{\sc \bf Andreas Goudelis}

\vskip1cm

Sujet:

\vskip0.2cm

\Large{\bf  Detection of WIMP-like dark matter in some }\\
\Large{\bf  extensions of the Standard Model}

\end{center}
\vskip0.75cm
\begin{flushright}

\vskip0.75cm
$\begin{array}{lll}
\mbox{Prof.}	& \mbox{Geneviève B\'elanger}  & \mbox{Rapporteur}\\
\mbox{Prof.}	& \mbox{Lars Bergström}      & \mbox{Rapporteur}\\
\mbox{Prof.}	& \mbox{Abdelhak Djouadi}    & \mbox{Directeur de Thèse}\\ 
\mbox{Prof.}	& \mbox{Ulrich Ellwanger}    & \mbox{Président du jury}\\  
\mbox{Prof.}	& \mbox{Yann Mambrini}       & \mbox{Membre invité du jury}\\  
\mbox{Prof.}	& \mbox{Aldo Morselli} 	     &\mbox{Membre du jury}\\
\mbox{Prof.}	& \mbox{Carlos Mu\~noz}	     &\mbox{Membre du jury}\\
\end{array}$

\end{flushright}

\thispagestyle{empty}

%%%%%%%%%%%%%%%%%%%%%%%%%%%%%%%%%%%%%%%%%%%%%%%
%%%%%%%%%%%%%%%%%%%%%%%%%%%%%%%%%%%%%%%%%%%%%%%
%%%%%%%%%%%%%%%%%%%%%%%%%%%%%%%%%%%%%%%%%%%%%%%
\newpage

\thispagestyle{empty}
\begin{flushright}
 \textit{Aux batailles qu'on a donn\'e}\\
 \textit{et \`a celles qui viennent.}
\end{flushright}
~\\

%%%%%%%%%%%%%%%%%%%%%%%%%%%%%%%%%%%%%%%%%%%%%%%
%%%%%%%%%%%%%%%%%%%%%%%%%%%%%%%%%%%%%%%%%%%%%%%
%%%%%%%%%%%%%%%%%%%%%%%%%%%%%%%%%%%%%%%%%%%%%%%
\newpage
\thispagestyle{empty}

\vspace*{-2cm}
\includegraphics[width=50mm]{beginning/psud.eps}
\hfill
\includegraphics[width=20mm]{beginning/LPT.eps}
\vspace{1cm}

\begin{center}
\fbox{{\bf UNIVERSITÉ PARIS-SUD 11}}

\vskip1cm

{\bf THÈSE DE DOCTORAT}
\vskip1cm

Spécialité: {\bf  PHYSIQUE THÉORIQUE}\\
\vskip0.3cm
Présentée\\
 pour obtenir le grade de
\vskip0.75cm
\large {\bf Docteur de l'Université Paris - Sud}

\vskip0.75cm

par

\vskip0.5cm

{\sc \bf Andreas Goudelis}

\vskip1cm

\Large{\bf  Detection of WIMP-like dark matter in some }\\
\Large{\bf  extensions of the Standard Model}

\vskip0.2cm

\end{center}
\vskip0.75cm
\begin{flushright}

\vskip0.75cm
$\begin{array}{lll}
\mbox{Prof.}	& \mbox{Geneviève B\'elanger}  & \mbox{Rapporteuse}\\
\mbox{Prof.}	& \mbox{Lars Bergström}      & \mbox{Rapporteur}\\
\mbox{Prof.}	& \mbox{Abdelhak Djouadi}    & \mbox{Directeur de Thèse}\\ 
\mbox{Prof.}	& \mbox{Ulrich Ellwanger}    & \mbox{Président du jury}\\  
\mbox{Prof.}	& \mbox{Yann Mambrini}       & \mbox{Membre invité du jury}\\  
\mbox{Prof.}	& \mbox{Aldo Morselli} 	     &\mbox{Membre du jury}\\
\mbox{Prof.}	& \mbox{Carlos Mu\~noz}	     &\mbox{Membre du jury}\\
\end{array}$

\end{flushright}

%%%%%%%%%%%%%%%%%%%%%%%%%%%%%%%%%%%%%%%%%%%%%%%
%%%%%%%%%%%%%%%%%%%%%%%%%%%%%%%%%%%%%%%%%%%%%%%
%%%%%%%%%%%%%%%%%%%%%%%%%%%%%%%%%%%%%%%%%%%%%%%
\newpage

\thispagestyle{empty}

~\\

\end{titlepage}

\tableofcontents

%...Introductions are not to be numerated and appear in plain style
\pagestyle{empty}

\chapter*{Introduction et sommaire}
\addcontentsline{toc}{chapter}{Introduction et sommaire}

\vspace{-0.7cm}
Ce travail traite la d\'etection de la mati\`ere noire dans quelques extensions du Mod\`ele
Standard de la physique des particules. Avant de se lancer dans la partie la plus technique, 
il serait peut - \^etre pertinent de r\'esumer l'id\'ee principale derri\`ere le concept de
la mati\`ere noire.

Aujourd'hui les physiciens disposent d'une s\'erie des lois physiques qui peuvent
fournir des cadres de description des int\'eractions entre des objets massifs. Ces lois
sont la Relativit\'e G\'en\'erale et la m\'ecanique Newtonienne. Ces th\'eories peuvent, entre autres,
relier trois choses: la masse d'un objet, celles des objets de son entourage ainsi que 
son comportement cin\'ematique (la mani\`ere dont il bouge).
Si on peut mesurer deux parmi ces trois quantit\'es, nous pouvons en principe pr\'edire la
troisi\`eme. Si nous pouvons mesurer toutes les trois quantit\'es, 
nous pouvons en tirer des conclusions concernant la validit\'e de la th\'eorie.

Ces deux th\'eories ont effectivement eu un succ\`es \'enorme pour expliquer toute une s\'erie
de ph\'enom\`enes, comme le mouvement des objets plan\'etaires dans le syst\`eme solaire, le mouvement
des objets au voisinage de la terre, bref, un grand nombre des ph\'enom\`enes o\`u la gravitation
est suffisament forte pour qu'elle puisse jouer un r\^ole important.

Supposons pourtant qu'il s'av\`ere que des situations apparaissent o\`u si on mesure la masse d'un objet
et celle des objets qui l'entourent, le comportement cin\'ematique pr\'evu par la th\'eorie est
en d\'esaccord avec celui qui est observ\'e. Dans un tel cas, il nous reste 
deux solutions:
\begin{itemize}
 \item On pourrait admettre que nous sommes arriv\'es aux limites de validit\'e de notre 
th\'eorie et essayer de modifier sa formulation ou son context conceptuel.
 \item Ou il faut admettre que les observations sont elles-m\^emes, dans un sens, 
probl\'ematiques: Ça peut \^etre que nous avons mal mesur\'e la masse de l'objet observ\'e, 
la masse des objets qui l'entourent ou sa trajectoire dans le ciel. Autrement dit, 
nous avons peut \^etre utilis\'e ``les bonnes \'equations'' mais pas les bonnes quantit\'es
dans ces \'equations.
\end{itemize}
Si nous ne souhaitons pas modifier la forme des th\'eories de la gravitation
(malgr\'e le fait qu'une telle approche est souvent adopt\'ee par plusieurs chercheurs et groupes, avec
des resultats tr\`es int\'eressants), il nous reste que la deuxi\`eme option. Il y a pourtant 
plusieures raisons pour croire que les mesures de masse et de comportement cin\'ematique sont
assez fiables 
\footnote{Il est pourtant assez int\'eressant que la premi\`ere mesure relative \'etait, en fait, 
fausse!}.
On pourrait donc supposer avec une certaine confiance que les mesures concernant les objets
qu'on voit ne sont pas fausses. Il semble alors qu'on arrive \`a une impasse. On pourrait dans
un tel cas supposer qu'il existe quelques objets qu'on ne peut pas voir effectivement
mais qui devraient \^etre pris en compte en tant qu'objets ``d'entourage'' afin de pr\'edire 
correctement le comportement cin\'ematique de l'objet qu'on observe. Cela a \'et\'e le premier
argument pour le postulat de la mati\`ere noire.

Il s'av\`ere qu'adopter une telle solution n'est pas quelque chose de nouveau en physique:
la plan\`ete de Pluto (qui n'a aujourd'hui rien de mystique pour les astrophysiciens) a \'et\'e
pr\'edite pour la premi\`ere fois sans \^etre visible \`a l'\'epoque gr\^ace \`a des anomalies 
dans le mouvement des objets plan\'etaires. L'existence des neutrinos (exp\'erimentalement
bien - \'etablie aujourd'hui) a \'et\'e postul\'ee car la lois de la conservation de l'\'energie 
totale d'un syst\`eme \'etait menac\'ee dans les observations des d\'esint\'egrations radioactives 
de quelques noyaux. La mati\`ere noire a \'et\'e introduite de fa\c{c}on similaire, \`a la
suite d'observations astronomiques \`a differentes \'echelles.

Mais que se passe-t-il s'il se trouve que cette mati\`ere noire ne peut pas \^etre constitu\'ee 
par la forme de mati\`ere connue? Dans ce cas, on pourrait dire que l'histoire devient
encore plus interessante! Pourrait-on esp\'erer observer cette nouvelle mati\`ere un jour?
Cela d\'epend fortement de sa nature, et est un des sujets principaux de ce travail.

Il existe plusieurs fa\c{c}ons (compl\'ementaires) de traiter la mati\`ere noire. Dans cette
th\`ese, nous nous focaliserons sur l'approche de la physique des particules, en introduisant
pourtant des \'elements d'astrophysique et de cosmologie (cette division est d\'ej\`a assez
brutale).

Le premier chapitre du travail ci-dessous pr\'esente bri\`evement quelques \'elements de base
qui sont indispensables pour travailler sur la mati\`ere noire. Nous allons pr\'esenter
quelques \'el\'ements de formalisme de Relativit\'e G\'en\'erale et de cosmologie, puis d\'ecrire
plus en d\'etail quelques arguments pour le postulat de la mati\`ere noire. Une petite
discussion suit sur les caract\'eristiques possibles de la mati\`ere noire et l'incapacit\'e
du Mod\`ele Standard de fournir un candidat viable.

Dans le deuxi\`eme chapitre nous discutons sur les modes de d\'etection de la mati\`ere noire.
Apr\`es une pr\'esentation des modes principaux de d\'etection
(au moins ceux qui existent aujourd'hui), nous expliquons quelques incertitudes
qui entrent dans chaque canal de d\'etection et nous pr\'esentons bri\`evement la situation
experimentalle actuelle. Ensuite nous d\'ecrivons quelques resultats originaux sur la capacit\'e
des experiences \`a reconstruire quelques propri\'et\'es des candidats pour la mati\`ere
noire.

Au troisi\`eme chapitre nous d\'ecrivons une solution minimale au probl\`eme de la mati\`ere
noire. Le Mod\`ele Standard de la physique des particules est l\'eg\`erement \'etendu pour
inclure une particule qui pourrait constituer la masse manquante de l'univers. Apr\`es une
description du mod\`ele et une \'etude de quelques unes de ses propri\'et\'es et contraintes, 
nous discutons ses perspctives de d\'etection en montrant qu'il pourrait 
\^etre test\'e aupr\`es des exp\'eriences courrantes ou \`a venir.

Dans le quatri\`eme chapitre, nous pr\'esentons une deuxi\`eme solution au probl\`eme de la mati\`ere
noire. Cette fois, l'approche est beaucoup moins minimale et \'economique et les mod\`eles
qui en d\'ecoulent sont consid\'erablement plus compliqu\'es. Pourtant, cette approche 
(appell\'ee \textit{sypersym\'etrie}) pr\'esente l'avantage d'avoir \'et\'e motiv\'ee par des questions
tout \`a fait diff\'erentes, l'existence d'un candidat de mati\`ere noire \'etant une cons\'equence 
des consid\'erations plus g\'en\'eriques. Nous pr\'esentons quelques \'el\'ements de formalisme 
supersym\'etrique et puis la version minimale de cette sorte des th\'eories, ainsi 
qu'un des candidats qu'elle propose. Ensuite, nous discutons d'un probl\`eme
sp\'ecifique qui appara\^it dans ce mod\`ele et nous \'etudions la mati\`ere noire dans
deux mod\`eles qui essayent de resoudre ce probl\`eme.

Finalement, un resum\'e et la conclusion forment un cinqui\`eme chapitre. Trois appendices suivent,
qui contiennent quelques \'el\'ements techniques ainsi qu'un certain nombre des points qui
renforcent quelques arguments qui sont donn\'es dans le corps du texte.\\
\begin{flushright}
 Paris, printemps/\'et\'e 2010
\end{flushright}

\newpage
\chapter*{Introduction and summary}
\addcontentsline{toc}{chapter}{Introduction and summary}

This work treats the detection of dark matter in some extensions of the Standard Model
of particle physics. Before starting with the more technical part of this work, it
could perhaps be pertinent to summarize the main idea behind the concept dark matter.

Today physicists have at their disposal a series of ``physical laws'' that provide
frameworks to describe the interactions between massive objects. These laws are General Relativity
and Newtonian mechanics. Among others, these physical theories can relate
three things: the mass of an object, that of its surrounding ones, as well as the 
kinematic behavior that we would expect this object to follow (how it should move).
If we can measure two of these quantities, we can in principle predict the third
one or, seen a bit differently, if we can measure all three of them we can draw
conclusions on the validity of these theories.

Indeed, these two theories have been extremely successful in explaining a very large amount
of phenomena, such as the motion of planets in the solar system, the motion of objects
close to the earth, in short, a huge variety of phenomena where gravitational
interactions are strong enough play a significant role.

Suppose however that it turns out that there are some cases where if we somehow measure
the mass of an object and of its surrounding ones, the predicted kinematic behavior is
in conflict with the observed one. In this case, we are left with more or less two
options:
\begin{itemize}
 \item We could admit that the limits of validity of the theories at our disposal
have been reached, and try to modify our theories' formulation, or conceptual
content, 
 \item Or we must admit that the observations are themselves, in some sense, problematic, 
where by this we mean that we could have mismeasured the mass of the observed object, 
the mass of its surrounding objects or its trajectory in the sky.
In other words, it could be that we used ``the correct formulae'' but we did not plug the
correct quantities in these formulae.
\end{itemize}
In practice, if we wish not to alter the form of current theories of gravitation (although such
an approach is also adopted by several researchers and groups, with very interesting results), 
we are stuck with the second option. Now, there are several reasons that allow one to believe
that mass and kinematic behavior measurements can be quite reliable
\footnote{It is though quite interesting that the first relevant measurement was actually
false!}.
So, it could be supposed quite safely that the measurements concerning the things we see are
not false. The situation seems thus to reach some sort of impass. We could then assume that
there are some objects that we cannot currently see and which should be taken into account
as ``surrounding objects'' in order to correctly predict the kinematic behavior of the
object under observation. This was the first argument that lead to the dark matter postulate.

It turns out that resorting to such a solution is not something new in physics: 
The planet Pluto (which today has nothing mystical for astrophysicists) was first predicted
without being visible at the time due to anomalies in the movement of planetary objects.
The existence of neutrinos (which is experimentally well - established today) 
was postulated because the law of conservation
of the total energy of a system was put in hazard when observing radioactive decays of
some nuclei. It is in a similar manner that dark matter was introduced, receiving 
as we shall see in the following increasingly strong support from other observational
evidence.

But what if it turns out that this dark matter cannot be constituted by the known 
matter forms? Well, in this case, one could say that things get even more interesting!
Could we hope one day to actually see this new matter? This depends strongly on its
nature, and this is one of the main topics in this work.

There are actually many (complementary) ways of tackling dark matter. In this work, we shall focus
on the approach of particle physics, introducing however elements of astrophysics and
cosmology (this division is already quite brutal).

The first chapter of this work briefly presents some background material needed to work with
dark matter. We shall present some elements of formalism of General Relativity and cosmology, 
then describe some motivation that lead to the dark matter postulate. A brief discussion follows
on the possible characteristics of dark matter and the incapacity of the Standard Model of particle
physics to accommodate some ``viable candidate''.

In the second chapter we start discussing how dark matter could be detected. After
presenting the main methods of dark matter detection (at least those existing today), we
further explain some of the uncertainties entering each detection mode and briefly present the
current status of the field. We then present some original results on the capacity of
the corresponding experiments to reconstruct some properties of dark matter candidates.

In the third chapter we describe a minimal approach towards the dark matter problem. The
Standard Model of particle physics is slightly extended so as to incorporate a 
particle that could account for the missing matter content of the universe. After describing the
model and examining some aspects of its constraints and phenomenology, we discuss its detection perspectives showing
that we could expect it to be tested in present or oncoming experiments.

In the fourth chapter we describe another solution to the dark matter problem. This time the
approach is much less minimal and economic, with the corresponding models being considerably
more complicated. However, this approach (called \textit{supersymmetry}) presents the merit of 
having been motivated from completely different questions, with the dark matter candidate
particle arising quite naturally as a consequence of more generic considerations. We present 
some elements of supersymmetric formalism and then the minimal version of such theories along
with one of the potential dark matter candidates it proposes. Then, we discuss a particular 
problem arising in this model and treat dark matter in two examples of models trying to tackle
this issue.

Finally, we summarize and conclude in the fifth chapter. Three appendices contain some technical
elements used throughout this work as well as some points aiming at corroborating certain arguments
given in the main body of the text.
\\ \\
\begin{flushright}
 Paris, spring/summer 2010
\end{flushright}

\newpage

%...Switch to fancy page layout
\pagestyle{fancy}

\chapter{The two Standard Models}

One of the main issues in today's high energy physics is the almost complete absence of 
experimental evidence of new physics. It turns out, however, that in trying to combine our
knowledge in the field of physics at very small scales (particle physics) and physics at very large
ones (cosmology), the necessity to go beyond one of them arises. In this chapter, we shall 
introduce the two ``standard pictures'' for the respective fields and explain how they turn
out to be insufficient in order to explain today's observations.

%%%%%%%%%%%%%%%%%%%%%%%%%%%%%%%%%%%%%%%%%%%%%%%%%%%%%%%%%%%%%%%%%%%%%%%%%%%%%%%%%%%%%%%%%%%%%%%%%%%%%
%%%%%%%%%%%%%%%%%%%%%%%%%%%%%%%%%%%%%%%%%%%%%%%%%%%%%%%%%%%%%%%%%%%%%%%%%%%%%%%%%%%%%%%%%%%%%%%%%%%%%
%%%%%%%%%%%%%%%%%%%%%%%%%%%%%%%%%%%%%%%%%%%%%%%%%%%%%%%%%%%%%%%%%%%%%%%%%%%%%%%%%%%%%%%%%%%%%%%%%%%%%
\section{A basic observation}

Looking at the sky and asking questions about ``everything there is'' is not a new habit. 
People have been engaging themselves into this enterprise since a very long time. It is in fact partly by
observing motions of extraterrestrial objects that we understood that everything in this
world seems to be moving with respect to other things. In the beginning, of course, at least to our
knowledge, there is one object which was considered to be completely inert, the earth.
With time, when the world came to position of being able to overcome the Aristotelean approach towards
nature, it was gradually understood that in fact our planet is also in constant motion.

Despite this fundamental understanding, though, it took quite some more time for another basic breakthrough
in our way of thinking to be established. Even with the rise of Newtonian (and then Lagrangian and 
Hamiltonian) mechanics, the mechanistic approach towards reality suggested that extremely complex phenomena
can take place in our world, but which actually happen in an otherwise inert, constant, eternal environment,
space: a field is something that propagates through space, even perhaps with a limited velocity, but space
itself does not play a crucial role in the process. It could perhaps, like in the Aristotelean theory,
acquire some properties, but its form always remained the same.

It is not until the twentieth century that scientists started wondering about the role of space (-time)
in physical phenomena. And it is not until the introduction of General Relativity that spacetime started
being conceived as something which is actually bound to the phenomena taking place ``inside'' it:
The action of a gravitational field is now understood as a deformation of spacetime.

At the same time, the development of quantum mechanics contributed further to the demolition of this
mechanistic view of our world. More on this shall be said in the following, when we describe the standard
model of particle physics.

Once we start believing that spacetime is a dynamical object, which can change and participate actively
in physical processes, a question that could rise is whether spacetime is something finite. And, according
to our current perceptions, it actually is. A number of observations have contributed to this evolution 
in our concepts, which are actually often said to consist the basis of modern cosmology.

The first observation is that the universe (including spacetime) seems to be actually expanding with time. 
If this is the case (and today we are quite confident that it is), this means that if we try to trace
its evolution throughout time it is necessary to develop the appropriate notions and formalism in order to be able
to treat phenomena taking place in such an environment. And, if the universe has been expanding up to its
current size for quite some time, an immediate question is how could it look at much earlier times, when
its size was much smaller than today. The belief that the total amount of energy and matter (unified already
in special relativity) stays constant, i.e. that there is no creation or eradication of these two in
the universe (since the ``universe'' is a closed thermodynamical system, it's simply ``everything there is'')
could probably make us hope that we can trace back what the place we live in looked like in past times.
In the next section, we shall introduce some of these notions and formalisms.

%%%%%%%%%%%%%%%%%%%%%%%%%%%%%%%%%%%%%%%%%%%%%%%%%%%%%%%%%%%%%%%%%%%%%%%%%%%%%%%%%%%%%%%%%%%%%%%%%%%%%
%%%%%%%%%%%%%%%%%%%%%%%%%%%%%%%%%%%%%%%%%%%%%%%%%%%%%%%%%%%%%%%%%%%%%%%%%%%%%%%%%%%%%%%%%%%%%%%%%%%%%
%%%%%%%%%%%%%%%%%%%%%%%%%%%%%%%%%%%%%%%%%%%%%%%%%%%%%%%%%%%%%%%%%%%%%%%%%%%%%%%%%%%%%%%%%%%%%%%%%%%%%
\section{Elements of General Relativity}

One of the main questions that arise in an expanding universe (but which has much more far-reaching
consequences) is how to relate physical and coordinate distances 
\cite{Schutz:1985jx, Dodelson:2003ft, Kolb:1990vq}. 
This is actually achieved through
the introduction of the metric of the space, $g_{\mu\nu}$, through:
\begin{equation}
 ds^2 = g_{\mu\nu} dx^\mu dx^\nu
\label{metric}
\end{equation}
where summation over repeated indices is implied.\\
The distance $s$ is invariant in every coordinate system, whereas in the above equation the metric
connects the values of coordinates to the more physical measure of the interval $ds$.

As we shall shortly describe, General Relativity relates the presence of matter to deformations
of the spacetime metric. In the special case where there is no matter in space spacetime
is flat, the metric is the Minkowski one and particles move in straight lines. Now, in a curved
manifold, particles moving freely will no longer follow straight lines but rather more complex
trajectories, called \textit{geodesics}. The equations of motion in curved spacetime acquire, in
their turn, also a more complicated form, which is
\begin{equation}
 \frac{d^2 x^\mu}{d \lambda^2} = -\Gamma^\mu_{\alpha\beta} \frac{dx^\alpha}{d\lambda}
\frac{dx^\beta}{d\lambda}
\label{geodesic}
\end{equation}
where $\lambda$ is some monotonically increasing parameter along the particles' path (which can always be
eliminated in actual calculations) whereas $\Gamma^\mu_{\alpha\beta}$ are the Christoffel symbols, 
defined as
\begin{equation}
 \Gamma^\mu_{\alpha\beta} = \frac{1}{2} g^{\mu\nu} (\partial_\beta g_{\alpha\nu} + 
\partial_\alpha g_{\beta\nu} - \partial_\nu g_{\alpha\beta})
\label{christoffel}
\end{equation}

Now, since we said that the universe is expanding, an important question is the way this should be
reflected in the metric. Actually, for a flat spacetime, the metric in an expanding universe is 
given by the famous Friedmann-Robertson-Walker-Lemaître (FLRW) formula
\begin{equation}
 g_{\mu\nu} = \left( \begin{array}{cccc}
               -1 & 0 & 0 & 0\\
		0 & a^2(t) & 0 & 0\\
		0 & 0 & a^2(t) & 0\\
		0 & 0 & 0 & a^2(t)
              \end{array} \right) 
\label{FLRWmetric}
\end{equation}
where $a(t)$ is a function of time, called the \textit{scale factor}, which describes the evolution
of distances with time: if the comoving distance between two points at some time is $1$, at later times
it will be $a(t) > 1$. Its present value is set to $1$ and it increases with time.

There is a fundamental consequence of the FLRW metric. If we start with a particle of $4$ - momentum
$p^\alpha = (E, \vec{p})$ and define the $\lambda$ parameter through $p^\alpha = dx^\alpha/d\lambda$
then we can easily demonstrate that the energy of the particle decreases with time (i.e. with the
expansion of the universe). This observation finds its physical interpretation in the fact that
as the universe expands, and all distances become larger, the wavelength of the particle also
stretches decreasing its energy.
\\ \\
At the center of General Relativity lies the observation that the metric can account for
gravitational phenomena. This is done through the Einstein equation
\begin{equation}
 G_{\mu\nu} = R_{\mu\nu} - \frac{1}{2} g_{\mu\nu} R = 8\pi G T_{\mu\nu}
\label{Einstein}
\end{equation}
where $G_{\mu\nu}$ is called the Einstein tensor, $R_{\mu\nu}$ is the Ricci tensor defined as
\begin{equation}
 R_{\mu\nu} = \Gamma^\alpha_{\mu\nu,\alpha} - \Gamma^\alpha_{\mu\alpha,\nu}
+\Gamma^\alpha_{\beta\alpha} \Gamma^\beta_{\mu\nu}-
 \Gamma^\alpha_{\beta\nu}\Gamma^\beta_{\mu\alpha}
\end{equation}
with $,\alpha$ denoting derivation with respect to $x^\alpha$, $R$ is the Ricci scalar defined as the contraction
of the Ricci tensor with the metric, $G$ is the Newton constant whereas $T_{\mu\nu}$ is 
the energy-momentum tensor. Eq.(\ref{Einstein}) introduces a clear correlation among the matter-energy
content of spacetime and its geometry: what would be called in Newtonian terms ``gravitational
force'' is actually the effect of the deformation of spacetime and the consequent modification in
the particles' trajectories.

In the LHS of Eq.\eqref{Einstein} one can also add a term of the form $\Lambda g_{\mu\nu}$, called
the cosmological constant term, with enormous consequences that we shall not be examining here.
Let us just note for the moment that this term can be used in order to explain the accelerating
expansion of the universe and is responsible for the so-called ``dark energy'' which seems to
be the main component of the matter-energy content of our universe.
\\ \\
Now, observations at large scales seem to favor a couple of very fundamental assumptions, namely that the
universe is homogeneous and isotropic. The first assumption means that for sufficiently large
volumes (and we shall comment on that in the following) the density of the universe is the same 
no matter in which region we consider the volume. The second assumption means
that there should be no privileged direction in the universe.

It turns out that these two basic assumptions constrain enormously the possibilities to
consider different metrics as solutions to the Einstein equations. The most general
form of the metric compatible with the two assumptions can actually be written, in spherical
coordinates, as:
\begin{equation}
 ds^2 = -dt^2 + a(t)^2\left[ \frac{dr^2}{1 - kr^2} + r^2 (d\theta^2 + \sin^2\theta d\phi^2)\right]
\label{FLRWspherical}
\end{equation}
where $k$ takes the values $-1, 0, 1$ for a universe with negative, zero or positive curvature respectively.
We note that the spatial part of the metric has a dependence on time. This is actually a generalization
of the FLRW metric that we wrote down before.
\\ \\
A question arises on the form of the energy-momentum tensor. In cosmology, the matter-energy content
of the universe (encoded exactly by the energy-momentum tensor) is usually taken to behave (at 
sufficiently large scales) as a perfect fluid. In this case, $T_{\mu\nu}$ takes the form
\begin{equation}
 T_{\mu\nu} = \mbox{diag}(\rho, p, p, p)
\end{equation}
where $\rho$ is the fluid's density and p is its pressure.

It is not that difficult to see that by taking the time component of the Einstein equation, we should
get a relation between the matter-energy density of the universe and the scale factor.
In fact, this relation is
\begin{equation}
 H^2(t) \equiv \left( \frac{\dot{a}(t)}{a(t)}\right)  = \frac{8\pi G \rho}{3} - \frac{k}{a^2(t)}
\label{FriedmannTime}
\end{equation}
which is called the Friedmann equation.
\\ \\
Furthermore, the continuity relation for the fluid yields
\begin{equation}
 \frac{d(\rho a^3)}{dt} = -p \frac{d (a^3)}{dt}
\end{equation}
\\
Now, define the \textit{critical density} of the universe as
\begin{equation}
 \rho_c \equiv \frac{3 H^2}{8 \pi G}.
\end{equation}
The value of this parameter determines whether the universe expands forever or eventually 
collapses to a singularity once more
\footnote{The presence of the cosmological constant term complicates the situation, but this
detail goes beyond the scope of the present work.}.

Then, by further defining normalized densities $\Omega \equiv \rho/\rho_c$, $\Omega_\Lambda
\equiv \Lambda/(8\pi G \rho_c)$ and $\Omega_k \equiv -k/\dot{a}^2$ for the energy-matter, dark
energy and curvature densities respectively, we can bring the Friedmann
equation to a very simple form
\begin{equation}
 \Omega + \Omega_\Lambda + \Omega_k = 1
\end{equation}
\\
With these tools at hand, we can in principle examine the evolution of all of the universe's 
components with time during its expansion. We should stress at this point that since the 
Einstein equations are obviously linear with the energy-momentum tensor, one could (at least
as far as gravitational interactions are concerned) separate the various contributions and
examine their evolution in an uncorrelated manner.

Finally, so far we have considered that the cosmic fluid is perfectly homogeneous. However, 
global homogeneity does not impose that locally there cannot be small inhomogeneities.
The usual approach is to actually consider the ``vacuum'' state as being the
perfectly isotropic and homogeneous one, then adding perturbations around the ground state
metric and examining the evolution with time. The very existence of structures in our universe
(planets, stars, galaxies, clusters of galaxies) is a witness that inhomogeneities must
have existed. These do actually exist and we can today see their footprints in the Cosmic 
Microwave Background (CMB) radiation anisotropies. 

%%%%%%%%%%%%%%%%%%%%%%%%%%%%%%%%%%%%%%%%%%%%%%%%%%%%%%%%%%%%%%%%%%%%%%%%%%%%%%%%%%%%%%%%%%%%%%%%%%%%%
%%%%%%%%%%%%%%%%%%%%%%%%%%%%%%%%%%%%%%%%%%%%%%%%%%%%%%%%%%%%%%%%%%%%%%%%%%%%%%%%%%%%%%%%%%%%%%%%%%%%%
%%%%%%%%%%%%%%%%%%%%%%%%%%%%%%%%%%%%%%%%%%%%%%%%%%%%%%%%%%%%%%%%%%%%%%%%%%%%%%%%%%%%%%%%%%%%%%%%%%%%%

\section{Evidence for Dark Matter}
\subsection{Galactic Rotation Curves}
According to our previous discussion, once one knows the matter content of some gravitating
system, one should be able to write down its equations of motion and predict its kinematic
behavior. Actually, in the cases of vanishing curvature (as for example when examining a
galaxy as a whole) then even simple Newtonian mechanics should be sufficient to fulfill this
task. Inversely, by observing the kinematic behavior of a gravitating system, we should be able
to determine its matter content.
Departures from the behavior predicted by the theory should either be explained by a 
modification in the relation among the dynamics and the matter content of the system, or by a
modification in the matter content itself.

This was actually one of the first arguments supporting the existence of some form of (yet, as
we shall argue in the following chapters) invisible matter. 
\footnote{Several reviews on (particle) dark matter exist. See, for example, 
\cite{Jungman:1995df, Bertone:2004pz, Munoz:2003gx, Feng:2010gw}}
According to Newtonian mechanics, the
circular velocity of the gas and the stars comprising a galaxy as a function of the distance from
the galactic center is given by the very simple formula
\begin{equation}
 v(r) = \sqrt{\frac{G M(r)}{r}}
\end{equation}
where $v$ is the velocity at distance $r$ from the center of the galaxy and $M$ is the mass enclosed
in a sphere of radius $r$. From this simple relation, even accepting that in reality things could be
slightly more complex, one would expect that once we move far enough from the galactic center, where
we can say quite confidently from observations of the luminous components that we have included in 
$M(r)$ the essential part of the galaxy's mass, we would expect a velocity falling roughly as 
$1/\sqrt{r}$. It turns out that very often this is not at all the case.

\begin{figure}[hbtp]
\begin{center}
\vspace{2cm}
\includegraphics[width=6cm]{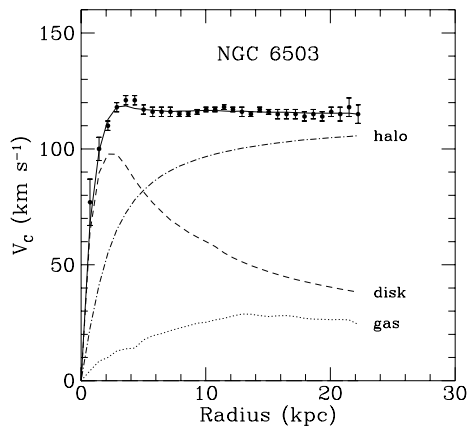}\\
\end{center}
\caption{Rotation curve of NGC 6503. The dotted, dashed and dash-dotted lines are
the contributions of gas, disk and dark matter, respectively. Figure taken from \cite{Bertone:2004pz}.}
\label{NGC6503}
\end{figure}

In fig.\ref{NGC6503} we see the rotation curve for the spiral galaxy NGC $6503$ and the various 
velocity distributions as a function of the distance from the galaxy's center. 
We clearly see
an unexpected behavior, with the the overall velocity following a flat distribution at large
distances from the luminous disk. This behavior suggests that either there should be some modification 
of the laws of gravity at the galactic scales, or that there is some important (in fact, dominant)
quantity of some sort of matter that has persistently evaded detection. We shall be referring,
according to habits, to such a matter as being ``dark'', i.e. non-luminous.

We should of course notice that both explanations seem equally reasonable at this stage.
Although throughout this work we shall be examining some possible consequences of the latter
possibility, there is extensive research since quite some time in the field of potential 
modifications of gravity: Modified Newtonian Dynamics (MOND), Tensor-Vector-Scalar (TeVeS)
theories as relativistic versions of the former, $f(R)$ theories etc. And, certainly, one cannot
exclude that the solution lies somewhere in between. Finally, it has also been argued that 
once we depart from the fluid approach for the various components of the universe and start
taking into account granular structure, results could be severely altered (although such
approaches have not been examined as thoroughly) \cite{Carati:2009us}.

\subsection{Gravitational Lensing}
Important evidence for the existence of either some form or non-luminous matter or some modification
of gravity comes equally from gravitational lensing experiments. Here, we observe the way light bends
under the influence of gravitational potentials as it passes by massive objects. So, for example, an
object ``behind'' (with respect to us) a galaxy cluster emitting light shall be seen distorted 
with respect to its original shape. By observing this distortion effect, we can infer information on the
distribution of matter in space. It turns out that quite often, by only taking into account luminous
matter, it is not possible to explain such effects.

\begin{figure}[hbtp]
\begin{center}
\vspace{1cm}
\includegraphics[width=6cm]{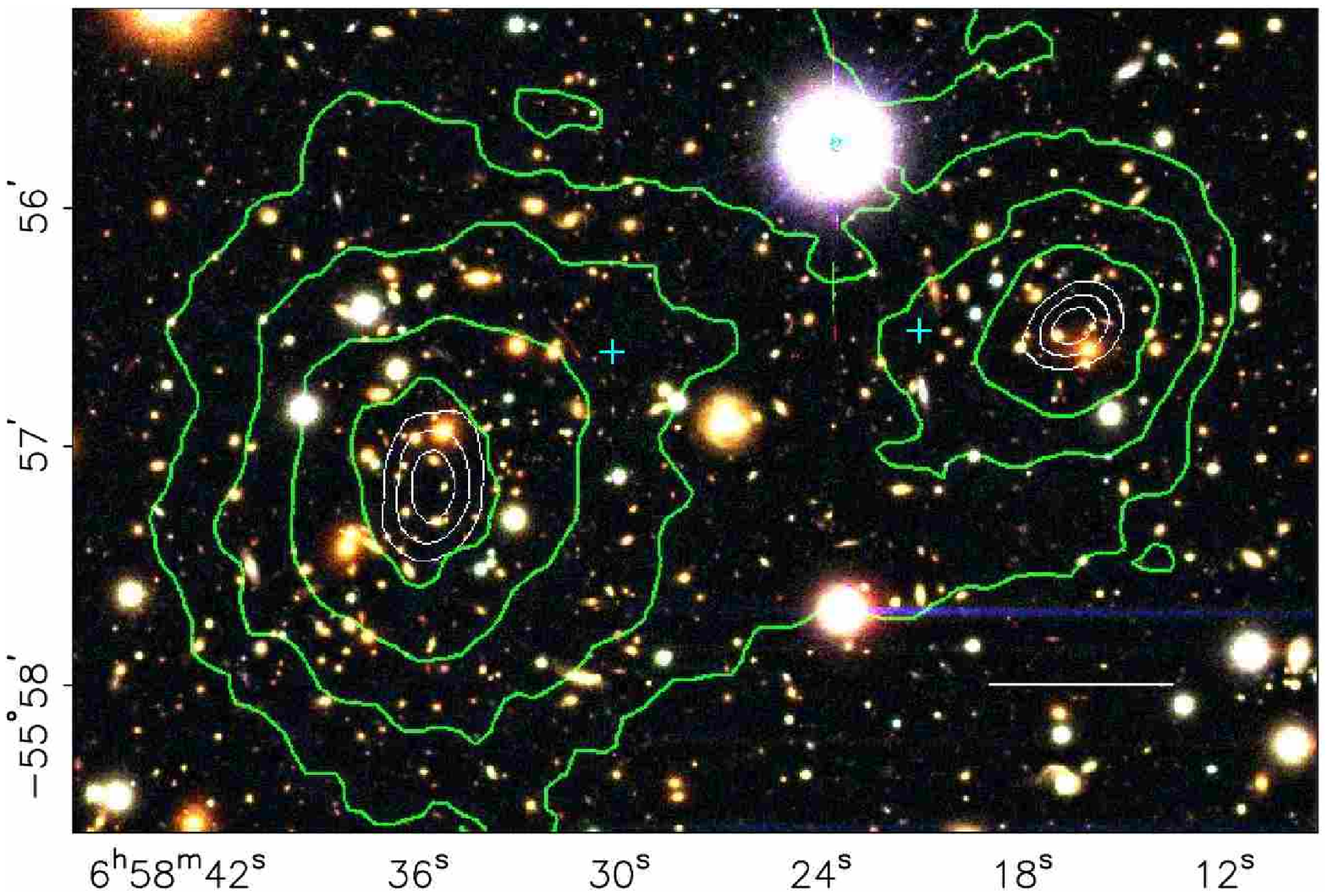} \hspace{2cm}
\includegraphics[width=6cm]{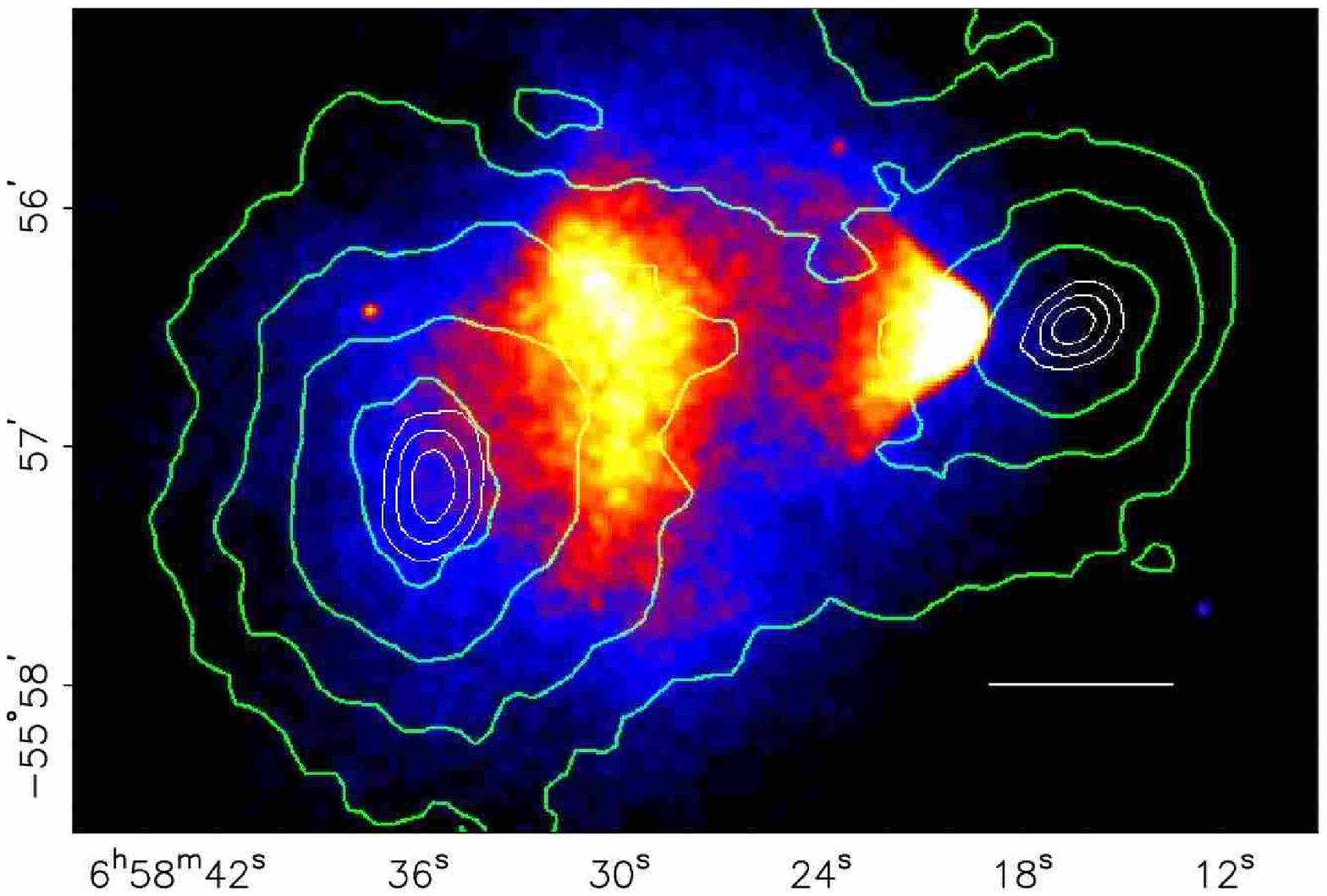}\\
\end{center}
\caption{The collision of the two galaxy clusters within the Bullet cluster. Left: The 
distribution of the star components of the cluster along with mass density isocontours.
Right: The distribution of dust as seen in X-rays against mass density isocontours.
Figure taken from \cite{Clowe:2006eq}.}
\label{BulletCluster}
\end{figure}

One of the most compelling evidence for the existence of dark matter came recently with weak
lensing  observations of the Bullet galaxy cluster (1E 0657-56) \cite{Clowe:2006eq} by the Hubble
space telescope. The Bullet cluster consists of two colliding galaxy clusters. These, in their
turn, have two main components, namely stars and dust/gas.
One would expect that upon collision the star components of the two clusters would not be significantly
slowed down, since they practically do not interact among them. On the other hand, the intergalactic
gas, which is the major component of the cluster's luminous matter, does interact electromagnetically
and should thus be significantly slowed down during the collision. This is actually the case. This can
be seen in fig.\ref{BulletCluster}. In the left hand-side image we see the star content of the two
clusters which is well separated after the collision. 
The dust component on the other hand, which can be seen in the right hand-side figure
in $X$-rays (mostly concentrated around the yellow regions), seems not to follow the
same behavior, being concentrated mostly around the ``collision center''. Then, since the
gas constitutes the main component of the two clusters, the overall matter distribution for
the Bullet cluster, depicted in the images by the mass density contours in green,  
should follow its behavior. However, we see that this is clearly not the case. On the contrary, 
the matter density seems to be following a behavior that resembles significantly to the one of stars.
This means that there should be some important, actually largely dominant, amount of matter that
is unobservable for the moment and drives the matter density distribution in the cluster. This
matter should also be quite collisionless, since if it interacted strongly (apart from the fact that
it would have probably been observed as such), it would at least behave in a way qualitatively 
similar to the gas. We note that this is considered to be one of the main observations favoring
the dark matter interpretation against the modified gravity one, since gravity (at least according
to our current perceptions) does not distinguish among matter forms.

We should note that at the scale of galaxy clusters, many more observations advocate for the
existence of dark matter, although quite often alternative approaches can also explain relevant
data. In these cases, a quite large number of observations can be made, mainly relevant to 
the observed temperature of the clusters, and compared against the expected temperature as a
function of the cluster's mass. Alternatively, one can compare the mass of the cluster as observed
in all wavelengths against data from weak gravitational lensing. In most cases, a severe discrepancy
is found among calculation and observations, supporting the existence of a significant amount
of dark matter.

\subsection{The Cosmic Microwave Background}
Many more observations support the idea that there is something wrong in the Einstein equations, 
either concerning their general form (modifications of gravity) or the matter
content we should plug in these equations in order to perform calculations. It is not our goal
to list all of this evidence at this point. We could not, however, omit one of the most important 
observations of modern cosmology, namely the anisotropies of the Cosmic Microwave Background radiation.

The existence of the CMB was first predicted by Gamow in 1948 and further 
established experimentally by the work of Penzias and Wilson in 1965. 
Giving a more or less complete description of the CMB-related theory and phenomenology
goes well beyond the scope of the present work. The main idea is that once the universe
cooled down sufficiently, the photons could no longer interact with matter (matter - radiation
decoupling) in an efficient way. Then, as we saw, expansion tends to ``stretch'' the wavelengths
of photons, leading to an overall cooling of this ``relic'' radiation. 

At first, what was observed is a practically perfect blackbody radiation, consisting exactly
of these photons as predicted. With the measurements of COBE and especially the 
WMAP mission, it was realized that although the 
CMB appears to \textit{globally} isotropic and homogeneous, at smaller scales there are 
small but measurable anisotropies in this radiation.

The standard picture today is that small inhomogeneities - perturbations around the 
ground state metric result in gravitational wells and hills. Then, the rest of the matter-energy
content of the universe feels these gravitational wells and tends to cluster around them.
As time passes by, the -initially small- perturbations grow and structures (gravitationally
bound systems) begin to form. And, according to our current beliefs, this process follows a 
``bottom - up'' procedure, that is, small structures are the first to form and then cluster
to produce large scale structures. 

At the same time, these inhomogeneities in the distribution of matter have an impact on the
local temperature of the relic photons, since the former undergo energy losses under the 
influence of the local gravitational wells. In this sense, the anisotropies in the CMB
constitute a snapshot of these initial matter-energy inhomogeneities in the early universe.
The observed temperature anisotropies in the sky can be expanded in terms of spherical harmonics
\begin{equation}
 \frac{\delta T}{T}(\theta, \phi) = \sum_{l=2}^{\infty} \sum_{m=-l}^{l} a_{l m} Y_{l m}(\theta, \phi)
\label{CMBexpansion}
\end{equation}
We can calculate exactly the variance of $a_{l m} Y_{l m}$ through
\begin{equation}
 C_l \equiv \left\langle|a_{l m}|^2 \right\rangle \equiv \frac{1}{2 l + 1} \sum_{m=-l}^{l}|a_{l m}|^2.
\end{equation}
Then, we can plot $C_l$ as a function of $l$ (in practice, what is plotted is $l(l + 1)C_l /2\pi $
against $l$) and fit a cosmological model to this data. The CMB analysis is a rather complex one,
which we do not intend to present here.

In the end, we can infer a number of values for the model parameters and calculate
many others. Among the most important ones, and the ones of interest for this work, are the 
density of baryonic matter and the \textit{overall} matter content in the universe. These values
are found to be, according to the WMAP $5$-year results \cite{Dunkley:2008ie}
\begin{equation}
 \Omega_b h^2 = 0.02267 \pm 0.00058, \ \ \Omega_m h^2 = 0.1326 \pm 0.0063
\end{equation}
with the discrepancy between the two quantities being more than obvious. Under the light
of the CMB measurement data, we can infer the overall dark matter density of the universe
as
\begin{equation}
 \Omega_{DM} h^2 = 0.1131 \pm 0.0034
\end{equation}
which shows clearly that the dark matter component should be the dominant matter component of
our universe.

The CMB is actually one of the most important pillars of modern cosmology, with consequences that
are much more far-reaching than described here. Up to the writing of this work, the only known
cosmological model that is consistent with all observations, including the CMB, is the so-called
$\Lambda$CDM model, that is, a model based on general relativity including a non-zero cosmological
constant term in the Einstein equations, along with Cold Dark Matter. We shall return to this point
in the following, but we note already that in this case ``Cold'' means non-relativistic.
%%%%%%%%%%%%%%%%%%%%%%%%%%%%%%%%%%%%%%%%%%%%%%%%%%%%%%%%%%%%%%%%%%%%%%%%%%%%%%%%%%%%%%%%%%%%%%%%%%%%%
%%%%%%%%%%%%%%%%%%%%%%%%%%%%%%%%%%%%%%%%%%%%%%%%%%%%%%%%%%%%%%%%%%%%%%%%%%%%%%%%%%%%%%%%%%%%%%%%%%%%%
%%%%%%%%%%%%%%%%%%%%%%%%%%%%%%%%%%%%%%%%%%%%%%%%%%%%%%%%%%%%%%%%%%%%%%%%%%%%%%%%%%%%%%%%%%%%%%%%%%%%%
\section{A small parenthesis: Spatial distribution of Dark Matter}
\label{SpatialDistribution}
Before setting off to examine what could be the nature of dark matter, it would be useful to shortly 
comment on how we could expect dark matter to be distributed in space.

Interestingly, our current notions on the distribution of dark matter rely mostly on theoretical models and
computer simulations. The most popular approach towards determining the distribution function of
dark matter in space (often called \textit{halo profile}) is by means of large $N$ - body simulations.

In this approach, a number of dark matter ``particles'' is placed in a confined volume, under some 
initial conditions, and then let to evolve according to well-specified non-linear gravitational dynamics.
Two of the most crucial parameters in $N$ - body simulations is the mass and length resolution, namely
the smallest dark matter ``particle'' considered as well as the smallest distance defined in the simulations
(which has to be finite in any case since gravitational interactions diverge at very small distances).

The results of most recent $N$ - body simulations tend to agree, at least qualitatively, at large scales:
dark matter forms extended halos that tend to be more dense in the centers of galaxies, their density decreasing
as one moves towards the outskirts of the galaxy, the halo itself being nevertheless much more extended
than the luminous part of the galaxy (for comparison, if one takes the Milky Way to have roughly a $20$ kpc
radius, its halo should extend up to more than $100$ kpc). However, as our computational capacities are
limited, the innermost regions of galaxies fall far beyond current resolutions. In order for one to sketch
the entire halo profile up to distances of a few pc from the galactic center, very strong assumptions and
extrapolations must be made. Hence, up to this day, the behavior of density profiles in the innermost 
regions of galaxies remains an issue of debate and controversy. We should add that a long-standing challenge
for such simulations is the incorporation of baryons in the setup (up to this day most simulations concern
pure dark matter halos), since they are expected to behave in a much different way than dark matter, which
is practically collisionless. Furthermore, there is quite some debate on the existence of substructures within
the halo, that is, inhomogeneities in the density function. Despite these controversies, there exist today 
some reference models against which all other $N$ - body simulations 
(but also analytical computations) are usually compared. 

The first reference model is the so-called \textit{modified isothermal} sphere model, that predicts a spherically
symmetric density distribution of the form
\begin{equation}
 \rho(r) = \frac{\rho_0}{1 + (r/a)^2}
\label{IsothermalSphere}
\end{equation}
where $\rho_0$ is a normalization parameter, that can be for example fixed in order to reproduce the
local density of dark matter which can be quite constrained from observations, $r$ is the distance 
from the galactic center and $a$ is a characteristic length.

An important benchmark model came in 1996, when Navarro, Frenk and White conducted a very 
important simulation \cite{Navarro:1996gj} and came up with a halo profile of the form
\begin{equation}
 \rho(r) = \frac{\rho_0}{(r/a)^\gamma [1 + (r/a)^\alpha]^{(\beta - \gamma)/\alpha}}
\label{NFWgeneral}
\end{equation}
The NFW simulation gave the values
$(\alpha, \beta, \gamma) = (1, 3, 1)$, with the resulting profile being called today the NFW profile. The
isothermal sphere model corresponds to choices $(\alpha, \beta, \gamma) = (2, 2, 0)$.

Some time later, Moore \textit{et al} \cite{Moore:1999nt} conducted a new simulation which gave quite different
values, namely $(\alpha, \beta, \gamma) = (1.5, 3, 1.5)$. By extrapolating the NFW and Moore \textit{et al}
profiles up to distances of a few kpc from the Galactic Center (GC), one can easily calculate that the two functions
present a difference of roughly an order of magnitude concerning the value they yield for the DM density.

However, both of these models (and actually, all models following the parametrization in Eq.(\ref{NFWgeneral}))
present a common feature: towards the inner regions of the galaxy, the dark matter density should rise
significantly, forming some sort of cusp. The findings of the recent Via Lactea II  simulation \cite{Diemand:2008in}
seem to confirm the general tendencies appearing in the NFW profile.

On the other hand, the Aquarius Project simulation \cite{Springel:2008cc} finds a quite different result,
with their halos being best reproduced by a completely different parametrization, the so-called Einasto
profile which can be written as
\begin{equation}
 \rho(r) = \rho_s \exp \left[ -\frac{2}{\alpha} \left( \left( \frac{r}{r_s}\right)^\alpha - 1 \right) \right] \ , \ \alpha = 0.17
\end{equation}
where $r_s = 20$ kpc is a characteristic length, whereas $\rho_s$ is, once again, a normalization factor.
The Einasto profile is significantly less steep than the NFW - like ones and does not diverge at very
small distances.

Finally, a number of simulations have tried to integrate baryons in the analysis. Such an example is
\cite{Prada:2004pi}. In this work, the authors find that in the presence of baryons dark matter could adiabatically
collapse in the inner region of the galactic center, forming an even more spiked profile, with 
$(\alpha, \beta, \gamma) = (0.8, 2.7, 1.45)$ if one takes as an initial distribution the NFW one.
This effect is often referred to as \textit{adiabatic compression}. In the following, we shall
be denoting this profile by NFW$_c$.

The general tendencies at larger distances are, less debated upon. One important issue is,
however, the existence of substructures in the galactic halo. These substructures appear in all high - 
resolution $N$ - body simulations in the forms of filaments and clumps and could in principle play
a significant role in indirect detection of dark matter that we shall discuss in the following chapters.
Further reference to this point will be made in the appropriate sections.

As computational capacity increases, simulations with rising resolutions are ran and it is expected
that in the few years to follow we might have a much better understanding of the way dark matter halos
form and evolve in a realistic environment.

%%%%%%%%%%%%%%%%%%%%%%%%%%%%%%%%%%%%%%%%%%%%%%%%%%%%%%%%%%%%%%%%%%%%%%%%%%%%%%%%%%%%%%%%%%%%%%%%%%%%%
%%%%%%%%%%%%%%%%%%%%%%%%%%%%%%%%%%%%%%%%%%%%%%%%%%%%%%%%%%%%%%%%%%%%%%%%%%%%%%%%%%%%%%%%%%%%%%%%%%%%%
%%%%%%%%%%%%%%%%%%%%%%%%%%%%%%%%%%%%%%%%%%%%%%%%%%%%%%%%%%%%%%%%%%%%%%%%%%%%%%%%%%%%%%%%%%%%%%%%%%%%%
\section{Thermal relics and WIMPs}
Once the existence of some -seemingly dominant- quantity of non-luminous matter is established, 
a natural question to ask would be what is its nature. We already saw that it seems very problematic
to assume that dark matter is of baryonic nature. Dark baryons are indeed expected to exist, in the
form of dust, gas but also dark compact objects, but the overall matter content of the universe 
cannot be baryon-dominated. 
The main evidence for the existence of dark matter comes from gravity-related
observations. But as we pointed out, gravity does not distinguish among different ``textures''
of matter: gravitational interactions only depend on the mass of the objects/particles under 
consideration.

Nevertheless, there is another issue that we have not examined so far, and this is kinematics
(which at the end of the day is indeed also related to the dark matter particles' masses).
According to our current beliefs, structure formation in the universe proceeds starting from
small-scale structures which followingly cluster into larger ones. Then, it turns out that
this assumption is incompatible with a dark matter that is moving at relativistic velocities.
The reason is simply that relativistic particles cannot easily cluster to form structures.
Instead, they would tend to form large-scale structures at first and then perhaps leave
some space for smaller ones to form. So, this could be a first important constraint:
dark matter must be cold, or at least not hot. Indeed, the possibility of Warm Dark Matter 
(semi-relativistic or cold/hot mixture, with the cold component being dominant) has been invoked
in order to explain a series of weaknesses of the simplest $\Lambda$CDM model such as
the rotation curves of some dwarf galaxies around the Milky Way or problems in the Lithium 
abundance calculation. We shall not be dealing with such models in the following. We shall
stick to the assumption that dark matter should be strictly non-relativistic.

Then, where does dark matter come from? It is clear that the number of different mechanisms
that could be invoked for the massive production of a non-relativistic species of particles 
(not to mention \textit{many} of them) could be huge. Indeed, many mechanisms have been invoked
in order to explain the observed dark matter abundance. To cite just two among the most 
common ones, dark matter could be the decay product of a heavier particle. This is 
actually the case for a large class of ``dark matter candidates'' such as the gravitino, 
superpatner of the graviton, arising always (but not only) in models of gauge mediated 
supersymmetry breaking. 

The most popular picture, however, is the one of ``thermal relics''. Suppose some stable species $\chi$
of mass $m_\chi$ in the early universe. For simplicity, also assume that all heavier particles
have been wiped away or decayed. Hence, the only species existing at the time (at least participating
in relevant interactions) are (some of the) standard model particles along with the stable species.
Suppose also for simplicity that the candidate can annihilate only with itself, producing standard
model particles. While the temperature of the universe is high enough, the standard model particles can 
annihilate in order to produce our candidate. If the candidate is heavy enough, at some point the
temperature of the universe could fall to such a point that the standard model particles are
no longer energetic enough so as to efficiently produce the $\chi$ particle. At this point, 
the density of the candidate starts falling tending asymptotically to zero.
Now, the question is whether we can actually reproduce the observed relic abundance of dark matter.
This question is obviously related in one hand to the mass of the $\chi$ particle and on the other
hand to the interactions in which it participates.

This process should in principle be described by some continuity relation, 
namely the Boltzmann equation. In the general case, the former is written as
\begin{equation}
 {\cal{L}}[f] = {\cal{C}}[f]
\end{equation}
where $f$ is the quantity under examination (in our case the density of the species), 
${\cal{L}}[f]$ is the Liouville operator describing the evolution of a phase-space volume
and ${\cal{C}}[f]$ is the collision operator describing all possible processes amounting to
production or destruction of $f$.
\\
The general relativistic form of the Liouville operator is given by
\begin{equation}
 {\cal{L}}_{\mbox{\begin{tiny}GR\end{tiny}}} = p^\alpha \frac{\partial}{\partial x^\alpha} - 
\Gamma^\alpha_{\beta \gamma} p^\beta p^\gamma \frac{\partial}{\partial p^\alpha}
\label{Boltzmann}
\end{equation}
For the sake of simplicity, let's assume that we are actually only dealing with a process of the
form
\begin{equation}
 1 + 2 \longrightarrow 3 + 4
\end{equation}
Then, the Boltzmann equation for species 1 in an expanding universe should take the form
\begin{eqnarray}
 a^{-3}\frac{d(n_1 a^3)}{dt} & = &
\int\frac{d^3 \vec{p}_1}{(2\pi)^3 2 E_1}
\int\frac{d^3 \vec{p}_2}{(2\pi)^3 2 E_2}
\int\frac{d^3 \vec{p}_3}{(2\pi)^3 2 E_3}
\int\frac{d^3 \vec{p}_4}{(2\pi)^3 2 E_4}\\ \nonumber
& \times & (2\pi)^4 \delta^3(\vec{p}_1+\vec{p}_2+\vec{p}_3+\vec{p}_4) \delta(E_1+E_2-E_3-E_4)
|{\cal{M}}|^2\\ \nonumber
& \times & [f_3 f_4 (1 \pm f_1)(1 \pm f_2) - f_1 f_2 (1 \pm f_3)(1 \pm f_4)]
\label{BoltzmannSimple}
\end{eqnarray}
where $n_i$ is the number density of species $i$, $(E_i, \vec{p}_i)$ is the $i$-th particle
species' four-momentum, ${\cal{M}}$ is the interaction's matrix element and $f_i$ is the occupation
number of species $i$. This equation is valid under the assumption of the reversibility of the
process.
\\ \\
To proceed, we make a series of assumptions:
\begin{itemize}
 \item Kinetic equilibrium: scattering takes place so rapidly that the
distributions of the various species are either Fermi-Dirac (FD) or Bose-Einstein (BE).
This means that the only uncertainty on the species' distributions lies in their chemical
potentials $\mu(t)$
 \item Furthermore, scattering takes place at a temperature well below $E-\mu$. Then,
the FD or BE nature of the species becomes indistinguishable. Statistics is simply 
Maxwell-Boltzmann.
\end{itemize}
Under these assumptions, the number density $n_i$ of a species $i$ is
\begin{equation}
 n_i = g_i \ e^{\mu_i/T} \int\frac{d^3\vec{p}_i}{(2\pi)^3} e^{-E_i/T}
\end{equation}
$g_i$ being the degeneracy of the species. Define further the number density at 
equilibrium as
\begin{equation}
  n_i^{\mbox{\begin{tiny}eq\end{tiny}}} = g_i \int\frac{d^3\vec{p}_i}{(2\pi)^3} e^{-E_i/T}
\end{equation}
and finally, define the thermally averaged cross-section for the reaction as
\begin{eqnarray}
 \left\langle\sigma v\right\rangle & \equiv & 
\int\frac{d^3 \vec{p}_1}{(2\pi)^3 2 E_1}
\int\frac{d^3 \vec{p}_2}{(2\pi)^3 2 E_2}
\int\frac{d^3 \vec{p}_3}{(2\pi)^3 2 E_3}
\int\frac{d^3 \vec{p}_4}{(2\pi)^3 2 E_4}\\ \nonumber
& \times & e^{-(E_1 + E_2)/T}\\ \nonumber
& \times & (2\pi)^4 \delta^3(\vec{p}_1+\vec{p}_2+\vec{p}_3+\vec{p}_4) \delta(E_1+E_2-E_3-E_4)
|{\cal{M}}|^2
\label{sigmav}
\end{eqnarray}
Then, the Boltzmann equation becomes
\begin{equation}
 a^{-3}\frac{d(n_1 a^3)}{dt} = 
n_1^{\mbox{\begin{tiny}eq\end{tiny}}}
n_2^{\mbox{\begin{tiny}eq\end{tiny}}}
\left\langle\sigma v\right\rangle
\left[ \frac{n_3 n_4}{n_3^{\mbox{\begin{tiny}eq\end{tiny}}} n_4^{\mbox{\begin{tiny}eq\end{tiny}}}} - 
\frac{n_1 n_2}{n_1^{\mbox{\begin{tiny}eq\end{tiny}}} n_2^{\mbox{\begin{tiny}eq\end{tiny}}}} \right]
\label{BoltzmannEquilibrium}
\end{equation}
In order to proceed further, we must make another assumption, namely that the annihilation
products $3$ and $4$ go very quickly into equilibrium with the thermal background. In this case, 
the first term in the brackets is simply equal to $1$, since we can replace $n_3, n_4$ by
$n_3^{\mbox{\begin{tiny}eq\end{tiny}}}, n_4^{\mbox{\begin{tiny}eq\end{tiny}}}$ respectively.
And then, we have already made the assumption that particle $1$ $\equiv$ particle $2$. So, suppose
that the initial state particles are described by some density $n_\chi$.
Under these assumptions, Eq.(\ref{BoltzmannEquilibrium}) becomes
\begin{equation}
 a^{-3} \frac{d (n_\chi a^3)}{dt} = \left\langle\sigma v\right\rangle [(n_\chi^{\mbox{\begin{tiny}eq\end{tiny}}})^2 - n_\chi^2]
\label{MasterRelic}
\end{equation}
From the last equation it becomes quite manifest how we can treat different final states. Because
clearly, it is not impossible that a pair of $\chi$ particles might be able to annihilate in a
whole series of final states. The idea then is just to replace $\sigma$, the partial self
annihilation cross-section with the total one (which we shall from now on call $\sigma$, separating
it from the partial ones which shall hereafter be referred to as $\sigma_i$). We also note that
in case particles $1$ and $2$ have a particle-antiparticle relation, then a factor of $1/2$ should
be included in front of the cross-section, since the density of the annihilating particles 
will be half the one of majorana-like particles. We also note that, as pointed out for example
in \cite{Gondolo:1990dk}, the velocity $v$ appearing in the total thermally averaged annihilation
cross-section is not the relative velocity among the two particles, but rather the Moller 
velocity, defined as:
\begin{equation}
 v_{\mbox{\begin{tiny}Mol\end{tiny}}} = 
\left[ |\vec{v}_1 - \vec{v}_2|^2 -  |\vec{v}_1 \times \vec{v}_2|^2 \right]^{1/2}
\label{MollerVelocity}
\end{equation}
Eq.\eqref{MasterRelic}  is, in principle and in the simplest of cases, the equation one has to solve in order to compute
the relic abundance for a dark matter species. One of the most tedious parts in this computation
is, of course, expected to be the calculation of $\left\langle\sigma v\right\rangle$. To perform these tasks a number of
numerical codes have been developed, like micrOMEGAs 
\cite{Belanger:2001fz, Belanger:2004yn, Belanger:2006is, Belanger:2007zz, Belanger:2008sj, Belanger:2010pz} 
or DarkSUSY \cite{Gondolo:2000ee, Gondolo:2002tz, Gondolo:2004sc, Gondolo:2005we}, 
offering a high level of automation to the whole process. Further complications such as coannihilations shall
be referred to in the following chapters, when relevant. For the moment, we limit ourselves to pointing
out that it can be the case that the dark matter candidate shares some quantum numbers with another
particle in the theory and that the two particles can annihilate with each other.

Going a little further, it would be interesting to briefly sketch what kinds of particles could 
in principle satisfy Eq.(\ref{MasterRelic}) according to the assumptions we have made.
First of all, it is not an absurd assumption to stick to stable particles. We saw that unstable
particles with very long lifetimes can also very well constitute viable candidates. But in our
case we shall be only considering stable ones. Let's define a couple of new variables, namely
$Y = n/s$, where $s$ is the total entropy density of the universe, as well as $x = m/T$.
Then, Eq.(\ref{MasterRelic}) can be recast into the form
\begin{equation}
\frac{dY}{dx} = \frac{1}{3 H} \frac{ds}{dx} 
\left\langle\sigma v_{\mbox{\begin{tiny}Mol\end{tiny}}}\right\rangle (Y_{\mbox{\begin{tiny}eq\end{tiny}}}^2 - Y^2)
\end{equation}
For heavy states, $\left\langle\sigma v_{\mbox{\begin{tiny}Mol\end{tiny}}}\right\rangle$ can be expanded with respect
to the Moller velocity as
\begin{equation}
 \left\langle\sigma v_{\mbox{\begin{tiny}Mol\end{tiny}}}\right\rangle = 
a + b \left\langle v^2 \right\rangle + {\cal{O}}(\left\langle v^4 \right\rangle) \approx a + 6 (b/x)
\end{equation}
Then, the evolution equation can be written as
\begin{equation}
\frac{dY}{dx} = -\left( \frac{45}{\pi} G\right)^{-1/2} \frac{g_*^{1/2} m}{x^2} 
(a + 6 (b/x)) (Y^2 - Y_{\mbox{\begin{tiny}eq\end{tiny}}}^2)
\end{equation}
with
\begin{equation}
g_*^{1/2} = \frac{h_{eff}}{g_{eff}^{1/2}} \left( 1 + \frac{T}{3 h_{eff}}\frac{d h_{eff}}{dT}\right) , \ \
g_{eff} = \frac{30 \rho}{\pi^2 T^4}, \ \ 
h_{eff}(T) =  \frac{45 s}{2 \pi^2 T^3}
\end{equation}
Eventually, and under some further assumptions, one can actually obtain an (not always valid) approximate
relation for the relic density
\begin{equation}
 \Omega_{\chi} h^2 \approx \frac{3 \times 10^{-27} \mbox{cm}^3 \mbox{sec}^{-1}}{\left\langle\sigma v\right\rangle}
\end{equation}
It is interesting that this equation yields the correct orders of magnitude for cross-section values
characterizing typically the weak interactions. This is the starting point for a very large class of
dark matter candidates within the category of thermal relics, namely Weakly Interacting Massive
Particles (WIMPs).
\\ \\
We shortly commented on the nature of dark matter and described in some detail a possible
mechanism that could in principle give rise to the observed cosmic abundance of dark matter. Then, 
the simplest idea would be to look for a dark matter candidate within the zoo of already known 
elementary particles. The world of elementary particles is today described at a very good level
by the Standard Model of particle physics, which we shall briefly describe in the following
section.
%%%%%%%%%%%%%%%%%%%%%%%%%%%%%%%%%%%%%%%%%%%%%%%%%%%%%%%%%%%%%%%%%%%%%%%%%%%%%%%%%%%%%%%%%%%%%%%%%%%%%
%%%%%%%%%%%%%%%%%%%%%%%%%%%%%%%%%%%%%%%%%%%%%%%%%%%%%%%%%%%%%%%%%%%%%%%%%%%%%%%%%%%%%%%%%%%%%%%%%%%%%
%%%%%%%%%%%%%%%%%%%%%%%%%%%%%%%%%%%%%%%%%%%%%%%%%%%%%%%%%%%%%%%%%%%%%%%%%%%%%%%%%%%%%%%%%%%%%%%%%%%%%
\section{The Standard Model of particle physics}
The Standard Model of particle physics 
\footnote{Numerous excellent textbooks on the Standard Model, Quantum Field Theory and gauge theory
exist. See, for example, 
\cite{Ryder:1985wq,Peskin:1995ev,Rivers:1987hi,Veltman:1994wz,Lahiri:2005sm,Cheng:1985bj,Huang:1982ik,Srednicki:2007qs}}
is actually comprised of two models, quantum chromodynamics
(QCD) that was proposed in the 1960's as a theory of strong nuclear interactions 
\cite{GellMann:1964nj, Fritzsch:1973pi, Gross:1973id, Politzer:1973fx, 'tHooft:1972qz} 
and the electroweak model proposed again in the 1960's by Glashow, Salam and Weinberg 
\cite{Glashow:1961tr, Weinberg:1967tq, Salam:1968rm}.

Before briefly describing the model itself, we state a few introductory remarks concerning the
Standard Model with a somewhat more general content:
\begin{itemize}
 \item It is a four-dimensional renormalizable quantum field theory: Particles are described by
field operators acting on basis vectors of Hilbert spaces, creating and annihilating degrees of freedom
(other particles). The operator fields live in the four-dimensional spacetime. The model can be in principle
extrapolated to arbitrarily high energies.
 \item It is a gauge theory. The fields' interactions are described by transformations of the 
former according to specific irreducible representations of specific Lie groups.
\end{itemize}

\subsection{Symmetries of the Standard Model}
The Standard Model is based on the direct product group $SU(3)_C \times SU(2)_L \times U(1)_Y$.
The first factor, $SU(3)_C$, is related to the strong interaction (Quantum ChromoDynamics, 
QCD). $SU(3)_C$ can be obtained from the rank $2$ $SU(3)_C$ semi-simple Lie algebra by usual
exponentiation. The group is of dimension $8$, having $8$ generators which are associated with
the ``carriers'' of the strong interaction, called \textit{gluons}. It is, as is well known, a non-abelian group, QCD
being a Yang-Mills theory. The subscript $C$ stands for ``Color'', the conserved charge of the
interaction according to Noether's theorem.

The Second factor, $SU(2)_L$ is again a non-abelian group generated by the semi-simple rank $1$
algebra $SU(2)$. It has dimension $3$, with an equal number of generators. The subscript $L$ here
stands for ``Left'', since left- and right-handed fermions transform differently under the group.
Its associated charge is called Isospin.
Along with the last (abelian) factor, $U(1)_Y$, it forms the so-called ``electroweak'' gauge
group which is associated with weak and electromagnetic interactions (with the $Y$ standing for
``hypercharge'', the associated charge).

Finally, all fields are characterized by Lorentz invariance. Spacetime is flat (Minkowski) and
four-dimensional, possessing the symmetries of the Poincar\'e group.

\subsection{The particle content}
Matter fields are described by three ``generations'' of fermions, transforming as spinors under Lorentz
transformations. Assuming a Dirac spinor $f$, this can be decomposed into ``left'' and
``right'' components defined as
\begin{equation}
 f_{L,\,R}=\frac12\,(1\mp\gamma_5)\,f.
\end{equation}
A first classification can be done according to the representation properties under $SU(3)_C$
for the various fermions. Some, called ``leptons'', transform as singlets under $SU(3)_C$, 
whereas the remaining ones, ``quarks'', transform as triplets, i.e. the fundamental representation.
Then, fermions are also organised according to their $SU(2)$ representations as follows
\begin{eqnarray}
L_1=\left(\begin{array}{l}\nu_e\\e^-\end{array}\right)_L,\qquad
L_2=\left(\begin{array}{l}\nu_\mu\\\nu^-\end{array}\right)_L,\qquad
L_3=\left(\begin{array}{l}\nu_\tau\\\tau^-\end{array}\right)_L,\\
e_1=e^-_R,\hspace{2.2cm} e_2=\mu^-_R,\hspace{2.15cm} e_3=\tau^-_R\hspace{1.3cm}
\end{eqnarray}
for leptons, and
\begin{eqnarray}
Q_1=\left(\begin{array}{l}u\\d\end{array}\right)_L,\qquad
Q_2=\left(\begin{array}{l}c\\s\end{array}\right)_L,\qquad
Q_3=\left(\begin{array}{l}t\\b\end{array}\right)_L,\\
u_1=u_R,\hspace{1.9cm} u_2=c_R,\hspace{2.cm} u_3=t_R,\hspace{1.05cm}\\
d_1=d_R,\hspace{1.9cm} d_2=s_R,\hspace{2.cm} d_3=b_R\hspace{1.05cm}
\end{eqnarray}
for quarks.
We already note what we said before,
that left and right components of the spinor fields transform differently under $SU(2)$. In this
version of the standard model (the truly minimal one) we do not include right-handed components
for the neutrino fields, since the latter are considered to have zero mass, which is actually
not true. This is, in fact, one of the very few evidence that one should go beyond the minimal
Standard Model in order to have a fully realistic description of nature.

The third component of the isospin for the left- and right-handed components of the spinor fields is
\begin{equation}
I_{f_L}^3 =\left(\begin{array}{l}+\frac12\\-\frac12\end{array}\right) \ . \ \
I_{f_R}^3=0
\end{equation}
The hypercharge $Y_f$ of a fermion $f$ is defined in terms of $I^3$ and of the electromagnetic
charge $Q_f$ as follows
\begin{equation}
Y_f=Q_f-I^3_f \ , \ \
\end{equation}
with $Q_f$ being defined in units of the elementary proton charge $+e$.
More specifically, we find that
\begin{equation}
Y_{Q_i}=\frac16,\qquad Y_{u_i}=\frac23,\qquad Y_{d_i}=-\frac13,\qquad Y_{L_i}=-\frac12,\qquad Y_{e_i}=-1.
\end{equation}
\\
It can further be checked that
\begin{equation}
\sum_\text{fermions} Y^3_f=0,
\end{equation}
which actually ensures the cancellation of chiral anomalies \cite{Adler:1969er,Bilal:2008qx} and is 
an indispensable ingredient for the renormalizability of the theory \cite{'tHooft:1971fh,'tHooft:1972fi}.
\\

Gauge fields correspond to spin-$1$ bosons which are responsible for the mediation of interactions.
In the strong sector, the fields $G_\mu^{1\dots 8}$ correspond to eight gluons, as many as the
generators of the algebra. These generators are defined by means of the Gell-Mann matrices $T_3^a$
and obey the corresponding Lie algebra
\begin{equation}
\left[T_3^a,T_3^b\right]=i\,f^{abc}\,T_3^c,\qquad Tr\left[T_3^a\,T_3^b\right]=\frac12\,\delta^{ab},
\end{equation}
where the tensor $f^{abc}$ corresponds to the structure constants of the group.
\\

In the electroweak sector, the field $B_\mu$  corresponds to the generator $Y$ of $U(1)_Y$ and the
three fields $W_\mu^{1,\,2,\,3}$ to the generators $T^{1,\,2,\,3}$ of the isospin group $SU(2)$. 
The generators $T_2^a$ are given by
$T_2^a\equiv\frac12\,\tau^a$, where the $\tau^a$ are the Pauli matrices describing the rotations:
\begin{equation}
\tau^1=\begin{pmatrix}0&1\\1&0\end{pmatrix},\qquad
\tau^2=\begin{pmatrix}0&-i\\i&0\end{pmatrix},\qquad
\tau^3=\begin{pmatrix}1&0\\0&-1\end{pmatrix},
\end{equation}
verifying the commutation relations
\begin{equation}
\left[T_2^a,T_2^b\right]=i\,\epsilon^{abc}\,T_2^c
\end{equation}
where $\epsilon^{abc}$ is the totally antisymmetric Levi-Civita symbol.
\\ \\
Finally, we define the field strength tensors as: 
\begin{eqnarray}
B_{\mu\nu}&=&\partial_\mu\,B_\nu-\partial_\nu\,B_\mu, \nonumber \\
W_{\mu\nu}^a&=&\partial_\mu\,W_\nu^a-\partial_\nu\,W_\mu^a+g_2\,\epsilon^{abc}
\,W_\mu^b\,W_\nu^c, \nonumber \\
G_{\mu\nu}^a&=&\partial_\mu\,G_\nu^a-\partial_\nu\,G_\mu^a+g_3\,f^{abc}\,G_\mu^b\,G_\nu^c,
\end{eqnarray}
where $g_1$, $g_2$ and  $g_3$ are the coupling constants of $U(1)$, $SU(2)$ and $SU(3)$ respectively. 
We should note that (as in general in Yang-Mills theories), the non-abelian gauge fields also 
possess self-interactions. This is not the case for abelian groups.

\subsection{Interactions and Lagrangian before EWSB}
Matter and gauge fields couple to each other according to the minimal coupling recipe, 
namely the only means of interaction among them is through terms containing the covariant
derivatives $D_\mu$ defined as
\begin{eqnarray}
D_\mu\,(Q_i,\,u_i,\,d_i)&=&\left[\partial_\mu-i\,g_3\,T_3^a\,G^a_\mu-i\,
g_2\,T_2^a\,W^a_\mu-i\,g_1\,\frac{Y}{2}\,B_\mu\right]\,(Q_i,\,u_i,\,d_i),
\nonumber \\
D_\mu\,L_i&=&\left[\partial_\mu-i\,g_2\,T_2^a\,W^a_\mu-i\,g_1\,\frac{Y}{2}\,
B_\mu\right]\,L_i, \nonumber \\
D_\mu\,e_i&=&\left[\partial_\mu-i\,g_1\,\frac{Y}{2}\,B_\mu\right]\,e_i.
\end{eqnarray}
This covariant derivative generates interaction terms among fermions $\psi$
and gauge bosons $V_\mu$ of the form
\begin{equation}
-g_i\,\bar\psi\,V_\mu\,\gamma^\mu\,\psi.
\end{equation}
The interaction is thus minimally and uniquely determined once the gauge symmetry group and 
the coupling constant is given.\\
Before the breaking of the symmetry group, the Standard Model Lagrangian density is
\begin{eqnarray}\label{lms}
\mathcal{L}&=&-\frac14\,G^a_{\mu\nu}\,G_a^{\mu\nu}-\frac14\,W^a_{\mu\nu}\,
W_a^{\mu\nu}-\frac14\,B_{\mu\nu}\,B^{\mu\nu} \\
&& +i\,\bar L_i\,D_\mu\,\gamma^\mu\,L_i +i\,\bar
e_i\,D_\mu\,\gamma^\mu\,e_i+i\,\bar Q_i\,D_\mu\,\gamma^\mu\,Q_i+i\,\bar
u_i\,D_\mu\,\gamma^\mu\,u_i+i\,\bar d_i\,D_\mu\,\gamma^\mu\,d_i \qquad \nonumber
\end{eqnarray}
This Lagrangian is invariant under local transformations for matter fields:
\begin{eqnarray}
(Q_i,\,u_i,\,d_i)(x)&\rightarrow& e^{i\,\alpha_3^a(x)\,T^a_3+i\,\alpha_2^a(x)\,
T^a_2+i\,\alpha_1(x)\,Y}\,(Q_i,\,u_i,\,d_i)(x) \nonumber \\
L_i(x)&\rightarrow& e^{i\,\alpha_2^a(x)\,T^a_2+i\,\alpha_1(x)\,Y}\,L_i(x)
\nonumber \\
e_i(x)&\rightarrow& e^{i\,\alpha_1(x)\,Y}\,e_i(x)
\end{eqnarray}
as well as gauge fields:
\begin{eqnarray}
G_\mu^a(x)&\rightarrow& G_\mu^a(x)-\frac{1}{g_3}\,\partial_\mu\,\alpha_3^a(x)-\epsilon^{abc}\,\alpha_3^b\,G_\mu^c,
\nonumber \\
W_\mu^a(x)&\rightarrow& W_\mu^a(x)-\frac{1}{g_2}\,\partial_\mu\,\alpha_2^a(x)-\epsilon^{abc}\,\alpha_2^b\,W_\mu^c,
\nonumber \\
B_\mu^a(x)&\rightarrow& B_\mu(x)-\frac{1}{g_1}\,\partial_\mu\,\alpha_1(x).
\label{inv-jauge}
\end{eqnarray}

We should note that the Lagrangian \eqref{lms} does not contain any mass term for the moment.
In fact, adding a mass term of the form $\frac12\,M_V^2\,V_\mu\,V^\mu$ for the gauge bosons would
explicitly violate gauge invariance. This can be easily seen in the case of an abelian gauge field.
Including a mass term would mean
\begin{eqnarray}
\frac{1}{2}M_B^2 B_\mu B^\mu \to \frac{1}{2}M_B^2 \left(B_\mu - \frac{1}{g_1} 
\partial_\mu \alpha_1\right) \left(B^\mu - \frac{1}{g_1} \partial^\mu \alpha_1\right) \neq
\frac{1}{2}M_B^2 B_\mu B^\mu.
\end{eqnarray} 
Furthermore, a mass term for a fermion $\psi$ would be of the form
\begin{eqnarray}
m_f\,\bar{\psi}\, \psi  = m_f\, \bar{\psi}\, \bigg( \frac{1}{2} (1-\gamma_5)
+\frac{1}{2}(1+\gamma_5) \bigg)\, \psi= -m_f (\bar{\psi}_R\, \psi_L+ \bar{\psi}_L\,
\psi_R) 
\end{eqnarray}
which is not invariant under isospin transformations, given that left-handed fermions are
doublets under $SU(2)$ whereas right-handed fermions are singlets.

But it is clear (and experimentally verified) that both fermions and bosons should be able
to have mass terms, since the masses of all particles (apart from neutrinos) have been 
experimentally measured.

The Higgs-Brout-Englert mechanism \cite{Higgs:1964pj,Higgs:1966ev,Englert:1964et} 
proposes a way to generate masses both for bosons and fermions
by breaking the electroweak symmetry spontaneously.

\subsection{The Higgs mechanism and mass generation}
In order to generate mass terms for quarks, leptons and gauge fields of the electroweak sector
(gluons are massless as well as the photon) we introduce a complex scalar $SU(2)$ doublet field $\Phi$, 
with a hypercharge $Y_\Phi=+1$ \cite{Djouadi:2005gi}
\begin{equation}
\Phi=\left(\begin{array}{c}\phi^+\\\phi^0\end{array}\right),
\end{equation}
The $SU(2)_L \times U(1)$ - invariant lagrangian density is given by
\begin{equation}
\mathcal{L}_H=\left(D_\mu\,\Phi\right)^\dagger\left(D^\mu\,\Phi\right)-\mu^2\,\Phi^\dagger\Phi-\lambda\,(\Phi^\dagger\Phi)^2,
\label{LHiggs}
\end{equation}
where the covariant derivative $D_\mu$ is given by:
\begin{eqnarray}\label{dercovl}
D_\mu\,\Phi = \left[\partial_\mu-i\,g_2\,T_2^a\,W^a_\mu-i\,g_1\,\frac{Y}{2}\,
B_\mu\right]\,\Phi
\end{eqnarray}
For a mass term $\mu^2<0$, the neutral component of $\Phi$ 
develops a non-zero vacuum expectation value
\begin{equation}
\langle\Phi\rangle_0=\langle 0|\Phi|0\rangle=\frac{1}{\sqrt{2}}
\left(\begin{array}{c}0\\ v \end{array}\right)\qquad\text{with}\qquad v\equiv 
\sqrt{-\frac{\mu^2}{\lambda}}.
\end{equation}
Note that the charged component of the field $\Phi$ should not acquire a VEV, since
we wish to conserve invariance under the group $U(1)_\text{Q}$ of electromagnetism: the
associated gauge boson, the photon, is massless.

It is possible to expand $\Phi$ around the minima $v$ in terms of real fields. At leading order,
we shall have:
\begin{equation}
\Phi=\left(\begin{array}{c}\theta_2+i\,\theta_1\\\frac{1}{\sqrt{2}}\,(v+H)-i\,\theta_3\end{array}\right)=e^{i\,\theta_a\,\tau^a}\,\left(\begin{array}{c}0\\\frac{1}{\sqrt{2}}\,(v+H)\end{array}\right).
\end{equation}
At this point, we can use the gauge invariance and fix ourselves at the unitary gauge, by
performing an $SU(2)$ gauge transformation on $\Phi$. This allows us to eliminate the $\theta_{1,\,2,\,3}$
degrees of freedom, which become non-physical:
\begin{equation}\label{junitaire}
\Phi(x)\rightarrow e^{-i\,\theta_a(x)\,\tau^a}\,\Phi(x)=\frac{1}{\sqrt{2}}\,\left(\begin{array}{c}0\\v+H\end{array}\right).
\end{equation}
In the unitary gauge, the kinetic term for $\Phi$ takes the form
\begin{eqnarray}\label{tcbh}
(D_\mu\,\Phi)^\dagger
(D^\mu\,\Phi)&\!=\!&\left|\left(\partial_\mu-i\,g_2\,\frac{\tau^a}{2}\,
W^a_\mu-\frac{i}{2}\,g_1\,
B_\mu\right)\Phi\right|^2 \\
&\!=\!&\frac12(\partial_\mu\,H)^2\!+\!\frac18\,g_2^2\,(v\!+\!H)^2\,\left|W_\mu^1
\!+\!i\, W_\mu^2\right|^2 \!+\!\frac18\,(v\!+\!H)^2\,\left|g_2\,W_\mu^3\!-\!
g_1\,B_\mu\right|^2. 
\nonumber
\end{eqnarray}
Then, if we define the physical gauge fields $W_\mu^\pm$, $Z_\mu$ and $A_\mu$
\begin{equation}
W_\mu^\pm\equiv \frac12\,\left(W_\mu^1\mp i\,W_\mu^2\right),   \,\,
Z_\mu\equiv \frac{g_2\,W_\mu^3-g_1\,B_\mu}{\sqrt{g_1^2+g_2^2}},\,\,
A_\mu\equiv \frac{g_1\,W_\mu^3+g_2\,B_\mu}{\sqrt{g_1^2+g_2^2}},
\end{equation}
equation \eqref{tcbh} can be recast into the form
\begin{equation}
|D_\mu\,\Phi|^2=\frac12(\partial_\mu\,H)^2+M_W^2\,W_\mu^+\,W^{-\mu}+\frac12\,M_Z^2\,Z_\mu\,Z^\mu+\frac12\,M_A^2\,A_\mu\,A^\mu,
\end{equation}
where the gauge boson masses will be
\begin{equation}
M_W=\frac{v\,g_2}{2},\qquad M_Z=\frac{v}{2}\,\sqrt{g_2^2+g_1^2},\qquad M_A=0.
\end{equation}

At this point, the $SU(2)_L \times U(1)_Y$ symmetry is no longer manifest: it has been 
spontaneously broken and the only residual symmetry is a $U(1)_\text{Q}$, which we identify
with the $U(1)$ abelian gauge symmetry of quantum electrodynamics. 
Among the degrees of freedom of $\Phi$, three have been absorbed by the
three vector bosons $W^\pm$ and $Z$ to give them longitudinal components and, thus, masses.
There remains a massless gauge boson $A$ which is identified with the photon: the residual
$U(1)_\text{Q}$ ``protects'' it from getting a mass term.

The fermion masses, in their turn, can be generated by means of the same field $\Phi$ along
with its conjugate $\tilde\Phi\equiv i\,\tau^2\,\Phi^*$ with a hypercharge of $Y_{\tilde\Phi}=-1$.
The mass terms are included in the Yukawa Lagrangian which is invariant under the SM group
gauge transformations
\begin{equation}
\mathcal{L}_F=-\lambda_{e_i}\,\bar L\,\Phi\,e_i-\lambda_{d_i}\,\bar Q\,\Phi\,d_i-\lambda_{u_i}\,\bar Q\,\tilde\Phi\,u_i+h.c.
\end{equation}

Once $\Phi$ acquires a non-zero VEV, and working always in unitary gauge \eqref{junitaire}, 
the Lagrangian density can be written as
\begin{eqnarray}
\mathcal{L}_F&=&-\frac{\lambda_{e_i}}{\sqrt{2}}\,H\,e_{iL}\,e_{iR}-\frac{\lambda_{d_i}}{\sqrt{2}}\,H\,d_{iL}\,d_{iR}-\frac{\lambda_{u_i}}{\sqrt{2}}\,H\,u_{iL}\,u_{iR}\nonumber\\
&&-m_{e_i}\,\bar e_{iL}\,e_{iR}-m_{d_i}\,\bar d_{iL}\,d_{iR}-m_{u_i}\,\bar u_{iL}\,u_{iR}+h.c.
\end{eqnarray}
The fermion masses are then identified with the terms:
\begin{equation}
m_{e_i}=\frac{\lambda_{e_i}\,v}{\sqrt{2}},\qquad
m_{d_i}=\frac{\lambda_{d_i}\,v}{\sqrt{2}},\qquad
m_{u_i}=\frac{\lambda_{u_i}\,v}{\sqrt{2}}.
\end{equation}
\\ \\
Finally, the remaining physical degree of freedom of $\Phi$ constitutes the so-called Higgs boson $H$.
After EWSB and since $v^2=-\mu^2/\lambda$ we can write down the Lagrangian for $H$.
\begin{eqnarray}
{\cal L}_{H}
= \frac{1}{2} (\partial^\mu H)^2 - \lambda v^2 \, H^2 - \lambda v \, H^3 - 
\frac{\lambda}{4} \, H^4 
\end{eqnarray}
from which we can immediately infer 
\begin{eqnarray}
M_H^2=2 \lambda v^2
\end{eqnarray}
as well as its self-couplings. As for the couplings of the $H$ boson to other particles, 
these can be read off the part of the Lagrangian which also contains the mass terms.
In the end, we find
\begin{eqnarray}
g_{H^3}= 3 \, \frac{M_H^2}{v} \ \ , \ g_{H^4} = 3 \frac{M_H^2}{v^2} \ \ , \
g_{Hff}=  \frac{m_f}{v} \ \ , \ 
g_{HVV}= -2  \frac{M_V^2}{v} \ \ , \ 
g_{HHVV}= - 2 \frac{M_V^2}{v^2}.
\end{eqnarray}
\\ \\
All of these parameters depend on the (non-zero) vacuum expectation value of the Higgs field $v$. 
This is in its turn related to the Fermi constant $G_F$, which has been measured with an extreme accuracy, 
as well as to the $W$-boson mass:
\begin{eqnarray}
M_W=\frac{1}{2} g_2v = \left( \frac{\sqrt{2} g^2}{8 G_F} \right)^{1/2} 
\Rightarrow v= \frac{1}{(\sqrt{2} G_F)^{1/2} } \simeq 246~{\rm GeV}.
\label{MW-vs-v}
\end{eqnarray}
The only free parameter of the Standard Model is, hence, the mass of the Higgs boson 
(since all other masses have already been measured experimentally).

%%%%%%%%%%%%%%%%%%%%%%%%%%%%%%%%%%%%%%%%%%%%%%%%%%%%%%%%%%%%%%%%%%%%%%%%%%%%%%%%%%%%%%%%%%%%%%%%%%%%%
%%%%%%%%%%%%%%%%%%%%%%%%%%%%%%%%%%%%%%%%%%%%%%%%%%%%%%%%%%%%%%%%%%%%%%%%%%%%%%%%%%%%%%%%%%%%%%%%%%%%%
%%%%%%%%%%%%%%%%%%%%%%%%%%%%%%%%%%%%%%%%%%%%%%%%%%%%%%%%%%%%%%%%%%%%%%%%%%%%%%%%%%%%%%%%%%%%%%%%%%%%%

\section{Candidates in the Standard Model?}
We saw the particle content of the Standard Model. Following the previous discussion, it is quite
logical to ask which of the particles listed in the previous section could constitute ``good''
candidates for dark matter. In order to evaluate this, first of all we should note that quarks
are by definition excluded, since the amount of baryons in the universe is bound by the WMAP data. 
As excluded are also gauge bosons and the Higgs particle, since they are very unstable.
Charged leptons are also strongly bound, since they interact electromagnetically and would have most
probably been observed. Furthermore, this would mean that the universe would be overall strongly 
electrically charged, something which is in contrast with observations.

The only possibility are, hence, neutrinos. Indeed, during the first days of interest around dark
matter neutrinos were considered to be the most plausible candidate. This was further supported
by the fact that as we saw, if one makes a series of assumptions, which are not that absurd, one can 
arrive to the conclusion that particles only involved in weak interactions should in principle 
constitute quite good candidates.

The main problem in the neutrino hypothesis (apart from the fact that today the total amount
of neutrinos in the universe is bound) is that when produced thermally, they are ultrarelativistic
due to their very small mass. They thus constitute a severely hot dark matter candidate, which
is an unacceptable feature according to our trends for structure formation. Indeed, 
neutrinos would free-stream in the early universe spoiling structures.

We thus see that the Standard Model itself is unable to provide a well-behaved dark matter
candidate. In order to accommodate such a feature, one has to extend the particle content.
This is, in our respect, a particularly interesting feature, especially in the absence of further
experimental evidence for the existence of Beyond Standard Model (BSM) physics. One indeed finds such evidence once
one tries to reconcile two very different fields, particle physics and cosmology.

Very large classes of extensions of the Standard Model offer viable candidates. In this work, we
shall be interested in particles falling in the class of Weakly Interacting Massive Particles 
(WIMPs), that is, candidates with masses and couplings falling roughly within the electroweak scale.
And, as we shall see, there are both minimal extensions of the SM as well as much larger ones, 
motivated from totally different arguments, that propose such candidates and actually quite
naturally.

The next question is whether we shall be able to probe some of these candidates: what kind of 
experimental techniques could we devise in order to actually detect dark matter? We shall 
be developing this point in the next chapter. For the moment, we just repeat that until now
the only evidence comes from gravitation-related data, which do not actually distinguish among
different kinds of particles: the intensity of the interaction only depends on the particle's mass.
If we should wish to determine the nature of the dark matter particles, we should rely upon their
-potential- capacity to interact through different forces than the gravitational one. It could
be of course, that this simply does not happen. But at least in the thermal relic picture that we
presented so far, other kinds of interactions are also expected to be relevant.

\newpage
\chapter{Detection of Dark Matter}
\label{Chapter2}
We closed the first chapter wondering whether it would be possible to envisage
techniques that could help us detect dark matter, especially when it comes to candidates 
falling in the WIMP category. We said that by definition these candidates should possess
properties similar to the electroweak sector of the Standard Model and be thus able
either to interact with ordinary matter or to annihilate and produce it.
\\ \\
From this last element, at least three ideas could arise quite straightforwardly: 
\begin{enumerate}
 \item The first one is that if WIMPs can
interact with ordinary matter, we could imagine building a detector on the earth and try
to detect the dark matter particles that -should in principle- continuously reach the
earth. Judging from the techniques that are used in order to detect other particles only interacting
through weak-scale forces, for instance neutrinos, the expected signal should be quite low. 
The detector should therefore be massive  so as to augment the probability
for a positive detection. Moreover, background sources should be understood and eliminated 
as much as possible. This technique is referred to as \textit{direct detection}.
 \item The second idea could emerge from the very mechanism invoked in the thermal relic scenario so as
to reproduce the correct DM abundance: WIMPs can annihilate into Standard Model particles.
If this was possible in the early universe, it should also be possible today. We could imagine
trying to detect exactly the annihilation products of this process. In fact, since such DM
annihilations are expected to take place throughout the galactic halo, it would only make sense 
to try and detect stable particles which might be either primary or secondary products of
WIMP annihilations: photons, electrons, neutrinos but also perhaps composite particles from
the hadronization of some of the annihilation products. This approach is called \textit{indirect detection}.
 \item Finally, since WIMPs have
roughly electroweak scale masses, it is possible that they could be produced in today's
high-energy colliders. Especially the CERN Large Hadron Collider \cite{LHCsite} and the 
Tevatron \cite{TEVATRONsite} are actually 
probing exactly the energies at which electroweak symmetry breaking is expected to take place. 
The same holds for oncoming or planned experiments, such as the International Linear Collider
\cite{ILCsite}.
\end{enumerate}
In the following paragraphs we develop the basic principles for these detection modes.
Then, we present some results concerning the capacity of these experiments to constrain some
WIMP properties, especially its mass.

%%%%%%%%%%%%%%%%%%%%%%%%%%%%%%%%%%%%%%%%%%%%%%%%%%%%%%%%%%%%%%%%%%%%%%%%%%%%%%%%%%%%%%%%%%%%%%%%%%%%%
%%%%%%%%%%%%%%%%%%%%%%%%%%%%%%%%%%%%%%%%%%%%%%%%%%%%%%%%%%%%%%%%%%%%%%%%%%%%%%%%%%%%%%%%%%%%%%%%%%%%%
%%%%%%%%%%%%%%%%%%%%%%%%%%%%%%%%%%%%%%%%%%%%%%%%%%%%%%%%%%%%%%%%%%%%%%%%%%%%%%%%%%%%%%%%%%%%%%%%%%%%%
\section{Direct detection}
The basic principle of direct detection is rather simple 
\cite{Goodman:1984dc,PhysRevD.33.3495,PhysRevD.33.2071,Munoz:2003gx,Cerdeno:2010jj}:
Since our galaxy is constituted primarily of dark matter, we expect that WIMPs constantly
reach the earth. As they interact weakly, most of the time they should just traverse it. But every now
and then, it could be that some of the WIMPs actually interact with the Earth's materials.
Then, if a large detector were built and exposed for a sufficiently large amount of time to WIMPs traversing it, 
some of the WIMPs might actually interact with the target material. There is a large number of experiments worldwide
that pursue this goal. They are typically built underground in order to achieve significant reduction of background.
Depending on the specific technique of every experiment, a large number of different observables
can be measured in order to detect and reconstruct a WIMP. In almost all cases however, 
the basic principle remains the same. WIMPs could interact with the nuclei and the electrons
of the target material, causing them to recoil, get excited or ionize and this is an in principle measurable effect.

\subsection{The event rate}
The event rate that one would expect in a detector depends on a certain number of parameters. 
Let us denote the total number of events by $N$. This number should be proportional to the number 
of target nuclei and WIMPs available for the interaction to take place. If we denote by $n_N$ 
the number of nuclei and $n_\chi$ the corresponding number for WIMPs, then the total number of 
events should be $N \propto n_N n_\chi$.

But not all WIMPs move at the same velocity: their velocities are rather dispersed according to
a certain \textit{velocity distribution} which should be known or calculable according to some
theoretical assumptions. More on this point will follow. In order now to get the total number of events 
for all velocities, we should integrate the distribution along with the relevant cross-section, 
which depends on the center-of-mass energy of the collision.
Then, the event rate per unit detector mass, time and energy should be
\begin{equation}
  \frac{dN}{dE_r} = \frac{\rho_0}{m_{Nucl} m_{\chi}} \int_{v_{min}}^{\infty} vf(v) 
\frac{d\sigma_{\chi Nucl}}{dE_r}(v, E_r) dv
\label{MasterDirect}
\end{equation}
where $N$ is the number of WIMP scatterings off target nuclei, $E_r$ is the nucleus recoil energy, 
$\rho_0$ is the local dark matter density near the earth, $m_{Nucl}$ is the nucleus mass, $m_\chi$ is the WIMP
mass, $v$ is the WIMP velocity, $f(v)$ is the WIMP velocity distribution in the detector rest frame 
and $\sigma_{\chi Nucl}$ is the
WIMP-nucleus scattering cross-section. The lower integration limit is $v_{min}$, the minimal velocity 
that can kinematically give rise to a scattering with recoil energy $E_r$.
\\ \\
This minimal velocity can be found to be
\begin{equation}
 v_{min} = \sqrt{(m_{Nucl} E_r)/(2 \mu_{Nucl}^2)}
\label{vmin}
\end{equation}
where $\mu_{Nucl} = m_\chi m_{Nucl}/(m_\chi + m_{Nucl})$ is the WIMP-nucleus reduced mass.

To calculate the total number of events per unit detector mass per unit time, one must integrate
Eq.(\ref{MasterDirect}) within the desired recoil energy region. 

Referring to Eq.\eqref{MasterDirect}, the BSM particle physics-related quantities
are just two: the WIMP - nucleus scattering cross-section and the WIMP mass. If the astrophysical
quantities are fixed, then one can extract bounds on the combination of these parameters in a more
or less model-independent way. Moreover, if it is clear how to pass from the nuclear level to the 
nucleonic one, the limits on the cross-section can be further translated into constraints on the
WIMP-\textit{nucleon} scattering cross-section, allowing for comparison among different experiments using
different target materials. This is actually the habit of experimental collaborations.

The situation is of course complicated by the fact that this equation is comprised of
several factors which we shall briefly analyze in the following. They are often
associated with uncertainties that can severely alter the interpretation of experimental data
or the theoretical predictions on event rates.

\subsubsection{The local density}
It is interesting that despite the strong uncertainties on the nature (as well as the spatial
distribution) of dark matter, there seems to exist quite some agreement (at least qualitatively)
concerning its density in the solar neighborhood. 
Experimental analyses usually use the -somehow- reference value of $0.3$ GeV cm$^{-3}$.

One of the latest and acknowledged calculations comes from 
ref.\cite{Catena:2009mf}. In this paper, the authors use observables related to the galactic
rotation curves in order to derive a local density of $0.385 \pm 0.027$  GeV cm$^{-3}$ for an Einasto halo 
profile and $0.389 \pm 0.025$ GeV cm$^{-3}$ for a NFW one.
This result is claimed by the authors to be quite robust, and the error bars lie 
around $7\%$ of the central value at $68\%$CL.
It is noteworthy that the authors' results do not change significantly
among different assumptions concerning the DM distribution in the galaxy.

It should be noted however that these estimates are practically always based on some assumptions. 
The authors of \cite{Weber:2009pt} for instance consider wider possibilities for halo profiles
finding a potential region for the local density from $0.2$ up to $0.4$ GeV cm$^{-3}$ at
$68\%$CL. 
In \cite{Salucci:2010qr} the authors do not make some particular assumption concerning the
halo profile finding a local density of $0.43 \pm 0.21$ GeV cm$^{-3}$ at the same CL.

Recently, a further study was performed in \cite{Pato:2010yq} that tries to estimate systematic
uncertainties in the local density calculation, as for example possible departures from perfect
sphericity. The conclusion of the authors is that systematic uncertainties can be more important
than statistical ones and their result for the local density value dispersion is
$0.466 \pm 0.033 \mbox{(stat)} \pm 0.077 \mbox{(syst)}$ GeV cm$^{-3}$ at $68\%$ CL and for an Einasto 
profile.

In any case, most studies are usually in agreement within roughly a factor of $2 - 3$ at $68\%$ CL. 
The recent results seem to be yielding values for the local density ranging between $0.2$ and $0.576$
GeV cm$^{-3}$ at $68\%$ CL.
We note that taking into account variations in the local density is 
quite straightforward, since the event rate depends just linearly on this parameter.

\subsubsection{The scattering cross-section and hadronic uncertainties}
So far we have omitted (and will continue doing so in the following)
possible interactions that could occur among WIMPs and the electrons in the target material. It 
has been shown that these interactions are much less frequent than interactions
with the nucleus (among others because of the huge mass difference between WIMPs and electrons).

Then, there can be inelastic interactions that excite the nucleus as a whole causing for a 
gamma-ray emission upon deexcitation. The typical lifetime of the excited states is of $O(\mbox{nsec})$. These
interactions produce a signal that is very similar to natural radioactivity, with the latter
providing a much stronger signal than the former and are, hence, ignored in analyses.

So, what we are left with is the elastic scattering cross-section between the WIMP and the 
nucleus. In order to calculate these, a series of steps must be taken:
\begin{itemize}
 \item First, one should compute the scattering cross-section at the partonic level, i.e. among
a WIMP and a quark/gluon. 
 \item Then, one must convolute this cross-section with Parton Distribution Functions (PDFs)
in order to pass from the partonic level to the WIMP - nucleon one.
 \item Finally, one must pass from the WIMP-\textit{nucleon} level to the WIMP-\textit{nucleus} one.
\end{itemize}
The WIMP-nucleus scattering cross-section $\sigma_{\chi Nucl}$ can usually be separated into two distinct 
parts, the spin-independent and the spin-dependent one
\begin{equation}
 \frac{d\sigma_{\chi Nucl}}{dE_r} = 
\left( \frac{d\sigma_{\chi Nucl}^{SI}}{dE_r} \right) + 
\left( \frac{d\sigma_{\chi Nucl}^{SD}}{dE_r} \right) 
\end{equation}
The spin-dependent part of the cross-section comes from axial current couplings appearing
in the interaction Lagrangian. On the other hand, the spin-independent comes from scalar-scalar
and vector-vector couplings. For heavy nuclei 
the spin-dependent contribution is quite a bit smaller than
the spin-independent one. Especially in cases of nuclei with an even number of neutrons and protons, 
this contribution vanishes since the total nuclear spin is zero.

The spin-independent cross-section is usually factorized in terms of the WIMP-nucleon one
and some form factor depending upon the structure of the target nucleus. More concretely,
we write
\begin{equation}
 \frac{d\sigma_{\chi Nucl}^{SI}}{dE_r} = \frac{m_{Nucl} \sigma_0}{2 \mu_{Nucl}^2 v^2} F^2(E_r)
\end{equation}
where $F$ is the nuclear form factor and $\sigma_0$ is the WIMP-nucleon scattering cross-section
at zero momentum transfer.
The latter, in turn, is the convolution of the parton-level cross-section with the
relevant parton distribution functions for protons or neutrons.

By substituting the last expression into Eq.(\ref{MasterDirect}) we get the final expression for
the event rate in the detector
\begin{equation}
   \frac{dN}{dE_r} = \frac{\rho_0 \sigma_0}{2 \mu_{Nucl}^2 m_{\chi}} F^2(E_R) \int_{v_{min}}^{\infty} \frac{f(v)}{v} dv
\label{PracticeDirect}
\end{equation}

The passage from the WIMP-\textit{nucleus} to the WIMP-\textit{nucleon} scattering cross-section
or vice-versa can also introduce some uncertainty. The most commonly used nuclear form factor today is the one proposed
by Engel in \cite{Engel:1991wq}. Previously, the common consideration included assuming that nuclear matter
follows a Gaussian distribution as a function of the distance from the nucleus' center. One example
work where several nuclear form factors are compared among them is \cite{Duda:2006uk}.

Furthermore, passing from the partonic cross-section to the nuclear one is not a so straightforward procedure, it involves
all of the aforementioned steps. So, PDFs come always with their respective uncertainties which, especially for
heavy flavors such as the $s$ - quark, can be significant. As described for example in \cite{Ellis:2008hf}, 
these errors can induce shifts in the predicted WIMP - nucleon scattering cross-sections reaching up
to an order of magnitude. We shall quantify this effect in the last chapter of this work.

Other uncertainties can appear in the scattering cross-section computation, depending on the specific particle 
physics framework under consideration. One such example are uncertainties in the Renormalization Group Equation (RGE)
evolution in GUT-scale models
\footnote{At this point, we of course mean uncertainties in the low-energy parameter values that can be
provoked by different treatments of RGEs. In other words, different RGE-solving codes can yield slightly
different low-energy results.}
. For the moment, and since estimation of such uncertainties demands the
definition of some particle physics framework, we shall ignore them.

\subsubsection{WIMP velocity distribution}

The velocity distribution of WIMPs in the detector frame $f(\vec{v})$ is one of the most uncertain elements
entering the event rate calculation.

The first obstacle to be tackled for an accurate determination of the distribution in the detector frame
is to pinpoint what is the distribution of WIMPs' velocities in a more ``natural'' reference frame, namely
the galactic one $f_1(v_1)$. Then, we expect that by means of Galilean transformations it will be able to convert this
distribution into $f(v)$.
\\ \\
A very common assumption is that WIMP velocity follows a Maxwell-Boltzmann distribution in the galactic rest
frame
\begin{equation}
 f_1(v_1) d^3 v_1 = \frac{1}{v_0^3 \pi^{3/2}} e^{-(v_1/v_0)^2} d^3 v_1
\end{equation}
around some central value $v_0$. Integration over the angular part of the distribution yields
\begin{equation}
 f_1(v_1) d v_1 =\frac{4 v_1^2}{v_0^3 \sqrt{\pi}} e^{-(v_1/v_0)^2} d v_1
\end{equation}
We note that if WIMPs have a velocity above some limit, let's denote it by $v_{esc}$, they can escape the galaxy
and are thus no longer gravitationally bound. In this respect, integrating the velocity distribution up to
infinite velocities does not make sense. Instead, we should limit the integration in eqs.\eqref{MasterDirect} and
\eqref{PracticeDirect} up to $v_{esc}$.

Then, it is necessary to determine the form this distribution takes in the detector rest frame. In this respect,
we should keep in mind that the earth participates in two additional motions with respect to the galactic frame, 
the distribution of interest should thus be determined by performing a Galilean transformation as 
\begin{equation}
 \vec{v}_1 \longrightarrow \vec{v} = \vec{v}_1 + \vec{v}_e (t)
\end{equation}
where $\vec{v}_e (t)$ is the earth's velocity in the galactic rest frame. The latter is comprised of two motions:
\begin{itemize}
 \item The motion of our solar system around the galactic center.
 \item The motion of the earth around the sun.
\end{itemize}
Let's define the galactic coordinates as a set of three vectors $(\vec{x}, \vec{y}, \vec{z})$ where the first
vector points towards the galactic center, the second to the direction of the galactic rotation and the third 
to the galactic north pole. In these coordinates, the sun's motion around the GC can be written as \cite{Cerdeno:2010jj}
\begin{equation}
 \vec{v}_\odot = (10.0 \pm 0.4, 5.2 \pm 0.6, 7.2 \pm 0.4) \ \mbox{km/sec}
\end{equation}
In its turn, the earth's motion around the sun can be expressed, in the same coordinates, as
\begin{equation}
 \vec{v}_e^{orb} = v_e [\vec{e}_1 \sin\lambda(t) - \vec{e}_2 \cos\lambda(t)]
\end{equation}
where the vectors $\vec{e}_1, \vec{e}_2$ are given by 
\begin{eqnarray}
 \vec{e}_1 & = & (-0.0670, 0.4927, -0.8676)\\
 \vec{e}_2 & = & (-0.9931, -0.1170, 0.01032)
\end{eqnarray}
In the end, we can express the velocity in the detector rest frame by substituting
\begin{equation}
 \vec{v}_e (t) = \vec{v}_\odot + \vec{v}_e^{orb}
\end{equation}
At this point, it is interesting to note the time dependence of the earth's velocity. In fact, this time
dependence is at the root of a class of direct detection experiments such as the DAMA/LIBRA 
\cite{Bernabei:2008yh,Bernabei:2008yi} 
experiment at the Gran Sasso national laboratory in Italy, as well as the KIMS experiment in Korea 
\cite{Lee:2006mz, Kim:2006xz}. 
Whereas the majority of direct detection experiments intend to detect the bulk of the signal generated by 
scatterings of dark matter on the detector nuclei, these experiments intend to measure the weak effect of
the signal's annual modulation expected by the harmonic time dependence of the velocity.
This modulation is exactly due to the fact that at some moment
every year, the earth's velocity has the same direction as the sun's rotation around the galactic center, whereas
at some other moment it has the opposite. We should thus expect a periodic fluctuation in the signal.
As weak as this effect might be, it is considered to be quite difficult to find another phenomenon which
could affect the number of events in the detector in a similar manner, something which is supposed to clearly distinguish among 
signal and background.

On the other hand, the presentation concerning the WIMP velocity distribution is somewhat simplistic. 
In the light of recent results from various experimental collaborations posing increasingly strong bounds
on the allowed $(m_\chi, \sigma_{\chi-N}^{SI})$ plane along with results revealing excesses that could
be interpreted as coming from dark matter collisions, a significant effort is being devoted to the study
of the impact of astrophysical assumptions on these bounds and the calculated event rates for various
models. A recent study in this direction is \cite{McCabe:2010zh} where it is found that especially for
low-mass WIMPs or candidates baring rather non-standard interactions, deviations from the behavior as predicted
by the aforementioned assumptions can be sizeable. Quantifying the overall uncertainties in direct dark
matter detection experiments is a very important work in order to better understand the behavior
that could be expected from different candidates as well as the possible nature of detected signals.

%%%%%%%%%%%%%%%%%%%%%%%%%%%%%%%%%%%%%%%%%%%%%%%%%%%%%%%%%%%%%%%%%%%%%%%%%%%%%%%%%%%%%%%%%%%%%%%%%%%%%
%%%%%%%%%%%%%%%%%%%%%%%%%%%%%%%%%%%%%%%%%%%%%%%%%%%%%%%%%%%%%%%%%%%%%%%%%%%%%%%%%%%%%%%%%%%%%%%%%%%%%
%%%%%%%%%%%%%%%%%%%%%%%%%%%%%%%%%%%%%%%%%%%%%%%%%%%%%%%%%%%%%%%%%%%%%%%%%%%%%%%%%%%%%%%%%%%%%%%%%%%%%
\section{Indirect Detection}
Indirect detection is based on the principle that  WIMPs should be able 
to annihilate in the same way as described in the previous 
chapter for the thermal relic mechanism. Of course, the thermally averaged annihilation cross-section might be different
than the one used in order to calculate the candidate's relic density, because since decoupling the average
WIMP velocity is expected to have significantly decreased. But the principle remains the same.
Then, the annihilation products should be (in principle) detectable. But which kinds of annihilation products
should we look for? First of all, it is clear that we should look for stable particles, and actually for particles
that we already know how to detect. In the framework of the standard model hence, the choices are rather limited:
we could look for gamma-rays, electrons or positrons, neutrinos, (anti)protons, as well as perhaps for some composite particles
such as (anti)deuteron and so on. 
And, of course, these annihilation products can be either primary
ones (i.e. produced directly by WIMP annihilation) or secondary (i.e. produced upon decay of unstable primary
products). It is important to note here the different nature of the various detectable annihilation products. 
This observation gives us an idea already that the physics entering their detection can be quite different.

Neutrinos
are extremely weakly interacting particles. In order to detect them, we should rely on techniques quite similar to 
the ones used for usual neutrino detection. Since they travel in straight lines throughout the galaxy, this means
that we can look at specific directions depending on where we expect the signal to be maximized. In this respect, it 
is quite common to look for DM annihilation-induced neutrinos towards the sun. Indeed, because of its mass, the
sun is expected to provoke capturing of WIMPs in its interior. As WIMPs annihilate, 
ultrarelativistic neutrinos can escape the sun's surface and reach the earth, giving a distinct 
contribution to the overall number of detected solar neutrinos. Although this detection mode is very interesting
especially for some classes of candidates, in the following we shall not be mentioning
it anymore.

Gamma-rays also traverse the galaxy without significantly interacting with the interstellar medium. Once again,
they are expected to be copiously produced at places with higher DM concentration, such as the center of the
galaxy where we believe there is a supermassive black hole. However, as we shall see later on, it can be that
other places in the galaxy with much fainter signals are also characterized by much lower backgrounds, hence 
they could offer even better detection perspectives.

Finally, charged matter such as positrons, antiprotons and antideuterons present a further complication, namely the
fact that being charged particles, they interact with the interstellar medium and can annihilate, undergo changes
in their propagation direction, or lose energy. Hence, while they are expected to be produced in places with large
dark matter densities, it can be that either they never reach us because these locations are too distant or that
they change direction.

In the three following paragraphs, we shall see the general features of each of these detection modes in more detail.

\subsection{Gamma-ray detection}
Suppose we are observing the sky along a line, forming an angle $\psi_0$ wrt the straight line connecting the sun and the GC. 
The differential gamma-ray flux coming from DM annihilations received on the earth in units 
GeV$^{-1}$ cm$^{-2}$ sec$^{-1}$ can be written as
\begin{equation}
 \frac{d \Phi_\gamma}{dE_\gamma} (\psi_0, E) = N_\chi
\frac{\left \langle \sigma v \right \rangle_{v\rightarrow 0}}{4 \pi m_\chi^2} 
\sum_f BR_f \frac{dN_\gamma^f}{dE} 
\int_{los} \rho^2\left(l(\psi)\right) dl(\psi)
\label{MasterGammas}
\end{equation}
where: 
$N_\chi$ depends on the nature of the annihilating particles, being $1/2$ for Majorana-like particles and $1/4$ for Dirac-like, 
$\left \langle \sigma v \right \rangle_{v\rightarrow 0}$ is the total thermally averaged self-annihilation cross-section
calculated for $v\rightarrow 0$, $m_\chi$ is the WIMP mass, $BR_f$ is the annihilation fraction into an $f$-th final state, 
$dN_\gamma^f/dE_\gamma$ is the differential yield of the $f$-th final state into $\gamma$'s, the sum runs over all 
possible final states, $\rho$ is the dark matter spatial distribution function, whereas the integral is performed along 
the line of sight (los) from us to the observed point. It is important to stress that in Eq.\eqref{MasterGammas}
we have neglected gamma-ray contributions that could come, for instance, from inverse Compton scattering
or synchrotron radiation of charged DM annihilation products. These contributions can actually turn out to be rather
sizeable. In this work, we shall nevertheless not be dealing with them.

In practice, no instrument can make observations along a 1-dimensional line. Instead, the flux on the earth
(or on a satellite) should be calculated within a cone centered around the angle $\psi_0$. The angle of the cone is
bound from below by the detector's angular resolution, that is, the minimal angular separation needed between two points in the sky
so that the detector can indeed distinguish these two points.
\\ \\
Following ref.\cite{Bergstrom:1997fj}, we define the dimensionless quantity $J$ as follows:
\begin{equation}
 J(\psi) = \frac{1}{R_0} \frac{1}{\rho_0^2} \int_{los} \rho^2 \left( l(\psi) \right) dl(\psi)
\label{Jay}
\end{equation}
where $R_0$ is the distance of the sun from the galactic center and $\rho_0$ is the local DM density 
that we mentioned before.
\\
To calculate the flux generated by DM annihilations within the cone, say of solid angle $\Delta\Omega$ and 
centered around $\psi_0$, 
we can calculate the average value of $J$ in the cone, then simply multiply the corresponding flux by 
$\Delta\Omega$. We therefore define
\begin{equation}
 \bar{J}_{\psi_0}(\Delta\Omega) = \frac{1}{\Delta\Omega} \int_{\Delta\Omega} J(\psi) d\Omega
\label{Jbar}
\end{equation}
where $d\Omega = \sin\theta d\theta d\phi$, with $\theta$ being the angle between the line connecting the 
sun to the GC and the observation line (varying in $[-\pi/2, \pi/2]$) and $\phi$ being the angle perpendicular
to the galactic disk (varying in $[0, 2\pi]$).

The most popular halo profiles are spherically symmetric around the Galactic Center, they are thus functions
of the distance $r$ from the GC only. From the cosine law, the distance $r$ can be expressed as 
\begin{equation}
 r = \sqrt{R_0^2 + l^2 - 2lR_0\cos(\psi)}
\end{equation}
we can hence express the halo profile as a function of $l$ and $\psi$. 

What are the limits of $l$? Normally, one should integrate from the observation point up to the 
end of the universe. Since in the following we shall be ignoring extragalactic contributions, and since in
any case the halo profile concerns the galaxy and not the extragalactic dark matter distribution, we 
integrate up to a maximal value for $l$ by defining a ``limit'' for our galaxy. Supposing the maximal radius
of the Milky Way is $r_{gal}$, we find $l_{max}$ to be
\begin{equation}
 l_{max} = \sqrt{r_{gal}^2 - R_0^2 \sin^2\psi} + R_0\cos\psi
\label{psimax}
\end{equation}
so the expression for $J$ becomes
\begin{equation}
 J(\psi) = \frac{1}{R_0} \frac{1}{\rho_0^2} \int_0^{l_{max}} \rho^2 \left( l(\psi) \right) dl(\psi)
\end{equation}
In the end, we find that
\begin{equation}
 \bar{J}_{\psi_0}(\Delta\Omega) = \frac{1}{\Delta\Omega} 
\int_{\phi_1}^{\phi_2} d\phi \int_{\psi_0}^{\psi_0 + \theta} J(\psi) \sin\psi d\psi
\label{FinalJbar}
\end{equation}
where $\phi$ varies between two integration limits of interest.

Finally, we can write down the expression for the differential flux that we would expect to detect on the
earth in units of time, detector surface, energy and solid angle as
\begin{equation}
 \frac{d\Phi}{dE} = N_\chi
\frac{\left \langle \sigma v \right \rangle_{v\rightarrow 0}}{4 \pi m_\chi^2} 
\sum_f BR_f \frac{dN_\gamma^f}{dE} \
R_0 \rho_0^2 \ \bar{J}
\label{GammaFluxCalc}
\end{equation}
It is very interesting to note that the particle physics part and the astrophysical part of the equation
are completely separated. This actually turns out to be very convenient, since astrophysics-related calculations
need only be done once. Then, they can be applied to any particle physics candidate.
In fact, this observation goes even further: the particle physics part itself is separated into known (Standard Model)
physics and BSM. 

In order to calculate the yields of the SM particles into gamma-rays we can employ well-known codes such as
PYTHIA \cite{Sjostrand:2006za} or SHERPA \cite{Gleisberg:2008ta}. Suppose the simplest case of a two-body SM
final state $f$, with the two particles having identical masses. Then, the energy of each particle in the final state
will be just $m_\chi$. And, since the decays of SM particles are known, the only factor that should change the spectrum
for a given final state should be the energy of the $f$ particles. For any energy of the primary final state products, 
we can use the usual Monte Carlo algorithms in order to compute the 
spectrum of the $f$ particles into photons. When scanning over large parameter spaces however, this turns out to 
be a quite CPU-consuming technique. In order to avoid this, we employ a trick: We only calculate the spectrum 
for a given mass value and then fit this
spectrum as a function of $E/m_\chi$. Then, changes in the WIMP mass can be taken into account straightforwardly.
It has been shown that this technique introduces an error typically of the order of less than $10\%$. We note that
since in the following chapters we shall be examining WIMPs with masses spanning about 2 orders of magnitude, in
practice we perform a small number of different fits for different WIMP masses and then use the appropriate function
according to the WIMP mass under examination. 

But it can be that the primary annihilation products are not SM particles but unstable ones that further decay
into multiple SM particle final states. We said that the fit is performed as a function of $E/m_\chi$. In reality,
one can generalize this into a fit of $E/E_f$, where $E_f$ is the energy of a final state particle. This shows us
that through the same method, we can also treat final states with non mass-degenerate particles, as well as final
states with more than two particles (although the latter case requires some more attention).

\begin{figure}[htb!]
\begin{center}
\includegraphics[width = 7cm, angle=270]{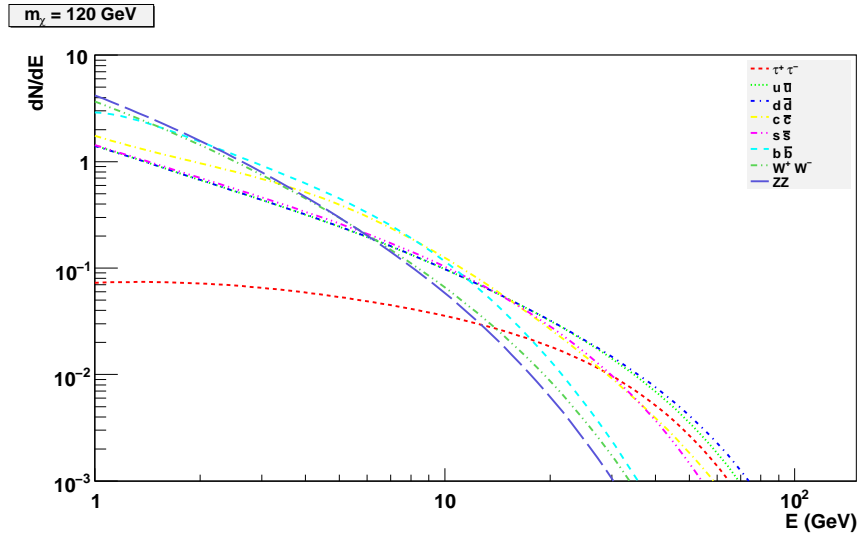}
\caption{{\footnotesize
Differential yield of various Standard-Model $2$-body final states into photons for a WIMP
mass of $120$ GeV.}}
\label{gammayieldmchi120GeV}
\end{center}
\end{figure}

In figure \ref{gammayieldmchi120GeV} we show an example of such a fit for Standard Model $2$-body final
states, assuming a WIMP mass of $120$ GeV. We can see that gauge bosons and heavy quarks tend to yield 
richer spectra at low-energies falling off rather quickly, whereas light quarks and especially leptons
give much harder spectra.

\subsubsection{The halo profile}
Upon simple inspection of Eq.\eqref{GammaFluxCalc} one can see that the gamma-ray flux expected on the earth depends
strongly on the assumptions adopted concerning the dark matter spatial distribution. We already mentioned in the
first chapter that while most halo profiles present a similar behavior at sufficiently large distances from the Galactic
Center, there does not seem to be much agreement on the corresponding form of the distribution close enough to the
center of the galaxy.

Since the galactic center is expected to be an area of important accretion of dark matter, it is quite natural 
to expect that the WIMP annihilation rate should be more significant in this region. It is thus quite
customary for calculations on expected fluxes to be performed with respect to the annihilation rates one would
expect close to the galactic center, actually at the very center of our galaxy.

Since however the flux for annihilating WIMPs depends quadratically on the halo profile (linearly for
decaying dark matter), the strong divergencies among different estimates should be expected to have an equally
strong impact on the results. This is indeed the case. Let us assume that we look for $\gamma$-rays from
dark matter annihilations within a conical region of $4 \cdot 10^{-3}$ sr around the GC. The relevant
parameters and values for the $\bar{J}$ quantity can be seen in Table \ref{tab:JbarMI} for several
different halo profiles discussed in section \ref{SpatialDistribution}: NFW, NFW with adiabatic compression, 
Moore \textit{et al} and Moore \textit{et al} with adiabatic compression. It is clear that
the possible values span several orders of magnitude and, since Eq.\eqref{GammaFluxCalc} depends linearly on
$\bar{J}$, so will the corresponding flux.
\begin{center}
\begin{table}
\centering
\begin{tabular}{|c|ccccc|}
\hline
&$a$ (kpc)&$\alpha$&$\beta$&$\gamma$
%&$\bar{J}(10^{-3} {\rm sr}) $
&$\bar{J}(4\cdot10^{-3} {\rm sr})$  \\
\hline
NFW & $20$ & $1$ & $3$ & $1$
%& $1.214 \cdot 10^3$
& $5.859\cdot10^2$\\
$\rm{NFW_c}$ & $20$ & $0.8$ & $2.7$ & $1.45$
%& $1.755  \cdot 10^5$
& $3.254\cdot10^4$\\
Moore et al. & $28$ & $1.5$ & $3$ & $1.5$
%& $1.603  \cdot 10^5$
& $2.574\cdot10^4$\\
$\rm{Moore_c}$ & $28$ & $0.8$ & $2.7$ & $1.65$
%& $1.242  \cdot 10^7$
& $3.075\cdot10^5$\\
\hline
\end{tabular}
\caption{{\footnotesize NFW and Moore et al.
density profiles without
and with
adiabatic compression ($\rm{NFW}_c$ and $\rm{Moore_c}$ respectively)
with the corresponding parameters, and values of $\bar{J}(\Delta\Omega)$.}}
%for $\Delta\Omega=10^{-3}, 10^{-5}\ {\rm sr}$.}
\label{tab:JbarMI}
\end{table}
\end{center}
It is straightforward to realize that a precise knowledge of the dark matter distribution is not only a crucial 
element in the calculation of the flux one would expect on the earth, but also a major uncertainty in 
gamma-ray detection. One potential wayout this problem could be to exclude the galactic center from the
analysis, since it is the most uncertain region. But then, the signal statistics is expected to significantly
decrease. At this point we should consider the fact that in any detection procedure, the important element
is not just the absolute magnitude of the signal but rather its relative magnitude to the relevant background.
As we shall see later on, looking at other directions than the GC can be an efficient way to optimize the signal-background relation.

\subsection{Charged Particle Detection}
Charged particles present the complication of propagating throughout the galactic medium. 
This effect results in the distortion of the spectrum produced at the source, a phenomenon which is
absent in the case of gamma-rays: the form of the charged particle spectrum received on the earth can 
be significantly different than the one produced at the source.
Numerous approaches have been proposed on how to treat such effects \cite{Baltz:1998xv,Strong:1998su,Lavalle:2006vb}.
The starting point for all these approaches is a continuity relation which encodes, according to each author's 
assumptions, the relevant physics. These methods vary from completely numerical, semi-analytical up to fully
analytical. Since analytical methods usually not only allow us to better understand the underlying physics
but are also computationally much more efficient, our choice for the following will be the two-zone diffusion
model and its semi-analytical solution as described in \cite{Lavalle:2006vb}. In this model, particle propagation
takes place in a cylindrical region (called the Diffusive Zone, DZ) of half thickness $L$. The propagating particles
can escape the DZ, in which case they are simply lost.

The physical processes involved in charged matter propagation in this framework, could be encoded as follows:
\begin{itemize}
 \item Charged particles scatter on irregularities of the galactic magnetic field (called Alfv\'en waves).
This is a diffusion process with a diffusion coefficient given by
\begin{equation}
K(E) = K_0\,\beta
\left( \frac{E}{E_0} \right)^\alpha
\label{DiffCoeff}
\end{equation}
with $\beta$ being the particle's velocity, $K_0$ the diffusion
constant, $\alpha$ a 
constant slope, $E$ the kinetic energy and $E_0$ a reference energy (which we take to be $1$ GeV)

\item They undergo a second order Fermi reacceleration due to the motion of scattering centers which can be
described by a coefficient
\begin{equation}
K_{EE} = \frac29\,V_{a}^{2}\,\frac{E^2\,\beta ^4}{K(E)}
\end{equation}

\item They lose energy at a rate $b(E)$ which depends on the specific final state particles

\item They are wiped away from the galactic disk due to the convective wind with velocity $V_c \approx (5 - 15)$ km/s

\item They can annihilate upon scattering on the InterStellar (IS) medium. In this study, we shall
consider the two primary components of the medium, namely Hydrogen and
Helium.
\end{itemize}
These effects can be encoded in a master equation which can be written as
\begin{equation}
\partial_t\psi+ \partial_z(V_c\,\psi) - \nabla(K\,\nabla\psi) - \partial_E \left[ b(E)\,\psi + K_{EE}(E)\,\partial_E\,\psi\right] = q \ , 
\label{masterProp}
\end{equation}
where we denote by $\psi = dn/dE$ the space-energy density of positrons or antiprotons.

Even under the simplifications that we mentioned before (such as cylindrical symmetry), Eq.(\ref{masterProp})
is impossible to solve analytically, perhaps even numerically. In order to overcome this difficulty, we have
to make further simplifications so as to try and bring the master equation in a form which is solvable.
These assumptions are not universal for different particle species, instead the specific nature of each final
state particle should be carefully taken into account and the errors brought about by these assumptions should 
be assessed.

Furthermore, since our universe is matter and not antimatter - dominated, whereas matter and antimatter are expected
to be produced at similar rates in dark matter annihilations, it is reasonable to expect that trying to detect 
antiparticles rather than particles should be a justified choice: the matter sector would simply suffer from much more
elevated background event rates than the antimatter one.

\subsubsection{Positrons}
As pointed out in ref.\cite{Delahaye:2008ua}, in the case of positron propagation convection and reacceleration can
be neglected up to a relatively good level of accuracy, with the relevant error being of the 
order of $10\%$ or less once one sticks to positrons
with energies above $10$ GeV. Moreover, above $10$ GeV one can safely ignore an additional effect called solar modulation.
In the following, we shall be examining positrons with energies above $10$ GeV, so these assumptions hold quite
well.

On the contrary, the main process that affects positron propagation is energy loss through synchrotron radiation and
inverse Compton scattering on CMB photons. To account for these losses, we shall be writing the energy loss rate
as
\begin{equation}
 b(E) = \frac{E^2}{E_0\,\tau_E}
\end{equation}
where $E$ is the positron energy and $\tau_E=10^{16}$s is the characteristic energy-loss time. Then, the master 
equation gets simplified to
\begin{equation}\label{masterposi}
\partial_t \psi - \nabla \left[ K(\vec{x},E)\,\nabla\psi \right]- 
\partial_E \left[b(E)\,\psi\right] = q(\vec{x},E)\,,
\end{equation}
where $K$ is the space diffusion coefficient if we assume steady state. 
This coefficient is  taken to be constant in space but depends on the energy as
\begin{equation}
K(E) = K_0\left(  \frac{E}{E_0}\right) ^\alpha.
\end{equation}
Here the diffusion constant, $K_0$, and the spectral index, $\alpha$, are propagation parameters.

This model includes thus three free parameters, namely $L,\,K_0$ and $\alpha$. Delahaye \textit{et 
al} have proposed three benchmark models for these parameters \cite{Delahaye:2007fr} which are usually
called MIN, MED and MAX. The first and the last ones correspond to parameter values giving minimal and
maximal fluxes respectively that are compatible with the B/C data.
The MED model, on the other hand, corresponds to  the parameters that best fit the B/C data.
The corresponding parameter values are given in table \ref{PropParametersPos}.
\\
The master equation for positron propagation (equation \eqref{masterposi}) gets simplified into the form
\begin{equation}
K_0\,\epsilon^\alpha \nabla^2 \psi  + 
\frac{\partial}{\partial \epsilon}\left( \frac{\epsilon^2}{\tau_E} \psi \right) + q = 0\,,
\label{masterPos}
\end{equation}
where $\epsilon\equiv E/E_0$. This is the expression that has to be solved in order to calculate the effects 
of positron propagation on a signal produced at some point in the galaxy.

\begin{table}
\begin{center}
\begin{tabular}{|c|ccc|}
\hline 
&$L$ [kpc]&$K_0$ [kpc$^2$/Myr]&$\alpha$\\
\hline 
MIN & $1$ & $0.00595$ & $0.55$ \\ 
MED & $4$ & $0.0112$ & $0.70$ \\
MAX & $15$ & $0.0765$ & $0.46$ \\
\hline 
\end{tabular}
\caption{{\footnotesize Values of positron propagation parameters
widely used in the literature and that roughly provide minimal and maximal positron fluxes,
or constitute the best fit to the B/C data.}}
\label{PropParametersPos}
\end{center}
\end{table}
The way to solve this equation has been described in detail in references \cite{Baltz:1998xv, Lavalle:2006vb} for 
example. In Appendix \ref{cosmicrayprop} we give some details on this calculation.

Following the method described there, it can be shown that under our assumptions, the positron flux on the
earth coming from dark matter annihilations is
\begin{equation}
 \Phi_{e^+} (E)= \frac{\beta_{e^+}}{4\pi} 
 \frac{\left\langle \sigma v \right\rangle}{2}  \left( \frac{\rho(\vec{x}_\odot)}{m_\chi} \right) ^2
\frac{\tau_E}{E^2}
\int_E^{m_\chi} f(E_s)\,\tilde{I}(\lambda_D)\,dE_s\,,
\label{PosFlux}
\end{equation}
where the detection and the production energy are denoted respectively by $E$ and $E_s$, $\vec{x}_\odot$ 
is the solar position with respect to the GC and $\beta_{e^+}$ is 
the positron velocity. $f(E_s)$ is the production spectrum for positrons, $f(E_s) = \sum_{i} dN_{e^+}^i/dE_s$,
 with $i$ running over all possible annihilation channels much like in the case of gamma-rays. 
The diffusion length,  $\lambda_D$, is defined by
\begin{equation}
\lambda_D^2 = 4\,K_0\,\tau_E \left(\frac{\epsilon^{\alpha-1} - \epsilon_s^{\alpha-1}}{1-\alpha} \right) .
\end{equation}
We should notice that the astrophysical dependence of the positron flux is nicely separated from the particle
physics of the problem. It is encoded in the so-called halo function, $\tilde{I}$, which is given by
\begin{equation}
\tilde{I}(\lambda_D) = \int_{DZ} d^3\vec{x}_s\,\tilde{G}\left(\vec{x}_\odot, E \rightarrow \vec{x}_s, E_s\right)\,
\left(\frac{\rho(\vec{x}_s)}{\rho(\vec{x}_\odot)}\right)^2\,,
\label{halofuncpos}
\end{equation}
where the integral is performed over the diffusive zone. The modified Green's function $\tilde{G}$ is 
given analytically in Appendix \ref{cosmicrayprop}.

The advantage of this method is that the halo function $\tilde{I}(\lambda_D)$ 
can be calculated (and either tabulated or fitted)
just once as a function of the diffusion length and then be easily used for performing parameter 
space scans which, as in our case, can be rather large. In the framework of the following analyses,
we developed dedicated FORTRAN codes in order to calculate the halo function and compute the
relevant positron fluxes.

\begin{figure}[htb!]
\begin{center}
\includegraphics[width = 7cm, angle=270]{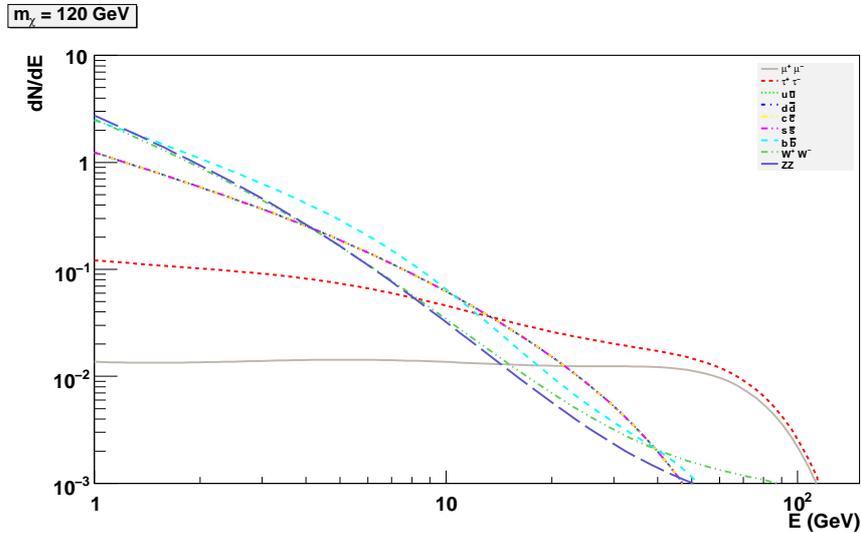}
\caption{{\footnotesize
Differential yield of various Standard-Model $2$-body final states into positrons for a WIMP
mass of $120$ GeV.}}
\label{positronyieldmchi120GeV}
\end{center}
\end{figure}

The decays of SM final-state particles into positrons can be computed as in the gamma-ray case.
In fig.\ref{positronyieldmchi120GeV} we can see the relevant yields, once again for a mass of $120$ GeV.
We note that in this figure we have made the simplification that all light quarks contribute similarly
in the total spectrum, which is a common assumption in the literature as well as in most public codes.

\subsubsection{Antiprotons}
Antiproton propagation is also governed by the master equation (\ref{masterProp}).
However, the dominant physical processes that affect antiprotons are quite different than in the positron case.
More specifically, it has been shown (see e.g. fig.2 of \cite{Maurin:2006hy}) that for antiproton energies 
above $10$ GeV energy losses, reacceleration as well as ``tertiary'' contributions can be neglected safely.
The main processes affecting propagation in this case are potential annihilations of antiprotons with the
interstellar medium and the fact that they can be wiped away from the galactic plane due to the galactic wind with
velocity $V_c$.

Let us denote by $\Gamma_{\overline{p}}^{\mbox{\tiny{ann}}} = \sum_{\mbox{\tiny{ISM}}} 
n_{\mbox{\tiny{ISM}}}\,v\,\sigma_{\overline{p} \ \mbox{\tiny{ISM}}}^{\mbox{\tiny{ann}}}$
the destruction rate of antiprotons in the interstellar medium, where $\mbox{ISM} = \mbox{H and He}$, 
$n_{\mbox{\tiny{ISM}}}$ is the average number density of ISM in the galactic disk, $v$ is the
antiproton velocity and $\sigma_{\overline{p} \ \mbox{\tiny{ISM}}}^{\mbox{\tiny{ann}}}$ is the
$\bar{p} - \mbox{ISM}$ annihilation cross-section. 
Implementing the aforementioned simplifications, 
the transport equation becomes:
\begin{equation}
\left[ -K\,\nabla + V_c\,\frac{\partial}{\partial z}
+2\,h\,\Gamma^{\mbox{ann}}_{\bar p}\,\delta(z) \right] \psi = 
q(r, t),
\end{equation}
with $h = 100$ pc being the half-thickness of the galactic disc.
Once again, some details on the solution of this equation can be found in Appendix \ref{cosmicrayprop}.
\\ \\
The final expression for the expected flux on the earth is
\begin{equation}
\Phi_{\odot}^{\bar{p}} (E_{\mbox{\tiny{kin}}}) = 
\frac{c\,\beta }{4\pi}
\frac{\langle\sigma v\rangle}{2}
\left(   \frac{\rho(\vec{x}_{\odot})}{m_\chi} \right)^2
\frac{dN}{dE}(E_{\mbox{\tiny{kin}}})
\int_{DZ} \left(\frac{\rho(\vec{x_s})}{\rho(\vec{x}_{\odot})} \right)^2
G^{\odot}_{\overline{p}}(\vec{x}_s)\,d^3x\,,
\label{PbarFlux}
\end{equation}
where none of the integrated quantities depends on the antiproton energy. 

Regarding  the propagation parameters $L$, $K_0$, $\alpha$, and  $V_c$, we take their values from 
the well-established MIN, MAX and MED models --see table \ref{PropParameters}. 
The former two models correspond to the minimal and maximal antiproton 
fluxes that are compatible with the B/C data. 
The MED model, on the other hand, corresponds to  the parameters that best fit the B/C data.
\begin{center}
\begin{table}
\centering
\begin{tabular}{|c|cccc|}
\hline 
&$L$ [kpc]&$K_0$ [kpc$^2$/Myr]&$\alpha$&$V_c$ [km/s]\\
\hline 
MIN & $1$ & $0.0016$ & $0.85$ & $13.5$\\ 
MED & $4$ & $0.0112$ & $0.70$ & $12.0$\\
MAX & $15$ & $0.0765$ & $0.46$ & $5.0$\\
\hline 
\end{tabular}
\caption{{\footnotesize Values of propagation parameters
widely used in the literature and that provide minimal and maximal antiproton fluxes,
or constitute the best fit to the B/C data.}}
\label{PropParameters}
\end{table}
\end{center}

\begin{figure}[htb!]
\begin{center}
\includegraphics[width = 7cm, angle=270]{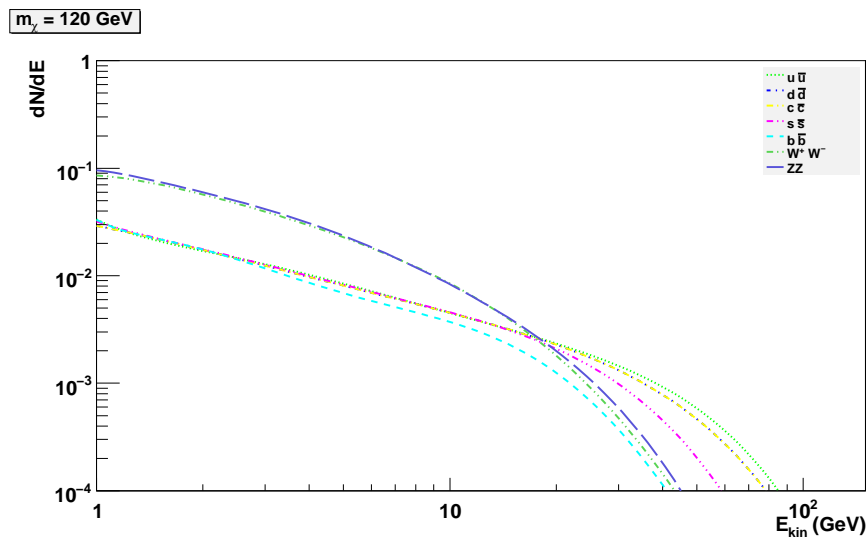}
\caption{{\footnotesize
Differential yield of various Standard-Model $2$-body final states into antiprotons for a WIMP
mass of $120$ GeV.}}
\label{antiprotonyieldmchi120GeV}
\end{center}
\end{figure}
As in the previous cases, the decay of SM 
particles into antiprotons can been calculated with {\tt PYTHIA}.  
In fig.\ref{antiprotonyieldmchi120GeV} we show the yields for SM 2-body final states into antiprotons
and for a $120$ GeV WIMP. The astrophysical factors and, eventually, the corresponding antiproton
fluxes are once again computed thanks to dedicated codes that we developed during this work.
Some more details on these codes are given in Appendix \ref{cosmicrayprop}.

\subsubsection{Uncertainties in antimatter detection}
Antimatter detection is unfortunately plagued by a very large number of uncertainties. It is interesting
that although the principle for their detection is quite similar to the gamma-ray case, the underlying physics
between the production at the source and the point of detection is so different that the dominant uncertainties
are very different.

Charged particles undergo more complex processes than gamma-rays, which make that they cannot originate
from very large distances. First of all, they can escape the diffusive zone and just get lost. Then, 
they can either lose energy, change direction on annihilate due to their interactions with the ISM. Even in
the antiproton case, where energy losses are quite irrelevant, the propagator cannot span too large a region
\footnote{In the case of positrons, the propagator expresses the probability that a particle produced at a point
$\vec{x_s}$ with energy $E_s$ reaches a point $\vec{x}$ with an energy $E$. In the antiproton case, the propagator
is a dimensionful quantity. Its interpretation is more tricky, but for our purposes we can say that it
is a measure of the same probability.}.
Since the earth is situated at a sufficiently large distance from the GC, and the most important ambiguity in the dark matter
distribution concerns actually the innermost regions of the galaxy, the halo profile is not expected to be among the
most determining factors in the flux calculation. This is indeed the case (see, for example, ref.\cite{Delahaye:2007fr}).

On the other hand, numerous other points of ambiguity exist. To give an example, assuming a well-defined 
production mechanism at the source, the most important factor that can modify the expected signal is
the propagation model, i.e. the values that should be used for the propagation parameters. These can indeed
give rise to fluxes differing not only in their normalization, but also in their form.
Also, the set of assumptions that we have made in order to arrive to a form of the diffusion equation that can be
solved analytically can be questioned. Although at large diffusion lengths different approaches 
seem to be in satisfactory agreement, there still exist important deviations among different methods for low
values of the diffusion length.

\subsubsection{Impact of substructures}
During the past years, much hope had been devoted on the possibility of an enhancement in the cosmic-ray
signal under the influence of granular structure in the halo. Qualitatively speaking, since $N$-body simulations
seem to favor the existence of a significant part of the DM halo within substructures (``clumps'') , 
some of which might follow even quite steep internal profiles, an arbitrarily large boost might be expected
in the annihilation rate. It has however been demonstrated that such large astrophysical boosts are
more or less excluded. In Appendix \ref{cosmicrayprop} we give some further details on this point.

It is not yet fully clear how clumpiness should be treated. The usual assumption was, until relatively
recently, that the impact of clumps should be included as an overall \textit{boost factor} in the total
flux. Lavalle \textit{et al} \cite{Lavalle:2006vb} showed that this is not at all the case and that we
should expect the boost factor to be (at least) a function of the propagated particle's energy. When
treating the possibility of clump-induced enhancement, we shall follow closely the treatment presented
by these authors. In this work, all effective boost factors are computed through FORTRAN programs that
we developed according to these prescriptions.
\\ \\
In recent years a new approach is being developed towards indirect detection, trying to combine as much information
as possible from as many different sources as possible in an effort to render observations and constraints more
robust. This ``multi-messenger, multi-experiment, multi-wavelength'' approach has indeed had some important successes
as we shall say in the following, rendering it an indispensable tool in our effort to better understand 
experimental results.

%%%%%%%%%%%%%%%%%%%%%%%%%%%%%%%%%%%%%%%%%%%%%%%%%%%%%%%%%%%%%%%%%%%%%%%%%%%%%%%%%%%%%%%%%%%%%%%%%%%%%
%%%%%%%%%%%%%%%%%%%%%%%%%%%%%%%%%%%%%%%%%%%%%%%%%%%%%%%%%%%%%%%%%%%%%%%%%%%%%%%%%%%%%%%%%%%%%%%%%%%%%
%%%%%%%%%%%%%%%%%%%%%%%%%%%%%%%%%%%%%%%%%%%%%%%%%%%%%%%%%%%%%%%%%%%%%%%%%%%%%%%%%%%%%%%%%%%%%%%%%%%%%

\section{Experiments, excesses and backgrounds}
\label{ExperimentsExcessesBkgs}
In the past decade there has been a very significant effort worldwide for the detection of dark matter 
by means of the aforementioned techniques (and not only). The listing we shall provide is by no means 
exhaustive, which actually demonstrates the important activity worldwide in the field of dark matter
detection. We should mention that we shall be focusing considerably more on experiments that will be
of interest in the following pages.

\subsubsection{Direct Detection}
Direct dark matter detection experiments consist typically of large detectors built underground in order
to minimize as much as possible unwanted background events. The experimental techniques vary significantly:
CDMS (and its upgrade CDMS II) \cite{Cooper:1993zg,Dixon:1997xi,Ahmed:2009zw} at the Sudan mine in the USA and EDELWEISS
in the Fr\'ejus underground laboratory in France \cite{deBellefon:1996da} use cryogenic detectors measuring phonons
and ionization induced from scattering of DM on the target material. 
The CDMS experiments use solid state semiconducting detectors made of Ge and Si whereas the EDELWEISS collaboration
has similarly chosen solid state Ge bolometers.
CRESST \cite{Bravin:1999fc} on the other hand uses solid state superconducting CaWO$_4$ heat detectors to measure phonons 
and scintillation whereas XENON \cite{Sorensen:2008ec,Aprile:2009yh} detects scintillation and ionization.
Some more details on this experiment will be mentioned in the following.
More specifically, upon interaction of a crystal detector with a WIMP, the crystal can get excited producing
phonons which can be measured. Ionization is the result of a WIMP - atom interaction where a detectable electron
is emitted by the ionized atom. Finally, scintillation occurs upon deexcitation of a nucleus who had previously
been excited through its scattering with a WIMP. Combining different techniques serves the purpose
of achieving good discrimination among signal and background: neutrons, for example, might yield
similar signals with a WIMP in one of the three channels, but a combined measurement (for example of
scintillation \textit{and} ionization) can discern among WIMPs and neutrons.

Constraints in direct detection experiments are usually given in the $(m_\chi, \sigma_{\chi N})$ plane.
Assuming there is a good modelization of the nuclear form factors for every experiment and an equally good
modelization of the astrophysics involved, unique bounds on the combination of these two parameters can be obtained
up to the uncertainties mentioned before.

Disagreement on specific issues let aside, 
CDMS II recently ended its functioning publishing its results \cite{Ahmed:2009zw} where the collaboration claims
the detection of two events passing all background rejection cuts. Although the statistical significance of their
signal is too low, there has been already quite some discussion on the meaning of the two events. An example
analysis can be found for the MSSM case in \cite{Bottino:2009km}.

A definitive answer to the CDMS II excess is expected to be given by the findings of the XENON $100$ kg detector 
which is currently running. Actually, the XENON collaboration also published recently its first results from their
new $100$ kg detector \cite{Aprile:2010um}. In this paper, the collaboration achieves the strongest limits ever
published, excluding lower cross-sections than every other apparatus in the world, especially in the intermediate
mass regime roughly between $10$ and $100$ GeV. We should note that the validity of this result is still under discussion
not only for the reasons we mentioned above, but also invoking arguments on the experimental setup (see, for example,
the discussion in \cite{Collar:2010gg,Collaboration:2010er}).

All of these experiments fall into the category of setups aiming to measure the ``bulk'' of the DM signal, in the
sense that they are not interested in the annual modulation that we mentioned. So far, letting aside the
CDMS II result, all of these experiments have only managed to set (increasingly strong) bounds on the 
$(m_\chi, \sigma_{\chi N})$ parameter space. Interestingly, one of the most controversial signals with a 
huge ($\sim 10 \sigma$) statistical significance has come from an experiment aiming to measure this marginal 
annual modulation effect, the DAMA experiment \cite{Bernabei:2008yh,Bernabei:2008yi} in Gran Sasso, Italy.
The DAMA observatory includes a whole series of different detectors. Among these setups, of particular
interest are DAMA/Libra and DAMA/NaI which use highly radiopure NaI(Tl) crystals in their detectors.
These experiments have indeed detected an annual modulation of the event rate exactly as predicted by the theory, 
which seems to point at low mass WIMPs if one attempts to interpret it through DM scatterings.
The DAMA results are going to be cross-checked by the oncoming KIMS experiment in South Korea, with the 
hope that if the modulation effect is real, it shall be confirmed. On the other hand, the dark matter interpretation of
the DAMA signal has been met with quite some skepticism from the community.

Another signal pointing possibly at low mass WIMPs came recently with the CoGeNT experiment \cite{Aalseth:2010vx}, 
which is mostly searching for light mass
WIMPs and reported the observation of an excess that cannot be associated to some known 
background source. The interpretation of CoGeNT data is known to require some caution, since the 
experiment does not discriminate among electron and nuclear recoils, hence controlling the background
can be slightly more subtle. Potential DM implications of the CoGeNT and DAMA results for dark matter, 
as well as ways to reconcile the two results with constraints coming from other DM detection experiments
(notably CDMS-II and XENON100) have been discussed, for example, in 
\cite{Kopp:2009qt, Chang:2010yk, Andreas:2010dz, Fitzpatrick:2010em, Mambrini:2010dq, Hooper:2010uy}. Especially in
the last of these references, the authors further manage to accommodate events recently announced by
the CRESST collaboration, through a $7.2$ GeV WIMP. The possibility that (some of) 
the three discrepancies could be due to dark matter scatterings is a particularly exciting one, that
actually proves to be quite challenging for our DM models, since usually WIMP candidates
tend to have relatively larger masses.

Finally, it should be mentioned that since a few years there has been increasing interest in yet
another mode of direct dark matter detection, called \textit{directional detection}
\cite{Vergados:2003pk, Morgan:2004ys, Vergados:2006gw, Alenazi:2007sy, Sciolla:2008vp, Green:2010zm}, 
in which the full variation of the event rate is reconstructed in an effort to obtain a full map
of dark matter in the earth's vicinity.

As for the backgrounds of direct detection, there cannot be a general treatment. The background rates largely
depend on the experiment and the specificities of each location, apparatus, etc. Where needed in the following, 
we shall explicitly state the background assumptions we make.

In the following, we shall be examining the sensitivity of the XENON experiment for a variety of different
dark matter candidates. We should thus provide a brief description of the experiment.
The XENON experiment aims at the direct detection of dark matter via its elastic scattering off
xenon nuclei. The detector is a mixture of liquid and gaseous xenon, allowing the simultaneous measurement 
of direct scintillation in the liquid and of
ionization, via proportional scintillation in the gas. In this way, XENON discriminates signal from
background for a nuclear recoil energy as small as $4.5$ keV. The main background for the experiment comes from 
neutron scatterings off xenon nuclei, due to natural radioactivity in the surrounding rock. 
Currently, the collaboration is working
with a 170 kg detector, but the final project is a detector containing 1 ton of xenon.

\subsubsection{gamma - ray detection}
There have been in the last decade several $\gamma$-ray detectors, either airborne 
(detecting directly gamma-rays from DM annihilations) or ground-based, in the form of
Atmospheric Cherenkov Telescopes. Some examples are:
\begin{itemize}
 \item The HESS ACT located in Namibia, which studies gamma-rays in the energy region $\sim [100, 10^5]$ GeV.
 \item The EGRET satellite, which has performed measurements at lower energies, roughly up to $10$ GeV
 \item The Fermi satellite mission, which is currently investigating the -very interesting- region
between $[0.3, \sim 300]$ GeV.
\end{itemize}

There have been excesses in the history of gamma-ray detection which have led to numerous efforts for their explanation
through dark matter annihilations.
In 2004, HESS announced the detection of some very high-energy gamma-rays originating from the galactic center
region at SgrA$^*$ \cite{Aharonian:2004wa}. During the first period after the announcement, numerous authors tried
to explain this signal as coming from dark matter annihilations. However, it became quite clear that the spectral form
could not easily be reconciled with dark matter annihilations and various other astrophysical mechanisms were
invoked to explain the excess \cite{Aharonian:2004jr}. Whatever the nature of this observation, the HESS collaboration
has given a fitting function for the detected flux
\begin{equation}
\phi^{\mathrm{HESS}}_{\mathrm{bkg}}(E) = F_0 ~ E_{\mathrm{TeV}}^{-\alpha},
\label{HESSpoint}
\end{equation}
with a spectral index
$\alpha=2.21 \pm 0.09$ and
$F_0=(2.50 \pm 0.21) \cdot 10^{-8} ~\mathrm{m^{-2} ~ s^{-1} ~ TeV^{-1}}$.
The data were taken during the second phase of measurements
(July--August, $2003$) with a $\chi^2$ of $0.6$ per degree of
freedom.

At the same time, HESS has measured the diffuse gamma emission at the area around the galactic center \cite{Aharonian:2006au}, 
with the corresponding spectrum being described by 
\begin{equation}
\phi^{\mathrm{diff}}_{\mathrm{bkg}}(E) = 1.1\cdot 10^{-4}\,E_{\mathrm{GeV}}^{-2.29}\,\mathrm{GeV^{-1} cm^{-2} s^{-1} sr^{-1}}\ .
\label{HESSdiffuse}
\end{equation}

EGRET in turn had previously announced the observation of a gamma-ray anomaly \cite{Hunger:1997we} below $10$ GeV
which exceeded by far the flux deduced by an extrapolation of the HESS measurements. Once again, several efforts
were made in order to explain the EGRET excess through dark matter annihilations. The recent results of the Fermi
Large Area Telescope \cite{Porter:2009sg} are incompatible with the EGRET excess.

Fermi is currently collecting data from various regions in the sky. Since Fermi is perhaps the
most promising gamma-ray detection experiment currently in operation, it would be useful to spend some time
describing it. The Fermi experiment \cite{Gehrels:1999ri,Peirani:2004wy} is 
a space satellite mission that was launched in June 2008 for a five-year run. Its instrument that is mainly of
interest for us, the Large Area Telescope (LAT) observes the whole sky
covering the energy range roughly from $30$ MeV up to $300$ GeV. The detector has a nominal effective area 
of $10 000$ cm$^2$ (which can actually vary, as we shall see later on) and an angular resolution of $0.1^\circ$, 
meaning that Fermi can distinguish two sources if they have a minimal separation of $0.1^\circ$ in the sky.
This setup allows Fermi to examine the inner regions ($\sim 7$ pc) of the galactic center.
Apart from that, Fermi shall also be looking for dark matter more or less all over the galaxy, at all longitudes
and latitudes.

\subsubsection{Antimatter detection}
Although highly challenging due to the important uncertainties in the particle propagation parameters, 
the area of antimatter detection has provided us with some of the most exciting results during the last few years.
Antimatter detection follows principles similar to the gamma-ray one. It is pursued by means of airborne detectors
either on satellites or on balloons.

One of the leading missions aiming at the detection of antimatter from dark matter annihilations today is the
PAMELA satellite \cite{Morselli:2000xg}. The PAMELA experiment is looking for positrons with energies lying in the
range $[0.1, 200]$ GeV, electrons with energies up to $1$ TeV as well as antiprotons of energy from $100$ MeV
up to $150$ GeV. The experiment has a geometric acceptance of $20.5$ cm$^2$ sr for positrons and antiprotons.
The collaboration, as is customary in cosmic ray detection, gives its results in the form of the positron
fraction, defined as the ratio of detected positrons over the total $e^+ + e^-$ number, and the antiproton 
results in the form of the $\bar{p}/p$ ratio. This is done in order to ``clean'' the results from the effects 
of solar modulation, which is a yet not quite well modeled effect.

Quite recently, the PAMELA satellite collaboration announced the observation of an substantial excess of cosmic rays
\cite{Adriani:2008zr}. This result is actually confirmed by the corresponding Fermi $e^+ + e^-$ measurements \cite{Abdo:2009zk}.
The observations of both missions are in straight contrast with the predicted backgrounds from the most popular 
cosmic ray propagation models which were used in order to calculate the backgrounds of these processes, such as
the background resulting from the so-called ``conventional'' propagation model \cite{Strong:2004de}. One of the most intriguing 
features of the PAMELA observations is that it is not accompanied, as would be natural in some of the most popular
models providing dark matter candidates, by a corresponding excess in antiproton measurements \cite{Adriani:2008zq}.
Perhaps the first possible thought in order to explain such an excess would be to enhance the DM self-annihilation
rate. However, it turns out that for candidates which have ``standard'' annihilation channels, doing so
would either require cross-sections falling largely out of the standard thermal cross-section needed in order to get
the correct relic densities, or astrophysical boosts that can most probably not appear (at least if one invokes standard
mechanisms such as clumps). Also, relevant excesses should have been observed in other channels as well, notably the
antiproton one as well as synchrotron emission. Various efforts have been made to overcome these constraints: 
non-perturbative effects that boost the
cross-section only at present times
\cite{ArkaniHamed:2008qn}, 
special leptophilic candidates with unusual properties that can only annihilate
or decay into leptons, superheavy candidates
and so on. Whether the PAMELA excess is indeed (partly) due to dark matter, 
is still under debate. A combination of different constraints has put severe bounds on DM interpretations of the
excess (see for example 
\cite{Cirelli:2008pk,Bertone:2008xr,Bergstrom:2008ag,Gogoladze:2009kv,
Cirelli:2009vg,Galli:2009zc,Pato:2009fn,Meade:2009iu,Profumo:2009uf,Huetsi:2009ex,Cirelli:2009bb,Cirelli:2009dv}
and references therein). 
It has nevertheless been argued that much more natural and standard astrophysical mechanisms such as pulsars
could account for this excess without invoking exotic physics \cite{Hooper:2008kg,Yuksel:2008rf,Profumo:2008ms}. 
If indeed the PAMELA excess in due to some non-DM related mechanism, it will certainly constitute a very 
important background for future dark matter searches in the positron channel. 

These questions are expected to be answered with the launch of the AMS-02 experiment \cite{AMSsite}. AMS-02 is again
an airborne mission to be placed onto the International Space Station for a three-year data acquisition. The
experiment has clearly among its scientific goals to detect cosmic rays from dark matter annihilations in the
energy ranges $[4, 300]$ GeV for positrons and $[16,300]$ GeV for antiprotons. The geometrical acceptance of the 
instrument is by far larger than the PAMELA one, namely an average of 420 cm$^2$ sr for positrons and 330 cm$^2$ sr 
for antiprotons \cite{Goy:2006pw}.

%%%%%%%%%%%%%%%%%%%%%%%%%%%%%%%%%%%%%%%%%%%%%%%%%%%%%%%%%%%%%%%%%%%%%%%%%%%%%%%%%%%%%%%%%%%%%%%%%%%%%
%%%%%%%%%%%%%%%%%%%%%%%%%%%%%%%%%%%%%%%%%%%%%%%%%%%%%%%%%%%%%%%%%%%%%%%%%%%%%%%%%%%%%%%%%%%%%%%%%%%%%
%%%%%%%%%%%%%%%%%%%%%%%%%%%%%%%%%%%%%%%%%%%%%%%%%%%%%%%%%%%%%%%%%%%%%%%%%%%%%%%%%%%%%%%%%%%%%%%%%%%%%
\section{WIMP mass determination and complementarity of different searches}
We discussed the basic principles of some approaches towards detection of dark matter
particles. Once something is detected, a next question is whether we can further
determine its properties: mass, couplings etc. Furthermore, it could be useful to try to compare
and combine different detection techniques in an effort to better constrain the DM properties.
We already mentioned that it was through the combination of data coming from very different sources that
it was possible to exclude and constrain several proposals trying to explain the PAMELA excess. Perhaps
we could even be more optimistic and hope that this combination could go further and be proven 
useful in the case of a positive detection.

But when could we characterize different experimental approaches as being complementary? Some comments
are perhaps in order at this point:
\begin{itemize}
 \item Given the significant uncertainties entering all dark matter detection modes, it is not absurd to
seek for an independent confirmation or cross-check of some experimental result (this is actually
sought for even in much more certain frameworks such as collider physics!). Going even further,
two experiments of the same kind might differ in many aspects, but the uncertainties remain the same.
In simple words, two gamma-ray detection experiments shall always have to face our limited knowledge 
of the dark matter distribution. Direct detection experiments or antimatter detection on the other
hand, are not so much plagued by this uncertainty.
 \item Different experiments might by sensitive to WIMPs with different characteristics. To give an
example, in the eventuality where the PAMELA excess is due to dark matter annihilations, and this
excess is due to some leptophilic candidate, direct detection efforts could be in vain. This is however
something which is not known in advance. Similarly, raising the self-annihilation cross-section for
a typical candidate should, as we said, induce a similar excess in other channels: $\gamma$'s, antiprotons
etc. The Fermi satellite could in principle help us probe some of these candidates. Moreover, even more
basic characteristics as the WIMP mass could be probed differently in different experiments.
 \item If an experiment or a detection mode is totally dominant in its detection capacity for some class of WIMPs, then 
indeed other detection modes could even be characterized as redundant. But if the sensitivities are comparable, 
then combining results could lead to a much better understanding of the properties of dark matter.
\end{itemize}
In ref.\cite{Bernal:2008zk} (see also \cite{Bernal:2008cu}), 
we attempted to examine at which point three completely different kinds
of dark matter detection could be complementary in determining some properties of WIMPs, especially
their mass. To do so, we looked into three kinds of detection modes: direct detection in the XENON
experiment, $\gamma$-ray detection in the Fermi satellite mission as well as WIMP production in a linear
lepton collider.

The question we shall try to answer is a rather optimistic one: suppose signals are detected in these three
experiments. We expect that given the uncertainties in statistics, astrophysics, systematics etc, the 
determination of the WIMP mass that could be responsible for these signals should present analogous 
uncertainties: as always, we do not get a point but rather a region in which the WIMP mass could lie.
What is the performance of each of these detection modes? In other words, given a positive signal, what
are the corresponding allowed regions for the WIMP mass (and perhaps other parameters)?
Furthermore, how robust are these results with respect to different candidates? Namely, can we perform
a model-independent analysis that does not enter the peculiar microphysics of different candidates?

\subsection{Statistical method}
In order to derive the allowed regions for all three detection modes, we employ a statistical method inspired
by Green's approach in \cite{Green:2007rb}. Let's take direct detection as an example. The two parameters that
we shall try to constrain are the WIMP mass as well as the WIMP-proton scattering cross-section (we assume that
the coupling of the WIMP to the proton and the neutron are the same).

Suppose a detected signal is generated by a WIMP of mass $m_\chi^{\mbox{\begin{tiny}real\end{tiny}}}$ and a 
scattering cross-section with the nucleon $\sigma_{\chi N}^{\mbox{\begin{tiny}real\end{tiny}}}$. Given these
two parameters and a well specified astrophysics, we can calculate the theoretically expected number of events
$N_{th}$ from equation \eqref{PracticeDirect}, integrating from a threshold energy up to the maximal observable energy.

It is well known that in real-life experiments the observed number of events can statistically fluctuate away from
the theoretical value, giving a number $N_{exp}$ of events. In order to account for this effect, an idea could be
to not actually try and analyze pseudo-data with the theoretical number of events, but rather something approaching
a more realistic situation. But how could we estimate the expected number of events starting from the theoretical one?
Usually, the statistical fluctuation of the signal is expected to follow a Poisson distribution. This is our choice 
in this case. We consider that the observed number of events follows a Poisson distribution, with mean value $N_{th}$. Then, 
we randomly pick $N_{exp}$ from this distribution and generate pseudo-events distributed over energies according to
equation \eqref{PracticeDirect}, which we numerically normalize to unity to render it a probability density function. 
The set of events, along with the corresponding $N_{exp}$ value, will in the
following be referred to as an ``experiment''.

Then, for every point in the $(m_\chi, \sigma_{\chi N})$ plane - we let the two parameters vary within
reasonable limits - we calculate the corresponding extended likelihood function
\begin{equation}
L = \frac{(N_{th}^{scan})^{N_{Exp}}}{N_{Exp}!}\exp{(-N_{th}^{scan})}
\prod_{i = 1}^{N_{Exp}} f(E;m_{\chi}, \sigma_{\chi-p})
\label{likelihood}
\end{equation}
where
\begin{equation}
f(E;m_{\chi}, \sigma_{\chi-p}) =
\frac{dN/dE(E;m_{\chi}, \sigma_{\chi-p})}
{\int_{E_{th}}^{E_{sup}} dN/dE(E;m_{\chi}, \sigma_{\chi-p})}
\end{equation}
is the normalized \textit{total} event rate (signal+background) and $N_{th}^{scan}$ is the theoretical number of events, 
expected from Eq.(\ref{PracticeDirect}), for the
given point of the parameter space. The normalization renders $f$ a probability
density function and, thus, suitable for use in a likelihood calculation.

The use of equation (\ref{likelihood}) presents the advantage that it takes into account the fact that the number of observed
events in an experiment can actually deviate from the expected behavior for several reasons. For the given
experiment, say $j$, we scan over the $(m_{\chi}, \sigma_{\chi-p})$ parameter space and calculate the value
$(m_{\chi}^{Est, j}, \sigma_{\chi-p}^{Est, j})$ that maximizes the expression (\ref{likelihood}). 
This is the estimation for our parameters for the $j$-th experiment.
We then calculate the mean value of all the estimations and find which experiment's estimation was closest to this mean value.
This experiment is considered to be the most representative of them all and is used to perform a final scan.
Finally, from the likelihood distribution we obtain through this scan we can plot discrimination capacity regions.

Direct detection experiments present the advantage of quite well-controlled background. The additional ambiguity that arises
in indirect detection and concerns uncertainties in the background will be dealt with in the relevant chapter.

As a final remark on the statistical treatment we used, let us say
that in order to be more precise, we would have to take into account (as is systematically
done in \cite{Baltz:2008wd}) the fact that the mass and cross-section precision are
themselves random variables and should, consequently, be given with their relevant
statistical variance. To do so, we would have to consider the actual distribution of
estimators for all experiments. However, 
such a treatment goes far beyond the scopes of this work, where we are interested
in a more qualitative comparison of different detection modes. In this respect, we
keep the experiment which averages the properties of a larger set of
experiments. Motivating this approach, our results are indeed in accordance with \cite{Baltz:2008wd} and
\cite{Green:2008rd}.

\subsection{Results for direct detection}
\begin{figure}[htb!]
\begin{center}
% \vspace{2cm}
\includegraphics[width=7cm, angle=-90]{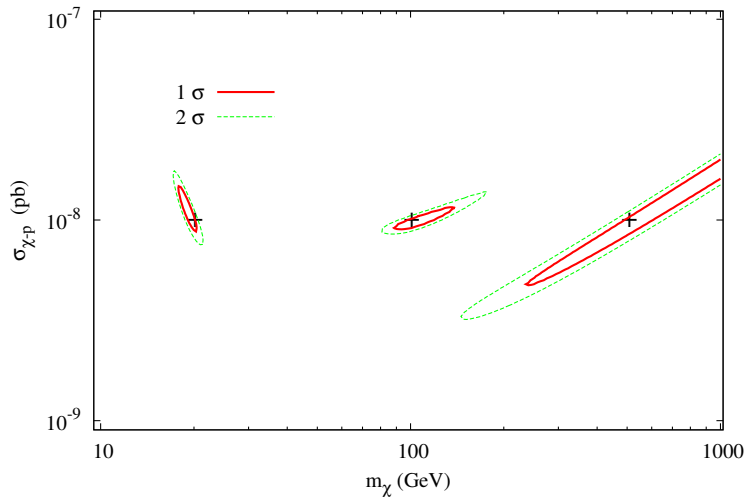}\\
\end{center}
\caption{{\footnotesize
Distribution of the maximum likelihood WIMP mass, $m_{\chi}$, and
cross-section, $\sigma_{\chi-p}$, for $3$ years of exposure in a $100$ kg
XENON experiment, for
$m_{\chi}=20,~100$, $500$ GeV and $\sigma_{\chi-p}=10^{-8}$ pb.
The inner (full) and outer (dashed) lines represent the $68\%$ and
$95\%$ CL region respectively.
The crosses denote the theoretical input parameters
($\sigma_{\chi-p}$, $m_{\chi}$).}}
\label{fig:MIdirectPlain}
\end{figure}

We consider a data acquisition period of $3$ years for the XENON experiment, with three different
detector masses, namely $10$ kg, $100$ kg and $1$ T. Following ref.\cite{Angle:2007uj} we take the energy
range from $4$ up to $30$ keV. The detector is taken to be ``perfect'', meaning that its efficiency has
been set to unity. It should be however noted that especially at low energies, this can be an important
issue.
In figure \ref{fig:MIdirectPlain} we show the capacity of the XENON experiment to reconstruct
the WIMP mass and spin-independent scattering cross-section assuming a $100$ kg detector and three
WIMP masses: $20$, $100$ and $500$ GeV. 

The
contours correspond to the allowed regions at $68\%$ and $95\%$ CL. We can clearly see a general 
tendency, namely that the mass reconstruction is much better for low WIMP masses, reaching the level
of a few percent for masses lower than $50$ GeV. The reason for that is quite simple: in general, the
event rate depends on the WIMP mass. However, upon closer inspection of Eq.(\ref{PracticeDirect}), one
can see that when the WIMP mass becomes quite larger than the nucleus mass (in our case this point
would be around $100$ GeV), the reduced mass $\mu_{Nucl}$ is roughly determined by the nucleus mass and the
event rate becomes more or less insensitive to the WIMP mass. This means
that for identical scattering cross-sections, two WIMPs with quite different masses can actually generate
a very similar recoil energy spectrum, i.e. the two WIMPs are indistinguishable. This is the reason why one sees that for
a $500$ GeV WIMP, only a weak lower mass bound can be extracted.

This result is however somewhat naive. The various uncertainties that enter the event
rate calculation have already been discussed. A good question to ask is how could (some of) these 
uncertainties influence the WIMP mass reconstruction capacity.

\subsubsection{Impact of some uncertainties}

As mentioned before, significant uncertainties can exist in the precise velocity distribution of
WIMPs in the detector rest frame. Among the parameters involved is $v_0$, the 
sun's circular velocity around the Galactic Center (GC). The relevant uncertainty is of the order
of $8-10$\% of the largely used value $220$ km/sec.

As far as background events are concerned, it is quite difficult to perform a general
study valid for every detector. Neutron backgrounds, which are in fact the most
difficult to distinguish from signal events, usually come from three sources
(see also \cite{Aprile:2002ef}):
\begin{itemize}
\item Cosmic muon - induced neutrons, which are not in general considered to cause
much nuisance.
\item Neutrons from the detector's surrounding rock.
\item Neutrons coming from contamination of the detector itself or surrounding materials.
\end{itemize}

It is difficult to model in general neutron backgrounds, as they are
mostly determined by the specific location in which every experiment is situated, as
well as by the specific shielding configuration adopted by each collaboration.
Two widely studied forms of neutron backgrounds are the case of a constant one,
which seems to be quite well-motivated by an experimental point of view and can
resemble to a heavy WIMP's signal, and an
exponential one which apart from its theoretical motivation is also interesting
as it gets to ``mimic'' (as pointed out in \cite{Green:2008rd}) the actual signal spectrum
for intermediate WIMP masses.
In this respect, we studied the impact of these two forms of background:

We consider firstly a constant background, with a value taken to be the
same as the maximal WIMP signal
in the first energy bin.
Throughout this paper, when examining the impact of uncertainties on
the mass determination accuracy, we will consider the case of a
somehow ``typical'' in many theoretical frameworks
case of a 100 GeV WIMP.
Then, we introduce an exponential background of the form
$\left (\frac{dN}{dE}\right )_{\mbox{bkg}}= A  \exp(-E/E_b)$,
where the slope of the exponential is fixed at $E_b = 25$ keV and
the $A$ factor is determined by demanding that the maximal
values of the signal and the background be the same. The
reason for this specific choice of parameters is that it is
for these values that the signal spectrum has a significant
resemblance to the background one, making it difficult to distinguish
from one another.

\begin{figure}[htb!]
\begin{center}
% \vspace{2cm}
\includegraphics[width=7cm, angle=-90]{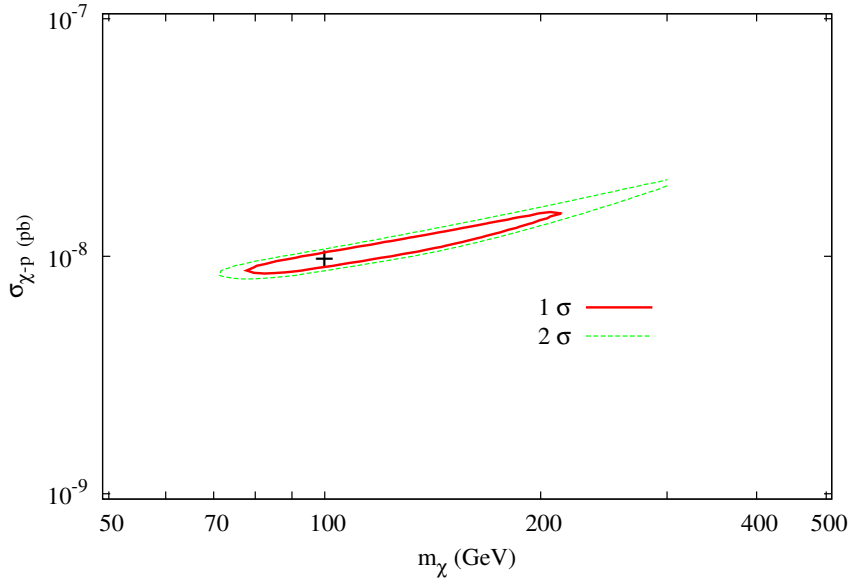}\\
\end{center}
\caption{{\footnotesize
$68\%$ and $95\%$ CL regions for the XENON $100$ kg experiment for a $100$ GeV WIMP
with a proton-WIMP scattering cross-section of $10^{-8}$pb
in the case where uncertainties in the $v_0$ parameter are taken into
account and, thus, included in the fitting procedure.}}
\label{fig:XENONv0}
\end{figure}

\begin{figure}[htb!]
\begin{center}
% \vspace{2cm}
\hspace{-1cm}
\includegraphics[width=5cm, angle=-90]{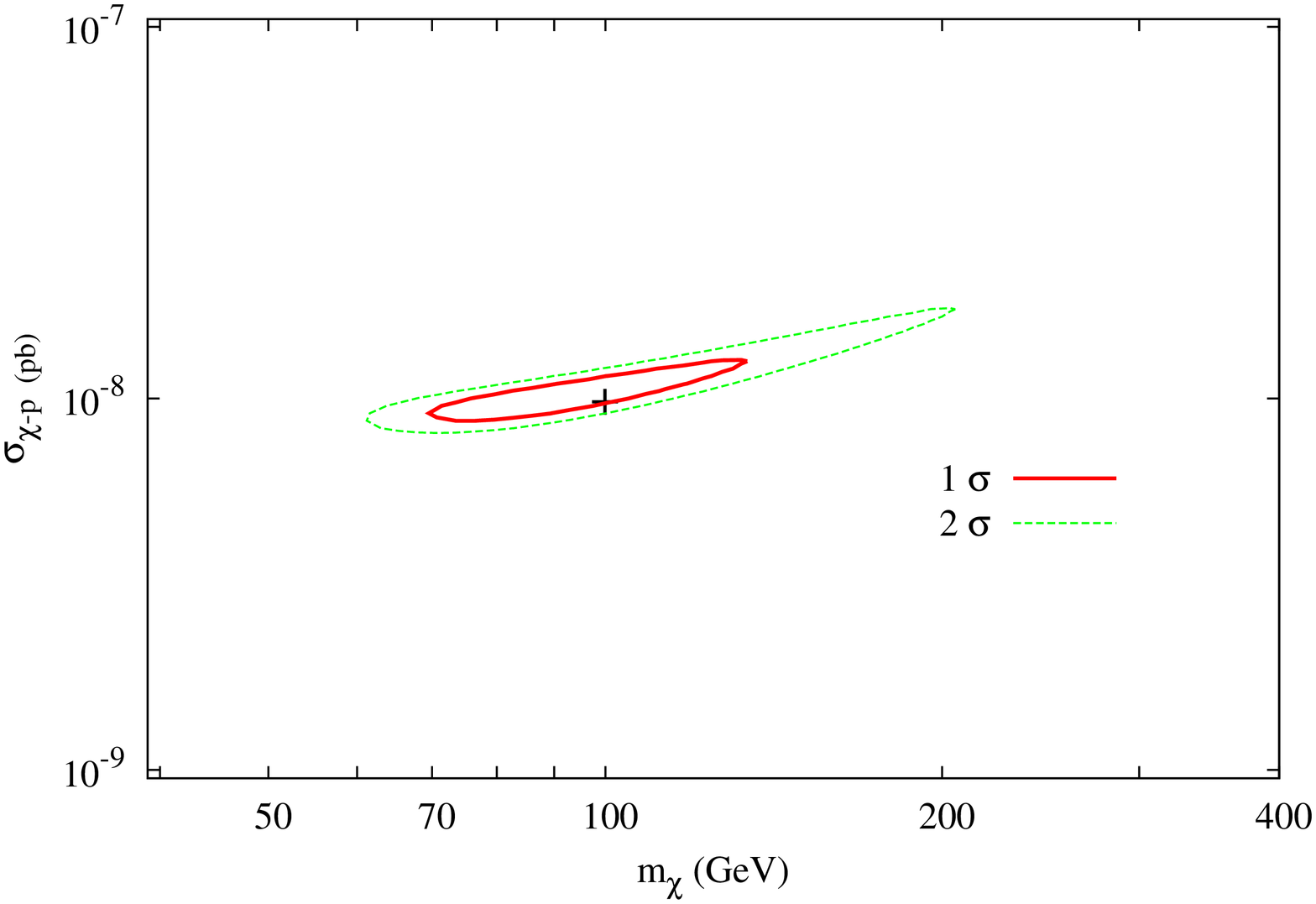} \hspace{1cm}
\includegraphics[width=5cm, angle=-90]{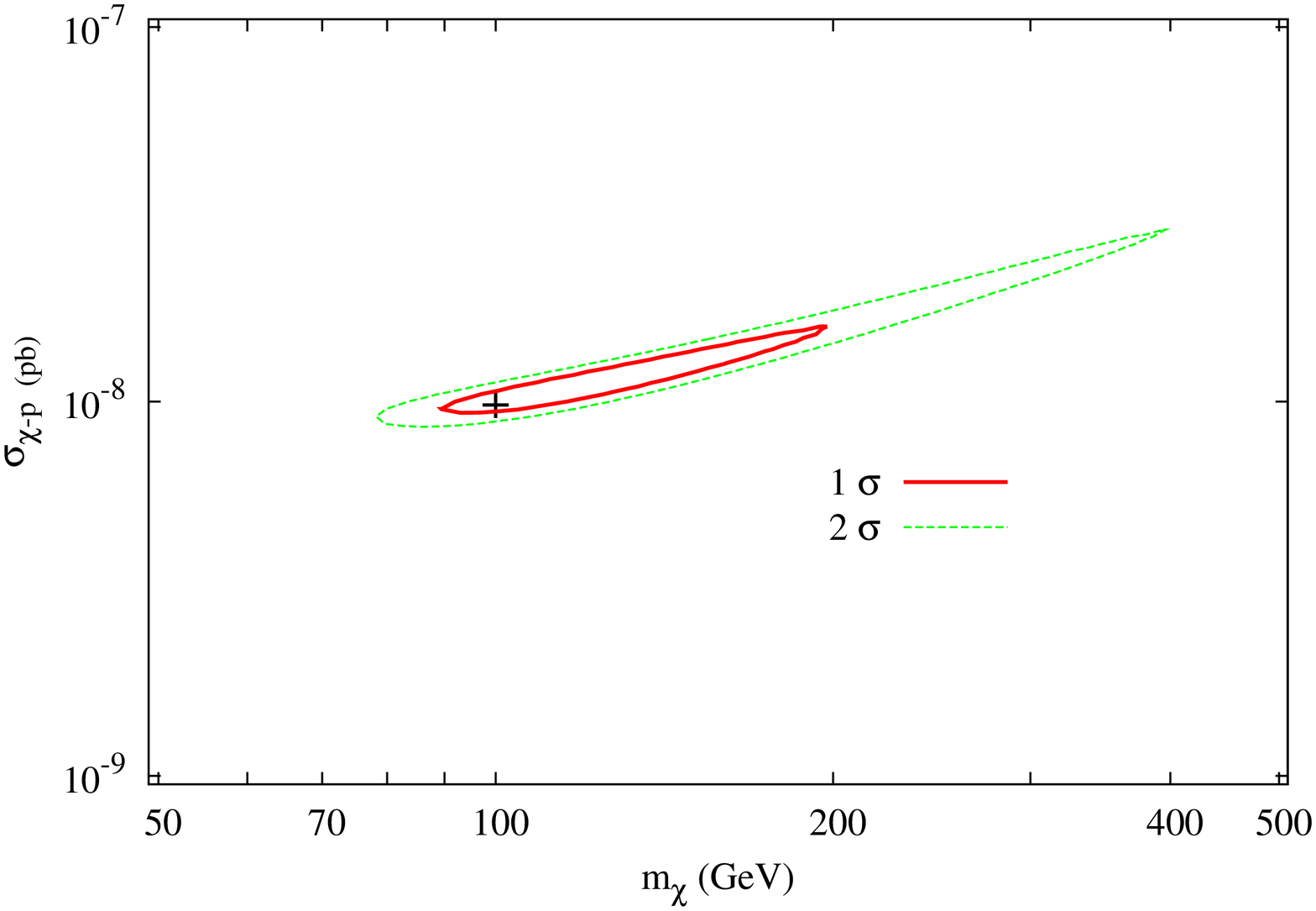}\\
\end{center}
\caption{{\footnotesize
$68\%$ and $95\%$ CL regions for the XENON $100$ kg experiment for a $100$ GeV WIMP
with a proton-WIMP scattering cross-section of $10^{-8}$pb
including a constant neutron background (left)
or an exponential one (right). The serious deterioration of accuracy in the
second case is due to the fact that the background parameters where chosen
in order to mimic the actual signal spectrum.}}
\label{fig:XENONbkg}
\end{figure}

Our results are shown in Figs.\ref{fig:XENONv0} and \ref{fig:XENONbkg}
for the cases of inclusion of $v_0$ in the fitting procedure and non-zero
backgrounds respectively.
The deterioration of the expected accuracy is obvious, when we compare these plots
to those of Fig.\ref{fig:MIdirectPlain}. Especially for the case of large uncertainties
in $v_0$ (we let it vary in the region between $200$ and $240$ km/sec) and of inclusion
of a background which is nearly degenerate with the signal, the expected precision
is dramatically aggravated. This shows, among others, the extreme importance of
a well-controlled environment and well-measured input parameters, other than the
WIMP mass.

\subsection{Gamma-ray detection}
Let's now repeat the previous exercise for the case of $\gamma$-ray detection from the galactic 
center. First things first, we should define the astrophysical assumptions for our study. We 
choose to examine the mass and $\left\langle \sigma v \right\rangle$ reconstruction capacity for the Fermi experiment,
assuming four different halo profiles: the standard Navarro, Frenk and White one, a modification of this
profile to include adiabatic compression, the Moore et al profile, as well as a modification of the latter to include,
again, the effect of adiabatic compression. Moreover, we consider a solid angle of $4 \cdot 10^{-3}$ sr around the
GC. The corresponding values for the $\bar{J}$ parameter defined in Eq.(\ref{Jbar}) can be found in Table \ref{tab:JbarMI}.

As background for our study, we consider the HESS measurements as presented in paragraph \ref{ExperimentsExcessesBkgs}.
Moreover, as this work was performed before the publication of the Fermi results, at energies below $10$ GeV
we also take into account the EGRET data. The resulting background is actually an interpolation between 
the HESS data at high energies and the EGRET ones at lower energies. This constitutes an important additional
background which is expected to deteriorate the precision. It would be interesting to repeat this analysis 
excluding the EGRET excess from the background. We leave this for future work. 
\footnote{The Fermi discovery potential has further been examined for example in \cite{Zaharijas:2006qb}.
An interesting recent treatment relevant to ours can be found in \cite{Bernal:2010ip}, 
where the authors further take into account non-prompt contributions.}
We consider a six-year mission run, assuming that the region of interest will be within the field-of-view
$50\%$ of the time.

\begin{figure}[htb!]
\begin{center}
% \vspace{2cm}
\includegraphics[width=7cm, angle=-90]{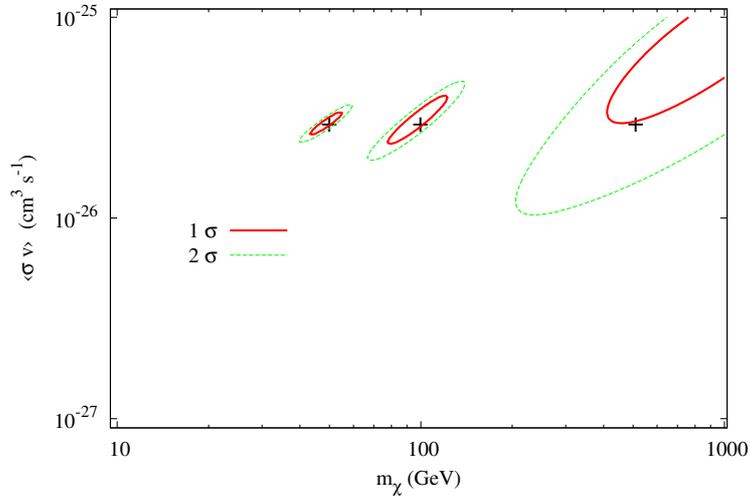}\\
\end{center}
\caption{{\footnotesize
Distribution of the maximum likelihood WIMP mass, $m_{\chi}$, and
annihilation cross-section, $\langle\sigma v\rangle$, after 6 years of observation
($50\%$ of time exposure)
of the galactic center with Fermi, with the hypothesis of a NFW halo profile,
for $m_{\chi}=50,~100$, $500$ GeV and
$\langle\sigma v\rangle=3 \cdot 10^{-26}$ cm$^3$s$^{-1}$.
The inner (full) and outer (dashed) lines represent the $68\%$ and
$95\%$ CL region respectively.
The crosses denote the theoretical input parameters
($\langle \sigma v \rangle$, $m_{\chi}$).}}
\label{fig:IndirectPlainMI}
\end{figure}

Let's begin with the simplest case: a WIMP of a given mass and total thermally averaged self-annihilation cross-section
which is the one naively suggested by the relic density arguments we gave in the previous chapter, $3 \cdot 10^{-26}$
cm$^3$ sec$^{-1}$. We follow a procedure similar to the one in direct detection, considering a two dimensional 
parameter space $(m_\chi, \left\langle \sigma v \right\rangle)$. Once again, we wish to see which regions would be
compatible with an excess provoked by the WIMP. For the sake of definitiveness, we take for the moment a perfectly known
$W^+ W^-$ final state and three candidate masses, $50$, $100$ and $500$ GeV. Obviously, this final state in inaccessible
for our first reference mass. Given however that the spectral shape is not that different than, for example, the 
$b \bar{b}$ one, this approximation is not influencing our main points. Following the extended likelihood approach
as before, we plot the non-discrimination regions in our two-dimensional parameter space. Our results can be seen in 
fig.\ref{fig:IndirectPlainMI} in the case of a NFW halo profile.

It is interesting that once again, the experiment is most sensitive to low-mass WIMPs.
The precision can easily reach the percent level
for Fermi for $m_{\chi}\lesssim 50$ GeV.
The gamma--ray spectrum will give more precise
measurements if the mass of the WIMP lies within the Fermi
sensitivity range. Indeed, the shape of the spectrum will be easily
reconstructed above the HESS/EGRET and diffuse background if
the endpoint of the annihilation spectrum
lies within the energy range reachable by Fermi.

\subsubsection{Uncertainties in gamma-ray Detection}

\begin{figure}[htb!]
\begin{center}
% \vspace{2cm}
\includegraphics[width=7cm, angle=-90]{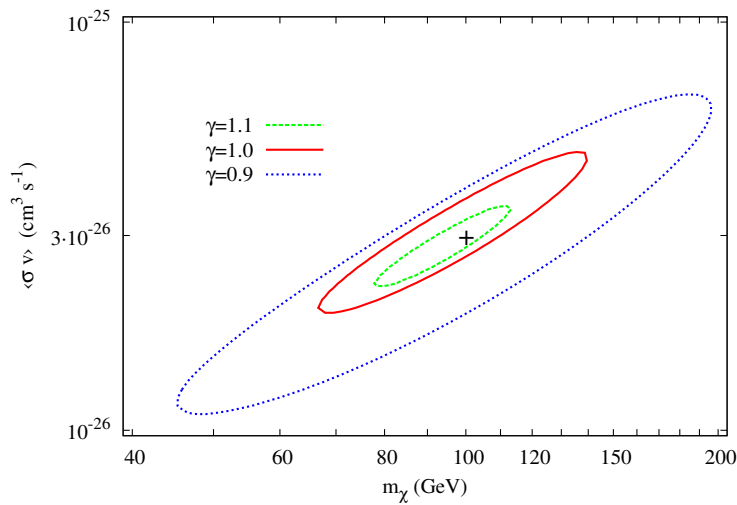}\\
\end{center}
\caption{{\footnotesize
NFW--like halo profile with $\gamma=0.9$, $1$ (NFW) and $1.1$ at $95\%$ confidence level.}}
\label{fig:IndirectGammaMI}
\end{figure}

Furthermore, we have studied the influence of the variation of
the inner slope of the halo profile on the resolution of the WIMP mass.
In addition to the NFW profile, we have considered some NFW--like profiles,
allowing the $\gamma$ parameter determining the inner slope of the profile
to vary from its original value by $10\%$. This is shown in
Fig.\ref{fig:IndirectGammaMI}, where in addition to the NFW halo profile ($\gamma=1$)
we also study profiles with
$\gamma=0.9, 1.1$.
As expected, the larger the $\gamma$ is, the more enhanced the galactic gamma ray flux becomes, and the better the
WIMP mass resolution turns out to be.
It is worth noticing here that we have confirmed that in the case of a compressed NFW profile
($\gamma \sim 1.45$), the precision of Fermi increases by two orders of magnitude.

\begin{figure}[htb!]
\begin{center}
% \vspace{2cm}
\includegraphics[width=7cm, angle=-90]{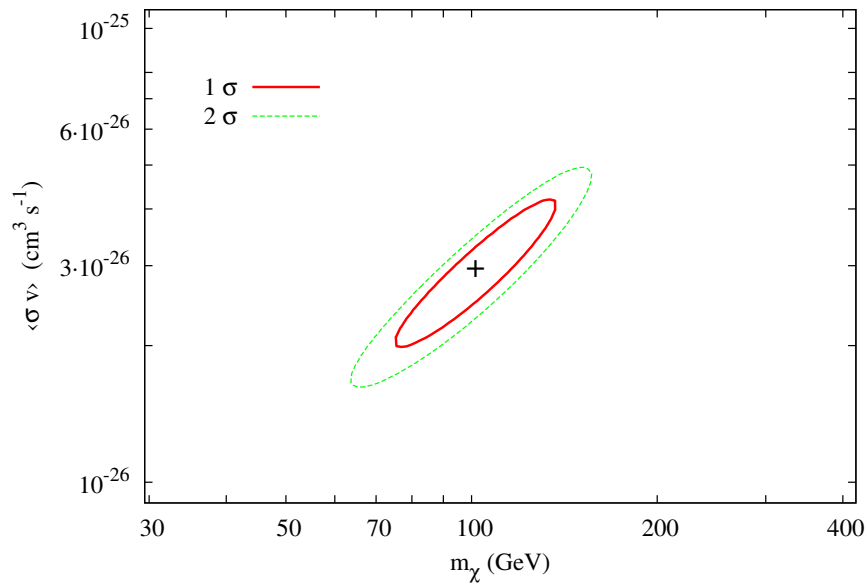}\\
\end{center}
\caption{{\footnotesize
$68\%$ and $95\%$ CL regions for a statistical treatment with the
overall background normalization included in the fitting procedure, $m_\chi = 100$GeV and
$\langle\sigma v\rangle = 3\cdot10^{-26}$ cm$^3$ s$^{-1}$  .}}
\label{fig:GammaBkgNormMI}
\end{figure}

So far, we have considered a perfectly known background.
Whereas this is a rather reasonable approximation in the case of direct detection,
it is less obvious
for the indirect one.  As it has been pointed out (see, for example,
\cite{Cesarini:2003nr, Jeltema:2008hf}), the uncertainties entering the calculation
of the backgrounds coming from the
galactic center region can considerably affect the results of any analysis.
More concretely, and especially for small WIMP masses and low energies
(where the performance of both direct
and indirect detection is maximal), the main background contribution comes from the aforementioned EGRET
source. However, both the overall normalization
and the spectral index characterizing
this source's spectrum contain
uncertainties. An interesting point would be to include
the overall background normalization as well as the spectral index in the fitting procedure.
In Fig.\ref{fig:GammaBkgNormMI} we show,
for the sake of comparison,  the result of a fitting procedure, where we also fit the background
normalization while simultaneously considering signals and backgrounds
with poissonian fluctuations. The original spectrum
is taken to be the full EGRET source plus the flux produced by a $100$ GeV WIMP annihilating in a
NFW halo. One could imagine discarding low-energy data which contain a maximal background
contamination. This, however, would significantly reduce the statistics and the corresponding
precision, since a major part of the signal would be discarded.
The fact that the inclusion of an uncertainty in the background normalization (i.e. its
inclusion in the statistical treatment) does not have a major impact on the results
can be explained from the fact that throughout this work we have used  the extended likelihood approach
in our statistical analysis, which already introduces a deviation from
the ideally expected situation. In this respect, our results are already quite conservative.

\begin{figure}[htb!]
\begin{center}
% \vspace{2cm}
\includegraphics[width=7cm, angle=-90]{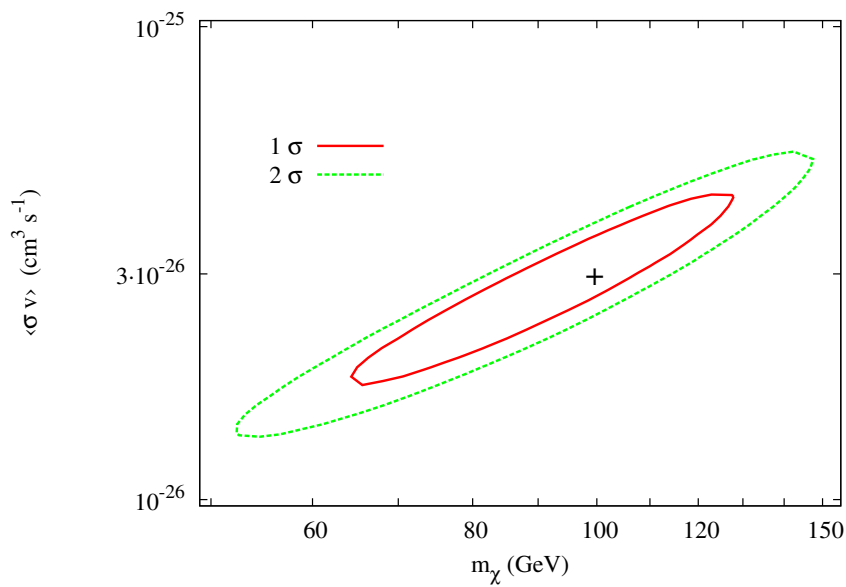}\\
\end{center}
\caption{{\footnotesize
$68\%$ and $95\%$ CL regions for the case where the uncertainties in the
EGRET point source spectral index are included in the fitting procedure,
$m_\chi = 100$GeV and
$\langle\sigma v\rangle = 3\cdot10^{-26}$ cm$^3$ s$^{-1}$  . An NFW halo profile
has been assumed.}}
\label{fig:specindexGammaMI}
\end{figure}

In the same way, in Fig.\ref{fig:specindexGammaMI} we show the corresponding results
where this time the spectral index is included in the fitting procedure instead.
The spectral index is left to vary in the region $[2.1, 2.4]$, which we find
to be a quite reasonable one as we verified that all over this region we obtain
reasonable fits of the EGRET data.

It is interesting to note that the variation of the background's spectral
index seems to have a larger impact on the precision that could be achieved,
with respect to the corresponding case of the background's overall
normalization.

This is somehow logical, first of all since by definition
the background depends linearly on the normalization factor,
but exponentially on the spectral index of the EGRET point source.
So, modifications in the latter bring along a much more drastic
modification of the background signal itself. Furthermore, variations of
the overall normalization have just the influence of "burying" the
signal a little more or a little less in a background which is already quite
elevated. On the contrary, by varying the spectral index we actually
change the \textit{shape} of the spectrum. This brings along a more important
uncertainty, since we could imagine much more numerous configurations in the
(spectral index, cross-section, mass) space that could satisfy selection criteria.

Although as we already mentioned the EGRET excess is now disproved by the latest Fermi results, 
the previous analysis can actually give us an idea, at least qualitatively, of the impact of
background uncertainties in the WIMP mass determination.

Proceeding to a different issue, until now we have considered a WIMP annihilating
into a pure $W^+ W^-$ final state. This is an assumption which is made to
simplify the overall treatment, but which at the same time somehow
restricts the generality of our results. It would be interesting to examine
what could be the impact of variations in the final state
on the WIMP mass determination capacity. Annihlation into $ZZ$ pairs
is not expected to seriously modify the results, since it resembles the $WW$ spectrum.
What would be more interesting
would be to see what happens when we consider (light or heavy) quark pairs and/or
leptons as WIMP annihilation products.

The only leptonic final state we consider is the
$\tau^+ \tau^-$ one, since annihilation into $\mu^+ \mu^-$ has a relatively small
contribution to the annihilation gamma-ray spectrum, whereas $e^+ e^-$ pairs
contribute through other processes, the examination of which exceeds the
purposes of the present treatment. We should, nevertheless, note here that
we do not take into account the effects of leptonic final state radiation which
can indeed become important, especially in the case of Kaluza-Klein dark matter
and in energy ranges lying near the WIMP mass.
The effect of such processes has been discussed in detail in ref.\cite{Birkedal:2005ep}
for the case of a generic WIMP and \cite{Bergstrom:2004cy} for the special case
of KK dark matter.
Obviously, this omission somehow restricts the generality of our results
as far as the impact of final states are concerned.

\begin{figure}[htb!]
\begin{center}
% \vspace{2cm}
\includegraphics[width=7cm, angle=-90]{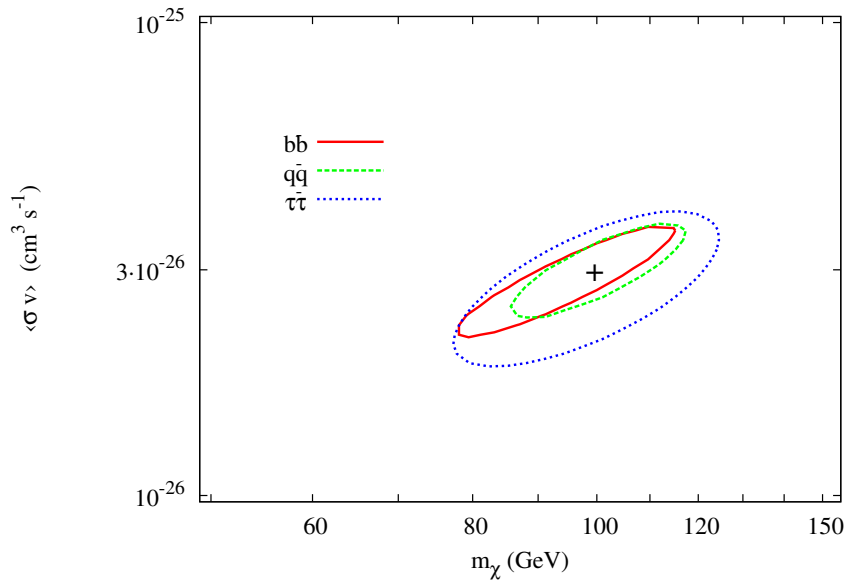}\\
\end{center}
\caption{{\footnotesize
$95\%$ CL regions for a $100$ GeV WIMP and different final
states. The total thermally averaged annihilation cross-section has been taken
to be $\langle\sigma v\rangle = 3\cdot10^{-26}$ cm$^3$ s$^{-1}$.
A NFW halo profile has been assumed.}}
\label{fig:OtherFinalStates}
\end{figure}

We performed two kinds of tests: The first one consists only
of modifying the annihilation products, considering a perfectly known final
state (meaning that the Branching Ratios are not included in the statistical treatment).
Our results can be seen in fig.\ref{fig:OtherFinalStates} for the case of pure
$b \bar{b}$, $q \bar{q}$ and $\tau^+ \tau^-$ final states and a $100$ GeV WIMP.

It is interesting to notice the relative amelioration of the mass resolution
with respect to the pure $W^+ W^-$ final state. A possible explanation could be that fermionic
final states tend to give more hard spectra with respect to bosonic ones (the extreme
case being leptonic ones), rendering the spectrum more easily distinguishable from
the background. The hardest spectrum is given by the $\tau^+ \tau^-$
final state. Nevertheless, in this case the characteristic spectral form is
somewhat compensated from the reduced statistics of the signal, due to the smaller
photon yield of leptons with respect to gauge bosons (or quarks). This is not
the case for annihilation into quarks, where the characteristic spectral form,
although less obvious than in the leptonic case,
is nevertheless combined with a more enhanced signal.

\begin{figure}[htb!]
\begin{center}
% \vspace{2cm}
\includegraphics[width=7cm, angle=-90]{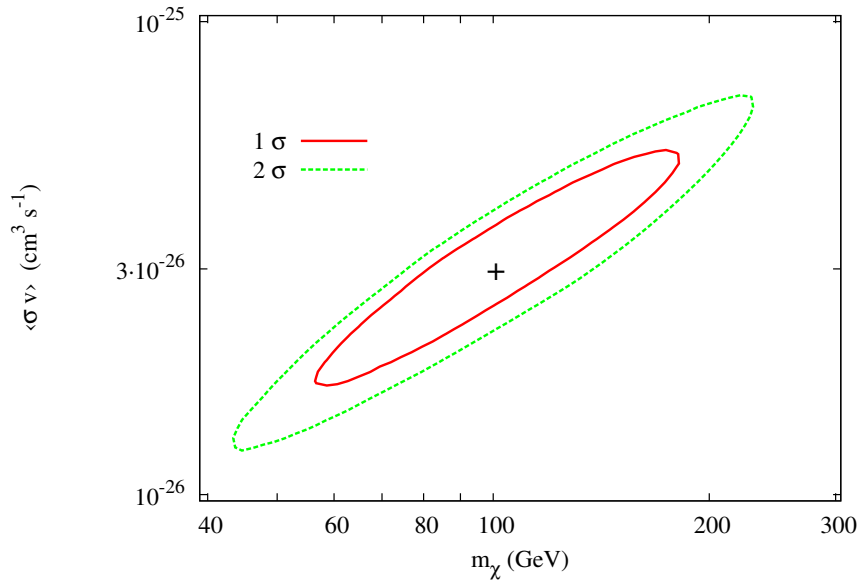}\\
\end{center}
\caption{{\footnotesize
$68\%$ and $95\%$ CL regions for a $100$ GeV WIMP with
$\langle\sigma v\rangle = 3\cdot10^{-26}$ cm$^3$ s$^{-1}$
considering a mixed final state consisting of $70\%$ $W^+ W^-$ and
$30\%$ $\tau^+ \tau^-$ and including the Branching Ratios in the statistical
treatment.}}
\label{fig:FitBRs}
\end{figure}

As a second test, we consider a mixed final state and fit the
BRs themselves along with the annihilation cross-section and the WIMP mass.
Our results can be seen in Fig.\ref{fig:FitBRs} where we have taken a 100 GeV WIMP
annihilating into a final state consisting of
$70\%$  $W^+ W^-$ and $30\%$  $\tau^+ \tau^-$. The sum of the two
branching fractions is obviously equal to $1$, so we only need to include one
further parameter in the statistical analysis.

In this case, we can clearly see that the mass resolution deteriorates
with respect to the case where a perfectly known final state is considered.
A possible explanation could be that a mixed lepton - gauge boson
final state yields a spectrum which presents neither the augmented statistics
of pure annihilation into gauge bosons (the gamma-ray yield of leptons is
significantly inferior to the one of gauge bosons) nor the characteristic hard spectral
form of annihilation into $\tau^+ \tau^-$ pairs.

\subsection{Direct vs Indirect detection}
It is interesting to remark from Figs.\ref{fig:MIdirectPlain} and \ref{fig:IndirectPlainMI},
how two completely different means of
observation, with completely different signal/background physics, are in fact
competitive (and hence complementary) in the search for the identification dark matter.

\begin{figure}[htb!]
    \begin{center}
\includegraphics[width=0.3\textwidth,clip=true,angle=-90]{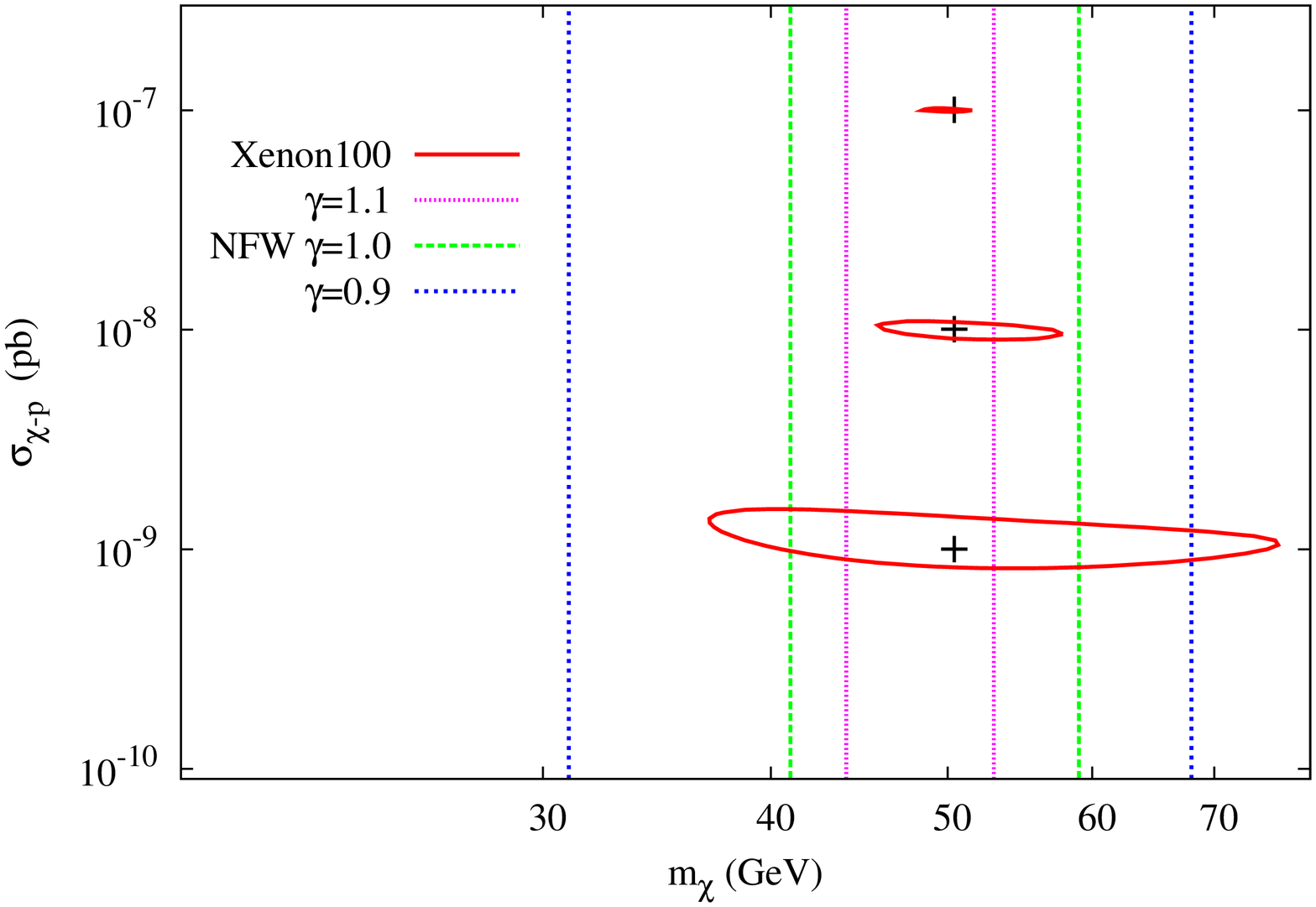}
\includegraphics[width=0.3\textwidth,clip=true,angle=-90]{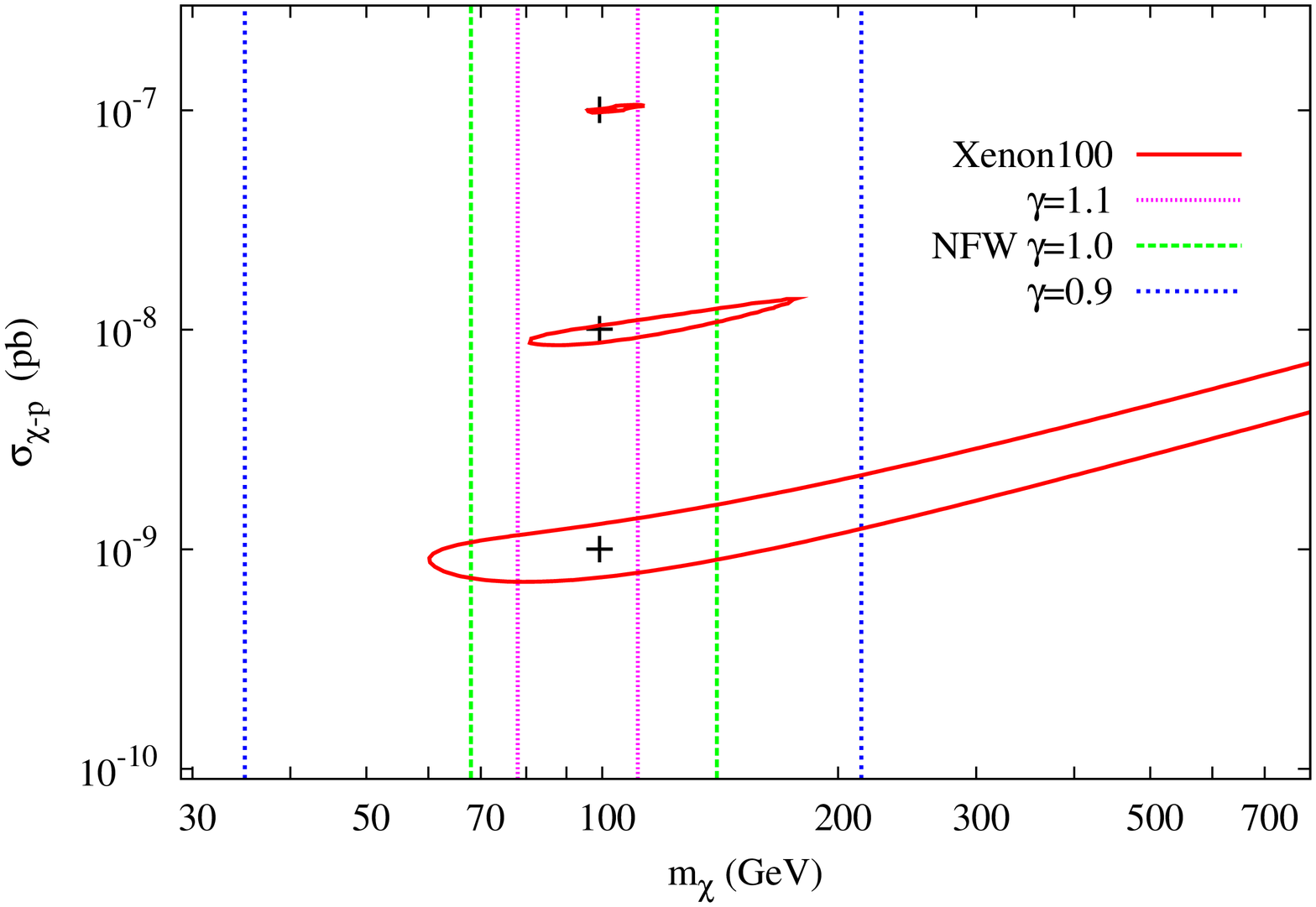}
\includegraphics[width=0.3\textwidth,clip=true,angle=-90]{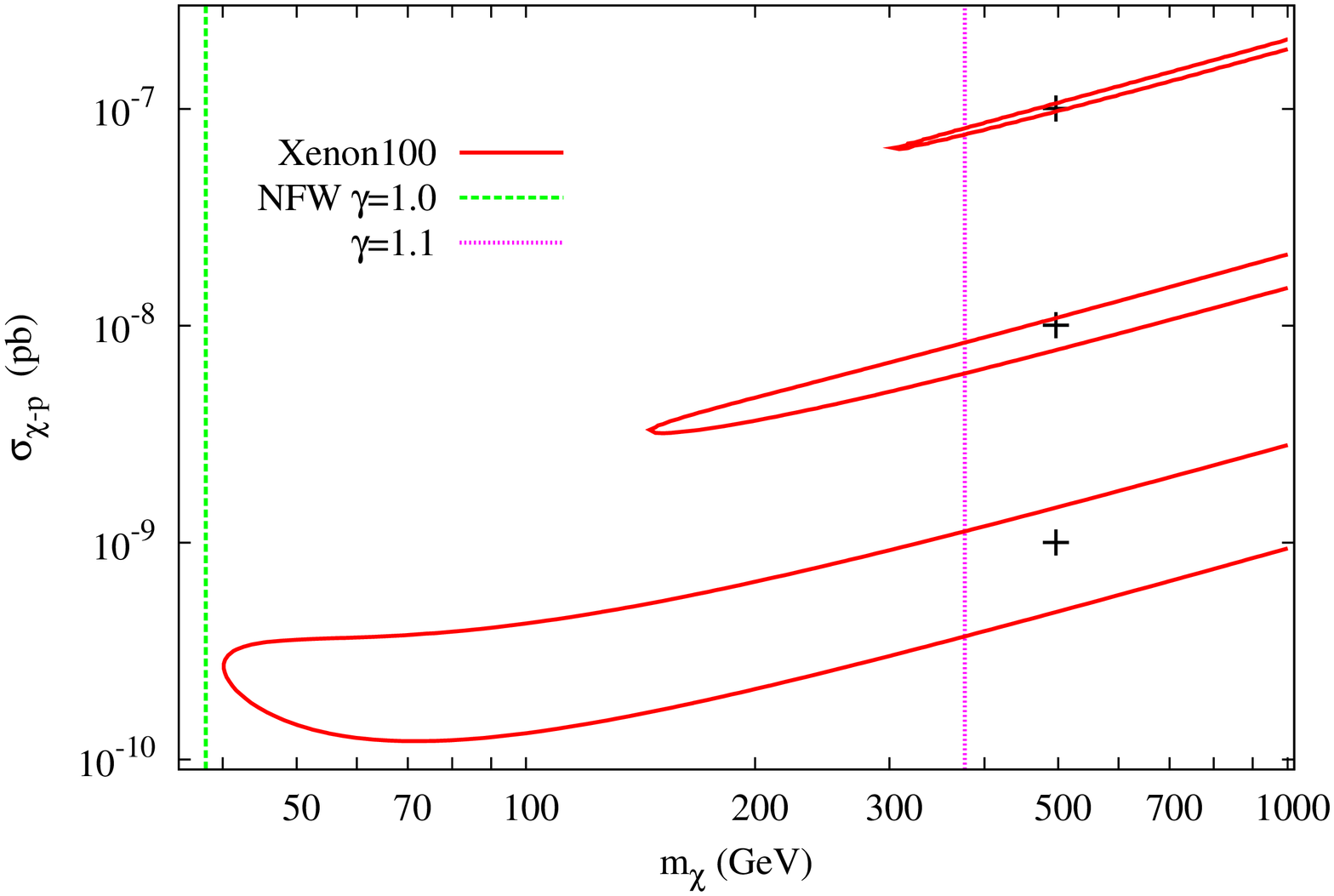}
          \caption{{\footnotesize
Comparison of the WIMP mass reconstruction 
capacity between a $100$ kg XENON experiment and the Fermi experiment
in the case of three different halo profiles (with different inner slopes $\gamma$),
at $95\%$ of confidence level, for several WIMP masses ($50$, $100$ and $500$ GeV)
and WIMP--nucleon cross-sections ($10^{-7}$, $10^{-8}$ and
$10^{-9}$ pb).
In each panel the cross denotes the input parameters.
}}
        \label{fig:comparison}
    \end{center}
\end{figure}
In Fig.\ref{fig:comparison} we compare the precision level for both experiments
as a function of the WIMP mass, for different values of the spin-independent
cross-section ($10^{-7}$, $10^{-8}$ and $10^{-9}$ pb) and for different
halo profiles.
For this treatment we minimize the impact of uncertainties discussed in
the previous paragraph, as we are mostly interested in examining the
\textit{a priori}, in some sense ``intrinsic'' sensitivity of the two detection techniques.
For example, in the case of direct detection, the necessity for minimization
of background noises and control of the noise sources has been repeatedly pointed out.
As for uncertainties entering the velocity distribution of WIMPs in the solar
neighborhood (or, why not, the form factor's functional form), these can,
\textit{in principle} be minimized by measurements exterior to the experiments
themselves.
The same holds for most uncertainties in the case of indirect detection.
As a small example, Fermi's overall sky survey capacity is hoped to contribute in
the minimization of uncertainties in non-DM annihilation sources, whereas other
observations in different energy regions can also contribute in this direction.
In this respect, for our comparative results,
we remove the extra parameters from the statistical treatment.
We see
that at $95\%$ of confidence level Fermi, after $3$ years of exposure
 will have an equivalent sensitivity to the $100$ kg XENON experiment
after $3$ years of running if $\sigma_{\chi-p}\lesssim 10^{-8}$ pb,
{\it independently of the WIMP mass}.
The indirect detection by Fermi will always be able to give an upper bound
on the WIMP mass for $m_{\chi}\sim 100$ GeV, whereas the XENON
$100$ kg experiment would only give a lower bound value if
$\sigma_{\chi-p} \lesssim 10^{-9}$ pb. In all cases, the lower
bounds given by Fermi for a NFW halo profile are similar to the
ones given by the XENON $100$ kg experiment for any WIMP mass if
$\sigma_{\chi-p} \lesssim 10^{-8}$ pb.

\begin{figure}[htb!]
\begin{center}
\hspace{-1cm}
\includegraphics[angle=-90,width=0.45\textwidth]{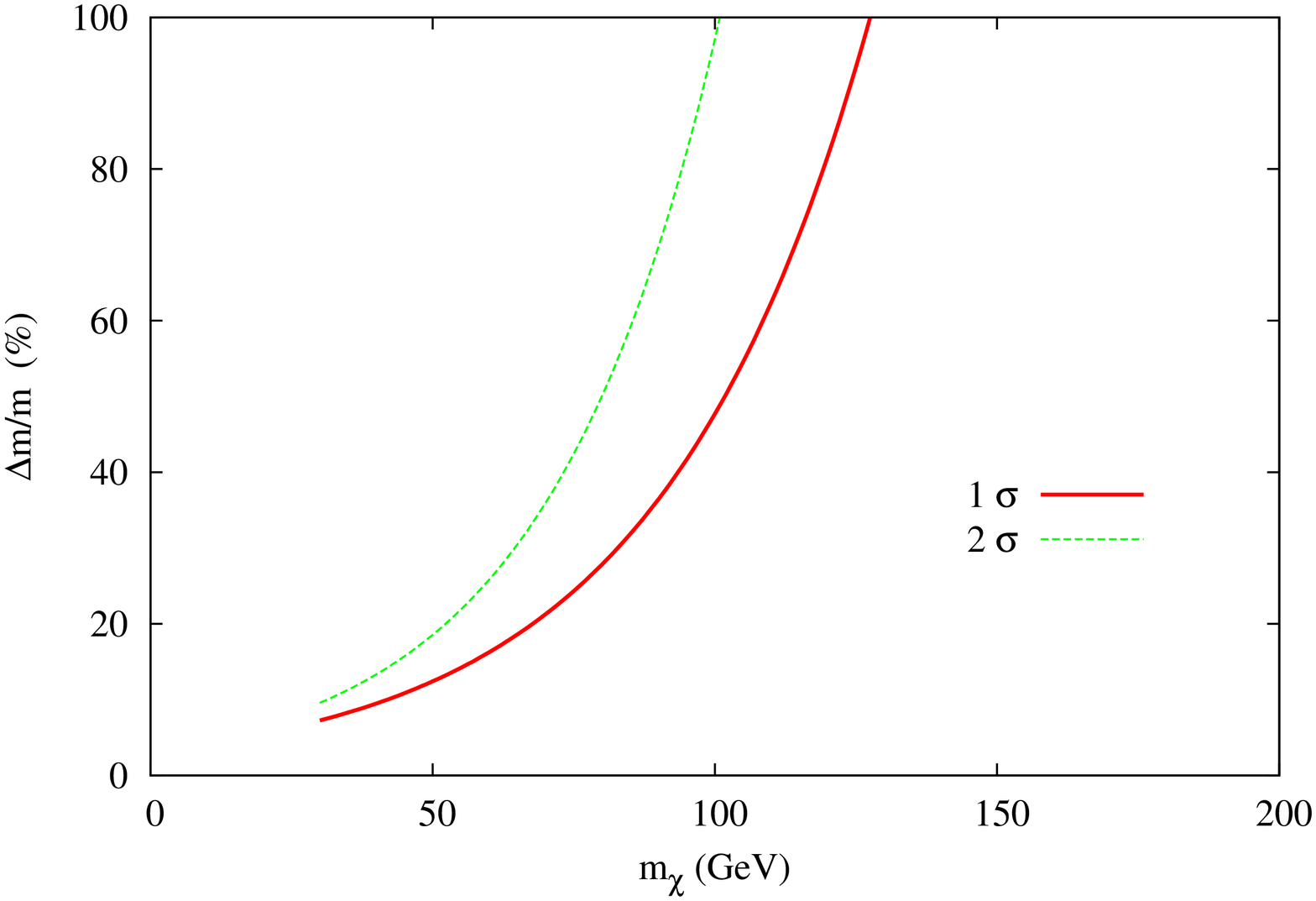} \hspace{1cm}
\includegraphics[angle=-90,width=0.45\textwidth]{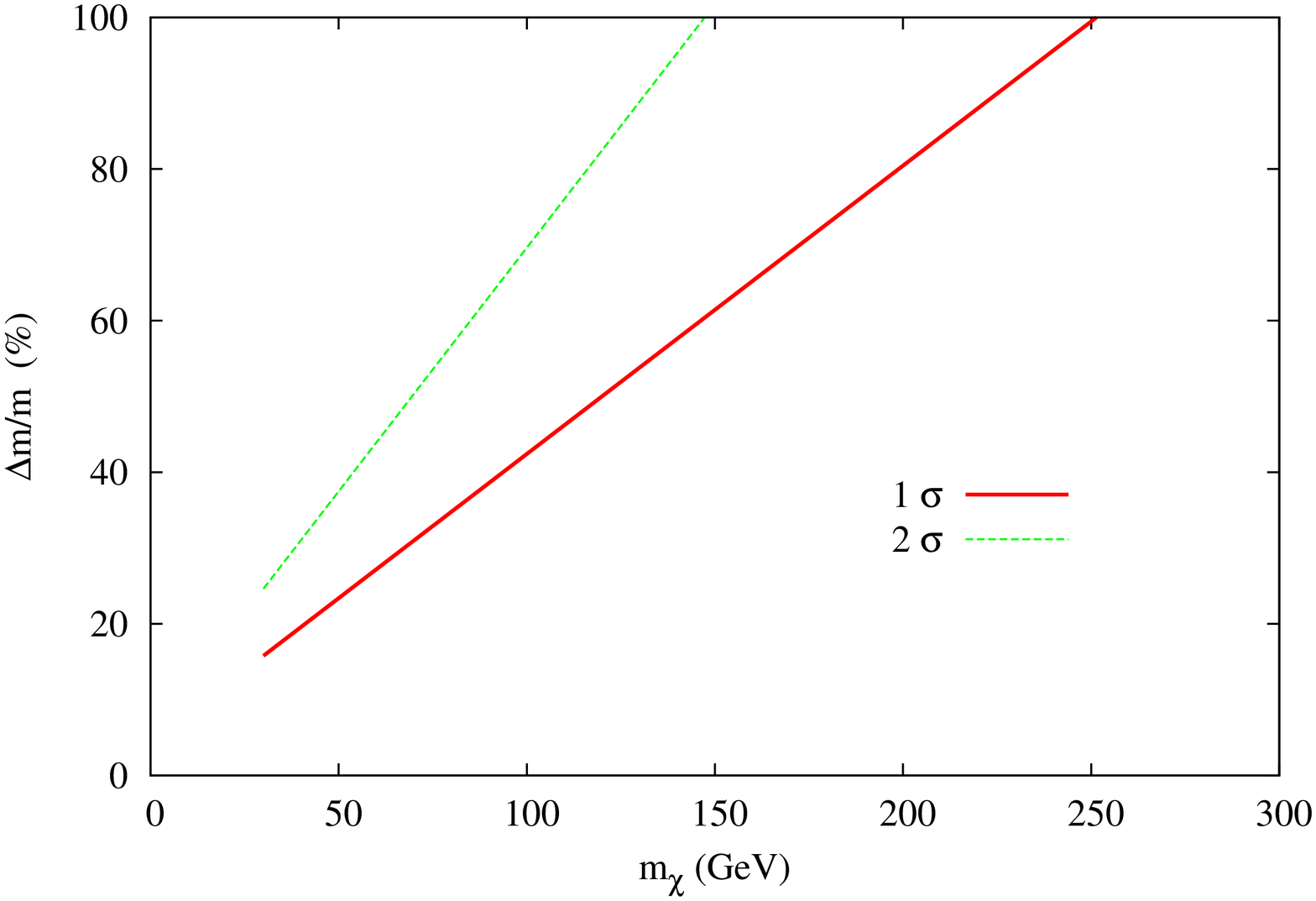}
\caption{{\footnotesize
Relative error $\Delta m_\chi/m_\chi$, at $68\%$ and $95\%$ CL, 
in the mass reconstruction for the XENON
$100$ kg experiment for $\sigma_{\chi-p}=10^{-8}$ pb (left) compared with the
Fermi experiment (right) for
$\langle\sigma v\rangle = 3 \cdot 10^{-26}$ cm$^3$s$^{-1}$ in the case of a NFW
halo profile.}}
\label{fig:DeltaM}
\end{center}
\end{figure}
To compare the uncertainties on the WIMP mass expected from direct
and indirect detection modes, we plotted in Fig.\ref{fig:DeltaM}
$\Delta m_\chi/m_\chi$
as a function of the WIMP mass for
$\sigma_{\chi-p} = 10^{-8} \mbox{pb}$ and a NFW halo profile.
One can clearly see in the figure
that
Fermi will be competitive with XENON $100$ kg to measure the WIMP mass
in the case of a NFW halo profile for $\sigma_{\chi-p} \lesssim 10^{-8}$ pb.
For heavy WIMPs, Fermi could even be more sensitive than XENON.

\subsection{WIMPs at present and future colliders}

So far we have completely neglected colliders as a potential source of information on 
dark matter properties. However, there is a quite
general agreement on the fact that despite the significant progress in
astroparticle physics experiments, which manage to impose more and more constraints on
various models, collider experiments remain an irreplaceable source of information
for particle physics. It is quite natural thus, to examine the potential of
colliders to constrain WIMP properties. We will examine the extent at which
astroparticle and collider experiments become competitive, trying at the same
time to stay as model-independent as possible.

This last point is, in fact, the major difficulty in treating collider experiments to extract astroparticle data:
most studies performed for new physics at colliders are very strongly model dependent. This is almost unavoidable
in the case of the LHC, due to the hadronic nature of the colliding particles. The fact that the initial energy of the
colliding particles/partons is not well-known, since it is determined by the parton distribution functions, renders it extremely difficult
(in fact, practically impossible) to look beyond the transverse plane. This obviously limits -up to a certain point- the
precision that could be obtained with respect to, for example, an $e^+e^-$ collider. As a result, it is quite difficult to make
predictions in a model-independent way, since a whole set of parameters must be taken into account in order to perform concrete
predictions. The cruciallity of these uncertainties will become clearer in the following.

\subsubsection{The Approach}

Recently, an approach was proposed in references \cite{Birkedal:2004xn, Birkedal:2005aa} which allows to actually
perform a model-independent study of WIMP properties at lepton colliders (such as the ILC project).
The goal we pursue is to extract constraints which are as stringent as possible for a generic dark matter candidate.
A generic WIMP can annihilate into pairs of standard model particles:
\begin{equation}
\label{annihilation}
\chi+\chi \longrightarrow X_i+\bar{X}_i\ .
\end{equation}

\noindent
However, the procedure taking place in a collider is the opposite one, having only
one species of particles in the initial state. The idea proposed in Ref. \cite{Birkedal:2004xn} is to
correlate the two processes in some way.
This can be done through the so-called ``detailed balancing'' equation, which reads:
\begin{equation}
\label{detbal}
\frac{\sigma(\chi+\chi \rightarrow X_i + \bar{X}_i)}{\sigma( X_i + \bar{X}_i \rightarrow \chi+\chi)} =
2\frac{v_X^2(2S_X+1)^2}{v_{\chi}^2(2S_{\chi}+1)^2}\ ,
\end{equation}
where $v_i$ and $S_i$ are respectively the velocity and the spin
of the particle $i$.
The cross-section $\sigma(\chi\chi \rightarrow X_i \bar{X}_i)$ is only averaged over spins.
\\ \\
The total thermally averaged WIMP annihilation cross-section can be expanded as
\begin{equation}
\label{expansion}
 \sigma_i v = \sum_{J=0}^{\infty}{\sigma_i^{(J)}v^{2J}}\ ,
\end{equation}
where $J$ is the angular momentum of each annihilation wave. Now, for low velocities, the lowest-order
non-vanishing term in the last equation will be dominant. So, we can express the total annihilation cross-section
as a sum of the partial ones over all possible final states for the dominant partial wave $J_0$ in each final state:
\begin{equation}
 \sigma_{an} = \sum_i{\sigma_i^{(J_0)}}\ .
\end{equation}

Next, we can define the ``annihilation fraction'' $\kappa_i$ into the standard model particle pair $X_i - \bar{X}_i$:
\begin{equation}
\label{anfrac}
\kappa_i = \frac{\sigma_i^{(J_0)}}{\sigma_{an}}\ .
\end{equation}

By combining Eqs. (\ref{detbal}) and (\ref{anfrac}) we can obtain the following expression for the WIMP pair-production cross-section:
\begin{equation}
\label{forwback}
\sigma(X_i \bar{X}_i \rightarrow 2\chi) =
2^{2(J_0 - 1)}\kappa_i\sigma_{an}\frac{(2S_{\chi}+1)^2}{(2S_X+1)^2}\left(1 - \frac{4M_{\chi}^{2}}{s}\right)^{1/2+J_0}\ .
\end{equation}

\noindent
A few remarks should be made about the validity of this formula:
\begin{itemize}
\item Equation (\ref{forwback}) is valid for WIMP pair-production taking place at
 center-of-mass energies \textit{just above} the pair-production threshold.
\item The detailed balancing equation is valid if and only if the process under
 consideration is characterized by time-reversal and parity invariance. It is well
 known that weak interactions violate both of them, up to some degree, which we
 ignore in this treatment.
\end{itemize}

A process of the form $X_i \bar{X}_i\longrightarrow \chi\chi$ is not visible in a
 collider, since WIMPs only manifest themselves as missing energy.
At least one detectable particle is required for the event to pass the
triggers.
An additional photon from initial state radiation (ISR) is required
to be recorded on tape: $X_i \bar{X}_i \longrightarrow \chi\chi\gamma$.
We can correlate the WIMP pair-production process to the radiative WIMP
pair-production for photons which are either soft or collinear with respect
to the colliding beams. In this case, the two processes are related
through \cite{Birkedal:2004xn}:
\begin{equation}
\label{correlation}
\frac{d\sigma(e^+ e^- \rightarrow 2\chi +\gamma)}{dx d\cos\theta} \approx {\cal{F}}(x,\cos\theta) \tilde{\sigma}(e^+e^-\rightarrow 2\chi)\ ,
\end{equation}
where $x = 2E_{\gamma}/\sqrt{s}$, $\theta$ is the angle between the photon direction and the
direction of the incoming electron beam, $\tilde{\sigma}$ is the WIMP pair-production cross-section produced at
the reduced center of mass energy $\tilde{s}=(1-x)s$, and $\cal{F}$ is defined as:
\begin{equation}
{\cal{F}}(x,\cos\theta)=\frac{\alpha}{\pi}\frac{1+(1-x)^2}{x}\frac{1}{\sin^2\theta}\ .
\end{equation}

\noindent
Now, by combining Eqs. (\ref{correlation}) and (\ref{forwback}) we get the master equation:
\begin{equation}
\label{ILCmastereq}
\frac{d\sigma}{dx d\cos\theta}(e^+e^- \rightarrow 2\chi+\gamma) \approx
\frac{\alpha\kappa_e \sigma_{an}}{16\pi} \frac{1+(1-x)^2}{x} \frac{1}{\sin^2\theta}2^{2J_0}(2S_\chi +1)^2
\left(1-\frac{4M_{\chi}^2}{(1-x)s}\right)^{1/2+J_0}\ .
\end{equation}
It would be useful here to note that although the value of $\kappa_e$ is here arbitrary, this parameter
acquires specific values in each model. As an example, in mSUGRA models $\kappa_e$ ranges roughly from
$0.2$ to $0.3$ \cite{Birkedal:2004xn}.

The problem now is that very collinear photons fall outside the reach of any detector, due to practical
limitations in the coverage of the volume around the beam pipe. Also, typically, lower cuts are included in the detected
transverse momentum of photons, $p_T = E_\gamma\sin\theta$, in order to avoid excessive background signals at low energies.
So, if we are to use this approach, we have to examine its validity outside the soft/collinear region. The accuracy
of the collinear approximation for hard photons at all angles has been discussed in the original paper \cite{Birkedal:2004xn},
with the conclusion that the approach works quite well.

However,
an important point should be taken into account here. From the previous discussion on the validity of the method, we have
to impose specific kinematic cuts on the detected photons. We consider the following conditions:
\begin{itemize}
\item We demand an overall condition $\sin\theta \geq 0.1$ and $p_T \geq 7.5$ GeV in order to assure the detectability of the photons.
\item In order to assure the fact that any photon under examination corresponds to non-relativistic WIMPs,
we demand $v_{\chi}^2 \leq 1/2$. This gives a lower kinematic cut, along with an upper cut corresponding just to
the endpoint of the photon spectrum:
\begin{equation}
\label{cuts}
\frac{\sqrt{s}}{2}\left(1-\frac{8M_{\chi}^2}{s}\right) \leq E_{\gamma} \leq \frac{\sqrt{s}}{2}\left(1-\frac{4M_{\chi}^2}{s}\right)\ .
\end{equation}
\end{itemize}

These conditions present a flaw: the energy limits depend on the mass we wish to constrain. On the other hand,
for the reasons explained before, we cannot treat the signals without imposing such kinds of cuts, if we do not
want either to misuse the method or stick to heavy WIMPs (which, for kinematic reasons, cannot be relativistic).
The only way to evade this problem is to suppose that other
dark matter detection experiments (or, eventually, the LHC in the framework of specific models) will have already
provided us with some sort of limits on the WIMP mass. In this case, having an idea of the region in which the WIMP
mass falls, we can also estimate the cuts that will safely keep us outside the relativistic region and only consider
photons within this region.

The main source of background events is the standard model radiative neutrino production,
$e^+e^-\longrightarrow \nu\bar{\nu}\gamma$. Apart from these background events, various models predict
additional signals of the form ``$\gamma$ + missing energy'', one of the most well-known examples being radiative
sneutrino production \cite{Dreiner:2006sb, Dreiner:2007qc}, predicted in the framework of several supersymmetric models.
In the spirit of staying as model-independent as possible, we will ignore all possible beyond standard model processes.

\subsubsection{Basic results for non-polarized beams}
\begin{figure}[htb!]
\begin{center}
\includegraphics[width=7cm,angle=270]{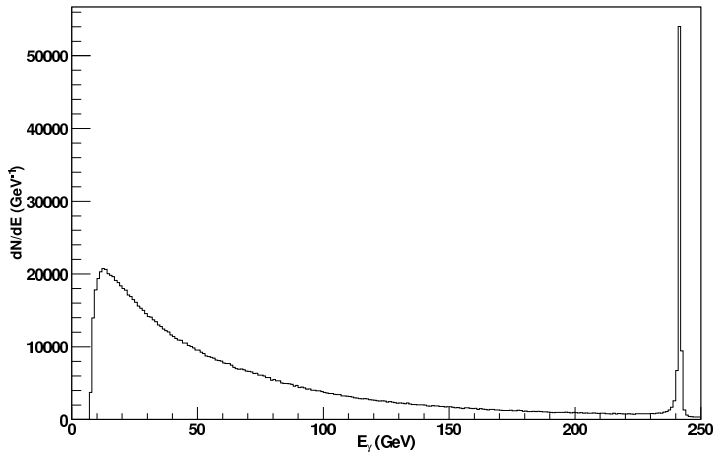}
\caption{{\footnotesize
Radiative neutrino production background $e^+e^-\rightarrow\nu\bar{\nu}\gamma$ for the ILC, for an unpolarized initial state.}}
\label{Fig.1}
\end{center}
\end{figure}
We place ourselves in the framework of the ILC
 project with a center-of-mass energy of $\sqrt{s} = 500$ GeV and an integrated
 luminosity of $500$ fb$^{-1}$. In order to estimate the background events, we
used the CalcHEP code \cite{Pukhov:1999gg, Pukhov:2004ca} to generate
$1.242.500$ $e^+e^-\longrightarrow \nu\bar{\nu}\gamma$ events, corresponding to
 the aforementioned conditions. The total radiative neutrino production
background can be seen in Fig.\ref{Fig.1}. The peak at
$E_\gamma=\sqrt{s}/2\cdot(1-M_Z^2/s)\simeq241.7$ GeV corresponds to the
radiative returns to the $Z$ resonance.

We generate a predicted ``observable'' spectrum for given values
of the WIMP mass and the annihilation fraction. During this study, we do not proceed to a (more realistic) full
detector simulation, as done for example in Ref. \cite{Bartels:2007cv}, but stick to prediction levels in order
to perform as thorough a scan as possible in the $(m_\chi,\kappa_e)$ parameter space and to have a picture of the
``a priori'' potential of the method.

\begin{figure}[htb!]
\begin{center}
\includegraphics[width=7cm,angle=270]{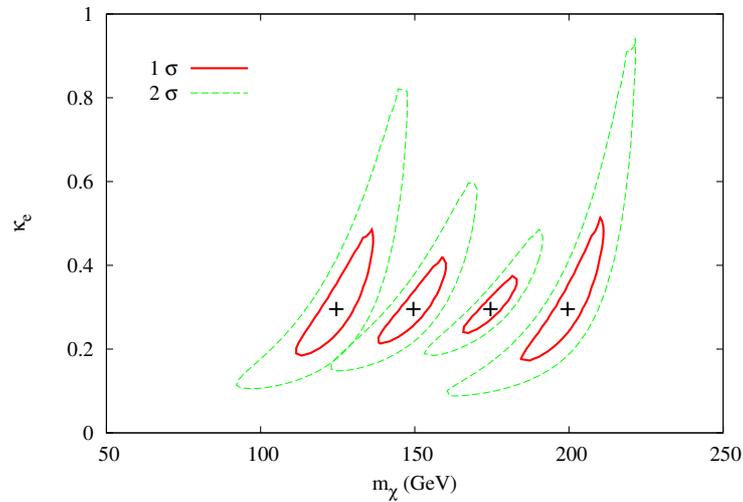}
\caption{{\footnotesize
Distribution of the maximum likelihood WIMP mass and annihilation fraction for the ILC in the  $(m_{\chi},\kappa_e)$ plane,
for $\kappa_e = 0.3$ and $m_\chi = 125, 150, 175$ and $200$ GeV. The inner (full lines) and outer (dashed lines) represent
the $68\%$  and $95\%$ CL region respectively.}}
\label{Fig.2}
\end{center}
\end{figure}

Figure \ref{Fig.2} shows the predicted ability of the ILC to determine WIMP masses and annihilation
fractions for four points in the $(m_\chi, \kappa_e)$ parameter space for a $68\%$ and $95\%$ Confidence Level.
These results concern WIMPs with spin $S_\chi = 1/2$ and an angular momentum $J_0 = 1$ which corresponds to an
annihilation cross-section $\sigma_{an} = 7$ pb \cite{Birkedal:2004xn}. As can be seen, we are able to constrain quite
significantly the WIMP mass ($20\% - 40\%$ precision), while constraints on $\kappa_e$ are weaker.

\begin{figure}[htb!]
\begin{center}
\includegraphics[width=7cm,angle=270]{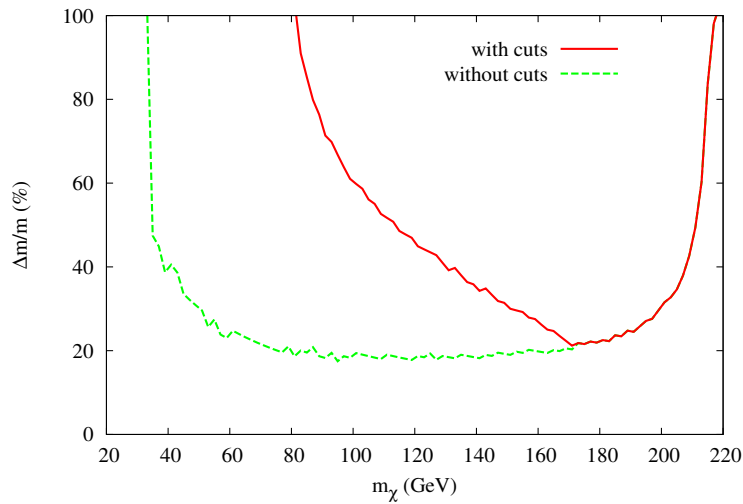}
\caption{{\footnotesize
Relative error in a generic WIMP mass determination, for $\kappa_e =0.3$ and at a 
95\% confidence level.
The solid line corresponds to the results obtained after imposing the proper kinematic cuts, whereas the dashed
line to the case where we do not take these limits in consideration.}}
\label{Fig.3}
\end{center}
\end{figure}
Figure \ref{Fig.3} shows the relative error ($\Delta m_\chi/m_\chi$) for the mass
 reconstruction as a function of $m_\chi$, for $\kappa_e = 0.3$ and a $95\%$
 confidence level.  The solid line corresponds to the proper treatment including
 kinematic cuts. For indicative reasons, we also show the abused results obtained
 if we do not impose kinematic cuts on the photon energy (dashed line). The
 amelioration of the method's efficiency is obvious, although this is after all a
 false fact, since we include regions in which the approach is not valid. Above
$m_\chi \simeq 175$ GeV the two lines become identical, since the WIMPs cannot be
 relativistic. The capacity of the method peaks significantly for masses of the
 order of $175$ GeV because around this value we reach an optimal
combination of phase space volume and available spectrum that passes the kinematic cuts. As we
move away from this value the accuracy tends to fall.

Let us make a final remark on the possibility of adopting a similar approach in the case of the LHC.
As we argued before, the large
uncertainty in the collision energy affects significantly the precision of the whole procedure
(which is, already, based on approximations). Formally, in order to perform such a study for the LHC,
the computed cross-sections must be convoluted with the proton parton distribution functions. As an additional element,
the photon background in the LHC is expected to be much greater than in the ILC. The possibility of determining
WIMP properties through a model-independent method at the LHC has been addressed to in Ref. \cite{Feng:2005gj},
where the authors conclude that WIMP detection will be extremely difficult, if even possible.

\subsubsection{Polarized beams}
\begin{figure}[htb!]
\begin{center}
\includegraphics[width=7cm,angle=270]{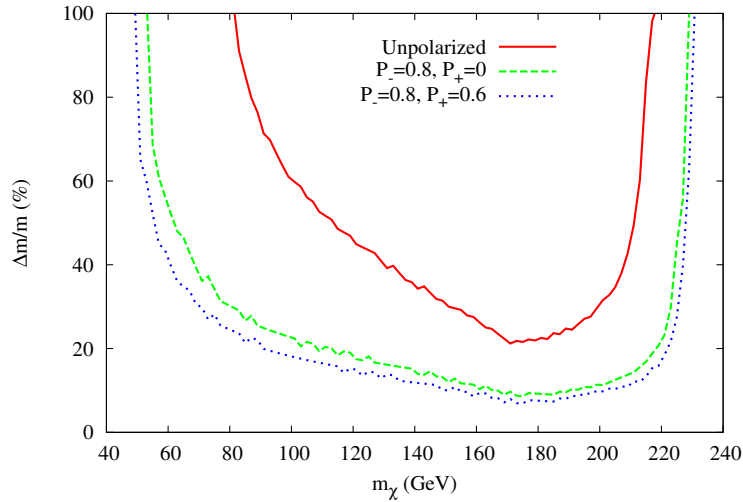}
\caption{{\footnotesize
Relative error in a generic WIMP mass determination, for three cases of beam polarization, including all proper kinematic cuts.}}
\label{Fig.4}
\end{center}
\end{figure}
The reach of the ILC can be further increased by polarizing the beams.
For polarized beams, the signal cannot be fully characterized by $\kappa_e$; instead, four independent annihilation
fractions are needed, corresponding to the four possible $e^+e^-$ helicity configurations.
\\ \\
To apply Eq. (\ref{ILCmastereq}) to this case, we make the replacement:
\begin{eqnarray}
\kappa_e&\rightarrow&\frac{1}{4}(1+P_-)\,\left[(1+P_+)\,\kappa(e_-^Re_+^L)+(1-P_+)\,\kappa(e_-^Re_+^R)\right]\nonumber\\
&+&\frac{1}{4}(1-P_-)\,\left[(1+P_+)\,\kappa(e_-^Le_+^L)+(1-P_+)\,\kappa(e_-^Le_+^R)\right]\ ,
\end{eqnarray}
where $P_\pm$ are the polarizations of the positron and the electron beams.
As in ref \cite{Birkedal:2004xn,Bartels:2007cv}, let us assume that the WIMP couplings to electrons conserve both
helicity and parity: $\kappa(e_-^Re_+^L)=\kappa(e_-^Le_+^R)=2\,\kappa_e$ and $\kappa(e_-^Re_+^R)=\kappa(e_-^Le_+^L)=0$.

In Fig.\ref{Fig.4} we show the relative error for the mass reconstruction for $\kappa_e=0.3$ and $95\%$ confidence level,
for the unpolarized scenario and for two different polarizations: $(P_-,P_+)=(0.8,0)$ and $(0.8,0.6)$.

%%%%%%%%%%%%%%%%%%%%%%%%%%%%%%%%%%%%%%%%%%%%%%%%%%%%%%%%%%%%%%%%%%%%%%%%%%%%%%%%%%%%%%%%%%%%%%%%%%%%%%%%%%%%%%%%%%%%%%%%%%
\subsection{Complementarity and Conclusions}

\begin{figure}[htb!]
\begin{center}
\includegraphics[width = 7cm, angle=270]{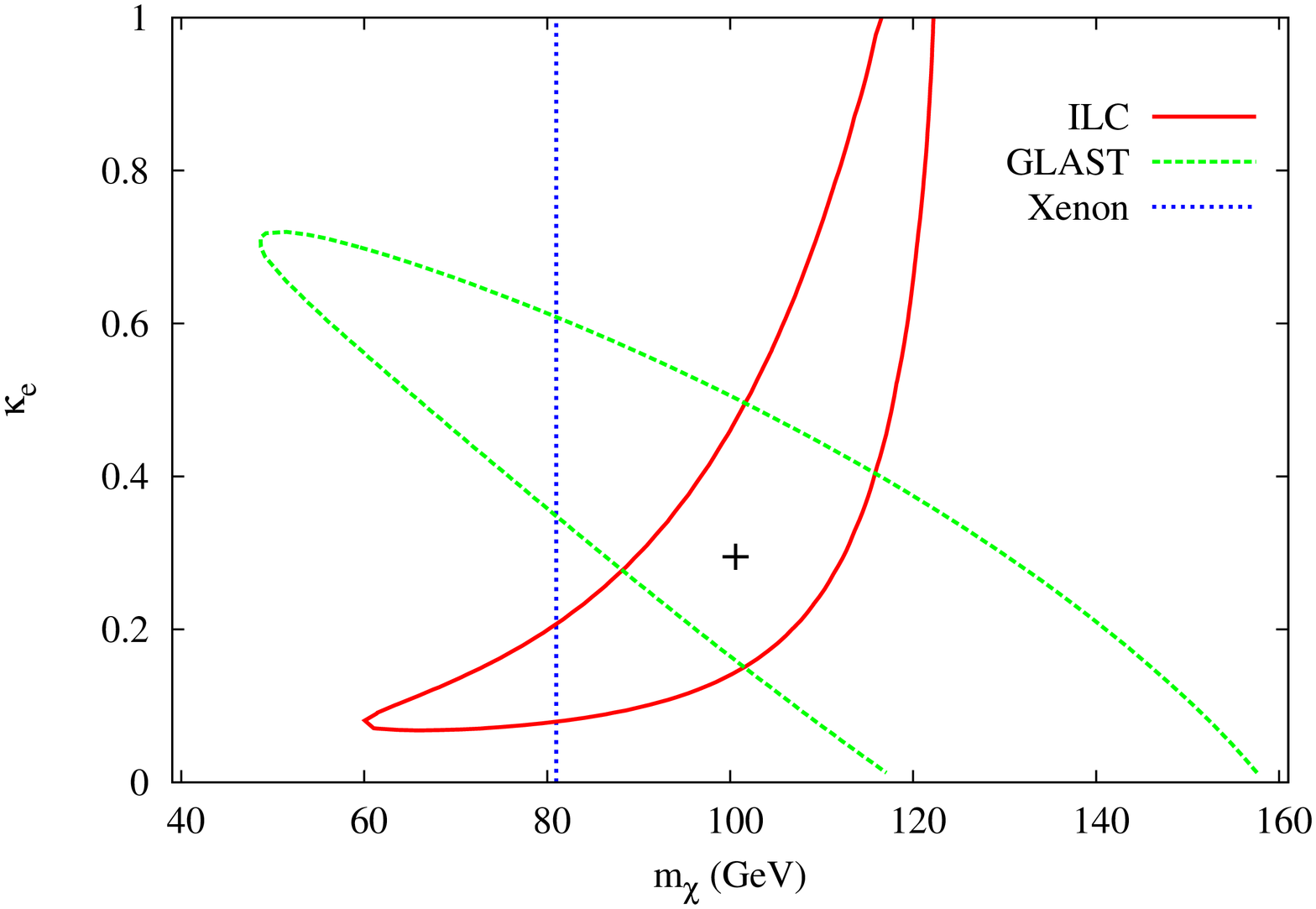}
\includegraphics[width = 7cm, angle=270]{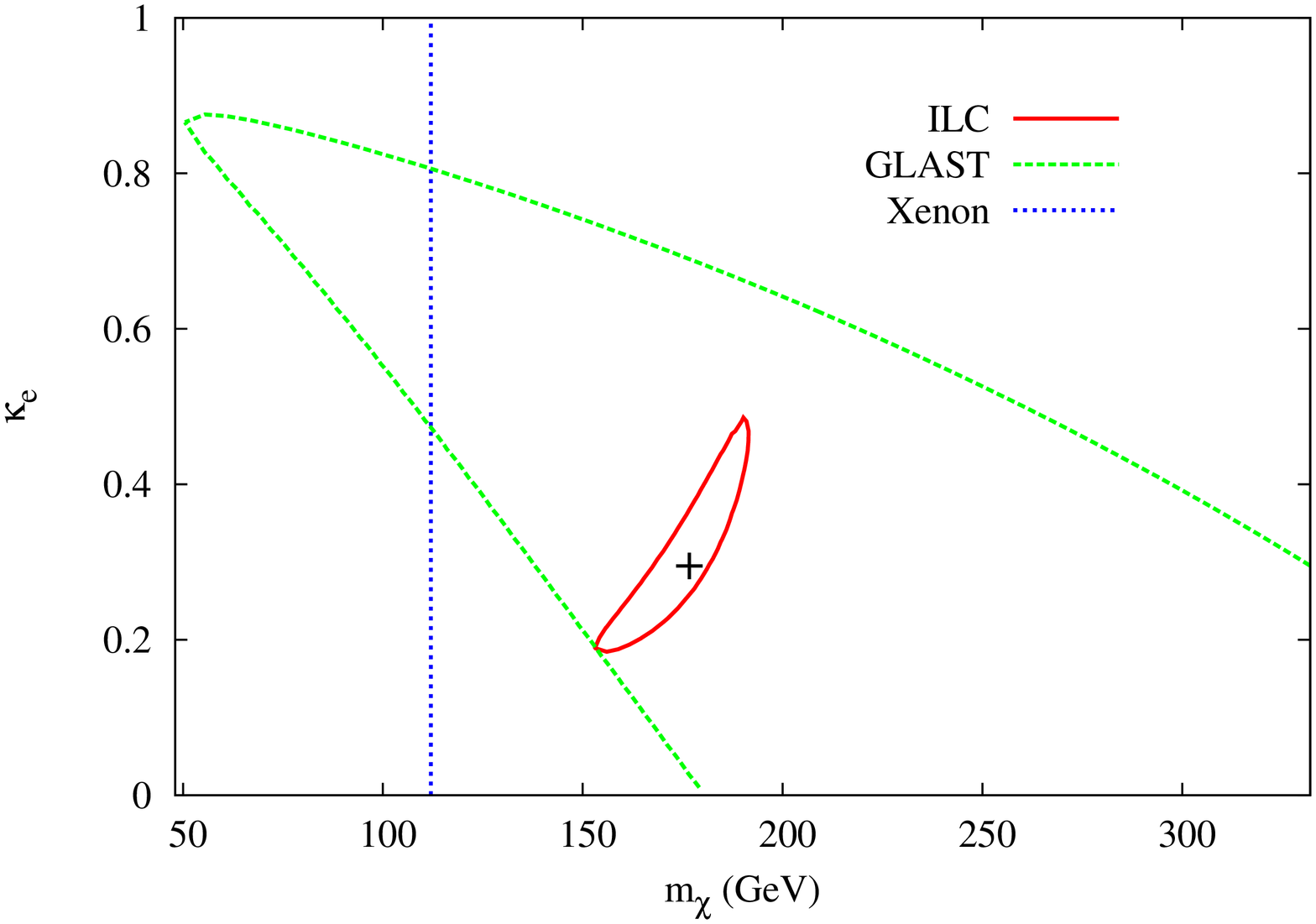}
\caption{{\footnotesize
Comparison between a $100$ kg XENON experiment (dotted line) with
$\sigma_{\chi-p}=10^{-8}$ pb, Fermi (dashed line) in the case of an NFW halo
 profile with $\langle\sigma v\rangle=3\cdot10^{-26}$ cm$^3$s$^{-1}$, and
 unpolarized ILC sensitivity (solid line) at $95\%$ of confidence level, for
 different WIMP masses $m_\chi=100$ and $175$ GeV, and $\kappa_e=0.3$.}}
\label{complementabananes}
\end{center}
\end{figure}
In Fig.\ref{complementabananes} we compare the precision levels for direct and
 indirect detection experiments, along with the corresponding results of the
method we followed for the ILC for two cases of WIMPs masses,
 $m_\chi = 100$ GeV and $175$ GeV, and $\kappa_e = 0.3$. We plot the
results in the $(m_\chi, \kappa_e)$ plane. This is done as the $\kappa_e$
parameter entering the ILC treatment presented before is, in fact,
the same parameter as the corresponding
branching ratio
$Br_i= \frac{\langle \sigma_i v \rangle}{\langle \sigma v \rangle}$
appearing in Eq. (\ref{MasterGammas}) for $i=e$.

The blue-dotted line corresponds to a $100$ kg XENON experiment, where the
WIMP-nucleus cross-section has been assumed to be $10^{-8}$ pb. The
green-dashed line
 depicts the results for a Fermi-like experiment assuming a NFW halo profile.
The total annihilation cross-section into standard model particles has been
taken to be $\langle\sigma v\rangle=3\cdot10^{-26}$ cm$^3$s$^{-1}$. The red-plain
line represents our results for an ILC-like collider, with non-polarized beams.
All the results are plotted for a $95\%$ confidence level.

We can see that for different regions of the WIMP mass, the three kinds of
 experiments that we have
used as prototypes can act in a highly complementary way. For example, in
the case of a 100 GeV WIMP, indirect detection
or an ILC-like experiment alone can provide us with limited precision
both for the WIMP mass (of the order of $60\%$) and the $\kappa_e$ parameter
 (where the results are even worse). Combined measurements can dramatically
 increase the precision, reaching an accuracy of $25\%$ in mass. If we
additionally include direct detection measurements,
we can further increase the precision.

In the case of a $175$ GeV WIMP, a point where the unpolarized ILC sensitivity peaks, we see that the
dominant information comes from this source. Nevertheless, even if we only combine direct and indirect detection
experiments, we see that we can, in fact, acquire non-negligible constraints on the dark matter candidate mass.

\begin{center}
\begin{table}
\centering
\begin{tabular}{|c|c|c|c|}
\hline
$m_{\chi}$ & XENON & Fermi & ILC   \\
\hline
 $50$ GeV & $-5/+7$   GeV & $\pm 12$ GeV   & $-$ \\
$100$ GeV & $-19/+75$ GeV & $-50/+60$ GeV  & $-40/+20$ GeV\\
$175$ GeV & $-65/$    GeV & $-125$ GeV & $-20/+15$ GeV\\
$500$ GeV & $-$           & $-$            & $-$ \\
\hline
\end{tabular}
\caption{{\footnotesize Precision on a WIMP mass expected from the
different experiments at a $95\%$ CL after 3 years of exposure,
$\sigma_{\chi-p}=10^{-8}$ pb a NFW profile and a
500 GeV unpolarized linear collider with an integrated luminosity of 500$\mbox{fb}^{-1}$}}
\label{Tab:summary}
\end{table}
\end{center}
To summarize the analysis, we show in Table \ref{Tab:summary} the precision expected for
several interesting dark matter masses.
Whereas a light WIMP (50 GeV) can be reached by both types of dark matter experiments
with a relatively high level of precision,
our analysis fails in the ILC case because of the relativistic nature
of the WIMP. On the contrary, the ILC would be particularly
efficient to discover and measure a WIMP with a mass of about 175 GeV.
Concerning a 500 GeV WIMP, which is kinematically unreachable at
the linear collider, it would be difficult to be observed by Fermi
or XENON. Only a lower bound could
be determined experimentally.

We have discussed the possibility of identifying WIMP properties in a model-independent way. For that we have
considered direct and indirect searches, and in particular the interesting cases of a XENON $100$ kg 
experiment and the Fermi satellite.
We have shown that
whereas direct detection experiments will probe efficiently
light WIMPs given a positive detection (at the $10\%$ level for
$m_\chi\lesssim 50$ GeV), Fermi will be able to confirm
and even increase the precision in the case of NFW profile, for a
WIMP-nucleon cross-section $\sigma_{\chi-p} \lesssim 10^{-8}$ pb.

Moreover, both XENON and Fermi are in principle complementary with a future ILC project,
and the measurements from the three experiments will be able to
increase significantly the precision that we can reach on the mass of
the WIMP.
\\ \\
In this chapter we introduced the basic formalism we shall be using in the following in order
to compute event rates and examine the detectability of various dark matter candidates 
in a series of experiments. We also presented an (somehow outdated we should say) analysis
of the capacity of three major detection modes to constrain dark matter properties.
But, so far, we have said very few on specific models that try to explain dark matter.
This will be the topic of the following chapters.

\newpage
\chapter{A minimal solution to the Dark Matter problem}

In the first chapter we explained why the Standard Model is unable to answer the Dark Matter
question. In short, it does not contain any electrically and color neutral particle that can
be produced non-relativistically. Therefore, if one wishes to try and explain the dark matter
problem, it is unavoidable to look for candidates in extensions of the Standard Model of
particle physics. In this chapter, we describe one of the simplest such extensions that
have been proposed in the literature, the singlet scalar model of dark matter.

\section{Some introductory remarks}
There is a plethora of theories beyond the standard model. They are often motivated by some 
experimental discrepancy with respect to the theory (dark matter, neutrino masses, LEP-2 excess).
In other cases, they try to resolve some problems appearing in the standard model itself from
a theoretical (someone could say almost aesthetic) point of view. Finally, some extensions are
even manifestly motivated by curiosity.

Although a full listing is quite difficult, we could cover a large variety of these extensions 
by categorizing them within three or four large classes: 
\begin{itemize}
 \item One can extend the particle content and/or
the gauge group. A simple example in this direction will be described in the present chapter.
 \item One can add spacetime dimensions which usually amount to new particles and
interactions in the four-dimensional spacetime. These theories, usually called extra-dimensional ones
shall not be dealt with in this work, despite their significant interest. Some notable references in
this class of models are 
\cite{Kolb:1983fm, Servant:2002aq, Cheng:2002iz, Servant:2002hb, Bergstrom:2004cy, Bergstrom:2004nr, Bergstrom:2006hk}.
 \item Finally, one can think of extending the Poincar\'e
symmetry characterizing current quantum field theories. This is the example of supersymmetry, that 
we shall examine in the next chapter.
\end{itemize}
It is interesting to note, as an introductory remark, that several among these approaches (actually,
all of them!) were not initially introduced in relation to the dark matter issue. The fact that they
can actually provide us with viable candidates is often an unexpected as well as impressive fact.

So far, we have implicitly made an assumption which obviously imposes a huge restriction to the potential
number of models that could solve the dark matter problem: that dark matter actually consists of a single
particle. This is a very common simplification. Although multi-candidate models \textit{do} exist, most
approaches are interested in finding exactly \textit{one} candidate in each model. A first reason is that
obviously it is easier to deal with a single particle rather than a certain number of them. Secondly, our
description of the so-called ``WIMP miracle'' presents us with an intriguing possibility: solving the
Boltzmann equation for a single particle can, indeed, amount quite naturally to searching for candidates
in an -in principle- experimentally reachable energy scale which can account for the entire quantity of
dark matter in our universe. Were we to suppose a large number of components, we would practically have
all the freedom in the world to introduce as many candidates as we want with couplings and masses in 
essentially all energy scales. Finally, we should say (and this shall be explicited in the following)
that as it turns out, finding a stable or quasi-stable particle in a model is not that easy a task, 
especially if we wish to couple it to some Standard Model sector in a non-purely gravitational 
manner. For example, if we were to consider a
candidate coupled to two lighter fermions, the candidate could easily decay, spoiling the picture of 
thermal relics. The stability or quasi-stability of a dark matter candidate is practically always
imposed by some discrete symmetry, rendering the lightest BSM particle (i.e. \textit{exactly one}) stable.

In the following, hence, we shall see that within all models we examine there is one and only one dark
matter candidate which should account for the entire amount of dark matter in our universe.
This has important repercussions, since in the case of multi-component dark matter, each individual
component need only respect the upper WMAP bound. On the contrary, in the case of single-component
dark matter, both the upper and the lower bound must be taken into account. So, the self-annihilation
cross-section of the candidate is not only bound from below but also from above.

\section{The singlet scalar extension of the Standard Model}
Among the three possibilities for extensions to the standard model we presented previously, the first is perhaps
the most straightforward one. If we leave the SM gauge group intact, we can add some more particle fields to the 
already existing particle content transforming according to our desired representations of $SU(3)_C\times SU(2)_L\times U(1)_Y$ 
and examine if they can constitute viable dark matter candidates. Evidently, describing the process as such is
an enormous oversimplification, since a whole number of experimental or theoretical constraints should
be respected. We shall see such examples in the paragraphs to follow.

Perhaps the simplest choice that can be made is the addition of a scalar field, being completely neutral under
the entire gauge group, but acquiring part of its mass through the usual Higgs mechanism. The field is hence only coupled
to the Standard Model through its interaction with the Higgs field. Although such a model had previously been 
considered in the literature, it was first introduced as a potential solution to the dark matter problem in
$1994$ \cite{McDonald:1993ex}. If we stick to the case of a single scalar singlet (in the original paper
the possibility for a larger number is also considered) coupled to the standard model only through the Higgs sector, 
the most general renormalizable tree-level Lagrangian that can be written is
\begin{equation}
 {\cal{L}} = {\cal{L}}_{SM} + {\cal{L}}_{S}
\end{equation}
where ${\cal{L}}_{SM}$ is the usual standard model Lagrangian and ${\cal{L}}_{S}$ is the Lagrangian involving the 
singlet field $S$ :
\begin{equation}
 {\cal{L}}_{S} = 
\frac{1}{2} \partial_\mu S \partial^\mu S - 
\frac{1}{2} m_0^2 S^2 - 
\lambda_1 S H^\dag H - 
\lambda S^2 H^\dag H - 
\frac{\lambda_3}{3} S^3 - 
\frac{\lambda_S}{4} S^4
\label{ScalarLagrangianGeneral}
\end{equation}
where for the moment, $S$ can be a complex field. In its present form, this Lagrangian does not guarantee the
stability of $S$: if for example the scalar were to be heavier than twice the Higgs boson mass, it could easily decay into
a Higgs pair due to the term $\sim S H^\dag H$ that induces a singlet-Higgs-Higgs vertex. 
To assure stability, we further impose a $Z_2$ symmetry (i.e. $S \rightarrow -S$) under which the singlet
is odd whereas all other particles are even. Referring to Eq.\eqref{ScalarLagrangianGeneral}, doing so eliminates the
terms proportional to $\lambda_3$ and $\lambda_1$. Furthermore, we make a simplifying assumption, that the $S$
field is real. We note that a linear term is forbidden by the $Z_2$ symmetry, but would be permitted in the original 
Lagrangian. Even if the $Z_2$ symmetry were absent though, such a term could be eliminated by a 
redefinition of the vacuum energy.
\\ \\
The previous assumptions amount to a Lagrangian
\begin{equation}
 {\cal{L}}_{S} = 
\frac{1}{2} \partial_\mu S \partial^\mu S - 
\frac{m_0^2}{2} S^2 - 
\frac{\lambda_S}{4} S^4 -
\lambda S^2 H^\dag H
\label{ScalarLagrangianFinal}
\end{equation}
The model introduces, hence, three new parameters: the scalar mass $m_0$, the scalar's self-coupling $\lambda_S$ and the
quartic coupling of the scalar to the Higgs field $\lambda$. The first two are parameters ``internal'' to the pure
$S$ sector. The third one determines the coupling strength of $S$ to the visible sector, through its 
interaction with the Higgs. We note that since $\lambda_S$ does by no means affect the visible sector,
it is difficult to directly constrain its value.
\\ \\
When the Higgs field acquires a non-zero vacuum expectation value, the scalar mass receives contributions from
terms of the form $\lambda v^2 S^2$ which shifts the actual tree-level mass to 
\begin{equation}
 m_S^2 = m_0^2 + \lambda v^2 \ , \ \
\label{ScalarMassTree}
\end{equation}
from which we can clearly see that we can change the free parameter basis from $(m_0, \lambda, \lambda_S)$
to $(m_S, \lambda, \lambda_S)$. At tree-level, the $\lambda_S$ parameter does not contribute to dark matter
phenomenology and is therefore an irrelevant parameter for most dark matter - related analyses. Taking into account the fact that
the Higgs boson has not yet been discovered, there are thus three free parameters of interest overall, namely
$(m_S, \lambda, m_h)$ with the latter being the Higgs mass, which is nevertheless bound by LEP and LEP-2 measurements, 
as well as theoretical constraints, having a minimal allowed value of $114.4$ GeV. If we further demand that at
the true vacuum of the theory $<S> = 0$, we can avoid mixing effects between the scalar and the Higgs boson.
The singlet's mass has \textit{a priori} no reason to lie in the electroweak scale: in fact, being a scalar
that is not protected by some symmetry (as is the case of gauge symmetry for gauge bosons and chiral symmetry for
fermions), its mass receives loop contributions depending quadratically on the theory's cutoff scale $\Lambda$ which can 
push the mass up to the Planck scale. However, for naturalness reasons, it is reasonable to expect 
that $m_S$ should fall roughly withing the electroweak scale.

Despite its simplicity, it turns out that this simple extension of the Standard Model can provide 
an interesting phenomenology and has been examined in the literature to quite some extent. Before going on to
discuss various constraints and phenomenological issues related to the model, we just note that even more
interesting effects can be introduced if one departures from the singlet scalar case and introduces, for 
example, a scalar doublet 
\cite{Majumdar:2006nt, LopezHonorez:2006gr, LopezHonorez:2007wm, Tytgat:2007cv, Lundstrom:2008ai, 
Agrawal:2008xz, Andreas:2009hj, Nezri:2009jd, Arina:2009um, Dolle:2009ft, Arina:2010zz, Honorez:2010re, 
Miao:2010rg, Krawczyk:2008zz, Gustafsson:2007pc}
or a fermion multiplet
\cite{Cirelli:2005uq, Cirelli:2007xd, Cirelli:2008id, Cirelli:2008jk, Cirelli:2009uv}
. But this is probably the truly simplest extension 
imaginable within the framework of renormalizable four - dimensional theories.

\section{Constraints and collider phenomenology}
Although this might seem as a rather simplistic approach, we can divide the constraints on the model
into two main categories, theoretical and experimental. Theoretical constraints can come from requirements
such as vacuum stability, perturbativity, unitarity and so on. Experimental constraints can come mainly
from precision electroweak measurements and direct searches at lepton colliders, or the requirement that
the model reproduces the correct dark matter relic density.

\subsection{Theoretical constraints}
Three types of theoretical constraints have been so far discussed in the literature: unitarity, perturbativity and
stability of the EWSB vacuum. In all cases, the authors stick to the case of a \textit{real} scalar field.

Perturbative unitarity constraints have been examined in \cite{Cynolter:2004cq}. Unfortunately, the authors 
conclude that their analysis does not restrict the scalar mass $m_S$. The main bounds that can be obtained
is that the Higgs mechanism contribution to the scalar mass cannot be larger than $900$ GeV and that for
$m_h \in (114.5, 251)$ GeV and $\lambda \in (0.21, 0.97)$ GeV the model parameters must satisfy
the relation 
\begin{equation}
 6 \lambda_S + \frac{4 \lambda^2}{8 \pi} \lesssim 8 \pi \ . \ \
\label{ScalarUnitarity}
\end{equation}

Vacuum stability and perturbativity constraints in their turn have been studied in detail in \cite{Gonderinger:2009jp}.
In this paper, the authors perform an analysis of the one-loop effective potential and find that not only one can
extract limits on the dark matter - related quantities, but interesting bounds can even be found on the value
of the scalar self-coupling especially when the relic density constraint is taken into account in combination
with others. The effective potential at $1$-loop order is calculated and its scale invariance imposes the introduction
of running parameters for the couplings and masses as usually in renormalization procedures. Due to the structure of
the model, the addition of the scalar does not contribute to the RGE evolution of the usual SM parameters other than
those that enter the scalar potential, which can be written in its turn as
\begin{equation}
 V_{eff}(h, S) = V_{tree}(h,S) + V_{1-loop}(h,S)
\label{EffectiveScalar}
\end{equation}
where the tree-level and 1-loop pieces of the scalar potential are, in our notations, 
\begin{equation}
 V_{tree}(h,S) = 
\mu^2 H^\dag H + 
\lambda_h (H^\dag H)^2 +
\frac{m_0^2}{2} S^2 - 
\frac{\lambda_S}{4} S^4 -
\lambda S^2 H^\dag H
\label{ScalarTreePotential}
\end{equation}
and
\begin{equation}
 V_{1-loop}(h,S) = 
\sum_j \frac{n_j}{64\pi^2} m_j^4(h,S)
\left[ \log \left( \frac{m_j^2(h,S)}{\mu_{r}^2} \right) - c_j \right]
\label{ScalarOneLoopPotential}
\end{equation}
where $h, S$ denote the classical fields defined as functional derivatives with respect to external sources of the
generating functional that only generates connected Green's functions and $\mu_{r}$ is the renormalization
scale. We should note a slight change in notation with respect to the first chapter, where the initial Higgs doublet was
denoted as $\Phi$ and the quartic Higgs coupling as $\lambda$, a symbol which at this point we reserve for
the $S$ scalar quartic coupling, denoting the corresponding Higgs coupling as $\lambda_h$.

Scale invariance of \eqref{EffectiveScalar} and previous knowledge of the standard model $\beta$ and $\gamma$
functions for the unaffected parameters allow the authors to fully determine the beta functions for all parameters
of the model (imposing, of course, appropriate boundary conditions).

Then, the conditions the authors demand are that no remote vacuum be formed below the cutoff $\Lambda$ of the
theory, to assure vacuum stability, as well as that no Landau pole appears below the cutoff in order to
ensure perturbativity. Many more details as well as explicit formulae can be of course found in the original 
paper.

Of particular interest for our work are the results obtained under the condition that the WMAP constraints be
satisfied in a saturated manner. Assuming a Higgs mass of $120$ GeV, the bounds on $(m_S, \lambda)$ combinations
depend strongly on one hand on the value of the scalar self-coupling (the bounds being alleviated as the
coupling increases) and on the other hand on the scale at which we consider that new physics enter the game.
For self-coupling of ${\cal{O}}(1)$, the model is practically unconstrained, especially if the cutoff is placed 
near the EW scale.

\subsection{Some notes on collider phenomenology}
In the general case where a real singlet scalar field is added to the Standard Model, it can mix with
the usual Higgs boson and affect not only the phenomenology of the latter, but also a whole series of
electroweak observables. This analysis has been performed in detail by Barger \textit{et al} in
\cite{Barger:2007im}. More specifically, the mixing is expected to bring about important changes in
the gauge boson propagator functions involved in several processes and quantities such as the $W$ mass, 
atomic parity violation or $Z$ - pole observables.

In our case, however, we already mentioned that for the scalar to be completely stable, we impose a
discrete symmetry to the Lagrangian. This eliminates all mixing effects with the Standard Model Higgs.
In this case, practically all experimental constraints are completely alleviated, at least from the 
point of view of collider physics. Nevertheless, even in this case it is possible to have interesting
effects especially on Higgs discovery physics at the Large Hadron Collider or the ILC. 

If the Singlet is light enough so that $m_h > 2 m_S$, the Higgs boson decay modes can be significantly
modified, since the possibility for invisible decays into a singlet pair is open. In this case, as
explained in the same paper, one should expect (depending on the singlet mass as well as on the
strength of the singlet-singlet-Higgs coupling) significant modifications on the Higgs discovery
potential of the LHC. More specifically, and especially in the Light Higgs scenario, fermionic decay
modes can by dominated by the invisible decays. Whereas the Higgs boson discovery potential thus 
deteriorates in the other channels, there can be discovery in the invisible channel through observation 
of events with missing energy in Vector Boson Fusion or Higgstrahlung processes. Furthermore, modifications can
be sizeable in the Higgs total decay width.

Interestingly, it turns out that low-mass singlets can further severely alter expectations on $b \rightarrow s$
transitions in the decays of $B$ mesons to Kaons along with missing energy \cite{Bird:2004ts}. 
These effects could be measurable in $b$ - factories and could constrain the very low mass regime of the
model, for $m_s < 2$ GeV. Current data can already put bounds for lower masses, excluding a significant
portion of that region of the parameter space.

The major constraints in the $Z_2 \times$SM case come from imposing dark-matter related limits, especially
the demand to reproduce the observed relic density as well as the imposition of bounds coming from direct detection
experiments.

\subsection{Relic density}
\begin{figure}[htb!]
\begin{center}
\includegraphics[width = 12cm]{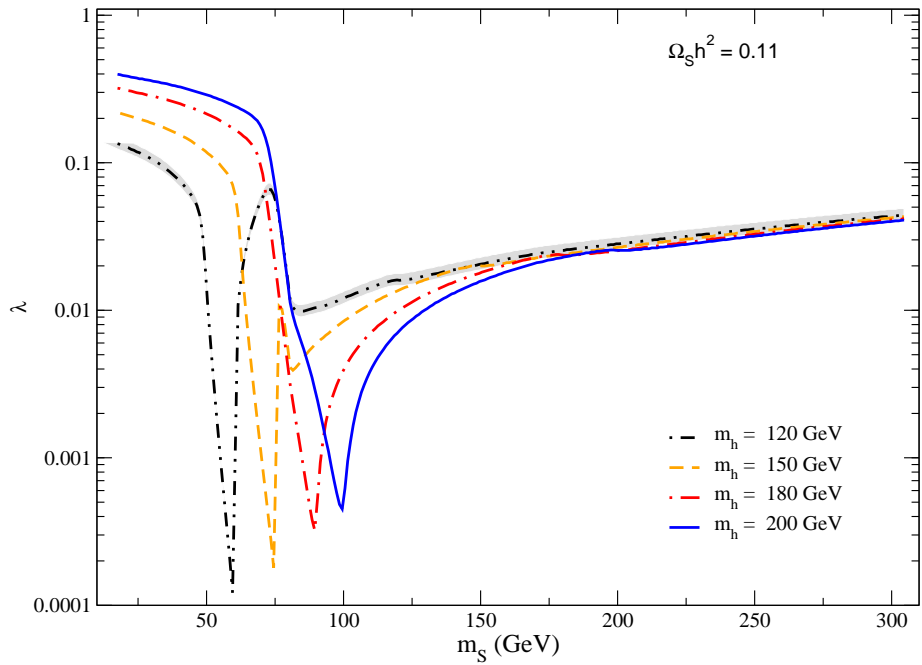}
\caption{{\footnotesize
Regions of the $(\lambda, m_s)$ satisfying the relic density constraint as given by WMAP. Figure taken
from ref.\cite{Yaguna:2008hd}
}}
\label{ScalarSingletRelic}
\end{center}
\end{figure}

The relic density constraint has been examined in a series of papers \cite{McDonald:1993ex, Gonderinger:2009jp, Barger:2007im, 
Burgess:2000yq, Kanemura:2010sh, Yaguna:2008hd} for different regions of the model's parameter space and
at times under different assumptions concerning the singlet itself (especially concerning its real or 
complex nature), depending also on the emphasis that each author wishes to give on specific aspects of the
model's phenomenology. The relic abundance calculation is actually really simple compared to more extended
models with several possible annihilation channels. The singlet coupling only to the Higgs sector, the only
diagram entering the calculation of $<\sigma v>$ is the $s$ - channel exchange of a Higgs boson between a pair
of singlets and the final state particles. The dark matter constraint can be satisfied in significant portions
of the parameter space $(\lambda, m_s, m_h)$. The $\lambda_S$ parameter, as we already mentioned, is 
irrelevant for the calculations and might only enter indirectly through the constraints described in the
previous sections. An instructive way to represent the parameter combinations that satisfy the constraint
can be found for example in \cite{Yaguna:2008hd}. We borrow fig.\ref{ScalarSingletRelic}
from this paper in order to explain the general parameter space behavior.

In this figure, the author plots the WMAP-compliant regions
in the $(m_s, \lambda)$ plane for various values of the Higgs mass, represented in fig.\ref{ScalarSingletRelic}
as lines or regions of different colors. In practice, the author scans the model's parameter space in order
to find points that satisfy the constraint $\Omega_S h^2 =  0.11$. For the case $m_h = 120$ GeV, the entire
region compatible with the $2 \sigma$ region allowed by WMAP is plotted.
It is particularly noteworthy that, as mentioned in the paper, for
reasonable values of $\lambda$ the model can account for the DM relic abundance by keeping the singlet's 
mass near the electroweak scale, so the corresponding DM candidate is a characteristic illustration of
WIMP dark matter. 

The most striking feature upon imposition of the WMAP limits is the steep decline in the
required $\lambda$ value for specific values of the scalar mass, varying according to the Higgs boson mass.
The first of these regions corresponds to the case where $m_s \approx {\cal{O}}(m_h/2)$, where annihilation is done
through a practically on-shell Higgs propagator. In this case, the squared propagator entering the
relevant amplitude explodes in value providing an enormous 
boost to the annihilation cross-section, hence the WMAP data
can be satisfied only by significantly decreasing the value of the relevant coupling. Seen inversely, by
keeping $\lambda$ constant, once the $h$-pole is reached the relic density falls dramatically and $S$ can 
no longer account for the observed relic abundance. The enhancement of $\sigma v$ depends of course
on the total width of the Higgs. This is the reason why as the Higgs mass increases, 
so does the required $\lambda$ value in order to satisfy WMAP. For the same reason, the increase in
the Higgs width, the resonance area tends to become less narrow for larger Higgs masses.

The second feature concerns the fall in the required $\lambda$ values in order to reproduce 
the DM relic density for moderate Higgs masses and slightly above the Higgs resonance. This feature
stays practically constant with changes in $m_h$. It corresponds to the $W^+ W^-$ pair-production
threshold, where $m_h \sim m_W$.
Once again the cross-section increases and the 
required singlet-singlet-Higgs coupling must be smaller so as not to underpoduce dark matter. This feature
is more striking for lower Higgs masses, whereas as $m_h$ increases the two resonances overlap.

A further feature is that for smaller singlet masses, the required $\lambda$ values in order to reproduce 
the DM relic density are comparably higher than for higher ones. Once again, this is a mainly kinematic effect.
Since the cross-section rises as the incident particle mass approaches the outgoing ones, once the gauge boson
channels open the singlet tends to annihilate mostly into $W^+ W^-$ and $Z Z$ pairs, which are evidently 
closer in mass than quarks or leptons. Furthermore, we see that as $m_S$ grows further beyond the $W$-resonance, 
the corresponding $\lambda$ values tend to increase once more.

The resonance regions let aside, for $m_h$ values of roughly $120$ - $200$ GeV, we see that the coupling values
in order to get the correct relic density are of the order of $10^{-1}$ for small masses and around $10^{-2}$
for masses up to $300$ GeV. These are the regions which are the most relevant for indirect detection, as we
shall see in the following.

Direct detection constraints have also been examined by a number of authors and under different points of view
\cite{Barger:2007im, Burgess:2000yq, Yaguna:2008hd, Bandyopadhyay:2010cc, Andreas:2010dz, Kanemura:2010sh}.
In most cases, the authors use existing data in order to impose bounds on the model parameters. After the 
excesses reported by DAMA but also CDMS-II and CoGeNT, there have been efforts to examine whether the singlet
scalar model can account for these excesses. The nature as well as the uncertainties entering some of the latest
reported excesses are still under discussion, but at first approach the model seems to be able to reconcile
them. Furthermore, as discussed in \cite{Andreas:2010dz} near-future measurements will be able to give a 
definitive answer on the point. In a recent review \cite{Guo:2010hq} the model is confronted with the latest
constraints coming from CDMS-II and XENON. It turns out that if standard assumptions are made, and for a Higgs
mass of $120$ GeV, the CDMS-II results alone exclude singlet masses roughly up to $80$ GeV. If the XENON-100
results are also taken into account, practically the entire parameter space for masses up to $200$ GeV are excluded.
The only region evading detection is the $h$-pole one, since the required $\lambda$ value is, as we explained, 
much smaller. We note that this region falls largely outside the reach of all planned direct detection experiments.

\section{Gamma-rays detection prospects}
Apart from its interesting relic density, direct detection and collider phenomenology, the singlet scalar model
of dark matter has implications for indirect detection experiments as well. This has been studied in \cite{Yaguna:2008hd}
for the case of gamma-rays coming from DM annihilations (excluding the Galactic Center) 
and in \cite{Goudelis:2009zz} for positrons and antiprotons. 
We note that in \cite{Guo:2010hq} the results of the second of these papers are summarized and extended as far as
the considered Higgs mass is concerned.

\begin{figure}[htb!]
\begin{center}
\includegraphics[width = 12cm]{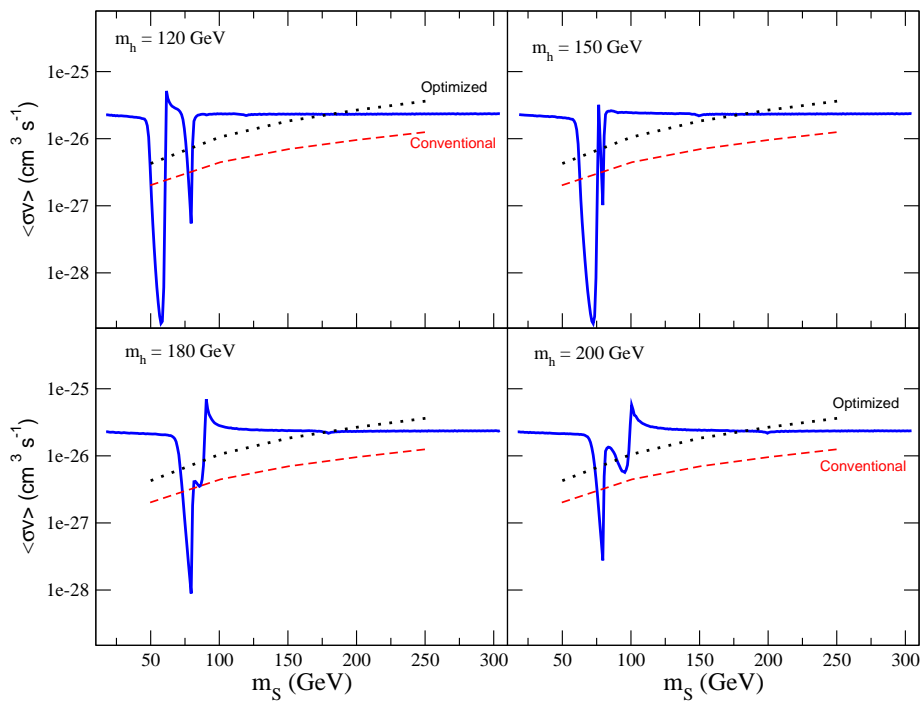}
\caption{{\footnotesize
Observable regions in the $(m_S, <\sigma v>)$ plane for four different values of the
Higgs mass and two different background considerations for the Fermi satellite mission and for gamma-rays
outside the Galactic Center. The blue line corresponds to the cross-section predicted by the singlet scalar
model. Figure taken from ref.\cite{Yaguna:2008hd}
}}
\label{ScalarSingletGammas}
\end{center}
\end{figure}

Let us first summarize the results concerning gamma-rays. For instructive reasons, once again, we
give in fig.\ref{ScalarSingletGammas} the plot from ref.\cite{Yaguna:2008hd} summarizing the author's findings.

In this figure, the blue lines follow the model's viable parameter space, where the singlet mass is varied
and the value of $\lambda$ is fixed for every mass by imposing the relic density constraint. We remind that, 
and this also concerns the following antimatter treatment we shall present, the dark matter relic density is
fixed to its central value (in the case of this paper this is $\Omega_{DM} h^2 = 0.11$) and not varied in the
whole WMAP-allowed region. The effect of allowing for such a variation would just result to the lines being
transformed to narrow regions around this central value. In this respect, the results are representative.
Then, the corresponding values of $<\sigma v>_{v\rightarrow 0}$ are calculated, which are relevant for 
indirect detection at present times when the WIMP velocity is small. The considered profile is the standard
Navarro, Frenk and White one, although since the GC is excluded from the analysis this is not such
and important factor.

The observed behavior is very smooth: $<\sigma v>_{v\rightarrow 0}$ remains practically constant for 
all singlet masses. Then, two important features appear. The first one is situated at the point where
$2 m_S \sim m_h$, where we have resonant annihilation into a Higgs propagator. The second one appears
for the case of on-shell production of a pair of $W$ vector bosons. Why does this fluctuation appear?
The answer lies in the fact that $<\sigma v>$ and $<\sigma v>_{v\rightarrow 0}$ are calculated for
different kinematic regimes. In order to obtain the correct relic density, and for typical Weakly Interacting
Massive Particles as in our case, a - more or less - standard value of the total thermally averaged self-annihilation
cross-section is needed. We already saw that in these two regions, this value is obtained through a combination
of low $\lambda$ values and the fact that annihilation takes place resonantly. The $\lambda$ value stays
practically the same at present times. The resonance conditions, however, do not. 
Since during decoupling WIMPs have a non-negligible
velocity, the resonance occurs \textit{slightly lower} than the points where $2 m_S = m_h$ or
$m_S = m_W$. What actually matters is the \textit{total} energy of the singlets at that time and not just
the one associated to their mass. Hence, for the specific points, the correct relic density was obtained 
for small $\lambda$ values but for resonance conditions which are no longer valid. 

To corroborate these comments, we should notice 
the rise in $<\sigma v>_{v\rightarrow 0}$ right after the Higgs resonance.
Being at the zero velocity limit, a small rise in the WIMP mass can reproduce resonant annihilation
which compensates the smallness of the scalar-scalar-Higgs coupling as soon as we are practically
exactly at $m_S = 1/2 m_h$. This is manifest in the cases of
$m_h = 120, 180$ and $200$ GeV but not in the $m_h = 150$ GeV one. The most plausible explanation 
for this is that since the two resonance regions get very close, the peak that would appear after the
$h$ resonance is immediately killed by the fall in $\lambda$ due to the $W^+ W^-$ one.

Now, as expected, the Fermi detection limits do not present any particular features. For the same
mass values the limits do obviously not depend on the model parameters as such. 
Small variations from one Higgs
mass scenario to another are the result of small differences in the final state composition, which
in any case is mostly $b \bar{b}$ below the W resonance and $W^+ W^-$ above. So, the detectability 
lines are practically the same among the different cases. We see that for masses up to roughly
$180$ GeV the perspectives are really good for both background models, whereas heavier singlets
are visible for the ``conventional'' background only, which is lower than the ``optimized'' one.

\section{Antimatter Detection}
In ref.\cite{Goudelis:2009zz} we studied the corresponding prospects for antimatter (positron and 
antiproton) detection at the AMS-02 mission as well as the constraints coming from the existing 
PAMELA collaboration data. We used the public code micrOMEGAs, which in its latest versions gives
the possibility to compute the relic density for WIMP-type candidates in generic models in order 
to define the viable parameter space. We examined singlets with masses lying in the region
$50 \leq m_S \leq 600$ GeV fixing the Higgs mass at $m_h = 120$ GeV, then varying $\lambda$ so as
to get a relic density of $\Omega_{DM} h^2 = 0.11$.

\begin{figure}[htb!]
\vspace{1cm}
\begin{center}
\includegraphics[width = 12cm]{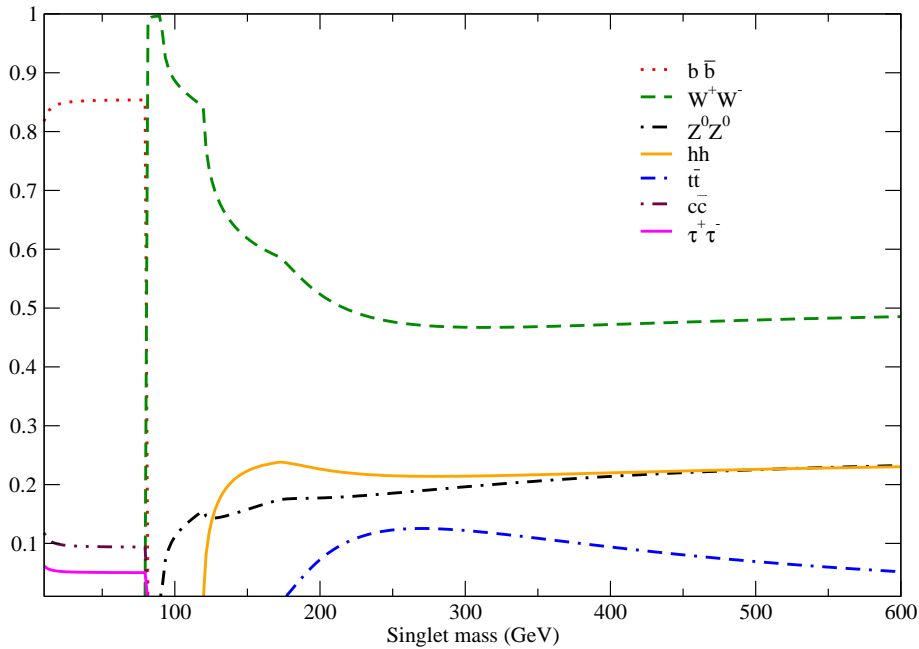}
\caption{{\footnotesize
Branching Fractions as a function of the dark scalar mass for the singlet scalar model.
}}
\label{ScalarSingletBRfractions}
\end{center}
\end{figure}

In fig.\ref{ScalarSingletBRfractions} we show the Branching Fractions (BRs) at zero velocity 
for our mass range and for the points satisfying the relic density constraint.

We see the behavior described in the previous, namely that for small masses the final state is
dominated by the $b \bar{b}$ component, whereas above the $W$ threshold the most important channel 
is the $W^+ W^-$ one. Smaller - but existing - contributions come from $hh, ZZ, c\bar{c}$ and
$t \bar{t}$ final states. At this point, it should be noted that a novel estimation of the final
state composition and the self-annihilation cross-section in \cite{Yaguna:2010hn} demonstrates
that in the low-mass regime, $3$ - body final state contributions can also be sizeable.

\subsection{Antiproton detection}
Our first study concerns the constraints coming from the PAMELA experiment and the detection perspectives
with AMS-02 in the antiproton channel. We compute our results for the three propagation models MIN, MED
and MAX as defined in the previous chapter assuming a smooth dark matter halo, as well as taking into
account potential substructure enhancing the annihilation rate.

In order to assess the background, we repeat that the relatively recent PAMELA data are well reproduced
by the conventional propagation model of Strong and Moskalenko. We thus consider that the experiment's 
results are essentially comprised of background events and just fit their spectrum, taking care so as
to maintain a good normalization to the low-energy data.

In this treatment, we shall characterize a point as being ``3-$\sigma$ excluded'' if the sum of the
signal produced by this point and the background events surpass the PAMELA measurements by more than
$3$ standard deviations, as stated by the collaboration itself. On the other hand, in order to assess
a point as being detectable by AMS-02 we employ a $\chi^2$ criterion. The $\chi^2$ is defined as
\begin{equation}
 \chi^2= 
\sum_{n=0}^N
\frac{(\phi^{tot}_n-\phi^{bkg}_n)^2}{(\phi^{tot}_n)}A\cdot T\ ,
\label{chi2}
\end{equation}
where $\phi^{tot}$ is the total antiproton flux, $\phi^{bkg}$ is the 
background flux, $N$ is the number of energy bins considered, $A$ is the
geometrical acceptance of the experiment, and $T$ is the data acquisition time. 
It is reminded that AMS-02 is expected to take data for  three years and features an antiproton  
geometrical acceptance of $330$ cm$^2$sr \cite{Goy:2006pw}. 
We consider $20$ energy bins evenly distributed in logarithmic scale between $10$ and $300$ GeV. 
A $95\%$ confidence level corresponds then to  $\chi^2>31$. 

We present our results in the same way as done in \cite{Yaguna:2008hd} for gamma-rays.
To obtain the excluded  regions in the plane ($m_S,<\sigma v>$), we first compute, 
for a model with the same branching ratios as the singlet scalar model (see figure \ref{ScalarSingletBRfractions}), 
the value of $<\sigma v>$ required to exclude the model, $<\sigma v>_{excl}$.  
By comparing $<\sigma v>_{singlet}$ with $<\sigma v>_{excl}$ we can then  determine whether the model is excluded or 
not at a given singlet mass and according to our exclusion criterion. 
An analogous procedure is followed to determine the detectable regions.

\begin{figure}[htb!]
\centering
      \includegraphics[width=0.40\textwidth,angle=-90]{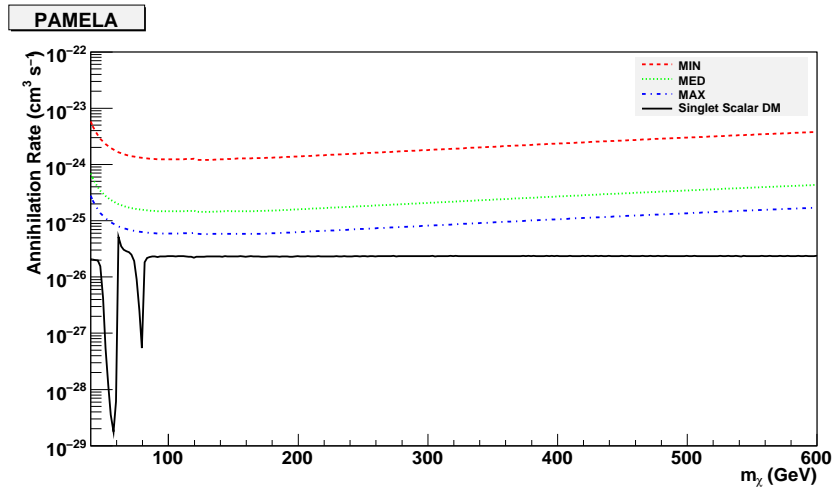}
      \caption{\footnotesize 
Regions of the parameter space that are  excluded by the antiproton data from the PAMELA experiment.
 The area above the MIN, MED, and MAX lines is excluded for the given propagation model. 
The solid (black) line shows the viable parameter space of the singlet scalar model of dark matter. }
       \label{ScalarPamelaAntiprotons}
\end{figure}

In fig.\ref{ScalarPamelaAntiprotons} we start the presentation of our results by comparing the minimal
cross-sections that would be required for each mass value and for the three propagation models against
the relevant singlet scalar's model predicted cross-sections in order to satisfy the relic density
constraint.

\begin{figure}[htb!]
\centering
      \includegraphics[width=0.40\textwidth,angle=-90]{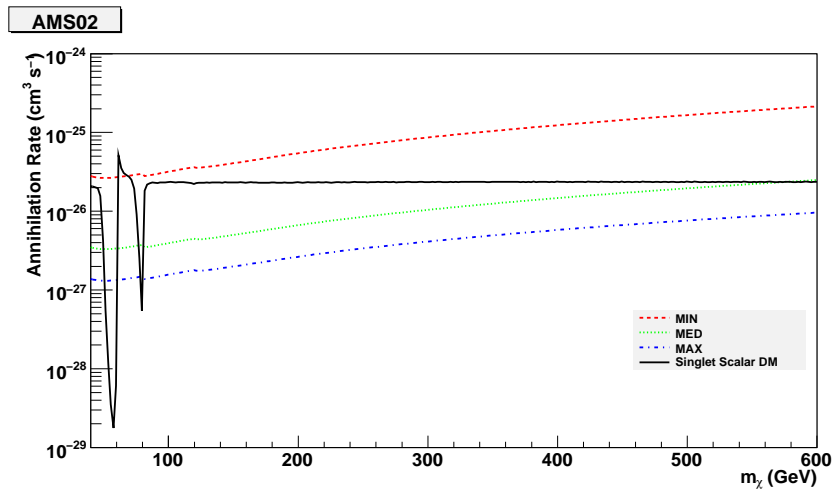}
      \caption{\footnotesize 
Regions of the parameter space that are within the sensitivity of the AMS-2 experiment. 
The area above the MIN, MED, and MAX lines is detectable by AMS-02. The solid (black) 
line shows the prediction of the singlet model. Notice that for MED and MAX, essentially 
the whole parameter space is detectable. }
       \label{ScalarAMSAntiprotons}
\end{figure}

In the figure, the red/dashed, green/dotted and blue/dotted-dashed curves concern the MIN, MED and MAX
models respectively, whereas the black solid line the model's prediction. We immediately see that no
parameter space point is yet excluded for any of the three propagation models. However, we observe that
the small region corresponding to the point where two singlets annihilate resonantly into a Higgs
propagator is at the verge of exclusion for the MAX model.
\\ \\
Then, in fig.\ref{ScalarAMSAntiprotons} we plot the corresponding minimal cross-sections that would
be needed in order to have a positive detection in AMS-02.

We see that the model has really good perspectives for antiproton detection. Some portion of its parameter
space will be probed for all three propagation models, whereas for the MED and the MAX one practically
the entire viable parameter space will be visible, apart from the points corresponding to the $h$-resonance
during decoupling as well as the corresponding $W^+ W^-$ resonance.

\begin{figure}[tbp!]
\centering
	\includegraphics[width=0.40\textwidth,clip=true,angle=-90]{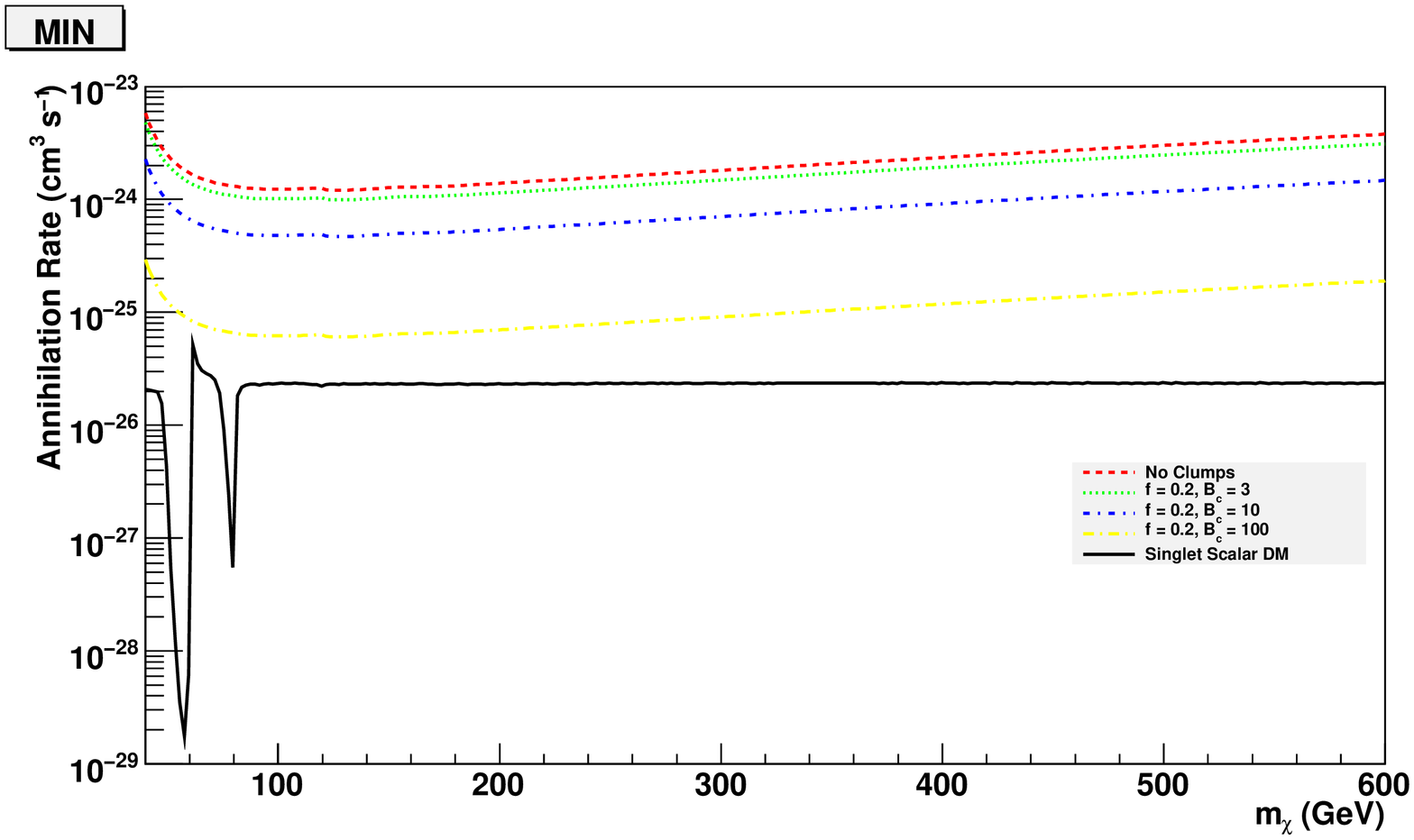}
	\includegraphics[width=0.40\textwidth,clip=true,angle=-90]{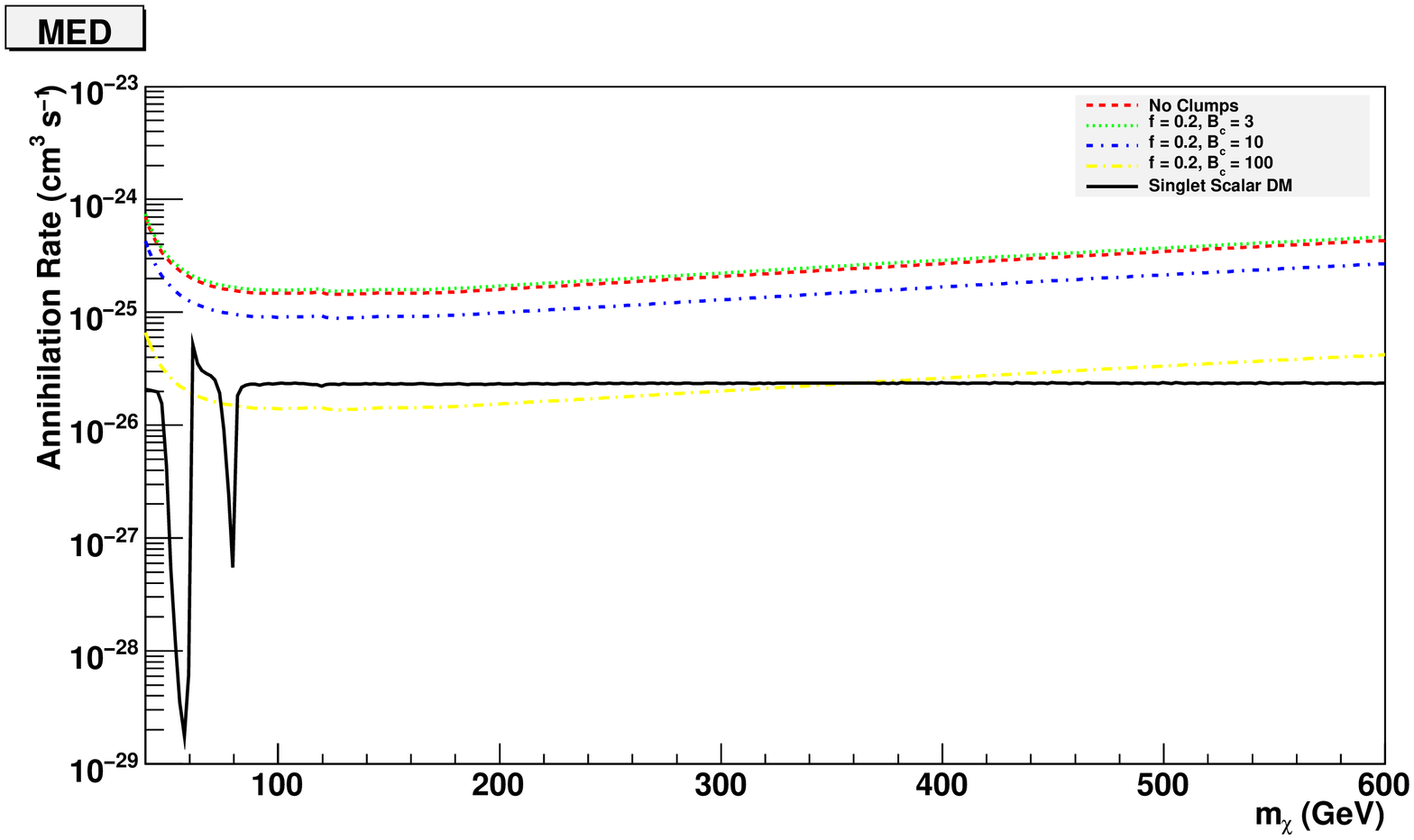}
	\includegraphics[width=0.40\textwidth,clip=true,angle=-90]{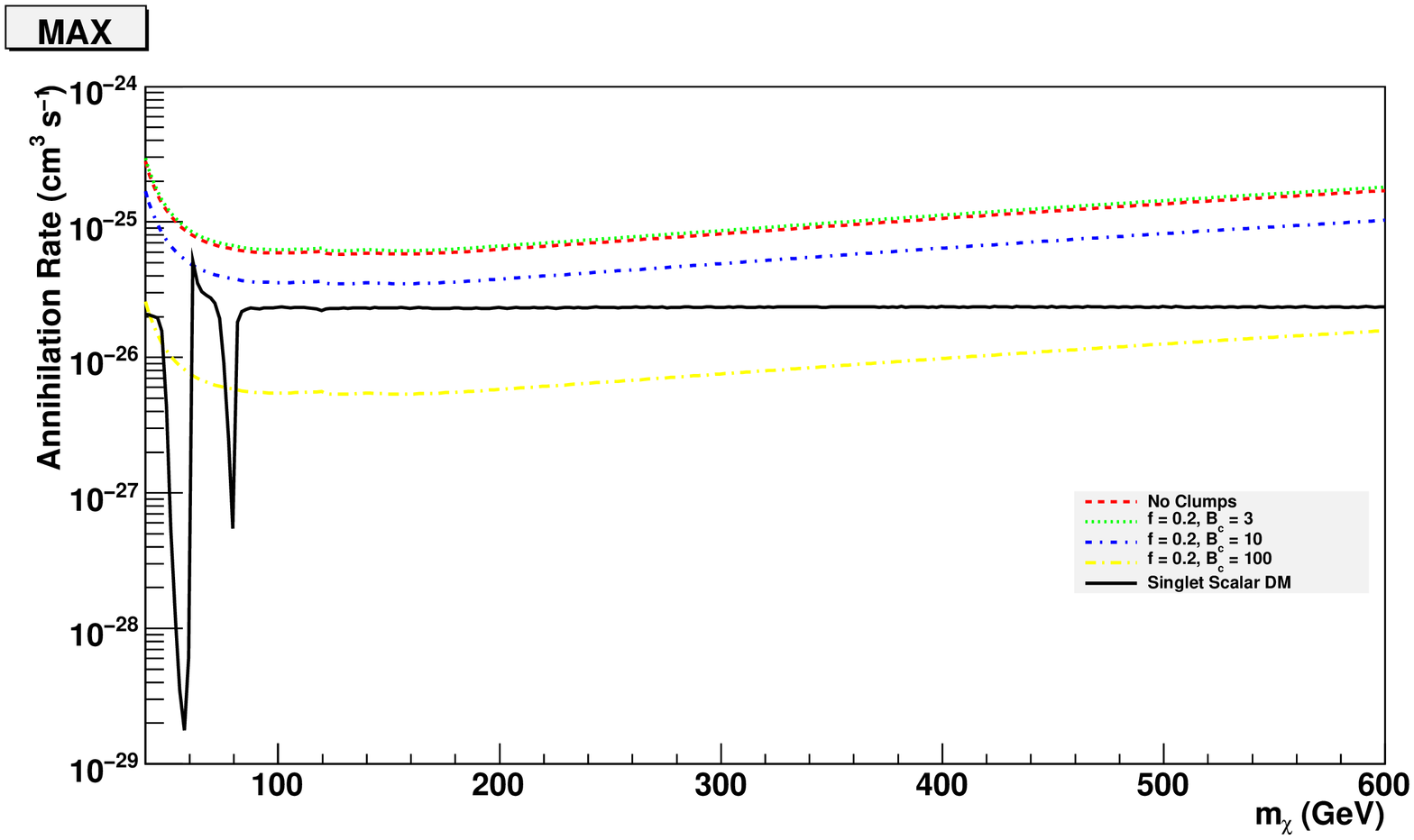}

          \caption{{\footnotesize
Regions excluded by the antiproton data from PAMELA including the possible effect of 
substructures in the DM halo. From top to bottom the figures correspond to the MIN, MED, 
and MAX propagation models. The solid (black) line shows the prediction of the singlet model. 
The area above the lines is excluded for the corresponding parameter values.}}
         \label{ScalarPamelaAntiprotonsClumps}
\end{figure}

The next step is to consider possible effects of substructure in the galactic halo. In 
fig.\ref{ScalarPamelaAntiprotonsClumps} we plot the exclusion limits for the three propagation models
as denoted in each plot's label, considering three different clump setups: No clumps (these lines are
the same as those in fig.\ref{ScalarPamelaAntiprotons} and are just given to facilitate
comparison), and three constant individual clump boost factors that we use to extract the effective 
one. In all cases and in what follows, we consider that $20\%$ of the DM halo is in clumps.

For all three propagation models, we see that the case $B_c = 3$ does not produce any essential 
modification with respect to the clump-less case. $B_c = 10$ produces an effective boost of the 
order of $3$ on average, which can also be seen in the graphs. Finally, $B_c = 10$ corresponds to
an average effective boost of at least ${\cal{O}}(10)$, which explains the decrease of the cross-section
needed for exclusion by around an order of magnitude. Let us examine the three propagation models
one by one: in the case of the MIN model, no viable parameter space point is excluded by current
data. The MED model is excluded if one assumes optimistic boost factors, especially in the
low/intermediate scalar mass regime. Finally, the MAX model is completely ruled out for large
boost factors, whereas even in the moderate boost case the $h$ - resonance at present times
(i.e. at zero velocity) is excluded. It should be noted that the the $h$ - pole region at decoupling
as well as the $W^+ W^-$ threshold are not excluded for any astrophysical setup.

\begin{figure}[tbp!]
\centering
	\includegraphics[width=0.40\textwidth,clip=true,angle=-90]{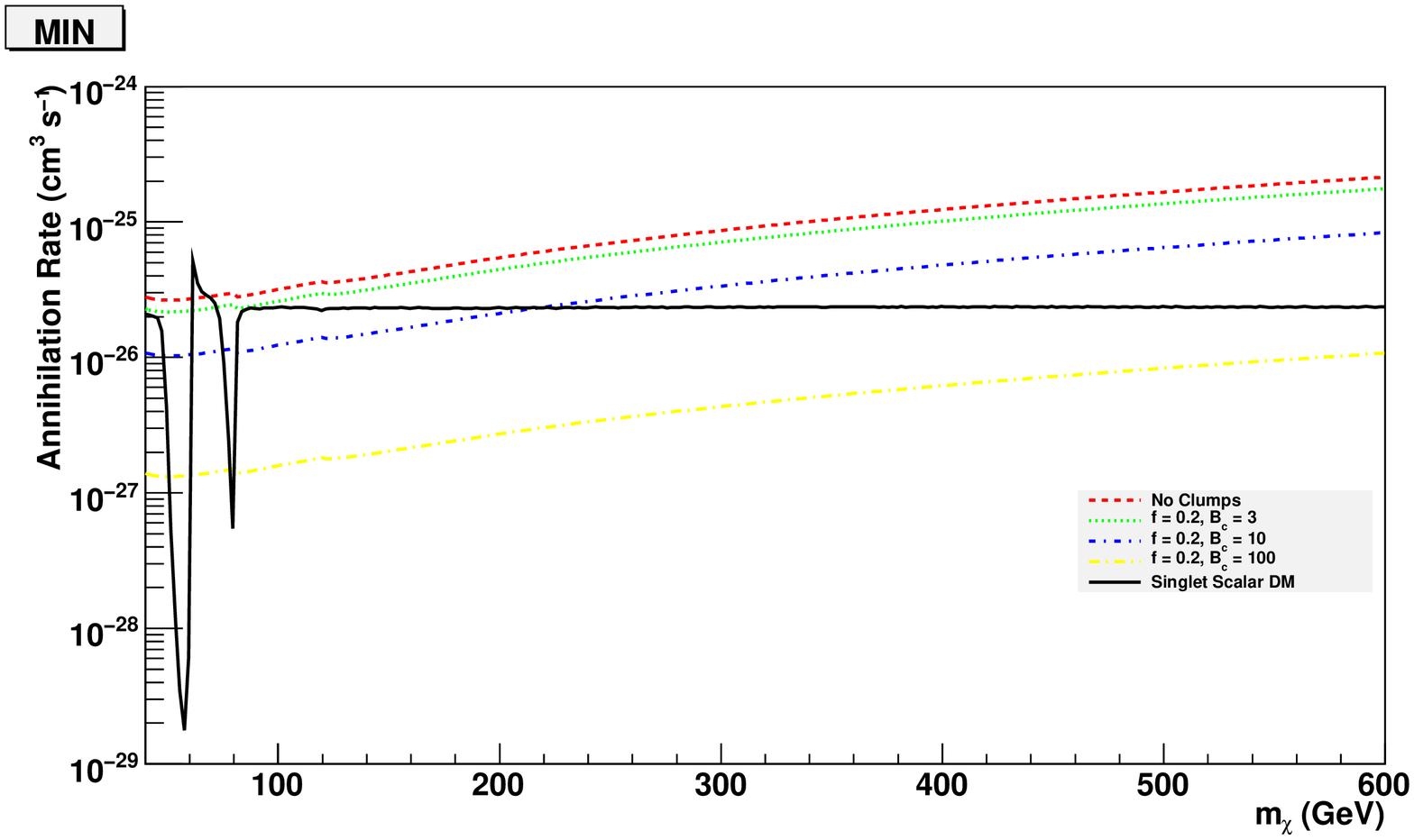}
	\includegraphics[width=0.40\textwidth,clip=true,angle=-90]{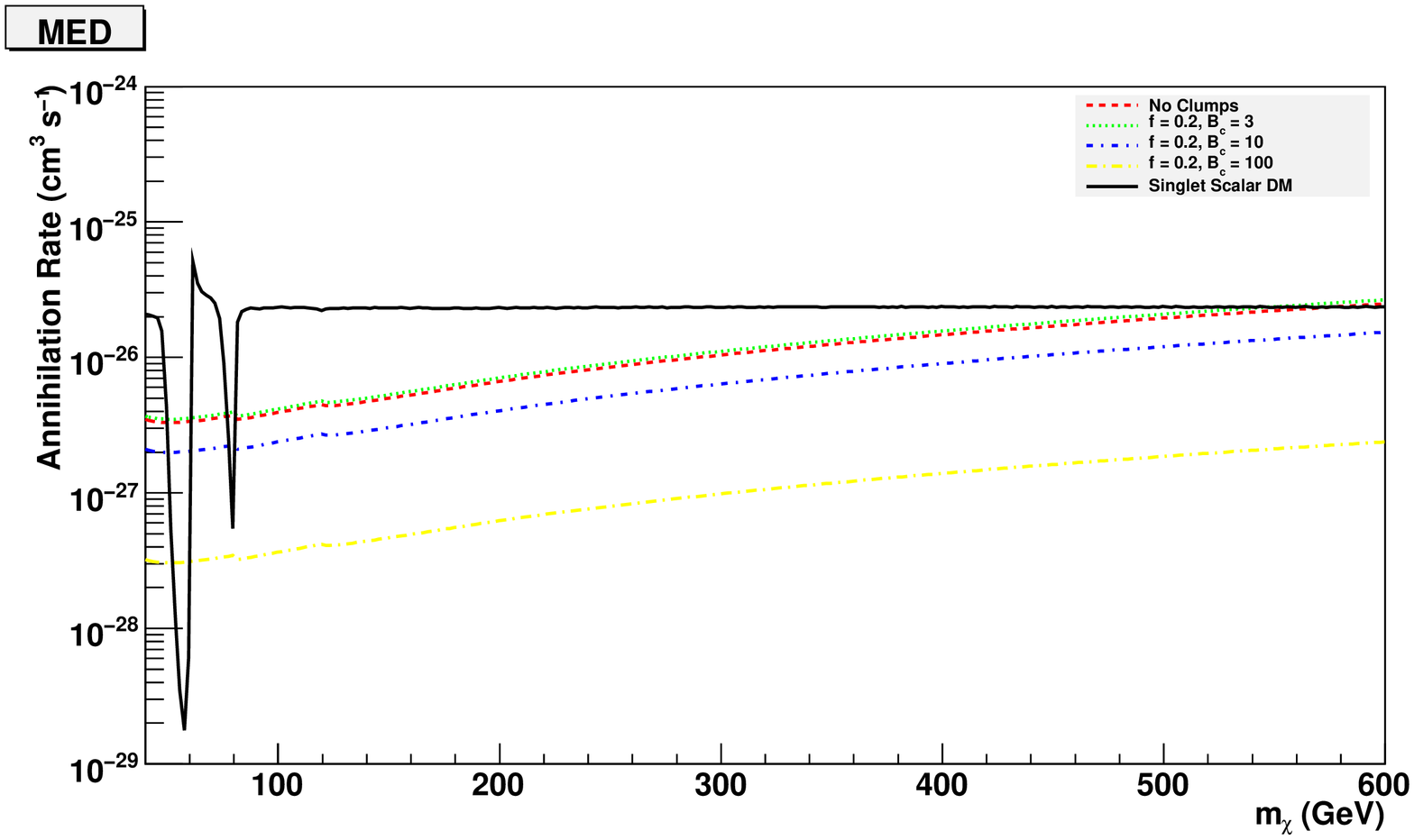}
	\includegraphics[width=0.40\textwidth,clip=true,angle=-90]{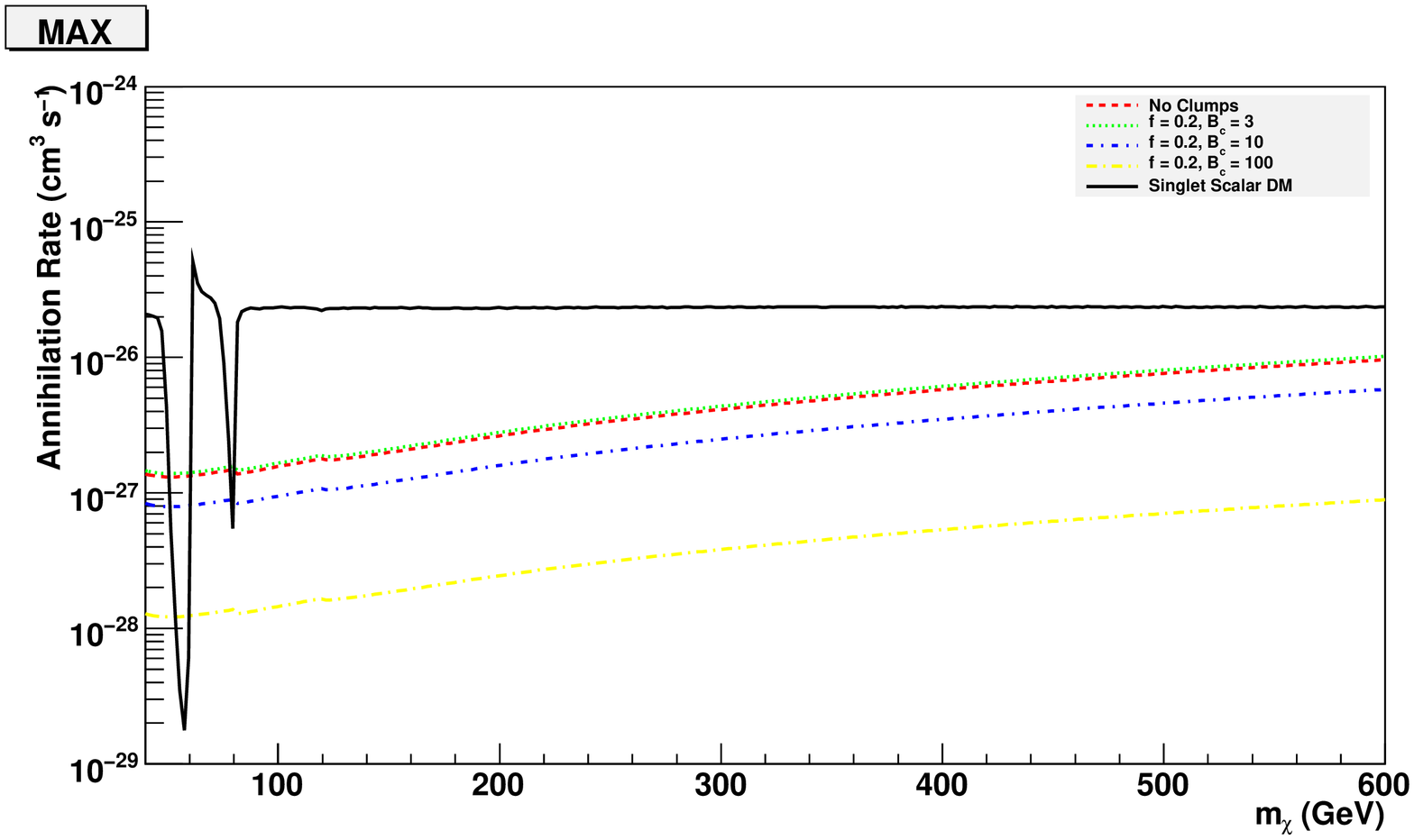}

          \caption{{\footnotesize
Detectable regions at AMS-02 including the possible effect of substructures. 
From top to bottom the figures correspond to the MIN, MED, and MAX propagation models. 
The solid (black) line shows the prediction of the singlet model. The area above the lines is detectable 
for the corresponding parameter values.  }}
        \label{ScalarAMSAntiprotonsClumps}
\end{figure}

In fig.\ref{ScalarAMSAntiprotonsClumps} we plot once again the minimal cross-sections that would
be required for AMS-02 to be able to detect singlet scalar DM assuming substructure boosts as in
fig.\ref{ScalarPamelaAntiprotonsClumps}. 

As expected, even for intermediate boost values good regions of the viable parameter space can be probed.
We note that for optimistic boost configurations, even the decoupling Higgs pole can be probed at a very
good level.

An interesting feature is that both the exclusion limits as well as the detection perspectives become
weaker as the singlet mass increases. This is actually due to two facts: first, 
an increase in the WIMP mass tends to have a negative impact on DM indirect
detection. Secondly, above the $W$ - resonance, there are major contributions from gauge boson final
states. But we saw that gauge bosons tend to underpoduce antiprotons with respect to quark final states.
The spectrum for such mass values is thus weaker than for smaller ones.

\subsection{Positron detection}
The next step in \cite{Goudelis:2009zz} was the calculation of the positron fluxes and corresponding
constraints from PAMELA and perspectives for AMS-02. Both the conditions we impose to characterize points as
being excluded or detectable and the way we present our results are the same as previously.

Before describing these results, we should comment upon one important point that will be
of determinant importance in this detection mode. We already described how the latest PAMELA and
Fermi-LAT data are in straight contrast to all previous expectations on the positron background, 
especially in our region of interest, $E \geq 10$ GeV and commented upon the fact that a huge effort
is devoted by numerous groups in order to find explanations for this excess through astrophysical mechanisms.
In this paper, we stuck to this approach, namely that the bulk of the PAMELA signal comes from background
events through some astrophysical mechanism. Therefore, we feel that the most conservative choice that
can be made (rendering results as robust as possible) is to take as background the entire PAMELA \cite{Adriani:2008zr} and
Fermi \cite{Abdo:2009zk} data. To this goal, we fit the positron fraction data from PAMELA 
as well as the $e^+ + e^-$ ones from Fermi. Since the two concern different energy regions, the
only solution is to extrapolate the fitting functions into our region of interest and multiply the
two fitting functions to obtain the relevant positron flux.

\begin{figure}[htb!]
\centering
      \includegraphics[width=0.40\textwidth,angle=-90]{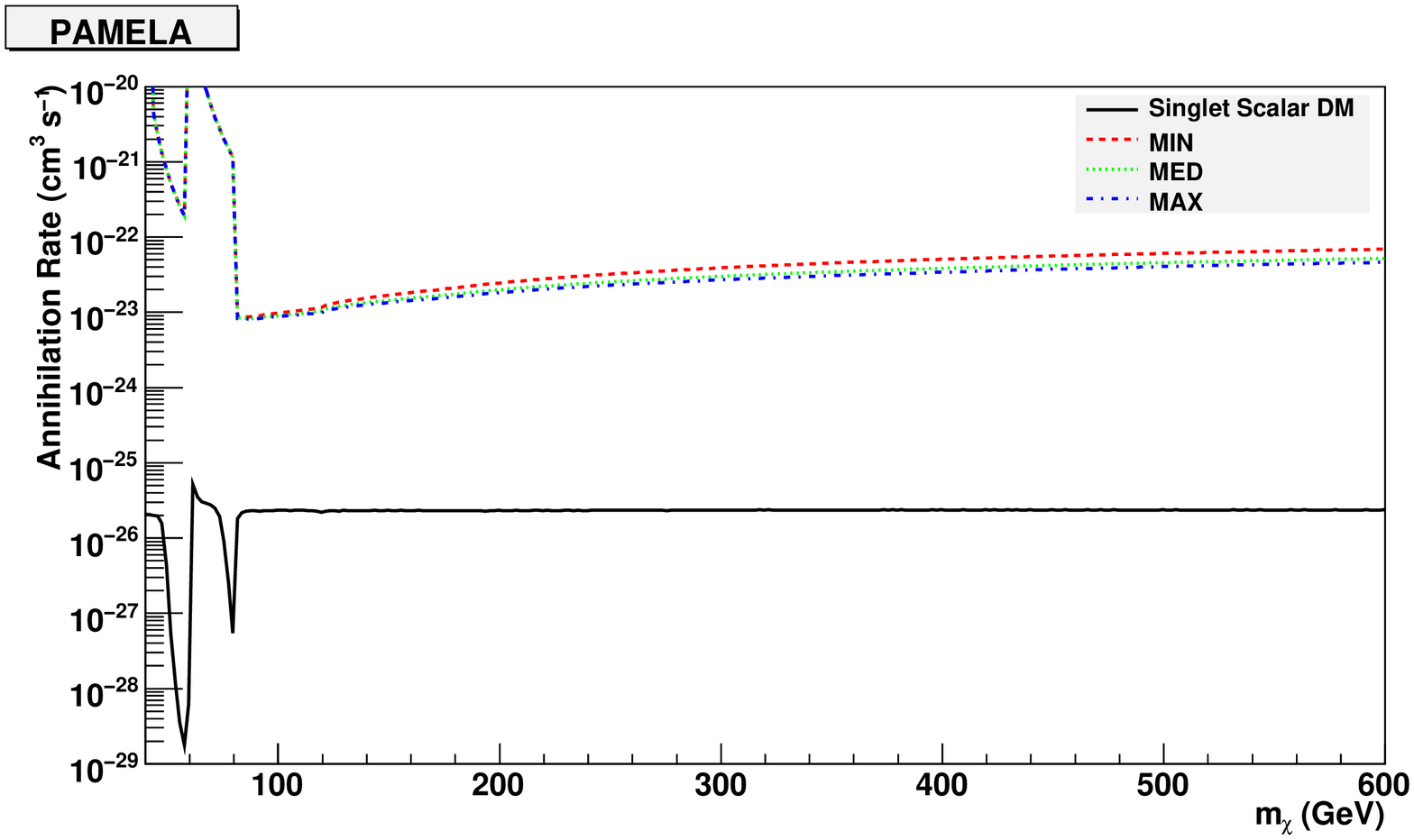}
      \caption{\footnotesize Regions of the parameter space that are  excluded by the  
recent positron data from the PAMELA experiment. The area above the lines is excluded for the 
corresponding propagation model. Notice that no region of the viable parameter space is currently ruled out.}
       \label{ScalarPamelaPositrons}
\end{figure}

In fig.\ref{ScalarPamelaPositrons} we demonstrate our results on the positron - excluded parameter space due
to the combined PAMELA/Fermi measurements.

It is clear from the figure that the viable parameter space falls largely outside the excluded region
by more than two orders of magnitude in the cross-section.

\begin{figure}[htb!]
\centering
      \includegraphics[width=0.40\textwidth,angle=-90]{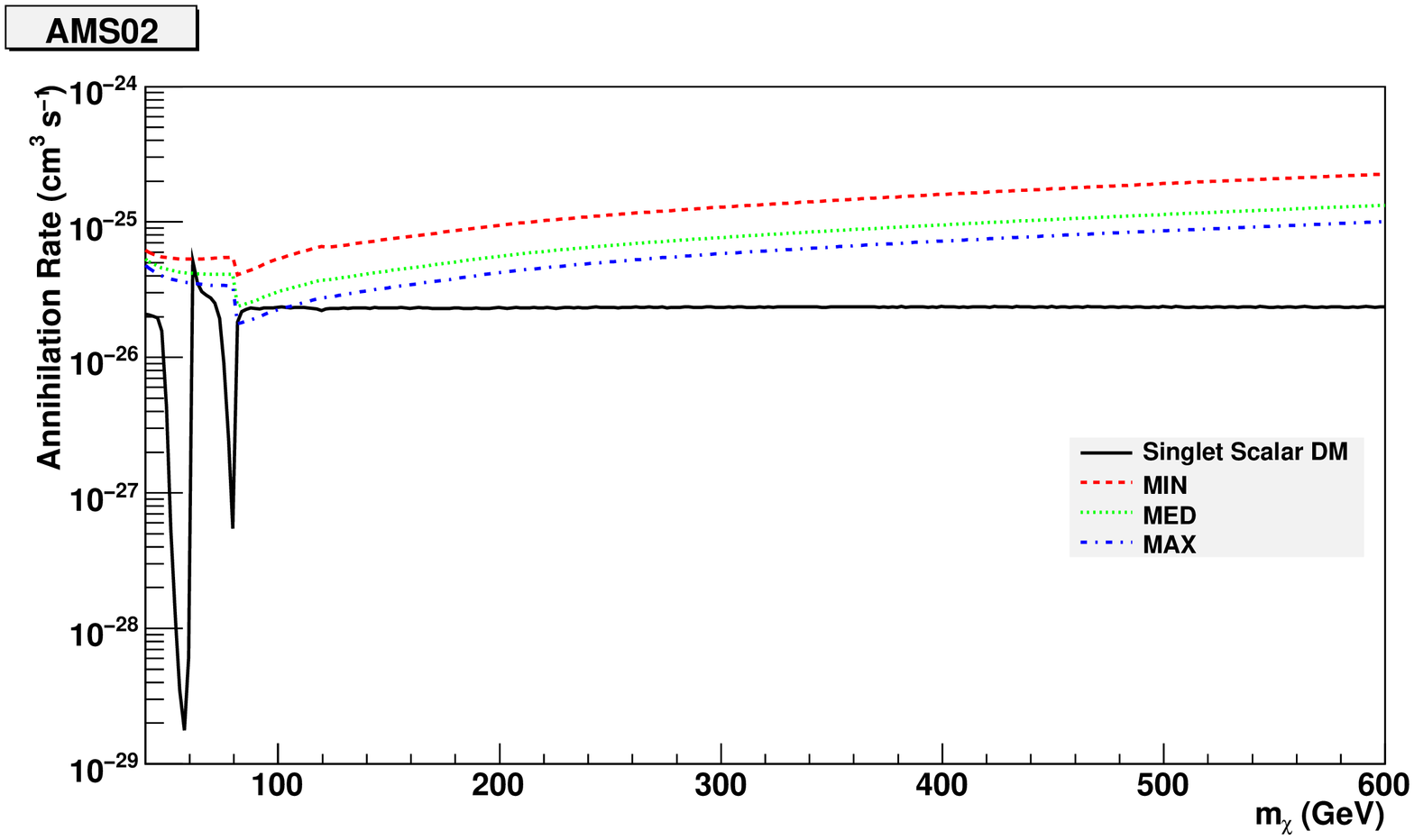}
      \caption{\footnotesize Regions of the parameter space that give a positron signal within 
the sensitivity of the AMS-02 experiment. The lines corresponding to the MIN, MED and MAX 
propagation models are shown. They must be compared to the actual prediction (solid line) 
of the singlet scalar model. }
       \label{ScalarAMSPositrons}
\end{figure}

Then, in fig.\ref{ScalarAMSPositrons} we present the corresponding results for the AMS-02 perspectives
in the same channel.
In this case the situation appears to be much better than in the case of PAMELA (remember already that
AMS-02 has a geometrical acceptance of roughly an order of magnitude larger than PAMELA and that
in this case only statistical errors are considered). The two most promising regions of the model
are the $h$ - resonance at present times, detectable in all three propagation models, as well as 
the $W$ - resonance again at present times. In the first case, i.e. the resonant annihilation
into a pair of $h$ bosons, the PAMELA or AMS-02 limits do not change in an important manner. It
is the rise in the total annihilation cross-section that renders these points visible, since the
$<\sigma v>_{singlet}$ values in this case are pushed into the detectable region. In the second case, on the
other hand, it is the detectability limits themselves that change due to the peculiar composition of the 
final state, which is almost entirely comprised of a $W^+ W^-$ pair. We saw that gauge bosons tend
to produce quite rich spectra in positrons, therefore a positive detection requires a smaller
cross-section in this case.

Note also that the changes among different propagation models
are much smaller than in the case of antiprotons.

\begin{figure}[tbp!]
\centering
	\includegraphics[width=0.40\textwidth,clip=true,angle=-90]{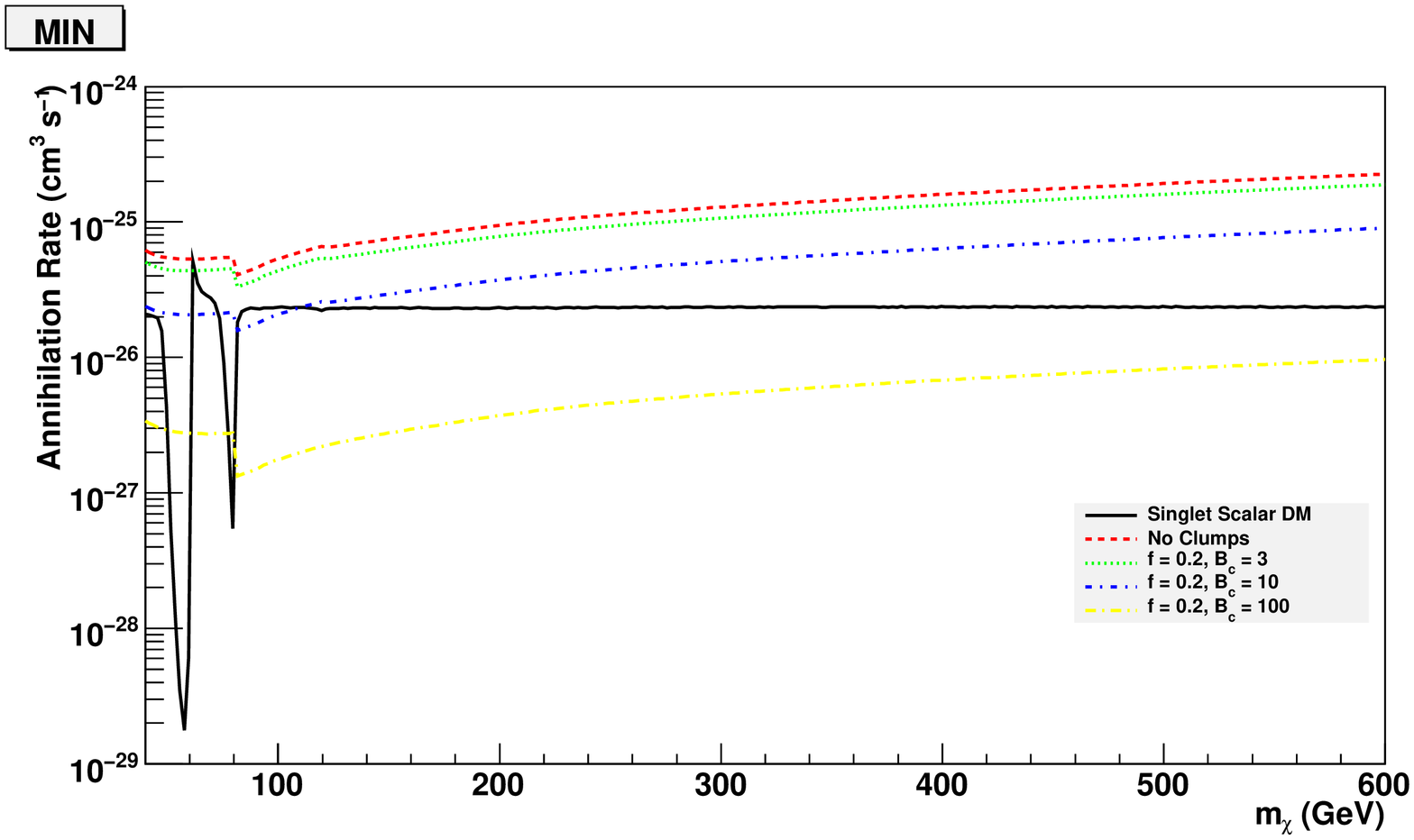}
	\includegraphics[width=0.40\textwidth,clip=true,angle=-90]{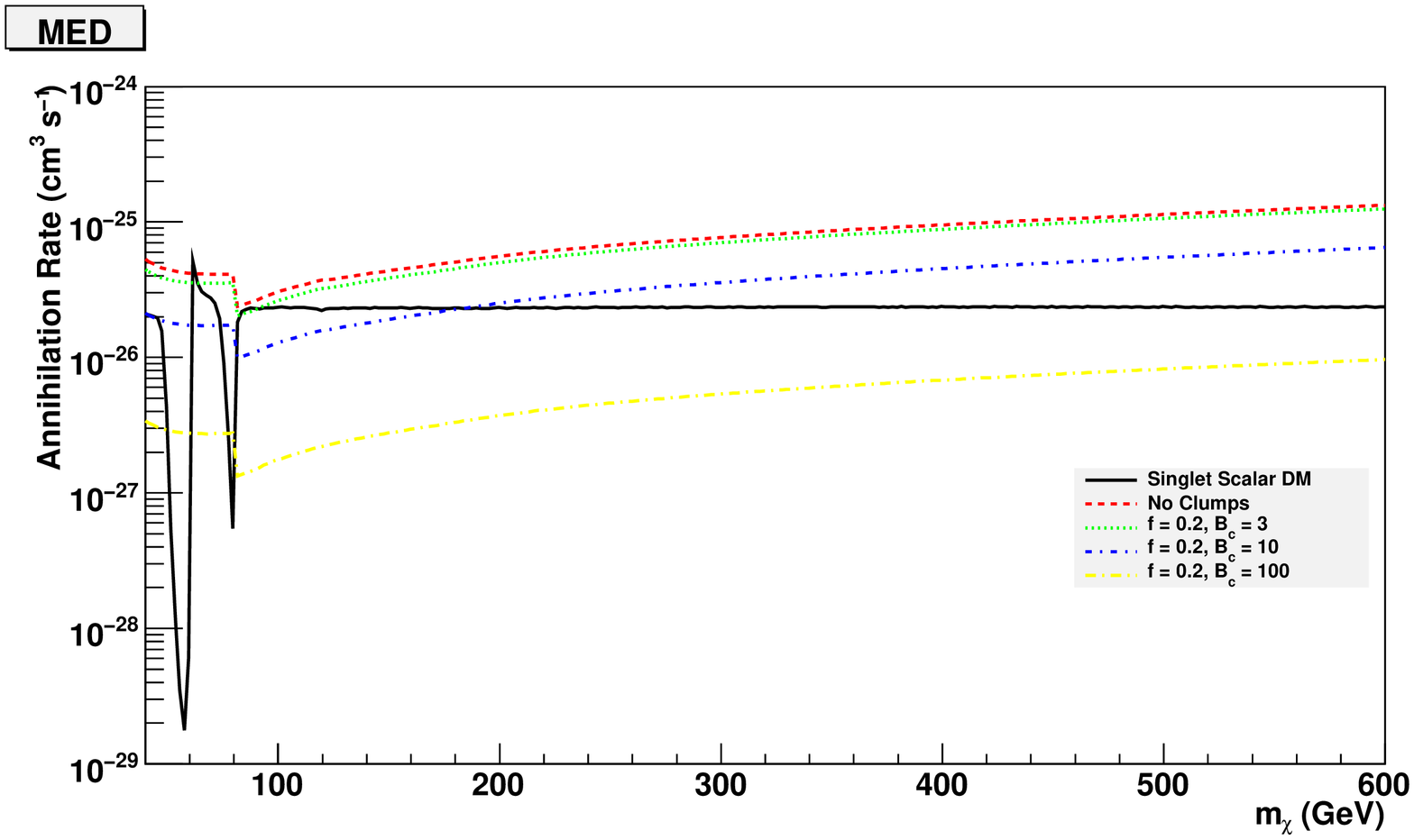}
	\includegraphics[width=0.40\textwidth,clip=true,angle=-90]{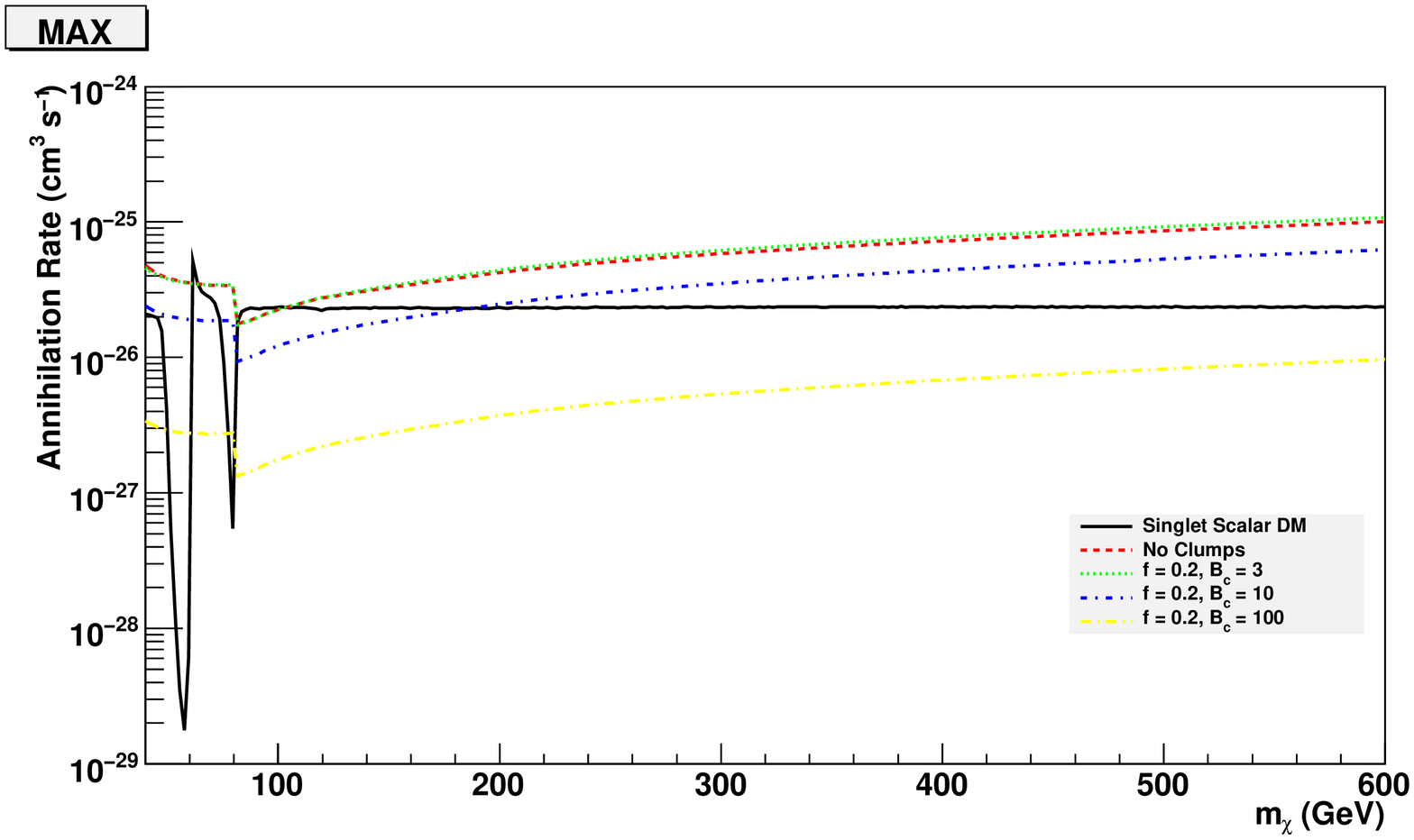}

          \caption{{\footnotesize
Detectable regions for the MIN, MED, and MAX propagation models in the presence of dark matter 
substructures. The region above the lines is detectable for the corresponding parameters. 
The solid (black) line shows the prediction of the singlet scalar model. 
}}
\label{ScalarAMSPositronsClumps}
\end{figure}

Finally, we calculate the constraints and prospects if clumpiness is taken into account. The results
can be seen in fig.\ref{ScalarAMSPositronsClumps} for the three propagation models and our three clump setups. In the case
of the MIN model, optimistic boost factor assumptions ($B_c = 100$) must be made in order to account
for a significant part of the parameter space to start falling into the region that can be probed.

In the cases of the MED and MAX models, even moderate astrophysical considerations can render an
important portion of the considered models detectable, especially in the low/intermediate mass regime.
For large boost factors, the entire viable parameter space is visible in all three models.
\\ \\
It is strange, at first sight, to see that for the same exclusion/detectability criteria, the
antiproton channel seems to offer much better possibilities than the positron one. One would
in fact expect the exact opposite situation, since the yield into positrons is for most final
states more than an order of magnitude larger than the yield into antiprotons.

However, at this point we should keep in mind that in all detection procedures the important 
factor is, in the end, not just the signal's magnitude but rather the relative magnitude of the signal 
to the background. In this sense, for example, if we imagine an experimental setup with strictly
zero background, even one or two events could constitute a clear proof for new physics. 
In the case of antimatter detection, since the discovery of the PAMELA excess, if we consider
- and this is the case in the present analysis - the whole dataset as being the result of some
astrophysical mechanism, then we are left with a very large background in the positron channel 
which would require much more elevated signal rates so as to produce a statistically significant 
excess. This is not the case in the antiproton channel though, where background rates are 
small enough so that even with relatively lower signal event rates it is easier to achieve
a statistically significant excess.
\\ \\
In this chapter, we examined the phenomenology of a simple extension of the standard model by a
real singlet scalar field. After presenting previous results on the various constraints and 
phenomenological aspects for the model, we presented our analysis concerning the constraints coming
from and the prospects for antimatter detection from dark matter annihilations in the galactic 
halo. We saw that for masses in the region $(50, 600)$ GeV, cross-sections of the order of 
$3 \cdot 10^{-26}$ cm$^3$ sec$^{-1}$ and for final states comprised mostly
of $b \bar{b}$ or $W^+ W^-$ pairs, there are no constraints coming, for example, from the existing
antimatter flux measurements. We nevertheless expect a significant amelioration with the launch
of the AMS-02 experiment. In the following chapter, we shall examine more complex models
falling into another of the classes we cited in the beginning of this chapter.

\newpage
\chapter{Supersymmetric solutions}

In the previous chapter we examined a really minimal way that tries to solve
the dark matter problem. We especially insisted on the fact that despite its
simplicity, the addition of just a singlet scalar in the theory can provide us
with a viable candidate for the missing matter content of the universe. We also 
saw that this candidate could be detected at present or oncoming 
experiments.

In this chapter we shall see how a dark matter candidate can arise within the context of
more complicated models. At first, we shall briefly describe some further issues that render the
Standard Model to be considered as a - probably - incomplete theory. Then, we shall see how 
an extension of the Poincar\'e symmetry that characterizes the SM Lagrangian can provide us
with a framework that can actually solve these issues. Eventually, the same framework can, 
under certain assumptions, also give rise to a stable neutral particle that can answer the
dark matter question. Furthermore, we shall discuss how despite its elegance, the models
that can be constructed by virtue of this new extended symmetry, called supersymmetry (SUSY), 
are not without issues. This shall especially be the case for the simplest model
that can be constructed in this way, called the Minimal Supersymmetric Standard Model (MSSM).
Eventually, we shall discuss dark matter in a context that tries to resolve some of these
issues, focusing on two specific examples: one going beyond the MSSM framework (BMSSM)
and another one admitting non-minimal versions of the former. All of this shall, of course, 
be clarified much more in the following.

%%%%%%%%%%%%%%%%%%%%%%%%%%%%%%%%%%%%%%%%%%%%%%%%%%%%%%%%%%%%%%%%%%%%%%%%%%%%%%%%%%%%%%%%%%%%%%%%%%%%%%%%%%%%%
%%%%%%%%%%%%%%%%%%%%%%%%%%%%%%%%%%%%%%%%%%%%%%%%%%%%%%%%%%%%%%%%%%%%%%%%%%%%%%%%%%%%%%%%%%%%%%%%%%%%%%%%%%%%%
%%%%%%%%%%%%%%%%%%%%%%%%%%%%%%%%%%%%%%%%%%%%%%%%%%%%%%%%%%%%%%%%%%%%%%%%%%%%%%%%%%%%%%%%%%%%%%%%%%%%%%%%%%%%%
\section{Some issues with the Standard Model...}
The Standard Model of particle physics is widely acknowledged as being extremely successful in its predictions, 
both at a qualitative and a quantitative level. All the new particles it predicted have been
discovered. It could be practically said that since the discovery of the 
top quark, its last missing ingredient before EWSB, there has been almost no compelling experimental evidence 
for new physics. Perhaps the two main experimental
issues that have arisen are neutrino oscillations, witnessing the existence of some non-zero neutrino masses, 
as well as the existence of dark matter, for which as we have said the Standard Model cannot account.
For the moment, its only missing ingredient is the mechanism breaking electroweak symmetry, or according
to our discussion until now the discovery of the Higgs boson. 

However, there are reasons to think that despite its success the Standard Model should probably not be the 
ultimate theory. We have of course already mentioned that the Standard Model is a renormalizable theory that
can be in principle extrapolated up to arbitrarily high energies. It is nevertheless quite reasonable 
to wonder what happens if it is not treated in this way, but it is instead taken to be just
the low-energy limit of some higher theory.

If this is the case, an important issue arises. It is known that every renormalization 
(more precisely: regularization) procedure
intrinsically introduces some cut-off scale, say $\Lambda$. If one just considers the pure Standard Model,
then at the end of this procedure the cut-off should be sent to infinity and finite results should be acquired.
This is actually the case for the Standard Model: it is a renormalizable theory. But if we instead accept that
the Standard Model has some specific region of validity, then the cut-off cannot be sent to arbitrarily high 
scales. This argument leads to one of the main concerns indicating that the SM needs some ultraviolet 
completion from a more fundamental theory, as we shall see in the following paragraph.
%%%%%%%%%%%%%%%%%%%%%%%%%%%%%%%%%%%%%%%%%%%%%%%%%%%%%%%%%%%%%%%%%%%%%%%%%%%%%%%%%%%%%%%%%%%%%%%%%%%%%%%%%%%%%
\subsection{The hierarchy problem}
\begin{figure}[ht]
\begin{center}
\includegraphics[width = 8cm]{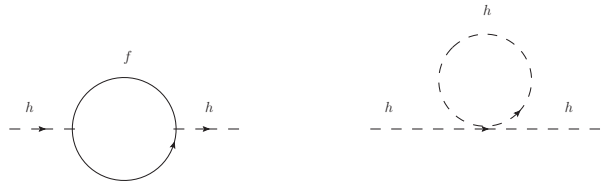}
\caption{{\footnotesize
Some 1-loop radiative corrections to the Higgs boson mass.}}
\label{fig:Higgs1loop}
\end{center}
\end{figure}

In fig.\ref{fig:Higgs1loop} we show two examples of one-loop contributions to the Higgs propagator, leading
to a redefinition of its physical mass according to usual renormalization procedures. 

To a good approximation (which shall be justified in a moment), one can ignore the light quark
loop contributions and only keep the one coming from the top quark. Then, the 
Higgs mass receives corrections like
\begin{equation}
 \delta_{m_h}^2 = \frac{3 \Lambda^2}{8 \pi^2 u^2} 
\left[ \left( 
4 m_t^2 - 2 M_W^2 - M_Z^2 - m_h^2) + {\cal{O}}\left( \log \frac{\Lambda}{\mu}\right)
\right) \right] \ , \ \
\label{Higgs1loopSM}
\end{equation}
where it can be seen that since fermion loop contributions scale as $m_f^2$, those coming from top quark 
loops are largely dominant. From Eq.\eqref{Higgs1loopSM} we can see that the standard model Higgs mass
receives corrections depending both logarithmically \textit{and} quadratically on the cutoff scale.
Logarithmic corrections are actually not that large, since even if we replaced the cutoff scale by the
Planck mass, where gravity is generally expected to manifest its quantum nature (hence rendering the theory
incomplete), then the corresponding contributions are numerically quite small. This is not at all the case,
however, for the quadratic corrections. There is, thus, a natural tendency for the Higgs mass to be pushed
towards the highest energy scale of the theory.

On the other hand, we expect that the Higgs mass should lie somewhere around the electroweak scale. This is not
only due to that ``logically'' it should be somewhere around the scale where electroweak symmetry breaking
takes place. It is also a requirement coming from a series of bounds, such as the triviality and vacuum stability
ones. Of course, one could argue that since there are both positive and negative contributions to the quadratic
terms, there could be some important cancellation among them. This calculation has been demonstrated to be
a viable solution, it requires though very precise cancellations among very large quantities in order to 
yield very small ones. This effect is usually referred to as \textit{fine tuning}, whereas the resulting
problem is known as the ``hierarchy problem`` \cite{I-hierarchie}. It would be more ``natural''
to think of a mechanism that precisely eliminates the quadratic divergencies or largely suppresses them. For the moment,
we just note that the usual way of eliminating terms in Lagrangians, perturbative expansions etc is the 
introduction of symmetries: the gauge symmetry ``protects'' gauge bosons from receiving quadratically divergent
contributions, whereas chiral symmetry does so for the case of fermions.

%%%%%%%%%%%%%%%%%%%%%%%%%%%%%%%%%%%%%%%%%%%%%%%%%%%%%%%%%%%%%%%%%%%%%%%%%%%%%%%%%%%%%%%%%%%%%%%%%%%%%%%%%%%%%
\subsection{Gauge coupling unification}
The Standard Model, apart from its endurance against experimental tests, is also a framework that allows
us to incorporate in the same lagrangian three out of the four fundamental forces in nature: the strong, 
weak and electromagnetic interactions. This unification is however incomplete. QCD on the one hand is just
a multiplicative factor for the electroweak sector, whereas even in the latter there are two distinct
gauge group factors resulting to two coupling constants.

This of course is by no means an original remark. Since the very early days of the standard model, a great
amount of research has been devoted to whether it would be possible to conceive a model with a simpler
gauge structure, perhaps based on a single semi-simple lie group, that can simultaneously describe all
three fundamental interactions (gravity is outside the scope of this work). Schematically,
one can imagine a large gauge group, having $SU(3)_C \times SU(2)_L \times U(1)_Y$ as one of its subgroups, 
breaking down at some scale in order to yield the SM along with some new matter and gauge bosons, 
which would probably be heavier than the ones we know.

\begin{figure}[ht]
\begin{center}
\includegraphics[width = 8cm]{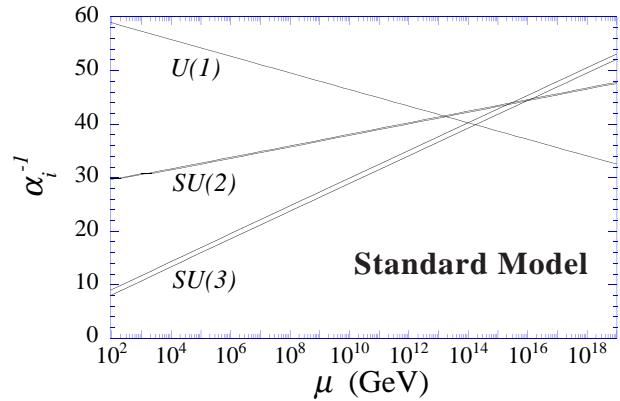}
\caption{{\footnotesize
Two-loop renormalization group evolution of the Standard Model gauge couplings. Figure taken from \cite{Martin:1997ns}.}}
\label{unif-SM}
\end{center}
\end{figure}

This point of view was encouraged by the findings of the LEP collider at CERN, at which a precise 
measurement of the three coupling constants was possible. 
The evolution of the coupling constants is described by the renormalization group equations. 
Setting $t = \ln \mu$, the general one-loop
form of the beta function $\beta_a$ associated to a generic group's coupling constant $g_a$ is
\begin{equation}
 \beta_a \equiv \frac{dg_a}{dt} = 
\frac{g_a^3}{16 \pi^2} 
\left( \sum_i l(R_i) - 3 C_2 (G) \right)
\label{RGEgeneral}
\end{equation}
where the summation is over the irreducible representations of the group $G$, $l(R_i)$ is the 
Dynkin index of the representation $R_i$ and $C_2 (G)$ is the quadratic Casimir of the adjoint
representation of $G$. In the case of the standard model, this expression becomes
\begin{equation}
 \beta_a = \frac{1}{16\,\pi^2}\,b_a\,g_a^3,
\end{equation}
with the $b_a$ coefficients being
\begin{equation}
\left(b_1,\,b_2,\,b_3\right)=(41/10,\,-19/6,\,-7)
\end{equation}
and the normalization of $g_1$ being chosen according to the one imposed by unification conditions
as the ones found in $SU(5)$. Then, it is customary to also define $\alpha_a\equiv g_a^2/(4\,\pi)$.
The evolution of $1/\alpha_a$'s can be 
seen in fig.\ref{unif-SM} taken from ref.\cite{Martin:1997ns}.

Starting from the precise LEP and LEP2 measurements, it has been rendered possible to perform
an accurate computation of the predicted behavior of the SM gauge coupling constants at high energy scales,
the result of this computation being shown in the figure. Very soon it was
understood that the three coupling constants have a manifest tendency to unify at a high energy scale, 
roughly from $10^{12}$ up to $10^{16}$ GeV. This tendency, although it could also clearly be an accident,
is however an important element supporting the idea of unification. It turns out however that the three
couplings do not exactly meet. Unification is incomplete within the framework of the Standard Model.

Supersymmetry (SUSY) (see, for example, \cite{Wess:1992cp, Bailin:1994qt, Mohapatra:1986uf, Martin:1997ns, Drees:1996ca}), 
and especially its low-energy variant, is perhaps the most popular way of addressing and
resolving the hierarchy and unification issues. 
In the following paragraphs we shall introduce the basic formalism needed and
see how SUSY can stabilize the Higgs mass and induce gauge coupling unification. Furthermore, we shall
focus on the solution SUSY can provide to the dark matter issue and examine the detection prospects
for one of the DM candidates appearing in supersymmetric frameworks, the lightest neutralino.

%%%%%%%%%%%%%%%%%%%%%%%%%%%%%%%%%%%%%%%%%%%%%%%%%%%%%%%%%%%%%%%%%%%%%%%%%%%%%%%%%%%%%%%%%%%%%%%%%%%%%%%%%%%%%
%%%%%%%%%%%%%%%%%%%%%%%%%%%%%%%%%%%%%%%%%%%%%%%%%%%%%%%%%%%%%%%%%%%%%%%%%%%%%%%%%%%%%%%%%%%%%%%%%%%%%%%%%%%%%
%%%%%%%%%%%%%%%%%%%%%%%%%%%%%%%%%%%%%%%%%%%%%%%%%%%%%%%%%%%%%%%%%%%%%%%%%%%%%%%%%%%%%%%%%%%%%%%%%%%%%%%%%%%%%
\section{Elements of SUSY formalism}
\subsection{Supersymmetric algebra and superspace}
The $S$ matrix describing transitions from one quantum state to another, possesses a certain number of
symmetries. In particular:
\begin{enumerate}
 \item It is invariant under the action of the elements of the Poincar\'e group, whose generators
are the translations and the Lorentz rotations $P_\mu$ and $M_{\mu\nu}$ respectively. These obey the
commutation relations
\begin{eqnarray}
\left[P_\mu,P_\nu\right]&=&0,\nonumber\\
\left[P_\mu,M_{\rho\sigma}\right]&=&i\,(\eta_{\mu\rho}\,P_\sigma-\eta_{\mu\sigma}\,P_\rho),\\
\left[M_{\mu\nu},M_{\rho\sigma}\right]&=&i\,(\eta_{\nu\rho}\,M_{\mu\sigma}-\eta_{\nu\sigma}\,M_{\mu\rho}-\eta_{\mu\rho}\,M_{\nu\sigma}+\eta_{\mu\sigma}\,M_{\nu\rho}).\nonumber
\end{eqnarray}
where $\eta_{\mu\nu}$ is the flat metric tensor.
 \item It is invariant under internal symmetries associated to some conserved quantum number (current)
according to Noether's theorem, such as the color, the electric charge etc. The generators of these
internal symmetries, say $T^a$, are Lorentz scalars and form Lie groups satisfying the relations
\begin{eqnarray}
\left[T^a,P_\mu\right]&=&0\nonumber\\
\left[T^a,M_{\mu\nu}\right]&=&0\\
\left[T^a,T^b\right]&=&i\,C^{abc}\,T^c\nonumber
\end{eqnarray}
where $C^{abc}$ are the group's structure constants.
\end{enumerate}
In 1967 Coleman and Mandula demonstrated a famous no-go theorem \cite{Coleman:1967ad} stating that apart from the Poincar\'e
group generators, the only conserved quantities in a local relativistic QFT with a mass gap can be a finite
number of Lorentz scalars associated with a Lie algebra of a compact Lie group 
(like the ones associated with the internal symmetries described above). This statement is valid if the theory
only contains commutators. In 1975, Haag, Lopuszanski and Sohnius demonstrated however that this restriction
can be evaded if one also includes anticommutators in the theory. Furthermore, they demonstrated that
the resulting extended Lie algebra (called \textit{supersymmetric algebra}) is the only one compatible 
with the symmetries of the $S$ matrix.

The most general supersymmetric algebra must contain the Poincar\'e group generators $P_\mu$ and $M_{\mu\nu}$
as well as a number $N$ of generators $Q^A$. It can be demonstrated that $N$ cannot
take arbitrary values, but should instead be $1, 2$ or $4$ in the case of global supersymmetry, or also
$8$ in the case of local supersymmetry. Despite the theoretical interest of theories with $N > 1$ 
(for example $N = 4$ SYM theories posess the maximal possible symmetry compatible with global supersymmetry
and are finite, i.e. their beta functions vanish; extended supersymmetry plays an important role in some
string constructions), since quite some time it has been known that they result
to non-chiral fermions. In this work, we shall hence only focus on $N = 1$ theories.
\\ \\
The algebra of the generators is \cite{Wess:1974tw}
\begin{eqnarray}\label{algebre}
\{\qa,\qbb\}&=&2\,\sigma^\mu_{\alpha\dot{\beta}}\,P_\mu,\nonumber\\
\{\qa,\qb\}&=&\{\qba,\qbb\}=0,\nonumber\\
\left[\qa,P_\mu\right]&=&\left[\qba,P_\mu\right]=0,\\
\left[\qa,M_{\mu\nu}\right]&=&\frac{1}{2}\,(\sigma_{\mu\nu})_\alpha^\beta\,\qb,\nonumber\\
\left[\qba,M_{\mu\nu}\right]&=&-\frac{1}{2}\,\qbb\,(\sigma_{\mu\nu})_{\dot{\alpha}}^{\dot{\beta}} \ , \nonumber
\end{eqnarray}
where the matrices $\sigma_{\mu\nu}\equiv\frac{1}{4}\,(\sigma_\mu\bar{\sigma}_\nu-\sigma_\nu\bar{\sigma}_\mu)$
are the generators of the special linear group of $2\times2$ complex matrices with unit determinant, $SL(2,\mathbb{C})$.
The first equation further shows us that the mass dimension of the $Q$ generators must be $1/2$.
The operators $Q_\alpha$ and $\bar{Q}_\alpha$ are Weyl spinors belonging to the representations 
$(1/2, 0)$ and $(0, 1/2)$ of the Lorentz group. Conventionally, we say that the former transforms as 
a left-handed Weyl spinor whereas the latter as a right-handed one.

An infinitesimal supersymmetric transformation of a field $\phi$ can be written by virtue of
two grassmann variables $\theta$ and $\bar{\theta}$ as
\begin{equation}
\delta_\xt\,\phi=(\ta\,\qa+\bar{\theta}_{\dot{\alpha}}\,\bar{Q}^{\dot{\alpha}})\,\phi.
\end{equation}

Now, since we are extending the symmetries of the $S$ matrix, it can be convenient to also introduce some
sort of ``extended'' space, called \textit{superspace}. Superspace in $N = 1$ supersymmetry
is described by the four usual spacetime coordinates
along with two Grassmann variables $\ta$ and $\tba$ which obey the anticommutation relations
 \begin{equation}
 \{\ta,\tb\}=\{\tba,\tbb\}=\{\ta,\tbb\}=0.
 \end{equation}
A global (non - position - dependent) supersymmetric transformation in superspace is then defined 
by ``exponentiation'' of the infinitesimal transformation as 
 \begin{equation}\label{transsusy}
 G(x^\mu,\xt,\bt)=\exp[i\,(-x^\mu\,P_\mu+\xt\,Q+\bt\,\bar{Q})].
 \end{equation}
where it is assumed that $\xt$ and $\bt$ are of dimension  $-1/2$.
The composition of two supersymmetric transformations in superspace is written as
\begin{equation}
G(x^\mu,\xt,\bt)\,G(y^\mu,\zeta,\bar{\zeta}) =
G(x^\mu+y^\mu+i\,\zeta\,\sigma^\mu\bt-i\,\xt\,\sigma^\mu\bar{\zeta},\xt+\zeta,\bt+\bar{\zeta}).
\end{equation}
In particular, the composition of two supersymmetric transformations one of which does not modify 
spacetime is 
\begin{equation}\label{g0g}
G(0,\zeta,\bar{\zeta})\,G(x^\mu,\xt,\bt)=G(x^\mu+i\,\xt\,\sigma^\mu\bar{\zeta}-i\,\zeta\,\sigma^\mu\bt,\xt+\zeta,\bt+\bar{\zeta}).
\end{equation}
Eq. \eqref{g0g} allows us to define a differential representation of the $\qa$ and $\qba$ operators, as
\begin{eqnarray}
\qa&=&\frac{\partial}{\partial\ta}-i\,(\sigma^\mu)_{\alpha\dot{\alpha}}\,\tba\,\partial_\mu,\nonumber\\
\qba&=&\frac{\partial}{\partial\tba}-i\,\ta\,(\sigma^\mu)_{\alpha\dot{\alpha}}\partial_\mu,
\end{eqnarray}
where $P_\mu\equiv -i\,\partial_\mu$.
Then, it is further possible to define covariant derivatives as
\begin{eqnarray}
\da&=&\frac{\partial}{\partial\ta}+i\,(\sigma^\mu)_{\alpha\dot{\alpha}}\,\tba\,\partial_\mu,\nonumber\\
\dba&=&-\frac{\partial}{\partial\tba}-i\,\ta\,(\sigma^\mu)_{\alpha\dot{\alpha}}\partial_\mu.
\end{eqnarray}
We should note that the supercharges and the covariant derivatives anticommute
\begin{equation}
\{D,Q\}=\{\bar{D},Q\}=\{D,\bar{Q}\}=\{\bar{D},\bar{Q}\}=0.
\end{equation}
%%%%%%%%%%%%%%%%%%%%%%%%%%%%%%%%%%%%%%%%%%%%%%%%%%%%%%%%%%%%%%%%%%%%%%%%%%%%%%%%%%%%%%%%%%%%%%%%%%%%%%%%%%%%%
\subsection{Superfields}
In a similar manner as we define fields as functions of spacetime, we can also define superfields $f(x,\xt,\bt)$ 
as functions of superspace. A superfield can in general be expanded in powers of $\xt$ and $\bt$.
This series cannot reach an arbitrarily high order, since the square of a Grassmann variable is zero
\begin{eqnarray}\label{schamp}
f(x,\xt,\bt)&=&z(x)+\xt\,\phi(x)+\bt\,\bar{\chi}(x)+\xt\xt\,m(x)+\bt\bt\,n(x)\nonumber\\
&+&\xt\,\sigma^\mu\,\bt\,A_\mu(x)+\xt\xt\bt\,\bar{\lambda}(x)+\bt\bt\xt\,\psi(x)+\xt\xt\bt\bt\,d(x),
\end{eqnarray}
with $\xt\xt\equiv\xt^a\,\xt_a=\xt^a\,\epsilon_{ab}\,\xt^b$ 
and $\bt\bt\equiv\bt_{\dot a}\,\bt^{\dot a}=\bt_{\dot a}\,\epsilon^{\dot a\dot b}\,\bt_{\dot b}$, 
where $\epsilon$ is an antisymmetric tensor defined as $\epsilon^{12}= \epsilon_{21}=1$.

A clarification is in order: the superfield $f$ should not be taken as a physical field, corresponding
to one specific particle. It is just a function of superspace, whose components form a \textit{supermultiplet}.
The components, on the other hand, can actually have physical meaning.

Now, $f$ contains both bosonic components, ($z$, $m$, $n$, $A_\mu$, $d$) as well as fermionic
ones ($\phi$, $\chi$, $\lambda$, $\psi$). The dimension of $z$ is the same as the superfield's $f$ one.
The dimensions of the other fields augment progressively with powers of $\xt$ and $\bt$ up to the
value $[f]+2$ for the field $d$.

The most general superfield as defined in \eqref{schamp} is a reducible representation of
the supersymmetry algebra. Next, it would be useful to construct the irreducible representations
by imposing conditions on $f$.

\subsubsection{The chiral superfield}
The chiral superfield takes its name by the chiral nature of the SM fermions. Since for chiral
fields their left- and right - handed components are independent, the superfield describing them 
should need two degrees of freedom in order to fully describe them. Left-handed chiral superfields 
$\Phi$ are defined as 
\begin{equation}\label{defchi}
\dba\Phi=0.
\end{equation}
If we define a bosonic coordinate $y^\mu\equiv x^\mu+i\xt\sigma^\mu\bt$, 
we notice that $\dba y^\mu=\dba\xt=0$. Hence, the chiral superfield can be written as
\begin{equation}\label{schampchi}
\Phi(y,\xt)=z(y)+\sqrt{2}\,\xt\,\psi(y)+\xt\xt\,F(y).
\end{equation}
If $\Phi$ is of dimension $1$, $z$ must be a physical complex scalar field, 
$\psi$ a left-handed Weyl spinor and $F$ and auxiliary field of dimension $2$.
Eq. \eqref{schampchi} can be written as
\begin{eqnarray}
\Phi(x,\xt,\bt)&=&z(x)+i\,\xt\sigma^\mu\bt\,\partial_\mu\,z(x)+\frac{1}{4}\,\xt\xt\bt\bt\,\square\,z(x)\nonumber\\
&+&\sqrt{2}\,\xt\,\psi(x)-\frac{i}{\sqrt{2}}\,\xt\xt\,\partial_\mu\psi(x)\,\sigma^\mu\,\bt+\xt\xt\,F(x).
\end{eqnarray}
It can further be demonstrated that the product of two chiral fields is a chiral field. In particular, 
the components $\xt\xt$ are invariant under SUSY transformations. They transform as a total derivative
\begin{eqnarray}\label{int}
\left.\Phi_i\,\Phi_j\right|_{\xt\xt}&=&z_i\,F_j+z_j\,F_i-\psi_i\,\psi_j,\\
\left.\Phi_i\,\Phi_j\,\Phi_k\right|_{\xt\xt}&=&z_i\,z_j\,F_k+z_k\,z_i\,F_j+z_j\,z_k\,F_i-\psi_i\,\psi_j\,z_k-\psi_k\,\psi_i\,z_j-\psi_j\,\psi_k\,z_i.\nonumber
\end{eqnarray}\s
\\ \\
In the same way as in Eq. \eqref{defchi}, we can define right-handed antichiral superfields
satisfying
\begin{equation}
\da\Phi^\dagger=0.
\end{equation}
Hence
\begin{equation}
\Phi^\dagger(y^\dagger,\bt)=z^*(y^\dagger)+\sqrt{2}\,\bt\,\bar{\psi}(y^\dagger)+\bt\bt\,F^*(y^\dagger).
\end{equation}

The product of a chiral and an antichiral superfield posess an interesting property: their $\xt\xt\bt\bt$
component contains the kinetic terms of $z$ and $\psi$
\begin{equation}\label{kin}
\left.\Phi_i\,\Phi_j^\dagger\right|_{\xt\xt\bt\bt}=F_i\,F^*_j+z^*_i\,\square\,z_j-\frac{i}{2}\,(\psi_i\,\sigma^\mu\,\partial_\mu\,\bar{\psi}_j-\partial_\mu\,\psi_i\,\sigma^\mu\,\bar{\psi}_j).
\end{equation}
Furthermore, this component transforms as a total derivative under supersymmetric transformations, 
it is therefore SUSY - invariant.

Using expressions \eqref{kin} and \eqref{int} we can construct the most general renormalizable Lagrangian
containing only chiral superfields
\begin{equation}\label{lchi}
\mathcal{L}=\left.\Phi_i^\dagger\,\Phi^i\right|_{\xt\xt\bt\bt}+\left[\lambda_i\,\Phi^i+\frac{m_{ij}}{2}\,\Phi^i\,\Phi^j+\frac{g_{ijk}}{2}\,\Phi^i\,\Phi^j\,\Phi^k+c.h.\right]_{\xt\xt}.
\end{equation}
The first factor corresponds to kinetic terms. The following terms correspond to the $\xt\xt$ component of the 
\textit{superpotential} $W$
\begin{equation}\label{spot}
W(\Phi^i)=\lambda_i\,\Phi^i+\frac{m_{ij}}{2}\,\Phi^i\,\Phi^j+\frac{g_{ijk}}{2}\,\Phi^i\,\Phi^j\,\Phi^k.
\end{equation}
An important remark is that the superpotential must be a holomorphic function of the superfields $\Phi^i$.
If we break it down to components, we can get the corresponding Lagrangian
\begin{eqnarray}
\mathcal{L}&=&i\,\partial_\mu\,\bar{\psi}_i\,\bar{\sigma}^\mu\,\psi^i+F^*_i\,F^i+z^*_i\square\,z^i\\
&+&\left[\lambda_i\,F^i+m_{ij}\,\left(z^i\,F^j-\frac{1}{2}\,\psi^i\,\psi^j\right)+g_{ijk}\,\left(z^i\,z^j\,F^k-\psi^i\,\psi^j\,z^k\right)+c.h.\right].\nonumber
\end{eqnarray}
The auxiliary fields $F$ and $F^*$ can be integrated out by means of their equations of motion
\begin{equation}
\frac{\partial\,\mathcal{L}}{\partial\,F^*}=0\qquad\text{et}\qquad\frac{\partial\,\mathcal{L}}{\partial\,F}=0
\end{equation}
which gives us an expression for $\mathcal{L}$ containing only the dynamical fields $z$ and $\psi$
\begin{equation}
\mathcal{L}=i\,\partial_\mu\,\bar{\psi}_i\,\bar{\sigma}^\mu\,\psi^i+z^*_i\square\,z^i+\frac{1}{2}\,\left(\frac{\partial^2\,W}{\partial\,z^i\,\partial\,z^j}\,\psi^i\,\psi^j+c.h.\right)-\mathcal{V}(z,z^*),
\end{equation}
where $\mathcal{V}(z,z^*)\equiv F^*_i\,F^i$ is the scalar potential.
This potential is manifestly positive: this is a consequence of supersymmetry. Its minimum
corresponds to $F^i\geq 0$.

\subsubsection{The vector superfield}
In order to describe the SM gauge bosons now, we introduce the vector superfields defined by their 
self-conjugation condition
\begin{equation}
V(x,\xt,\bt)=V^\dagger(x,\xt,\bt).
\end{equation}
In terms of components, according to Eq. \eqref{schamp}, a vector superfield is written as
\begin{eqnarray}
V(x,\xt,\bt)&=&C(x)+i\,\xt\,\chi(x)-i\,\bt\,\bar{\chi}(x)-\xt\,\sigma^\mu\,\bt\,v_\mu(x)\nonumber\\
&+&\frac{i}{2}\,\xt\xt\,\left[M(x)+i\,N(x)\right]-\frac{i}{2}\,\bt\bt\,\left[M(x)-i\,N(x)\right]\nonumber\\
&+&i\,\xt\xt\bt\,\left[\bar{\lambda}(x)+\frac{i}{2}\,\bar{\sigma}^\mu\,\partial_\mu\,\chi(x)\right]-i\,\bt\bt\xt\,\left[\lambda(x)+\frac{i}{2}\,\sigma^\mu\,\partial_\mu\,\bar{\chi}(x)\right]\nonumber\\
&+&\frac{1}{2}\,\xt\xt\bt\bt\,\left[D(x)+\frac{1}{2}\,\square\,C(x)\right];
\end{eqnarray}
where the fields $C$, $M$, $N$, $D$ and $v_\mu$ are real.
We should notice that the vector superfield is gauge invariant. The number of
degrees of freedom can be significantly reduced through gauge fixing. The Wess-Zumino gauge \cite{Wess:1974jb} is
a generalization of the usual unitary gauge and has the form
\begin{equation}\label{jwz}
V\rightarrow V+\Phi+\Phi^\dagger;
\end{equation}
where $\Phi$ is a non-physical chiral superfield that can be adjusted to eliminate $C$, $M$, $N$ and $\chi$.
This choice of gauge also implies that the fields $\lambda$ and $D$ are gauge-invariant, and that the vector
$v_\mu$ transforms as in the non-supersymmetric case:
\begin{equation}
v_\mu\rightarrow v_\mu-i\,\partial_\mu\,(z-z^*).
\end{equation}
\\ \\
So, in this gauge the vector superfield takes the form
\begin{equation}
V(x,\xt,\bt)=
-\xt\,\sigma^\mu\,\bt\,v_\mu(x)+
i\,\xt\xt\bt\,\bar{\lambda}(x)-i\,\bt\bt\xt\,\lambda(x)+
\frac{1}{2}\,\xt\xt\bt\bt\,D(x).
\end{equation}
This superfield is comprised of a gauge field $v_\mu$, a gaugino $\lambda$ and an auxiliary real field $D$.
The field strength tensor (which would be the equivalent of the usual 
$F^{\mu\nu}=\partial_\mu\,A_\nu-\partial_\nu\,A_\mu$)
is defined by means of gauge-invariant spinor fields
\begin{eqnarray}\label{vcin}
W_\alpha&=&-\frac{1}{4}\,\bar{D}\bar{D}D_\alpha\,V,\nonumber\\
\bar{W}_{\dot{\alpha}}&=&-\frac{1}{4}\,DD\bar{D}_{\dot{\alpha}}\,V.
\end{eqnarray}
We note that $\bar{D}_{\dot{\alpha}}\,W_\alpha=0$ and $D_\alpha\,\bar{W}_{\dot{\alpha}}=0$, 
which means that these fields are chiral and anti-chiral respectively.
In terms of components, we have
\begin{eqnarray}
W_\alpha&=&-i\,\lambda_\alpha(y)+\xt_\alpha\,D(y)-\frac{i}{2}
\,(\sigma^\mu\,\bar{\sigma}^\nu\,\xt)_\alpha\,v_{\mu\nu},\nonumber\\
\bar{W}_{\dot{\alpha}}&=&i\,\bar{\lambda}_{\dot{\alpha}}(y^+)+\bt_{\dot{\alpha}}\,\bar{D}(y^+)+
\frac{i}{2}\,(\sigma^\mu\,\bar{\sigma}^\nu\,\bt)_{\dot{\alpha}}\,v_{\mu\nu},
\end{eqnarray}
where $v_{\mu\nu}\equiv\partial_\mu v_\nu-\partial_\nu v_\mu$.
Since $W_\alpha$ is a chiral field, the component $\xt\xt$ of $W^\alpha\,W_\alpha$,
\begin{equation}
\left.W^\alpha\,W_\alpha\right|_{\xt\xt}=-2\,i\,
\lambda\,\sigma^\mu\,\partial_\mu\,\bar{\lambda}-
\frac{1}{2}\,v^{\mu\nu}\,v_{\mu\nu}+
\frac{i}{4}\,\epsilon^{\mu\nu\rho\sigma}\,v_{\mu\nu}\,v_{\rho\sigma},
\end{equation}
is SUSY-invariant, since it transforms as a total derivative.

From this last expression, we can construct the most general renormalizable Lagrangian containing only 
vector superfields
\begin{eqnarray}\label{lvec}
\mathcal{L}&=&\frac{1}{4}\,\left(\left.W^\alpha\,W_\alpha\right|_{\xt\xt}+\left.\bar{W}_{\dot{\alpha}}\,\bar{W}^{\dot{\alpha}}\right|_{\bt\bt}\right)\nonumber\\
           &=&\frac{1}{2}\,D^2-i\,\lambda\,\sigma^\mu\,\partial_\mu\,\bar{\lambda}-\frac{1}{4}\,v^{\mu\nu}\,v_{\mu\nu}.
\end{eqnarray}
where the auxiliary field $D$ can be eliminated by means of its equations of motion. This field
will also contribute to the scalar potential, without modifying its positivity.
%%%%%%%%%%%%%%%%%%%%%%%%%%%%%%%%%%%%%%%%%%%%%%%%%%%%%%%%%%%%%%%%%%%%%%%%%%%%%%%%%%%%%%%%%%%%%%%%%%%%%%%%%%%%%
\subsection{Interactions and supersymmetry breaking}
\subsubsection{Interactions}
Having so far examined theories with pure chiral or vector superfields, we can now construct 
a supersymmetric gauge - invariant Lagrangian describing interactions among chiral superfields $\Phi_i$ 
and vector superfields $V$.
\\ \\
Under the action of a non-abelian group $G$, a chiral superfield transforms as
\begin{equation}\label{tcin}
\Phi^i\rightarrow(e^{-i\,\Lambda})^i_j\,\Phi^j,\qquad\Phi^\dagger_i\rightarrow\Phi^\dagger_j\,(e^{i\,\Lambda^\dagger})^j_i,
\end{equation}
where $\Lambda^i_j\equiv (T^a)^i_ j\,\Lambda_a(x,\xt,\bt)$ is defined in terms of chiral superfields $\Lambda_a$.
The matrices $T^a$ are hermitian generators of $G$ in the representation in which $\Phi$ lives. In particular, 
in the adjoint representation
\begin{eqnarray}
Tr\left(T^a\,T^b\right)&=&C(r)\,\delta^{ab},\nonumber\\
\left[T^a,T^b\right]&=&i\,f^{abc}\,T^c.
\end{eqnarray}
Since the transformation law for the chiral superfield is \eqref{tcin}, the kinetic term $\Phi^\dagger_i\,\Phi^i$
appearing in the Lagrangian \eqref{lchi} is no longer gauge - invariant.
However, if we generalize the transformation law of $V$, Eq. \eqref{jwz}, for non-abelian groups as
\begin{equation}
e^V\rightarrow e^{-i\,\Lambda^\dagger}\,e^V\,e^{i\,\Lambda},
\end{equation}
we can construct a SUSY-invariant kinetic term as
\begin{equation}
\left.\Phi^\dagger_i\,\left(e^V\right)^i_j\,\Phi^j\right|_{\xt\xt\bt\bt}.\\
\end{equation}
\\ \\
The field strength tensor \eqref{vcin} for non abelian interactions must then be redefined as
\begin{eqnarray}
W_\alpha=-\frac{1}{4}\,\bar{D}\bar{D}\,e^{-V}\,D_\alpha\,e^V \ , \
%\nonumber\\
\bar{W}_{\dot{\alpha}}=-\frac{1}{4}\,DD\,e^{-V}\,\bar{D}_{\dot{\alpha}}\,e^V.
\end{eqnarray}
These field strength tensors in their turn transform as
\begin{equation}
W_\alpha\rightarrow W^\prime_\alpha=e^{-i\,\Lambda}\,W_\alpha\,e^{i\,\Lambda}.
\end{equation}
The Lagrangian for vector superfields is the same as in Eq. \eqref{lvec}, apart from the fact
that a trace must be taken over gauge indices.

Then, the most general renormalizable Lagrangian including gauge interactions among chiral and vector superfields
can be written as
\begin{eqnarray}
\mathcal{L}&=&\frac{1}{4\,C(r)}\,\left(\left.W^\alpha\,W_\alpha\right|_{\xt\xt}+\left.\bar{W}_{\dot{\alpha}}\,\bar{W}^{\dot{\alpha}}\right|_{\bt\bt}\right)+\left.\Phi^\dagger\,e^V\,\Phi\right|_{\xt\xt\bt\bt}\nonumber\\
&+&\left[\left.\left(\lambda_i\,\Phi^i+\frac{m_{ij}}{2}\,\Phi^i\,\Phi^j+\frac{g_{ijk}}{2}\,\Phi^i\,\Phi^j\,\Phi^k\right)\right|_{\xt\xt}+c.h.\right].
\end{eqnarray}

\subsubsection{Supersymmetry breaking}
A phenomenologically viable supersymmetric model should include terms breaking supersymmetry.
One possibility is that supersymmetry can be an exact symmetry of the theory, but which is spontaneously
broken by the vacuum choice. In this way, supersymmetry will not be manifest at low energies, especially
the electroweak scale which is of interest for us.
However, it turns out quite difficult to spontaneously break supersymmetry. 
% The definition of its
% algebra, equations \eqref{algebre}, implies that the Hamiltonian is defined in terms of the hypercharges
% \begin{equation}
% H=\frac{1}{4}\,\left(\qa\qba+\qba\qa\right).
% \end{equation}
% Supersymmetry is not broken once  $H|0\rangle=0$. This implies that the vacuum
% energy is zero. Inversely seen, if supersymmetry is broken spontaneously, the 
% vacuum state will have a positive energy. It can be demonstrated that SUSY is broken
% if $F_i$ or $D^a$ do not cancel in the vacuum. We thus search for models where the 
% equations $F_i=0$ and $D^a=0$ are not satisfied simultaneously.

There have been several proposals for spontaneous breaking of supersymmetry in the literature. 
These proposals always include new particles and interactions at some high energy scale.
The standard picture is that supersymmetry breaking occurs in some ``hidden sector'' which
does not communicate directly with the rest of the spectrum, the breaking being then ``mediated''
to the other parts of the Lagrangian through some \textit{messenger} sector.
The most well-known examples are gravity-mediated supersymmetry breaking \cite{Freedman:1976xh,Cremmer:1982en}, 
gauge-mediated breaking \cite{Giudice:1998bp}, anomaly-mediated breaking \cite{Randall:1998uk, Falkowski:2005ck}
or supersymmetry breaking induced by the existence of extra dimensions \cite{Quiros:2003gg}.
Combinations of these mechanisms can also be effective, often motivated by string constructions
\cite{Dudas:2008qf}.
However, there is no consensus on the mechanism that induces SUSY - breaking. It might be
that future experimental data shall help in this direction.

It is nonetheless possible to parametrize the effects of supersymmetry breaking at low
energies, introducing terms breaking it explicitly in the otherwise SUSY - invariant Lagrangian.
The new couplings must be \textit{soft}, so as to not reintroduce quadratic divergencies in the theory. 
In particular, we cannot introduce dimensionless couplings.

The most general renormalizable Lagrangian explicitly breaking supersymmetry, $\mathcal{L}_\text{soft}$
must include
\begin{itemize}
\item[\textbullet]  masses for the scalars $-m_{\phi_i}^2\,\left|\phi_i\right|^2$,
\item[\textbullet]  masses for the gauginos $-\frac{1}{2}\,m_{\lambda_i}\,\bar{\lambda}_i\lambda_i$,
\item[\textbullet]  trilinear scalar interactions $-A^{ijk}\,\phi_i\,\phi_j\,\phi_k+ c.h.$
\item[\textbullet]  bilinear scalar terms $-b^{ij}\,\phi_i\,\phi_j+c.h.$.
\end{itemize}

Fermionic terms could also in principle be included, but they can be absorbed by a 
redefinition of the superpotential, the scalar masses and the trilinear couplings.
We note that it has been rigorously demonstrated that a theory breaking supersymmetry
softly does not reintroduce quadratic divergencies in the perturbative expansion
\cite{Girardello:1981wz}.

The soft Lagrangian $\mathcal{L}_\text{soft}$ breaks supersymmetry, since it 
only contains scalars and gauginos without their superpartners. These soft terms
would induce a mass for the scalars and the gauginos, even in the absence of
mass terms for vector bosons and ordinary fermions.

%%%%%%%%%%%%%%%%%%%%%%%%%%%%%%%%%%%%%%%%%%%%%%%%%%%%%%%%%%%%%%%%%%%%%%%%%%%%%%%%%%%%%%%%%%%%%%%%%%%%%%%%%%%%%
%%%%%%%%%%%%%%%%%%%%%%%%%%%%%%%%%%%%%%%%%%%%%%%%%%%%%%%%%%%%%%%%%%%%%%%%%%%%%%%%%%%%%%%%%%%%%%%%%%%%%%%%%%%%%
%%%%%%%%%%%%%%%%%%%%%%%%%%%%%%%%%%%%%%%%%%%%%%%%%%%%%%%%%%%%%%%%%%%%%%%%%%%%%%%%%%%%%%%%%%%%%%%%%%%%%%%%%%%%%
\section{The Minimal Supersymmetric Standard Model}
\subsection{The MSSM}

The Minimal Supersymmetric Standard Model (MSSM)
\cite{Fayet:1976et,Fayet:1976cr,Fayet:1977yc,Nilles:1983ge,Martin:1997ns} 
is, as stated in its name, the simplest supersymmetric extension to the Standard Model.
It is minimal in the sense that it contains the smallest possible number of 
new fields.

The MSSM is based on the SM gauge group $\sut\times\sud\times
\uu$. Supersymmetry associates to every gauge boson a spin - $1/2$ fermion.
The gauge bosons belong to $8+3+1$ vector superfields, associated to the groups
$\sut$, $\sud$ and $\uu$. The superpartners of the gauge bosons are usually 
collectively called gauginos. In particular, the superpartners of the gluons are
called gluinos $\tilde g$, whereas the ones associated to the $W^\pm$, $W^3$ and
$B$ gauge bosons of the electroweak sector are called winos $\tilde W$ and 
binos $\tilde B$ respectively. In table \ref{svec} we indicate the quantum numbers
of the various vector superfields of the MSSM. As usually, the charge $Q_\text{em}$
associated with electromagnetism is given by the sum of the third component of the
isospin, $T_3$ and the hypercharge $Y$ of the particle.

\begin{table}[!h]
\centering
\begin{tabular}{|c||cc|ccccc|}
\hline
$\begin{array}{c}\mbox{Super-}\\\mbox{field}\end{array}$ & \multicolumn{2}{c|}{$\begin{array}{c}\mbox{Spin}\\1\qquad 1/2\end{array}$} & $\sut$ & $\sud$ & $T_3$ & $\uu$ & $Q_\text{em}$ \\
\hline\hline
$B$ & $B_\mu$ & $\tilde{B}$ & $1$ & $1$ & $0$ & $0$ & $0$ \\
$W$ & $\begin{array}{r}W_\mu^+\\W_\mu^3\\W_\mu^-\end{array}$ & $\begin{array}{r}\tilde{W}^+\\\tilde{W}^3\\\tilde{W}^-\\\end{array}$ & $1$ & $3$ & $\begin{array}{r}+1\\0\\-1\end{array}$ & $0$ & $\begin{array}{r}+1\\0\\-1\end{array}$ \\
$g$ & $g_\mu$ & $\tilde{g}$ & $8$ & $1$ & $0$ & $0$ & $0$ \\
\hline
\end{tabular}
\caption{Vector superfields of the MSSM}
\label{svec}
\end{table}

On the other hand, the matter content of the MSSM is described by chiral 
superfields. We follow the standard convention according to which the chiral
superfields are defined in terms of left-handed Weyl spinors, hence their charge
conjugates correspond to right-handed spinors. Two $\sud$ doublets ($Q_i$ and $L_i$) and three singlets 
($u^c_i$, $d^c_i$ and $e^c_i$) are needed to account for the different quarks and leptons.
The index $i=1,2,3$ corresponds to different fermion families. We note that
like the minimal SM, the MSSM does not contain right-handed neutrinos.

In the Standard Model, a single Higgs doublet field $H$ along with its conjugate is sufficient
to provide masses for all quarks and massive leptons.
In the MSSM, two Higgs doublets are instead necessary, often denoted as
$H_u$ and $H_d$. It is actually supersymmetry itself that imposes the introduction of
two doublets instead of one. This can be seen in two ways:

\begin{itemize}

\item In supersymmetry, the superpotential \eqref{spot} is a holomorphic
function in the superfields it contains. Hence, a Higgs supermultiplet
with isospin $Y = +1/2$ cannot give rise to Yukawa couplings that generate masses
for down-type quarks. The inverse applies to a $Y = -1/2$ Higgs supermultiplet, 
which can only generate masses for up-type quarks. The holomorphic nature of the
superpotential obliges us hence to introduce two distinct Higgs doublet fields.

\item The Higgs superpartners, called Higgsinos,  are expected (and actually \textit{do}) 
give rise to new contributions to the chiral anomaly. 
In order to achieve cancellation of these anomalies, one must have
\begin{equation}
\sum_\text{fermions}Y^3=\sum_\text{fermions}T_3^2\cdot Y=0.
\end{equation}
Now, whereas in the Standard Model the quark and lepton contributions cancel, 
in the MSSM case this is no longer valid. It is only after the introduction of
a second Higgs doublet that anomaly cancellation is actually achieved.
\end{itemize}

The chiral superfields and their quantum numbers are summarized in table \ref{schi}, which along
with table \ref{svec} sum up the particle content of the MSSM.

\begin{table}
\centering
\begin{tabular}{|c||cc|ccccc|}
\hline
$\begin{array}{c}\mbox{Super-}\\\mbox{field}\end{array}$ & \multicolumn{2}{c|}{$\begin{array}{c}\mbox{Spin}\\1/2\qquad 0\end{array}$} & $\sut$ & $\sud$ & $T_3$ & $\uu$ & $Q_\text{em}$ \\
\hline\hline
$Q$ & $\begin{array}{r}  u_L\\d_L\end{array}$ & $\begin{array}{r}  \tilde{u}_L\\\tilde{d}_L\end{array}$ & $3$ & $2$ & $\begin{array}{r}1/2\\-1/2\end{array}$ &  $1/6$ & $\begin{array}{r}2/3\\-1/3\end{array}$ \\
$u^c$ & $\bar{u}_R$ & $\tilde{u}_R^*$ & $\bar{3}$ & $1$ & $0$ & $-2/3$ & $-2/3$ \\
$d^c$ & $\bar{d}_R$ & $\tilde{d}_R^*$ & $\bar{3}$ & $1$ & $0$ &  $1/3$ &  $1/3$ \\
\hline
$L$ & $\begin{array}{r}\nu_L\\e_L\end{array}$ & $\begin{array}{r}\tilde{\nu}_L\\\tilde{e}_L\end{array}$ & $1$ & $2$ & $\begin{array}{r}1/2\\-1/2\end{array}$ & $-1/2$ & $\begin{array}{r}0  \\-1  \end{array}$ \\
$e^c$ & $\bar{e}_R$ & $\tilde{e}_R^*$ &       $1$ & $1$ & $0$ &    $1$ &    $1$ \\
\hline
$H_u$ & $\begin{array}{r}H_u^+\\H_u^0\end{array}$ & $\begin{array}{r}\tilde{H}_u^+\\\tilde{H}_u^0\end{array}$ & $1$ & $2$ & $\begin{array}{r}1/2\\-1/2\end{array}$ &  $1/2$ & $\begin{array}{r}1\\0 \end{array}$ \\
$H_d$ & $\begin{array}{r}H_d^0\\H_d^-\end{array}$ & $\begin{array}{r}\tilde{H}_d^0\\\tilde{H}_d^-\end{array}$ & $1$ & $2$ & $\begin{array}{r}1/2\\-1/2\end{array}$ & $-1/2$ & $\begin{array}{r}0\\-1\end{array}$ \\
\hline
\end{tabular}
\caption{The chiral superfields of the MSSM.}
\label{schi}
\end{table}

There is only one missing ingredient before writing down the MSSM Lagrangian density.
If supersymmetry where an exact symmetry, then every particle belonging to the same
supermultiplet would have the same mass as its supersymmetric partner. So, for example, 
there should exist selectrons $\tilde{e}_L$ and $\tilde{e}_R$ with masses $m_{\tilde e}=m_e\sim 0.51$ MeV.
However, it is quite apparent that if this was the case, such scalars should have been
observed. This means that if supersymmetry has anything to do with the physical world
at the electroweak scale, it must be broken.

The MSSM Lagrangian is thus comprised of two main pieces: The first
one is supersymmetric and contains all kinetic terms for chiral and vector superfields, 
as well as all terms that can be derived from the superpotential.
The second part contains all SUSY - breaking terms, which we argued should be soft.
So, we can write
\begin{equation}\label{mssm}
\mathcal{L}=\mathcal{L}_\text{kin}+\mathcal{L}_\text{W}+\mathcal{L}_\text{soft}.
\end{equation}
\\
Under the light of the previous discussion, the most general gauge-invariant superpotential of the MSSM is
\begin{eqnarray}\label{potentiel}
W=\sum_{i,j=1}^3\,\sum_{a,b=1}^2\left[\lambda^{ij}_u\,Q_{ai}\,\epsilon^{ab}\,H_{ub}\,u_j-\lambda^{ij}_d\,Q_{ai}\,\epsilon^{ab}\,H_{db}\,d_j\right.\nonumber\\
\left.-\lambda^{ij}_l\,L_{ai}\,\epsilon^{ab}\,H_{db}\,e_j+\mu\,H_{ua}\epsilon^{ab}\,H_{bd}\right].
\end{eqnarray}
where $\lambda_u$, $\lambda_d$ and $\lambda_l$ are complex $3\times 3$ matrices in
family space, corresponding to the Yukawa couplings, whereas the $\mu$ term
is a supersymmetric mass term for the Higgs doublets.
\\ \\
This superpotential is invariant under a discrete symmetry called $R$ - parity, defined as
\begin{equation}
R_p=(-1)^{2\,S+3\,(B-L)},
\label{Rparity}
\end{equation}
where $S$ is each particle's spin and $B$ and $L$ are the baryonic and leptonic numbers respectively.
$R$ - parity was first introduced to ensure baryon and lepton number conservation, 
so as to prevent rapid proton decay. Particles belonging to the same supermultiplet do not have
the same $R$ - parity: Standard Model particles have $R_p=1$, whereas their superpartners have
$R_p=-1$.
\\ \\
Apart from its initial motivation, $R$ - parity has very far-reaching consequences:
\begin{itemize}
\item  sparticles are forcedly produced in pairs.
\item  The Lightest Supersymmetric Particle (LSP) is completely stable.
\item  sparticles other than the LSP decay in an odd number of sparticles.
\end{itemize}

The structure of the MSSM Lagrangian is highly constraining for the parameters
it includes. However, the soft breaking terms generate a huge number of free
parameters. The relevant part of the Lagrangian is
\begin{eqnarray}\label{doux}
\mathcal{L}_\text{soft}=&-&\frac{1}{2}\left[M_1\,\tilde{B}\tilde{B}+\sum_{a=1}^3M_2\,\tilde{W}^a\tilde{W}_a+
\sum_{a=1}^8M_3\,\tilde{g}^a\tilde{g}_a+c.c.\right]\nonumber\\
&-&\sum_{i=1}^3\left[m_{\tilde{Q}_i}^2\,|\tilde{Q}_i|^2+m_{\tilde{L}_i}^2\,|\tilde{L}_i|^2+m_{\tilde{u}_i}^2\,
|\tilde{u}_i|^2+m_{\tilde{d}_i}^2\,|\tilde{d}_i|^2+m_{\tilde{e}_i}^2\,|\tilde{e}_i|^2\right]\nonumber\\
&-&m_{\tilde{H}_u}^2\,|\tilde{H}_u|^2-m_{\tilde{H}_d}^2\,|\tilde{H}_d|^2-(B\,\mu\sum_{a=1}^2\tilde{H}_u^a\,\tilde{H}_{da}+c.c.)\\
&-&\left[A_u^{ab}\,\tilde{Q}_{ai}\,\epsilon^{ij}\,\tilde{H}_{uj}\,\tilde{u}_b-A_d^{ab}\,\tilde{Q}_{ai}\,
\epsilon^{ij}\,\tilde{H}_{dj}\,\tilde{d}_b-A_e^{ab}\,\tilde{L}_{ai}\,\epsilon^{ij}\,\tilde{H}_{dj}\,\tilde{e}_b+c.c.\right].\nonumber
\end{eqnarray}
where in Eq. \eqref{doux} $M_1$, $M_2$ and $M_3$ correspond to the masses of the bino, the winos and the gluinos respectively.
The terms in the second line correspond to the mass terms for squark and sleptons. In the third line, 
there are new contributions to the Higgs potential. Finally, the fourth line corresponds to trilinear couplings 
among three scalars. We note that the $A_{u,d,l}$ factors are $3\times 3$ complex matrices in 
family space.
\\ \\
After having presented some basic elements of formalism, we shall now see how supersymmetry (and, notably, the MSSM)
provides solutions to the Standard Model issues mentioned previously: the Higgs mass hierarchy problem, the
unification of gauge couplings, as well as the dark matter problem.

%%%%%%%%%%%%%%%%%%%%%%%%%%%%%%%%%%%%%%%%%%%%%%%%%%%%%%%%%%%%%%%%%%%%%%%%%%%%%%%%%%%%%%%%%%%%%%%%%%%%%%%%%%%%%
\subsection{SUSY to the rescue!}
\subsubsection{Solution to the hierarchy problem}

Supersymmetry can offer a solution to the hierarchy problem discussed previously.
In the same way as gauge symmetry ``protects'' the masses of vector bosons and
chiral symmetry the ones of fermions from receiving quadratic divergencies, SUSY
protects the masses of scalars. We saw that the problem arose from the contributions
of fermion and gauge boson loop correction to the Higgs mass. It is well-known that
fermion loops always carry an extra factor of $-1$ with
respect to the case where a boson ``circulates'' in the loop.

So, for example, if we suppose $N_S$ scalar particles of mass $m_S$ and with
trilinear and quartic couplings $v\,\lambda_S$ and $\lambda_S$ respectively, 
their $1$-loop contribution to the Higgs mass shall be of the form
\begin{equation}
\Delta\,M_H^2=\frac{N_S\,\lambda_S}{16\,\pi^2}\left[-\Lambda^2+2\,m_S^2\,\log\frac{\Lambda}{m_S}\right]
-\frac{N_S\,\lambda_S^2\,v^2}{16\,\pi^2}\left[-1+2\,\log\frac{\Lambda}{m_S}\right]+\mathcal{O}\left(\frac{1}{\Lambda^2}\right),
\end{equation}
which also contains quadratic divergencies.

However, if we suppose that the Higgs couplings to the scalar particles have some relation with its
couplings to fermions of the form $|\lambda_f^2|=\lambda_S$, and that the number of bosonic and 
fermionic degrees of freedom is equal ($N_S=N_f$), 
the total radiative corrections induced by the presence of fermions and
scalars is
\begin{equation}
M_H^2 =m_H^2 + \frac{N_f\,\lambda_f^2}{4\,\pi^2} \left[(m_f^2-m_S^2)\,\log\frac{\Lambda}{m_S}+3\,m_f^2\,
\log\frac{m_S}{m_f}\right]+\mathcal{O}\left(\frac{1}{\Lambda^2}\right).
\end{equation}

At this point, we see that the quadratic divergencies have disappeared.
We remark that logarithmic divergencies are still there, but even
if the theory's cutoff is pushed to the Plank scale, these remain quite moderate.
An important remark is, though, that in order to have exact cancellation of
the quadratic divergencies a very strong condition is imposed, namely that
$m_S=m_f$. This last condition is however in straight contradiction to our
discussion so far, since we said that supersymmetry must be broken and that
sparticles should receive additional contributions with respect to their
SM counterparts if we wish for a phenomenologically viable theory.

This last point motivates electroweak scale supersymmetry: the superpartners
must not be much heavier than the corresponding SM particles, since this would
destabilize the Higgs boson mass once more. As we shall see in the following, 
even more complications may appear, concerning mostly the mass of the lightest
Higgs boson, both from an experimental and a theoretical point of view.

\subsubsection{Gauge coupling unification}

Another point where the MSSM turns out to be successful is the unification of
coupling constants. The renormalization group equations for the three gauge couplings 
$g_1$, $g_2$ and $g_3$ are again given by Eq. \eqref{RGEgeneral}, but this time
the coefficients $b_a$ are different than in the Standard Model case:
\begin{equation}
\left(b_1,\,b_2,\,b_3\right)=\left\{\begin{array}{l}(41/10,\,-19/6,\,-7)
\qquad\text{MS}\\(33/5,\,1,\,-3)\hspace{2.0cm}\text{MSSM.}\end{array}\right.
\end{equation}
\begin{figure}[!h]
\centering
\includegraphics[width=0.55\textwidth,clip=true]{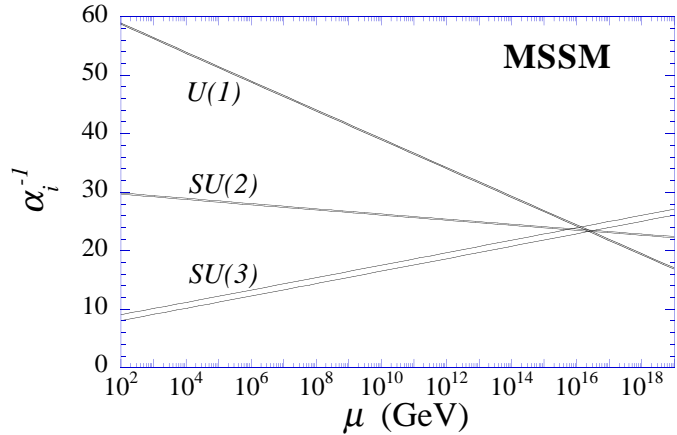}
\vspace*{-2mm}
\caption{\footnotesize{
Gauge coupling constant evolution in the MSSM. 
One and two loop corrections are included in the running. Figure taken from ref.\cite{Martin:1997ns}.
}}
\label{unificag-MSSM}
\vspace*{-3mm}
\end{figure}
The coefficients in the MSSM case are larger, due to the larger number of particles 
contributing to the beta-functions. In fig.\ref{unificag-MSSM} we can see a
comparison of the RGE evolution of the $\alpha_a^{-1}$'s, at two loops
in perturbation theory. Contrary to the Standard Model case, the MSSM contains
the right number of fields to ensure unification of the gauge couplings at some
large scale, called the Grand Unification scale $\mgut\sim 2\cdot 10^{16}$ GeV.
Unification of coupling constants at some scale could indicate the existence of 
some Grand Unified Theory (GUT) at the scale where the coupling constants acquire
the same value.

This is a further motivation for weak-scale supersymmetry: it has been
demonstrated that gauge coupling unification is not a general feature of supersymmetric
theories and models. In order to achieve it, the superpartner contributions must enter
the RGE running at a scale being sufficiently low so as to leave enough ``space'' for the
coupling constants to evolve in this way.

%%%%%%%%%%%%%%%%%%%%%%%%%%%%%%%%%%%%%%%%%%%%%%%%%%%%%%%%%%%%%%%%%%%%%%%%%%%%%%%%%%%%%%%%%%%%%%%%%%%%%%%%%%%%%
\subsection{The physical particles of the MSSM}
Before setting off to examine dark matter in the framework of the MSSM, it is useful to discuss the
physical particle spectrum of the model, i.e. the mass eigenstates appearing in the Lagrangian. 
Tables \ref{svec} and \ref{schi} summarize the various gauge eigenstates of the model. We know
that after EWSB several particles of the Standard Model that are gauge eigenstates are no longer mass
eigenstates, since they appear in quadratic crossed terms in the Lagrangian. This is also the case
for the MSSM, with mixing effects being even more extended than in the SM. As a first remark, we note
that the SM particles' definitions are not altered at tree level (apart, of course, from the
Higgs sector). We shall now describe what happens
with the rest of the particle content. Our intention is not to provide a detailed description, but rather
to summarize a number of elements that shall be useful for the following, namely to just identify the
physical degrees of freedom.

\subsubsection{The Higgs fields}
We saw that in the MSSM two Higgs doublets are required, which we denote by $H_u \equiv H_2 = (H_2^+, H_2^0)^T$ 
and $H_d \equiv H_1 = (H_1^0, H_1^-)^T$. Upon EWSB, the neutral components of the two doublets acquire 
non-zero VEVs $v_1/\sqrt{2}$ and $v_2/\sqrt{2}$ for $H_1^0$ and $H_2^0$ respectively, with $(v_1 + v_2)^2 = 
v^2 = (246 \ \mbox{GeV})^2$. We further define the parameter
\begin{equation}
 \tan\beta = \frac{v_2}{v_1}
\end{equation}
As usually, the physical Higgs fields are obtained by expanding around the scalar potential's minimum
\begin{eqnarray}
 H_1 & = & (H_1^0, H_1^-)^T = \frac{1}{\sqrt{2}} (v_1 + H_1^0 + i P_1^0, H_1^-)^T\\ \nonumber
 H_2 & = & (H_2^+, H_2^0)^T = \frac{1}{\sqrt{2}} (H_2^+ , v_2 + H_2^0 + i P_2^0)^T
\end{eqnarray}
with the real parts corresponding to $CP$ - even Higgs bosons and the imaginary parts to $CP$ - odd ones
and the goldstone bosons.

Then, the physical fields/mass eigenstates of the model can be expressed as linear combinations of the
gauge eigenstates, as follows:
\begin{itemize}
 \item First, we can write down the expression for the neutral goldstone boson and the $CP$ - odd  
Higgs
\begin{equation}
\left( \begin{array}{c}   G^0 \\ A \end{array} \right) 
= \left( \begin{array}{cc} \cos \beta & \sin \beta \\
- \sin \beta & \cos \beta \end{array} \right) \ 
\left( \begin{array}{c}   P_1^0 \\ P_2^0 \end{array} \right)
\end{equation}
 \item Then, we have the charged Goldstone bosons and the charged Higgses
\begin{equation}
\left( \begin{array}{c}   G^\pm \\ H^\pm \end{array} \right) 
= \left( \begin{array}{cc} \cos \beta & \sin \beta \\
- \sin \beta & \cos \beta \end{array} \right) \ 
\left( \begin{array}{c}   H_1^\pm \\ H_2^\pm \end{array} \right) 
\end{equation}
 \item Finally, we have the two physical $CP$ - even Higgs bosons
\begin{equation}
\left( \begin{array}{c}   H \\ h \end{array} \right) 
= \left( \begin{array}{cc} \cos \alpha & \sin \alpha \\
- \sin \alpha & \cos \alpha \end{array} \right) \ 
\left( \begin{array}{c}   H_1^0 \\ H_2^0 \end{array} \right) 
\end{equation}
\end{itemize}
where $\alpha$ is a rotation angle. In principle, the Higgs sector should contribute 
six new free parameters to the theory. It turns out however that there are several relations among
them, which amount to only two free parameters, often taken to be $\tan\beta$ and $M_A$.

\subsubsection{The sfermions}
Sfermions can also mix, since they share quantum numbers. The mass matrices have the general form
\begin{equation} 
\label{sqmass_matrix}
{\cal M}^2_{\tilde{f}} =
\left(
  \begin{array}{cc} m_f^2 + m_{LL}^2 & m_f \, X_f  \\
                    m_f\, X_f    & m_f^2 + m_{RR}^2 
  \end{array} \right) 
\end{equation}
where
\begin{equation}
\begin{array}{l} 
\ m_{LL}^2 =m_{\tilde{f}_L}^2 + (I^{3L}_f - Q_f s_W^2)\, M_Z^2\, c_{2\beta} \\\
m_{RR}^2 = m_{\tilde{f}_R}^2 + Q_f s_W^2\, M_Z^2\, c_{2\beta} \\\
\ \ X_f  = A_f - \mu (\tan\beta)^{-2 I_f^{3L}} \ . \ \
\label{mass-matrix}
\end{array}
\end{equation}
These mass matrices are diagonalized by means of $2 \times 2$ unitary matrices 
\begin{equation}
 R^{\tilde f} =  \left( \begin{array}{cc}
     c_{\theta_f} & s_{\theta_f} \\ - s_{\theta_f} & c_{\theta_f}
  \end{array} \right)  \ \ \ \ , \ \ c_{\theta_f} \equiv \cos \theta_{\tilde f} 
\ \ {\rm and} \ \ s_{\theta_f} \equiv \sin \theta_{\tilde f} \ . \ \
\end{equation}
The mixing angle and sfermion masses are then given by 
\begin{equation}
s_{2\theta_f} = \frac{2 m_f X_f} { m_{\tilde{f}_1}^2
-m_{\tilde{f}_2}^2 } \ \ , \ \ 
c_{2\theta_f} = \frac{m_{LL}^2 -m_{RR}^2} 
{m_{\tilde{f}_1}^2 -m_{\tilde{f}_2}^2 }
\end{equation}
and
\begin{equation}
m_{\tilde{f}_{1,2}}^2 = m_f^2 +\frac{1}{2} \left[
m_{LL}^2 +m_{RR}^2 \mp \sqrt{ (m_{LL}^2
-m_{RR}^2 )^2 +4 m_f^2 X_f^2 } \ \right] \ . \ \
\end{equation}
The mixing is very strong in the stop sector for large values of the parameter
$X_t=A_t- \mu \cot \beta$ and generates a mass splitting between the two mass 
eigenstates which makes the state $\tilde{t}_1$ much lighter
than the other squarks and possibly even lighter than the top quark itself.
These points shall be of interest later on, when we discuss the so-called 
``little hierarchy problem''.

\subsubsection{Gaugino and Higgsino sector}
The two charged Winos as well as the two charged Higgsinos can mix to yield four fermionic
mass eigenstates called charginos. These eigenstates have a tree-level mass matrix
\begin{equation}
{\cal M}_\pm =
  \begin{pmatrix}
    M_2 & \sqrt{2} M_W s_\beta \\
    \sqrt{2} M_W c_\beta & \mu
  \end{pmatrix},
\label{mchar}
\end{equation}
where $M_W$ is the $W$ boson mass and $s_\beta = \sin\beta$.

Next, the four gauge eigenstates $(\tilde{B},\tilde{W}^3,  \tilde{H}_u^0, \tilde{H}_d^0)$ can also
mix, giving rise to four mass eigenstates collectively called neutralinos. At tree-level, the neutralino
mass matrix is given by
\begin{equation}
{\cal M}_0 =
 \begin{pmatrix}
    M_1 & 0 & -M_Z s_W c_\beta & M_Z s_W s_\beta \\
    0 & M_2 & M_Z c_W c_\beta & -M_Z c_W s_\beta \\
    -M_Z s_W c_\beta & M_Z c_W c_\beta & 0 & -\mu \\
    M_Z s_W s_\beta & -M_Z c_W s_\beta & -\mu & 0
  \end{pmatrix}.
\label{mneutra}
\end{equation}\\
where $s_\beta = \sin\beta$, $c_\beta = \cos\beta$, $s_W = \sin\theta_W$,  $c_W = \cos\theta_W$,
$\theta_W$ is the Weinberg angle and $M_Z$ is the $Z$ boson mass.

The matrix \eqref{mneutra} is complex symmetric and can thus be diagonalized analytically 
by a unitary matrix $Z_0$ as
\begin{equation}
 {\cal M}_0 = Z_0 D_0 Z_0^\dag
\end{equation}
The exact form of $Z_0$ is, in the general case, quite complex. In the end, one gets
four mass eigenstates that we shall hereafter denote $\chi_i^0 \ ,  \ i = 1...4$, with
$m_1^0 < m_2^0 < m_3^0 < m_4^0$.
\\ \\
Having presented the physical particle content of the MSSM, we can next wonder whether some
of these particles could account for the observed dark matter abundance.

%%%%%%%%%%%%%%%%%%%%%%%%%%%%%%%%%%%%%%%%%%%%%%%%%%%%%%%%%%%%%%%%%%%%%%%%%%%%%%%%%%%%%%%%%%%%%%%%%%%%%%%%%%%%%
%%%%%%%%%%%%%%%%%%%%%%%%%%%%%%%%%%%%%%%%%%%%%%%%%%%%%%%%%%%%%%%%%%%%%%%%%%%%%%%%%%%%%%%%%%%%%%%%%%%%%%%%%%%%%
%%%%%%%%%%%%%%%%%%%%%%%%%%%%%%%%%%%%%%%%%%%%%%%%%%%%%%%%%%%%%%%%%%%%%%%%%%%%%%%%%%%%%%%%%%%%%%%%%%%%%%%%%%%%%
\section{Dark Matter in the MSSM}
So far, we have mainly focused on the completely minimal supersymmetric extension of the Standard Model.
In short, for every bosonic or fermionic degree of freedom in the SM, $N = 1$ global supersymmetry introduces 
a fermionic or bosonic one. Nevertheless, there are obviously numerous other extensions that can be envisaged,
depending on each author's concerns. In any case, whatever the precise model, it is reasonable to look for
candidates which are neutral under color and electromagnetism, as well as most probably stable (or at least
with couplings weak enough to prevent their rapid decay).

In the MSSM framework we presented, the imposition of $R$-parity renders the Lightest Supersymmetric
Particle completely stable. Potential candidates could be (referring to the physical states now) sneutrinos 
(left-handed, since we have not written down terms for right-handed neutrinos) or neutralinos. 
Furthermore, we note that once supersymmetry is rendered local, an additional candidate can
be found, the graviton's superpartner called gravitino. Gravitino dark matter has been studied
in a series of different frameworks, such as the MSSM \cite{Luo:2010he}, extensions of the former
that solve the so-called ``$\mu$ problem'' and/or the neutrino mass problem 
\cite{Choi:2009ng,Munoz:2009qp, Buchmuller:2007ui, Covi:2009xn, Covi:2009bk} and so on.
Then, one could aim at introducing right - handed
neutrinos in order to achieve a see-saw mechanism yielding small neutrino masses, in which case the
right-handed neutrino or sneutrino could enter the game (see, for example, \cite{Cerdeno:2008ep}). 
Finally, we should also mention that
in theories trying to address the absence of $CP$ violation in the strong sector via the introduction
of a Peccei - Quinn symmetry, the associated boson called axion is a plausible candidate. Although today
axions are rather constrained, if such a model is rendered supersymmetric, the axion's superpartner called
axino can provide a good candidate \cite{Covi:2004rb}.

Among these candidates, the most well - known and widely studied is the lightest neutralino
(often also called just neutralino).

\subsection{Neutralino dark matter}
We already mentioned that all neutralinos are linear combinations of the superpartners of the SM
neutral gauge bosons and the neutral Higgs bosons. In a generic manner, we can write
\begin{equation}
 \chi_1^0 = Z_{11} \tilde{B} + Z_{12} \tilde{W}^3 + Z_{13} \tilde{H}_1^0 + Z_{14} \tilde{H}_2^0
\label{NeutralinoGeneric}
\end{equation}
where we assume that we have rearranged the neutralino matrix in order to have the lightest one
at the top row. Since neutralinos are comprised of four distinct contributions, all of which contribute
to the total couplings in different manners, it is customary to further define two quantities representing
the neutralino composition, namely the gaugino fraction and the Higgsino fraction
\begin{eqnarray}
 f_G & = & \left|Z_{11}\right|^2 + \left|Z_{12}\right|^2 \ , \ \ \\ \nonumber
 f_H & = & \left|Z_{13}\right|^2 + \left|Z_{14}\right|^2 \ . \ \
\end{eqnarray}
\\ \\
A huge amount of literature has been devoted to the study of neutralino dark matter. It is quite difficult
to present an overview of the relevant phenomenology in the case of the most general $R$ - parity conserving MSSM, 
since the number of free parameters in the general case is of the order of $120$, which is an uncontrollably
large parameter space to be probed efficiently.

The most usual approach towards supersymmetric phenomenology is to make simplifying assumptions, 
often (but not exclusively) motivated by some higher theory, aiming at the reduction of the number
of free parameters. One such example is the so-called Constrained MSSM (CMSSM). In this model, 
it is assumed that 
\begin{itemize}
 \item The three gauge couplings (properly normalized) meet at the Grand Unification Scale
 \item The Bino, Wino and gluino masses are universal at the GUT scale $M_1(\mgut) = M_2(\mgut) = M_3(\mgut) \equiv m_{1/2}$
 \item All scalar masses unify at the GUT scale
\begin{eqnarray}
m_{\tilde{Q}_i} (M_{GUT}) &=& m_{\tilde{u}_{Ri}} (M_{GUT}) =
m_{\tilde{d}_{Ri}}(M_{GUT})  =m_{\tilde{L}_i} (M_{GUT}) 
= m_{\tilde{\ell}_{Ri}} (M_{GUT}) \\ \nonumber
&=& m_{H_1}(M_{GUT}) =m_{H_2} (M_{GUT}) \equiv  m_0
\end{eqnarray}
 \item Trilinear couplings are universal at the GUT scale
\begin{equation}
A^u_{ij} (M_{GUT}) = A^d_{ij} (M_{GUT}) = A^\ell_{ij} (M_{GUT}) \equiv  A_0 \, 
\delta_{ij}
\end{equation}
\end{itemize}
If one further assumes a specific relation among the bilinear and trilinear soft breaking terms
as well as a relation between the gravitino and scalar masses, the resulting model is called
minimal supergravity (mSUGRA).

In the case of the CMSSM, the resulting model is described by only five free parameters:
$\tan\beta$, $m_{1/2}$, $m_0$, $A_0$ and $\mbox{sign}(\mu)$, where the last parameter is the
sign of the Higgsino mass parameter. It can be demonstrated that demanding radiative electroweak 
symmetry breaking, minimization of the Higgs potential yields
\begin{eqnarray}
 B \mu & = & \frac{1}{2} 
\left[ 
(m_{H_1}^2 - m_{H_2}^2)\tan 2\beta + M_Z^2\sin 2\beta
\right]\\
 \mu^2 & = & 
\frac{m_{H_2}^2 \sin^2\beta - m_{H_1}^2 \cos^2\beta}{\cos2\beta}
- \frac{M_Z^2}{2}
\end{eqnarray}
which demonstrates that while the absolute value of $\mu$ is fixed, its sign remains
unknown.

\begin{figure}[ht]
\begin{center}
\includegraphics[width = 8cm]{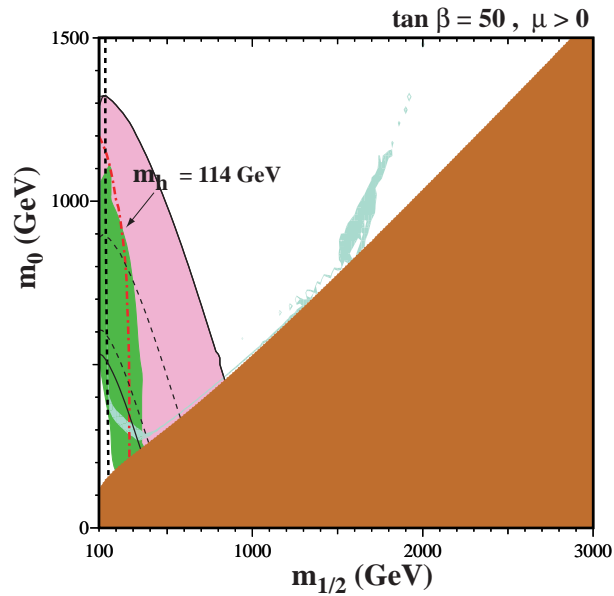}
\caption{{\footnotesize
$(m_{1/2}, m_0)$ plane for $\tan\beta = 50$, $\mu > 0$ and $A_0 = 0$. Red dot-dashed lines
correspond to $m_h = 114$ GeV and black dashed lines to $m_{\chi^\pm} = 104$ GeV. The brown 
region yields a stau LSP, the dark green region is excluded by $b \rightarrow s \gamma$ and
the pink region is favored by the muon anomalous magnetic moment measurements. The turquoise
regions yield the correct relic density. Figure taken from \cite{Ellis:2010kf}
}}
\label{fig:CMSSMdark}
\end{center}
\end{figure}

In order to describe some typical features of neutralino
dark matter in the CMSSM, we borrow figure \ref{fig:CMSSMdark} from \cite{Ellis:2010kf}.
Non-dark matter - related constraints appearing in the figure are explained in the caption.

We can clearly see from the figure that whereas the regions where the neutralino becomes the LSP
are quite significant,
the WMAP limits strongly restrict the viable parameter space. The WMAP - compliant regions
are represented in turquoise. It is interesting that despite 
the relatively small number of
surviving points, the processes contributing to getting the correct relic density vary from
one region to another. Let us start the description of these zones by stating that in most of
the parameter space, the CMSSM yields too large a relic density. The neutralino LSP is mostly
bino, and a bino LSP couples very weakly to both the $Z$ and the Higgs bosons. The couplings
of the neutralino to various MSSM particles are presented in Appendix \ref{NeutralinoCouplings}.
Some enhancement is thus needed
in order to obtain WMAP-compliant results. For small $(m_{1/2}, m_0)$ values, the correct 
relic density is obtained through crossed-channel sfermion exchange. This is called the 
\textit{bulk region}. As $m_{1/2}$ increases, the correct relic density is obtained near the region
where the stau becomes the LSP. In this case, the dominant process
enhancing neutralino annihilation is actually its coannihilation with the lightest $\tilde{\tau}$, 
usually called the NLSP (Next-to LSP). This region is called the \textit{coannihilation region}. At larger
$(m_{1/2}, m_0)$ values and away from the coannihilation region, the self-annihilation cross-section
gets enhanced kinematically, since in this region two neutralinos can annihilate efficiently
into a nearly on-shell Higgs propagator $H$ or $A$. This region is called the \textit{funnel region}.
Finally, there is a fourth region where the correct relic density can be obtained, called the
\textit{focus point/hyperbolic branch} region, where the neutralino obtains a significant Higgsino fraction, 
resulting in an enhancement of its couplings to gauge and Higgs bosons. 
\\ \\
For the moment, we pause our discussion on neutralino dark matter in order to present an issue 
of the CMSSM related to the Higgs boson mass, motivating the models that we shall be looking into
in the following.

%%%%%%%%%%%%%%%%%%%%%%%%%%%%%%%%%%%%%%%%%%%%%%%%%%%%%%%%%%%%%%%%%%%%%%%%%%%%%%%%%%%%%%%%%%%%%%%%%%%%%%%%%%%%%
%%%%%%%%%%%%%%%%%%%%%%%%%%%%%%%%%%%%%%%%%%%%%%%%%%%%%%%%%%%%%%%%%%%%%%%%%%%%%%%%%%%%%%%%%%%%%%%%%%%%%%%%%%%%%
%%%%%%%%%%%%%%%%%%%%%%%%%%%%%%%%%%%%%%%%%%%%%%%%%%%%%%%%%%%%%%%%%%%%%%%%%%%%%%%%%%%%%%%%%%%%%%%%%%%%%%%%%%%%%
\section{A parenthesis: the little hierarchy problem}
\label{LittleHierarchyProblem}
Starting from the discussion on the physical spectrum of the MSSM, it is quite straightforward to 
compute the model's tree-level prediction for the neutral $CP$ - even Higgs boson masses.
The result is
\begin{equation}
 m_{h, H}^2 = \frac{1}{2} 
\left[
m_Z^2 + m_A^2 \mp 
\sqrt{(m_A^2 - m_Z^2)^2 + 4 m_A^2 m_Z^2 \sin^2 2 \beta}
\right]
\label{TreeHiggs}
\end{equation}
From this equation we can see an immediate problem: denoting the lightest of the
two Higgses by $h$, we see that its mass is forcedly lower than the $Z$ boson mass.
LEP2 has posed the most stringent existing bounds on the lightest Higgs boson mass
for the Standard Model \cite{Barate:2003sz} and the MSSM \cite{Schael:2006cr}. In the former case, this
limit is very stringent $m_h > 114.4$ GeV. In the MSSM case, the situation turns out
to be slightly more complicated.

\begin{figure}[htb]
%\vspace{1cm}
\begin{center}
\includegraphics[width = 10cm]{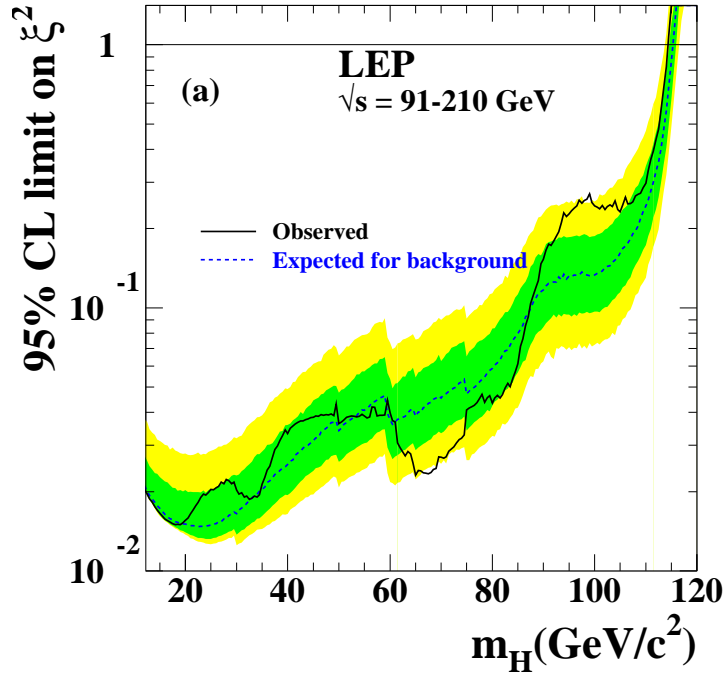}
\caption{{\footnotesize
$95\%$ CL upper bound on the ratio $\xi^2 = (g_{hZZ}/g_{hZZ}^{SM})^2$.
Figure taken from \cite{Schael:2006cr}.
}}
\label{fig:LEPHiggsBSM}
\end{center}
\end{figure}

These bounds come mainly from two direct search channels, namely Higgsstrahlung and associated 
production of $h$ and $A$. For later use, we note that the cross-sections for these processes
are, in comparison to the Standard Model Higgsstrahlung case
\begin{eqnarray}
 \sigma(e^+ e^- \rightarrow hZ) & = & g^2_{hZZ} \sigma_{\mbox{\begin{tiny}SM\end{tiny}}}(e^+ e^- \rightarrow hZ)\\ \nonumber
 \sigma(e^+ e^- \rightarrow hA) & = & g^2_{hAZ} \sigma_{\mbox{\begin{tiny}SM\end{tiny}}}(e^+ e^- \rightarrow hZ) \times
\frac{\lambda_{Ah}^2}{\lambda_{Zh} (\lambda_{Zh}^2 + 12 M_Z^2/s)}
\label{HiggsProdLEP}
\end{eqnarray}
where $m_Z$ is the $Z$ boson mass, $s$ is the center-of-mass energy of the collision, 
$\sigma_{\mbox{\begin{tiny}SM\end{tiny}}}(e^+ e^- \rightarrow hZ)$ is the Higgsstrahlung 
cross-section in the SM, $\lambda_{ij} = (1 - M_i^2/s - M_j^2/s)^2 - 4 M_i^2 M_j^2/s^2$ and
the two couplings $g_{ijk}$ are given by the Higgs mixing angles
\begin{eqnarray}
 g^2_{hZZ} & = & \sin^2(\beta - \alpha)\\ \nonumber
 g^2_{hAZ} & = & \cos^2(\beta - \alpha)\\ \nonumber
\end{eqnarray}
The LEP2 bound $m_h \gtrsim 114$ GeV is actually also applicable to the MSSM, in the 
case where $g^2_{hZZ} = {\cal{O}}(1)$.
\\ \\
In fig.\ref{fig:LEPHiggsBSM}, taken from \cite{Schael:2006cr}, the LEP collaboration 
calculates the limits on the squared ratio
$\xi^2 = (\sigma_{hZZ}/\sigma_{hZZ}^{SM})^2$ between a generic BSM $hZZ$ coupling and
the Standard Model one as a function of the Higgs boson mass. As we can see from
the figure, once the squared coupling becomes smaller than $1$, there is indeed much 
space for lighter $h$ masses. Referring to Eq.\eqref{HiggsProdLEP}, the $\xi$ factor
would be simply $\sin(\beta - \alpha)$.

The problem in CMSSM and mSUGRA-like models arises from the fact that in these models the suppression
factor is typically ${\cal{O}}(1)$. An example analysis is presented in \cite{Kane:2004tk}, 
where it can be seen that a Light Higgs Scenario (LHS) cannot be realized within the 
framework of mSUGRA without introducing tensions with other observables.
\\ \\
One the other hand, Eq.\eqref{TreeHiggs} is a tree-level prediction.
Once one starts taking seriously into account the quantum nature
of the MSSM and includes radiative corrections to the Higgs mass, these turn out to be quite sizeable
for some regions of the CMSSM parameter space. However,
even in this way, there are some requirements which often appear to be quite restrictive \cite{Djouadi:2005gj}:
\begin{itemize}
 \item From Eq.\eqref{TreeHiggs} we can see that the large $\tan\beta$ regime is rather favored, 
since then the lightest Higgs mass approaches its highest tree-level limit, the $Z$ mass.
 \item Even after the inclusion of radiative corrections, it turns out that the regions of the
parameter space where the LEP2 limits can be satisfied are somewhat restricted, requiring either
large stop masses (the stop sector gives the dominant contribution to the Higgs mass), either
substantial left-right stop mixing.
\end{itemize}

We thus see that in order to obtain a large enough Higgs mass, quite particular assumptions must be
made concerning the values of $\tan\beta$, the masses and mixings of the stop sector, or the
relation among the two Higgs mixing angles.
Furthermore, assuming large stop masses starts posing once again issues with the hierarchy 
problem, since we already mentioned that the superpartner masses should be as close to the
electroweak scale as possible in order to efficiently cancel the Higgs mass quadratic divergencies.
This problem is referred to as the \textit{little hierarchy problem}.
\\ \\
According to this discussion, two possible wayouts could be envisaged:
\begin{itemize}
 \item Taking into account contributions to the Higgs mass that could come from
physics beyond the MSSM (BMSSM), which are known to be present in several of its
extensions.
 \item Trying to find a framework which does not necessarily modify the particle
content of the MSSM, but instead focuses on reducing $\sin(\beta - \alpha)$.
\end{itemize}
In the remaining of this work we shall be examining dark matter in two supersymmetric 
frameworks trying to evade the little hierarchy problem in both ways.
%%%%%%%%%%%%%%%%%%%%%%%%%%%%%%%%%%%%%%%%%%%%%%%%%%%%%%%%%%%%%%%%%%%%%%%%%%%%%%%%%%%%%%%%%%%%%%%%%%%%%%%%%%%%%
%%%%%%%%%%%%%%%%%%%%%%%%%%%%%%%%%%%%%%%%%%%%%%%%%%%%%%%%%%%%%%%%%%%%%%%%%%%%%%%%%%%%%%%%%%%%%%%%%%%%%%%%%%%%%
%%%%%%%%%%%%%%%%%%%%%%%%%%%%%%%%%%%%%%%%%%%%%%%%%%%%%%%%%%%%%%%%%%%%%%%%%%%%%%%%%%%%%%%%%%%%%%%%%%%%%%%%%%%%%
\section{Beyond the MSSM}

\subsection{The framework}
One of the first thoughts that might come in mind in an effort to satisfy the LEP2
bounds could be to further extend the particle content of the MSSM. New contributions
to the lightest Higgs mass could in principle raise the tree-level (or even the loop-level)
prediction and reconcile it with the experimental constraints. Numerous such examples
are known in the literature. An important issue is, however, that nothing is
known about what could be the physics beyond the MSSM (\textit{if} the MSSM has something to do
with physical reality).

During the last few years, a series of papers 
\cite{Brignole:2003cm, Casas:2003jx, Pospelov:2005ks, Pospelov:2006jm, Dine:2007xi} 
have followed a somewhat alternative
approach: instead of examining all possible extensions of the MSSM, one could just assume that
new physics enters the game at some scale $M$, imposing a cutoff to the theory. Below this 
cutoff, the new degrees of freedom can in principle be integrated out of the theory resulting
in an effective Lagrangian near the EW scale. The new operators can then be organized according
to their suppression by the cutoff scale in the superpotential. 
Dimension - $5$ operators will be suppressed as $1/M$, 
dimension - $6$ as $1/M^2$ and so on.

At first sight, the number of potential operators that one could include in
order to write down the most general dimension five superpotential is huge, not to speak of
higher - dimensional operators. However, it turns out that employing superfield techniques and
identities, the total number of these operators can be significantly reduced by redefining 
a certain number of them. In fact, the most general dimension - $5$ \cite{Antoniadis:2008es} 
and dimension - $6$ \cite{Antoniadis:2009rn} MSSM
superpotentials were recently written down. Several new parameters and contributions should
be taken into account, but apparently much less than one would have initially expected.

In the light of the previous discussion on the little hierarchy problem, one could start with 
a little less ambition and strictly try to address the Higgs mass issue, for example by only
including dimension - $5$ operators only involving Higgs fields. This was done quite recently 
by Dine, Seiberg and Thomas in ref.\cite{Dine:2007xi}. Baryon and lepton number violating operators
are ignored, as is done for operators that could be added in the squark sector.

The authors found a remarkable result, namely that there are only two independent operators 
falling under the previous considerations that can be added to the MSSM superpotential.
The first of these operators is supersymmetric:
\begin{equation}
 W_5^{\mbox{\begin{tiny}SUSY\end{tiny}}} = \frac{\lambda_1}{M} (H_u H_d)^2 \ .
\label{BMSSMsusypres}
\end{equation}
Another contribution comes from supersymmetry breaking terms. This operator can be written as
\begin{equation}
 W_5^{\cancel{\mbox{\begin{tiny}SUSY\end{tiny}}}} = \frac{\lambda_2}{M} {\cal{Z}} (H_u H_d)^2
\label{BMSSMsusybreak}
\end{equation}
where ${\cal{Z}}$ is a supersymmetry breaking
spurion field, ${\cal{Z}} = m_{\mbox{\begin{tiny} SUSY\end{tiny}}} \theta^2$. Here, 
$m_{\mbox{\begin{tiny} SUSY\end{tiny}}}$ is the supersymmetry breaking scale.
\\ \\
The total superpotential for this model (dubbed Beyond the MSSM, BMSSM) is, of course, the
sum of the three contributions
\begin{equation}
 W_{\mbox{\begin{tiny}BMSSM\end{tiny}}} = 
W_{\mbox{\begin{tiny}MSSM\end{tiny}}} + 
W_5^{\mbox{\begin{tiny}SUSY\end{tiny}}} +
W_5^{\cancel{\mbox{\begin{tiny}SUSY\end{tiny}}}} \ .
\label{BMSSMsuperpotential}
\end{equation}
Differentiating the superpotential with respect to the theory's scalar fields and then
squaring, we can get the Higgs scalar potential. This shall include both $1/M$
and $1/M^2$ - suppressed terms, of which according to our discussion we only keep the
former, i.e. the crossed terms between the MSSM contribution and the new pieces.
The supersymmetric and supersymmetry breaking parts give us the following contributions
respectively
\begin{eqnarray}
 \delta V_1 & = & 2 \epsilon_1 (H_u H_d) (H_u^\dag H_u + H_d^\dag H_d) + h.c.\\
 \delta V_2 & = & \epsilon_2 (H_u H_d)^2 + h.c.
\label{BMSSMscalarcontr}
\end{eqnarray}
where we have defined the two new parameters that the model introduces
\begin{eqnarray}
 \epsilon_1 & \equiv & \frac{\mu^* \lambda_1}{M}\\
 \epsilon_2 & \equiv &  - \frac{m_{\mbox{\begin{tiny} SUSY\end{tiny}}} \lambda_2}{M} \ .
\label{BMSSMparameters}
\end{eqnarray}
Finally, the new operators introduce a new Higgs-Higgs-Higgsino-Higgsino interaction 
Lagrangian
\begin{equation}
 \delta{\cal{L}} = - \frac{\epsilon_1}{\mu^*}
\left[
2 (H_u H_d) (\tilde{H}_u \tilde{H}_d) + 
2 (\tilde{H}_u H_d) (H_u \tilde{H}_d) + 
  (H_u  \tilde{H}_d) (H_u \tilde{H}_d) + 
  (\tilde{H}_u H_d) (\tilde{H}_u H_d)
\right] + h.c.
\label{BMSSMinteractions}
\end{equation}
This contribution modifies the Higgsino annihilation process, which
can be relevant for dark matter phenomenology provided the neutralino has a significant Higgsino
component. At this point we should notice that this additional interaction Lagrangian
does not depend on the $\epsilon_2$ parameter. It is thus reasonable to expect that
the dark matter - related phenomenology should only depend indirectly on $\epsilon_2$, 
notably through the modifications of the various particles' masses.

At the same time, all of the above contributions modify the relation among the Higgs
mass and the stop sector, as well as the neutralino and chargino masses. Concerning the
lightest Higgs mass, the Non - Renormalizable (NR) corrections to the Higgs mass can become sizeable, allowing
to satisfy the LEP2 limits even at tree-level for moderate stop mass values and without
substantial left-right stop mixing. The modifications in the lightest Higgs mass, the neutralino
mass matrix, as well as in some useful couplings can be found along with some other useful
formulae in Appendix \ref{NeutralinoCouplings}.
\\ \\
Several aspects of the model have been studied in the literature
\cite{Blum:2008ym, Cheung:2009qk, Cassel:2009ps, Carena:2009gx, Bernal:2009hd}.
Since the model is much more extended and complex than the singlet scalar one, we shall
not describe in as much detail its various phenomenological consequences and constraints.
For the sake of brevity, we shall only focus on the dark matter - related phenomenology.
The first step is to describe the relic density constraints coming from the WMAP measurements.
Then, we shall present some work effectuated in \cite{Bernal:2009jc} (see also \cite{Bernal:2010uf}) 
concerning the dark matter detection prospects for the model.

\subsection{Relic Density}
The impact of the NR operators on the relic density calculation has been examined in detail in
references \cite{Berg:2009mq} and \cite{Bernal:2009hd}. In both of these papers, the authors
scan over the BMSSM parameter space (the conventions used in the two papers are different), 
finding WMAP - compliant regions that are absent in the MSSM case, either due to the elimination
of a number of them by ``external'' constraints, or because the MSSM just cannot produce the 
relevant phenomenology.

We shall be examining the dark matter phenomenology in the two benchmark scenarios
discussed in \cite{Bernal:2009hd}: the first one tries to compare the BMSSM phenomenology
with the low-energy phenomenology of a typical CMSSM model, whereas the second one 
begins with the definition of a low-energy model which in the BMSSM framework can lead
to light stops and heavy sleptons.

\subsubsection{Correlated stop - slepton masses}
\label{CorrSSmasses}
The first scenario of \cite{Bernal:2009hd}  begins with the typical set of the five free parameters present in
CMSSM - like models
\begin{equation}
 \tan\beta, \ m_{1/2}, \ m_0, \ A_0, \ \mbox{sign}(\mu)
\end{equation}
In such a framework, the low energy parameters can be approximately given by
\begin{eqnarray}
 m^2_{\tilde q}&\approx&m_0^2+6\,m_{1/2}^2,\nonumber\\
m^2_{\tilde\ell_L}&\approx&m_0^2+0.5\,m_{1/2}^2,\nonumber\\
m^2_{\tilde\ell_R}&\approx&m_0^2+0.15\,m_{1/2}^2,\nonumber\\
M_1&\approx&0.4\,m_{1/2},\nonumber\\
M_2&\approx&0.8\,m_{1/2},\nonumber\\
M_3&\approx&3\,m_{1/2}.
\end{eqnarray}
Three out of the five parameters of the CMSSM model are fixed, choosing
specifically $\tan\beta = 3$ or $10$, $A_0 = 0$, $\mbox{sign}(\mu) > 0$ and then 
the remaining $(m_0, m_{1/2})$ parameter space is scanned and the relic density as
well as a number of EW observables are computed.

The resulting low-energy model is next enriched with the addition of the NR operator
contributions and the observables are recalculated. We should stress at this point
that the new framework should \textit{by no means} be conceived as a generalized 
CMSSM model, since it is impossible to take into account the BMSSM physics effects
on the running of physical quantities from the GUT scale down to the electroweak scale:
the two models are compared \textit{only} with respect to their low-energy phenomenology.

\begin{figure}[htb]
\vspace{2cm}
\begin{center}
\hspace{-4cm}
\includegraphics[width = 5cm]{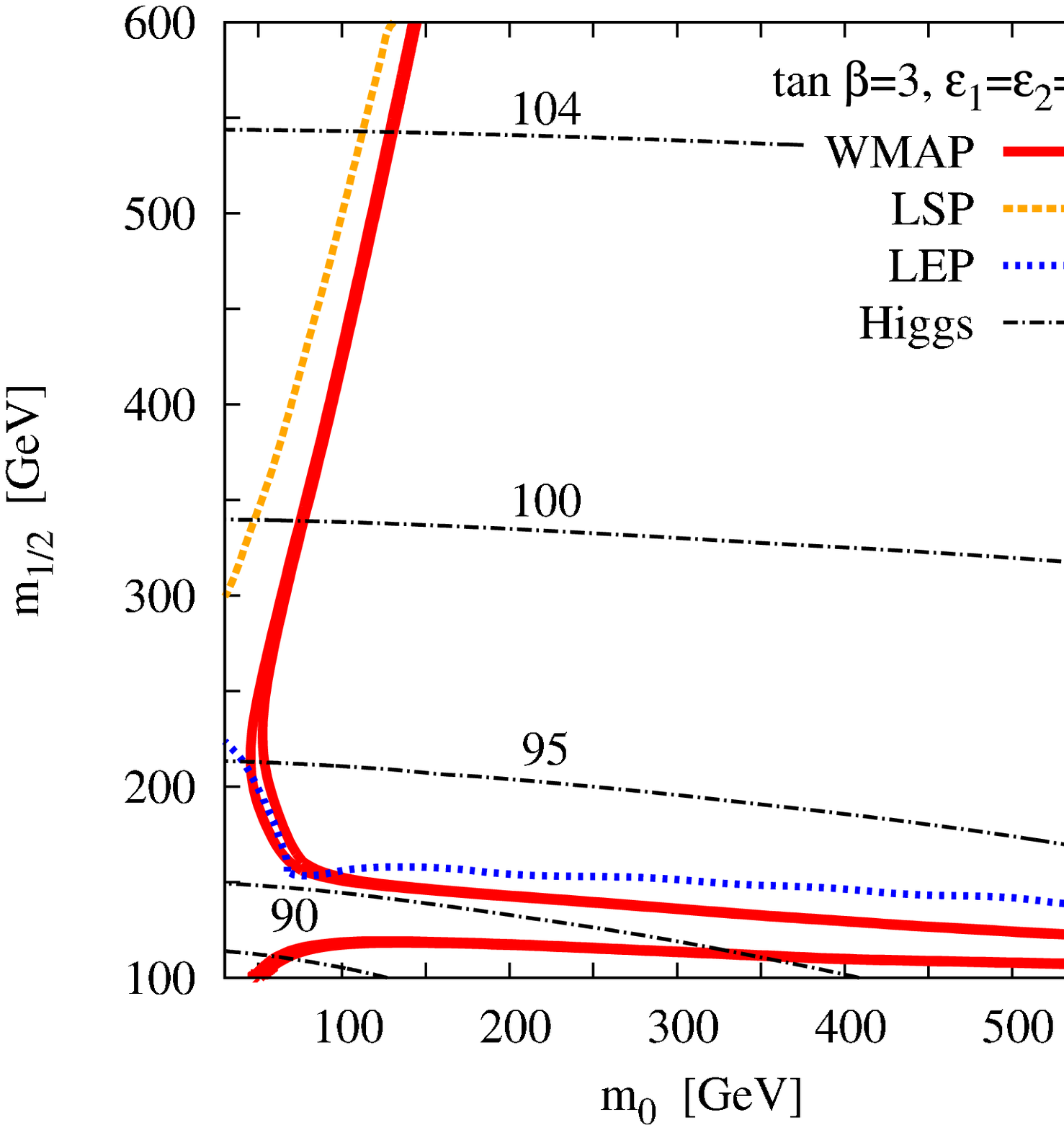}
\hspace{3cm}
\includegraphics[width = 5cm]{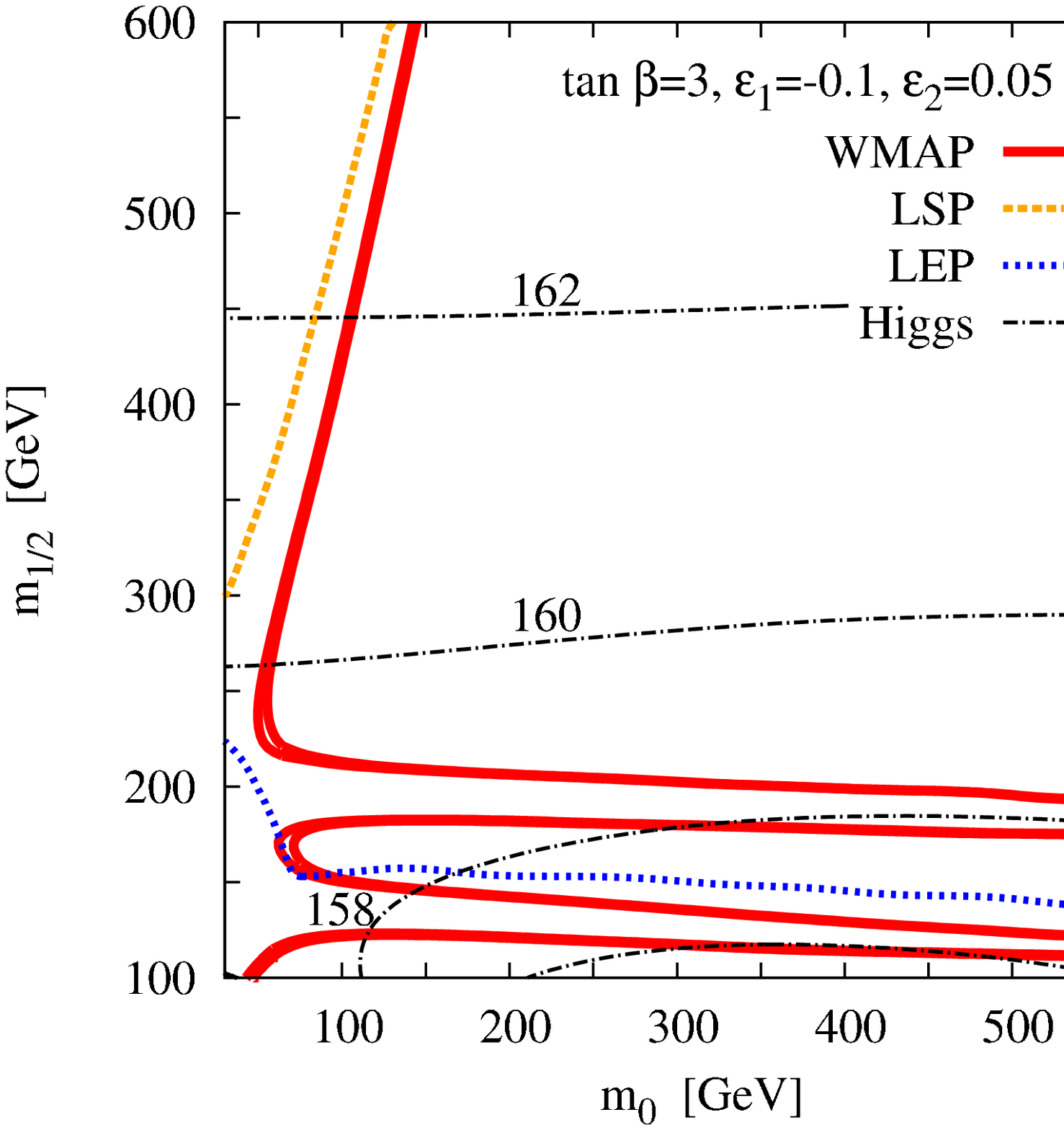}
\caption{{\footnotesize
Regions in the $(m_0, m_{1/2})$ plane where the WMAP constraints are satisfied (red solid regions)
in the case of a plain CMSSM (left panel) and a BMSSM (right panel) model. The regions below the blue
dotted lines are excluded by null searches for charginos at LEP. The regions to the left and above
the yellow dashed lines are excluded due to stau LSP. Finally, the black dotted-dashed lines are
light Higgs mass contours in GeV.}}
\label{fig:BMSSMrelic}
\end{center}
\end{figure}

In order to present the relic-density related phenomenology, we borrow fig.\ref{fig:BMSSMrelic}
from ref.\cite{Bernal:2009hd}. In this figure, the left-hand side plot corresponds to the
CMSSM benchmark with the BMSSM contributions set to zero.
The WMAP - compliant regions are delimited by red solid lines (in practice the regions seem
as lines in the plot). Furthermore, a series of constraints are also depicted in the figure:
the regions above and on the left of the yellow dashed lines correspond to a non-neutralino
LSP (in this case it is the stau), the regions below the blue dotted line are excluded by the null
chargino searches at LEP whereas the black dotted-dashed lines are light Higgs mass isocontours
as seen in the figure. We note that the entire parameter space depicted in the figure is 
in any case excluded due to the lightest Higgs mass constraint.

Once the low-energy spectrum and observables are computed, the next step is to turn on the
BMSSM contributions. In the original paper this is done in two steps: first, only the 
$\epsilon_1$ parameter is turned on choosing $\epsilon_1 = -0.1$ and $\epsilon_2 = 0$.
We point out that in our notations, a negative $\epsilon_1$ value yields a positive 
contribution to the lightest Higgs mass, as can be verified in Appendix \ref{NeutralinoCouplings}.
Everything is recalculated with this configuration. Then, both parameters are given
non-zero values, namely $\epsilon_1 = -0.1$ and $\epsilon_2 = 0.05$ and the process is
repeated. For the time being, we omit the intermediate step with $\epsilon_2 = 0$
and just give the results with both parameters turned on (right panel of fig.\ref{fig:BMSSMrelic}).
Later, when we present our original work on the BMSSM dark matter detection issue, 
we shall present all three
configurations. A first remark that could be already made for the BMSSM variant is that  the
LEP2 Higgs mass bound is immediately satisfied for the entire parameter space thanks
to the contributions from the new operators.

Let us now start the description of the relic density results with the left panel, 
i.e. the plain mSUGRA model (forgetting
for the moment that it is in any case excluded). As a general rule, we could
say that the model tends to yield too small neutralino self-annihilation cross-sections and thus
overproduce dark matter. At low $m_{1/2}$ values and almost parallel to the $m_0$ axis, we notice
two strips where the WMAP bounds are satisfied. These are around the region where the neutralino
mass is close to half the mass of the lightest Higgs boson or the $Z$ boson, 
$m_{\chi_1^0} \sim m_{h, Z}/2$. In this regime, the neutralino self-annihilation cross-section
gets enhanced because the light Higgs propagator in the $s$ - channel starts being nearly on-shell.
The second region of interest is the one almost parallel to the stau LSP constraint line, at low
$m_0$ values. In this region, the correct relic density is obtained not because of some enhancement
in the neutralino self-annihilation cross-section, but due to efficient $\chi_1^0$ - $\tilde{\tau}$ 
coannihilation. Finally, for small $(m_0, m_{1/2})$ values, we have the bulk region, where 
the relic density calculation is driven by crossed sfermion exchange.

The most striking feature brought about by the introduction of the NR operators (right-hand side plot) 
is the appearance
of a new, distinct zone where the relic density constraint is satisfied. This is the region again
almost parallel to the $m_0$ axis and above the chargino LEP exclusion limits. In fact, this region 
is not exactly new: the introduction of the dimension - $5$ operators has the effect (and the aim)
to raise the Higgs mass. Hence, whereas in the CMSSM case the $Z$ and $h$ pole regions are 
practically degenerate
\footnote{The degeneracy of the two poles is of course not a universal phenomenon in the CMSSM, 
it just occurs in the scenario considered here!}
, the new contributions uplift the Higgs mass causing for the separation
of the two resonant regions. And whereas the $Z$ pole region remains excluded by the LEP
chargino search limits, the $h$ pole region is now perfectly viable.

\subsubsection{Light stops, heavy sleptons}
\label{LSHS}
The second scenario introduced in \cite{Bernal:2009hd} makes no reference whatsoever to GUT-scale
conditions (even if in the previous case we are not interested in them either). The authors start
with a set of low-energy parameters, namely 
\begin{equation}
 M_2,\ \mu,\ \tan\beta,\ X_t,\ m_U,\ m_Q, \ m_{\tilde f},\ m_A,
\end{equation}
which correspond respectively to the Wino mass, the Higgsino mass parameter, the usual ratio 
of the Higgs vacuum expectation values, the trilinear coupling - dependent parameter $X_t=A_t- \mu \cot \beta$, 
the right stop mass parameter, a common mass parameter for the third generation 
left squarks, a common mass for the sleptons, the first two generation squarks and the 
right sbottom and, finally, the pseudoscalar mass. Six out these eight parameters are fixed as
$\tan\beta=3$ or $10$, $X_t=0$, $m_U=210$ GeV, $m_Q=400$ GeV,
$m_{\tilde f}=m_A=500$ GeV and a scan is performed in the $(M_2, \mu)$ parameter space.
The other two gaugino masses are fixed as $M_1 = 5/3 \tan^2 \theta_W M2 \approx 1/2 M_2$
whereas $M_3$ is irrelevant for our analysis.
Then, the $\epsilon_1$ and $\epsilon_2$ parameters are turned on as in the previous scenario
(choosing $\epsilon_1=-0.1$ and then also $\epsilon_2=+0.05$) and the scan is repeated.
\\ \\
This scenario is chosen in order to yield light stops
\begin{equation}
 m_{\tilde t_1}\lesssim 150\ {\rm GeV},\ \ \ \ 370\
{\rm GeV}\lesssim m_{\tilde t_2}\lesssim 400\ {\rm GeV}
\end{equation}
which, as argued in the paper, is a favorable scenario for electroweak baryogenesis.

\begin{figure}[htb]
\vspace{2cm}
\begin{center}
\hspace{-4cm}
\includegraphics[width = 5cm]{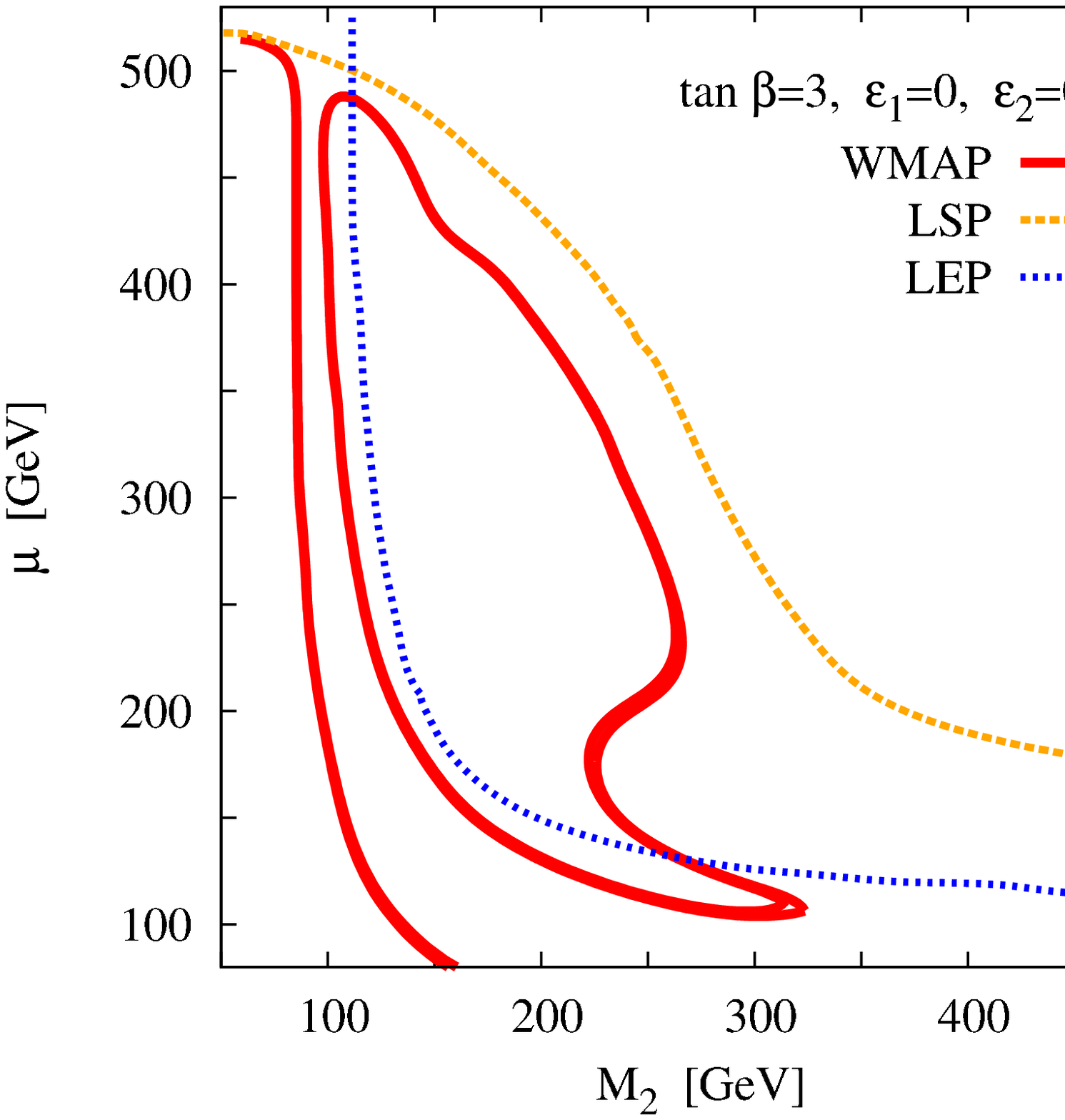}
\hspace{3cm}
\includegraphics[width = 5cm]{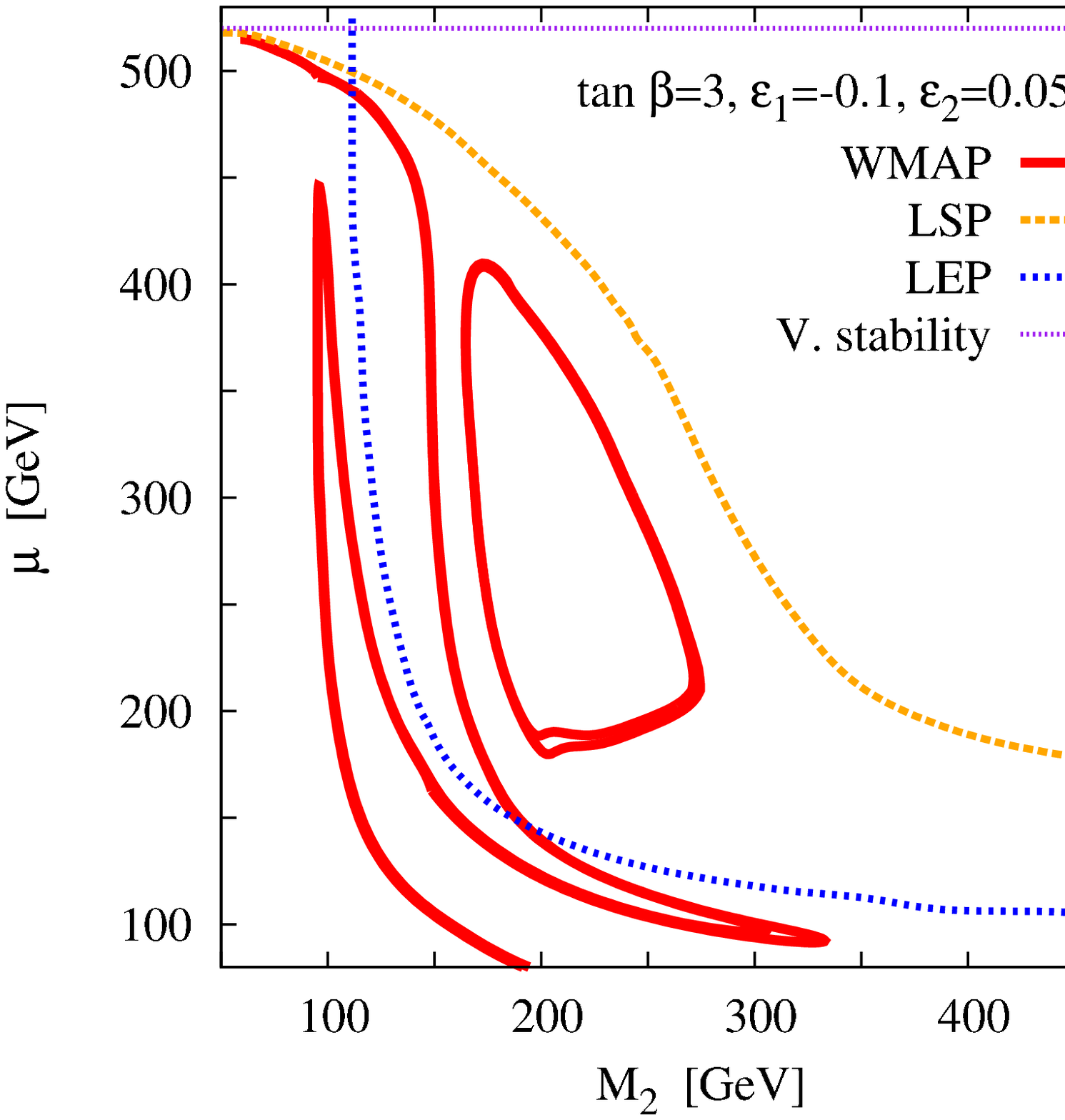}
\caption{{\footnotesize
Regions in the $(M_2, \mu)$ plane where the WMAP constraints are satisfied (red solid regions)
in the case of a plain MSSM (left panel) and a BMSSM (right panel) model. The regions below and on the 
left of the blue
dotted lines are excluded by null searches for charginos at LEP. The regions to the right and above
the yellow dashed lines are excluded due to stop LSP. The violet curve in the right panel
depicts constraints coming from vacuum stability.
}}
\label{fig:BMSSMrelicLSHS}
\end{center}
\end{figure}
We once again borrow fig.\ref{fig:BMSSMrelicLSHS} from \cite{Bernal:2009hd} in order to 
describe the relic density phenomenology.
We start with the plain MSSM model (left panel). The first regions where the WMAP constraints are satisfied
are the two red regions at roughly $M_2 \approx 100$ GeV and almost parallel to the
$\mu$ axis. These correspond to neutralino annihilation taking place near the $h$ and $Z$ poles
as before. Around these regions the relic density is too high, whereas between them it is
too low. The next region is the one quasi-parallel to the neutralino LSP constraint, where 
the driving mechanism enhancing the neutralino annihilation process is coannihilation with the 
NLSP, which in this case is the lightest stop. Finally, there is one last region around $M_2 \sim 200$ Gev
and $\mu \sim 100 - 250$ GeV, where the neutralino starts acquiring a significant Higgsino fraction, 
augmenting its couplings to the $Z$ boson and amounting mostly to gauge and Higgs boson final states.
Interestingly, for larger $M_2$ values the self-annihilation cross-section becomes too large, 
and the lower WMAP limit is violated.

Introducing the dimension - $5$ operators has, once again, mainly the effect of separating the $h$
and $Z$ pole regions, due to the rise in the lightest Higgs mass. Once again, the $h$ pole region
evades the LEP chargino search limits. Finally, we remark the appearance of a violet line at large
$\mu$ values. This corresponds to a further constraint, which is the requirement for
vacuum stability. Indeed, in the presence of the NR operators the scalar Higgs potential can
get destabilized, with a second remote vacuum forming, rendering the EW breaking vacuum of the theory
metastable.

Having discussed the two scenarios and their relic density - related phenomenology, we can now 
proceed to examine the detection prospects for the model in various channels.
%%%%%%%%%%%%%%%%%%%%%%%%%%%%%%%%%%%%%%%%%%%%%%%%%%%%%%%%%%%%%%%%%%%%%%%%%%%%%%%%%%%%%%%%%%%%%%%%%%%%%%%%%%%%%
%%%%%%%%%%%%%%%%%%%%%%%%%%%%%%%%%%%%%%%%%%%%%%%%%%%%%%%%%%%%%%%%%%%%%%%%%%%%%%%%%%%%%%%%%%%%%%%%%%%%%%%%%%%%%
%%%%%%%%%%%%%%%%%%%%%%%%%%%%%%%%%%%%%%%%%%%%%%%%%%%%%%%%%%%%%%%%%%%%%%%%%%%%%%%%%%%%%%%%%%%%%%%%%%%%%%%%%%%%%

\section{Dark matter detection in the BMSSM}
In ref.\cite{Bernal:2009jc} we examined the detection perspectives for the two BMSSM models
presented previously. Finding first the WMAP-compatible regions (to be depicted in red
in the plots that follow), 
we estimated whether the parameter space points for the two scenarios
can be probed using four different detection techniques: direct detection in a XENON - like
experiment, gamma-ray detection at the Fermi satellite mission as well as positron and
antiproton detection coming from DM annihilations in the AMS-02 experiment.

The method we followed was to first compute the detection rates with $\epsilon_1 = \epsilon_2 = 0$, 
i.e. in the plain MSSM case. This shall correspond to the first row in all plots that follow
for this study. Then, we further examine two BMSSM variants for each scenario, 
turning on the values $(\epsilon_1 = -0.1, \epsilon_2 = 0)$ (second-row plots) and 
$(\epsilon_1 = -0.1, \epsilon_2 = 0.05)$ (third-row plots) in each one of them. Our computations were done assuming
two distinct values for $\tan\beta$, namely $3$ and $10$. The first case shall correspond to the left-hand
side plots, whereas the second one to the right-hand side ones.

Before presenting our results, it is useful to define what we shall be 
meaning in this treatment when we characterize
a parameter space point as being ``detectable''.
We employ a method based on the $\chi^2$ quantity.
Consider whichever mode of detection: direct or indirect in any of the three channels
($\gamma$-rays, $e^+, \bar{p}$) we shall be considering. In all four modes, what 
is finally measured is a number of events per energy bin.
Let us call $N^{sig}_i$ the number of signal (dark matter - induced) events in
the $i$-th bin, the nature of which depends on the specific experiment, 
$N^{bkg}_i$ the corresponding
background events in the same bin, and $N^{tot}_i$ the sum of the two.
The variance $\chi^2_i$ in every bin is defined as:
\begin{equation}
\chi^2_i =
\frac{(N_i^{tot}-N_i^{bkg})^2}{N_i^{tot}} \ .
\end{equation}
Then, the condition that we impose to characterize a point as detectable, is that at least 
in one energy bin 
$\chi^2_i\gtrsim 3.84$. In Gaussian error terms, this corresponds to a $95\%$ CL.

We shall now present the work effectuated for each of the four detection modes, introducing
our experimental and theoretical assumptions and simplifications, then describing the
results for each of our two models.

\subsection{Direct detection}
The first part of the work consists of calculating the regions of our parameter space that can
be probed in a XENON-like experiment for our twelve models (2 scenarios each with 6 sub-variants
as described before).
To this goal, we consider $7$ recoil energy bins 
between $4$ and $30$ keV. The background in this analysis is set
to zero. Furthermore, we assume three exposure values (time $\times$ detector mass) which we take
as $\epsilon = 30, 300$ and $3000$ kg$\cdot$year. These exposure values could correspond e.g. 
to a detector with $1$ ton of xenon and $11$ days, $4$ months or $3$ years of data acquisition, respectively.

Concerning astrophysics, in this analysis we take the standard halo model with neutralinos
following a Maxwell-Boltzmann velocity distribution in the galactic rest frame and neglecting the
motion of the earth around the sun, whereas the local DM density is set to $0.385$ GeV cm$^{-3}$ \cite{Catena:2009mf}. 
The sun's velocity around the GC is set to $220$ km sec$^{-1}$.

\subsubsection{Correlated stop-slepton masses}
\begin{figure}[tbp!]
\begin{center}
\vspace{-0.2cm}%\hspace{-2.5cm}
\includegraphics[width=5.5cm,angle=-90]{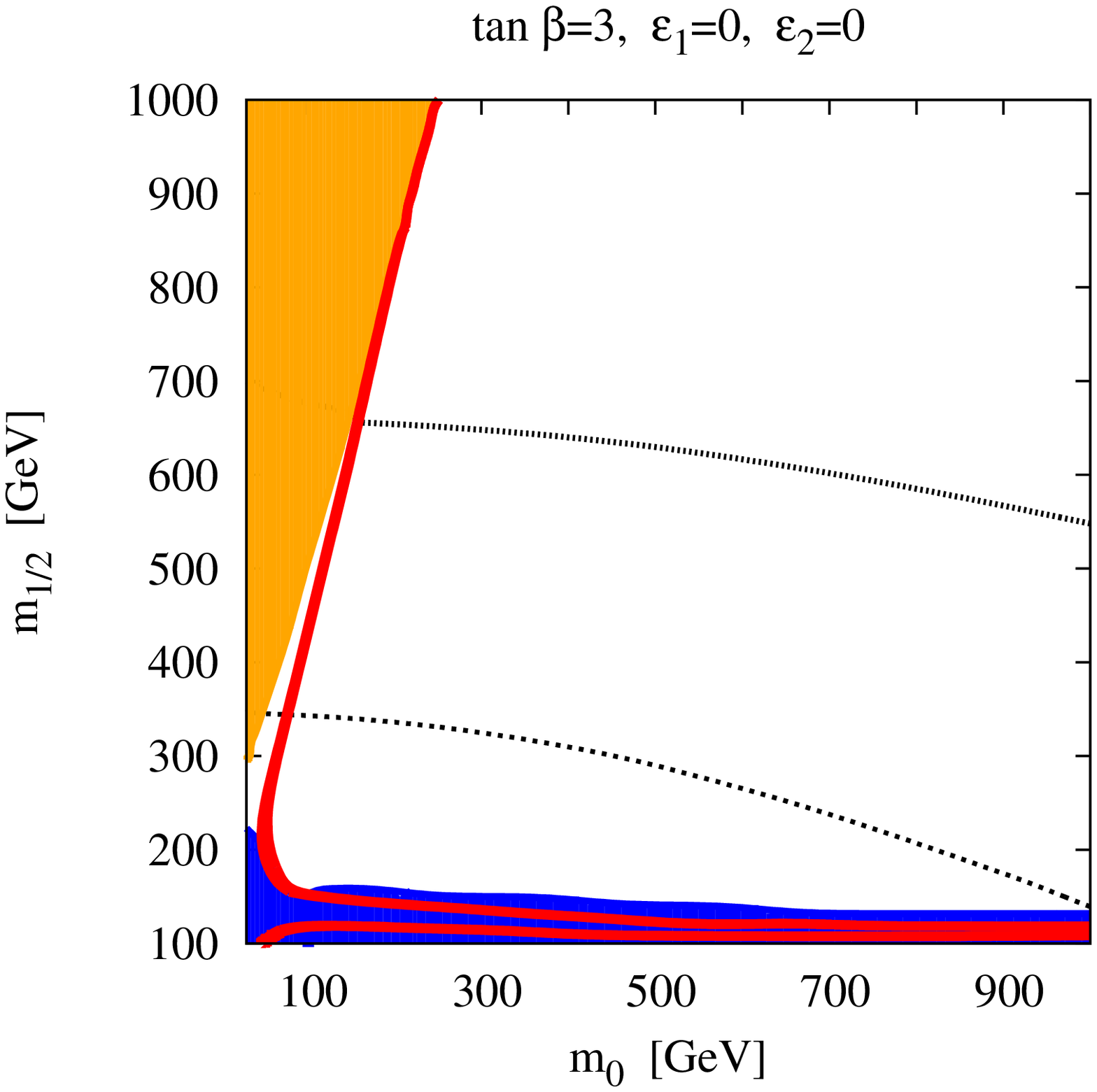}\hspace{0.2cm}
\includegraphics[width=5.5cm,angle=-90]{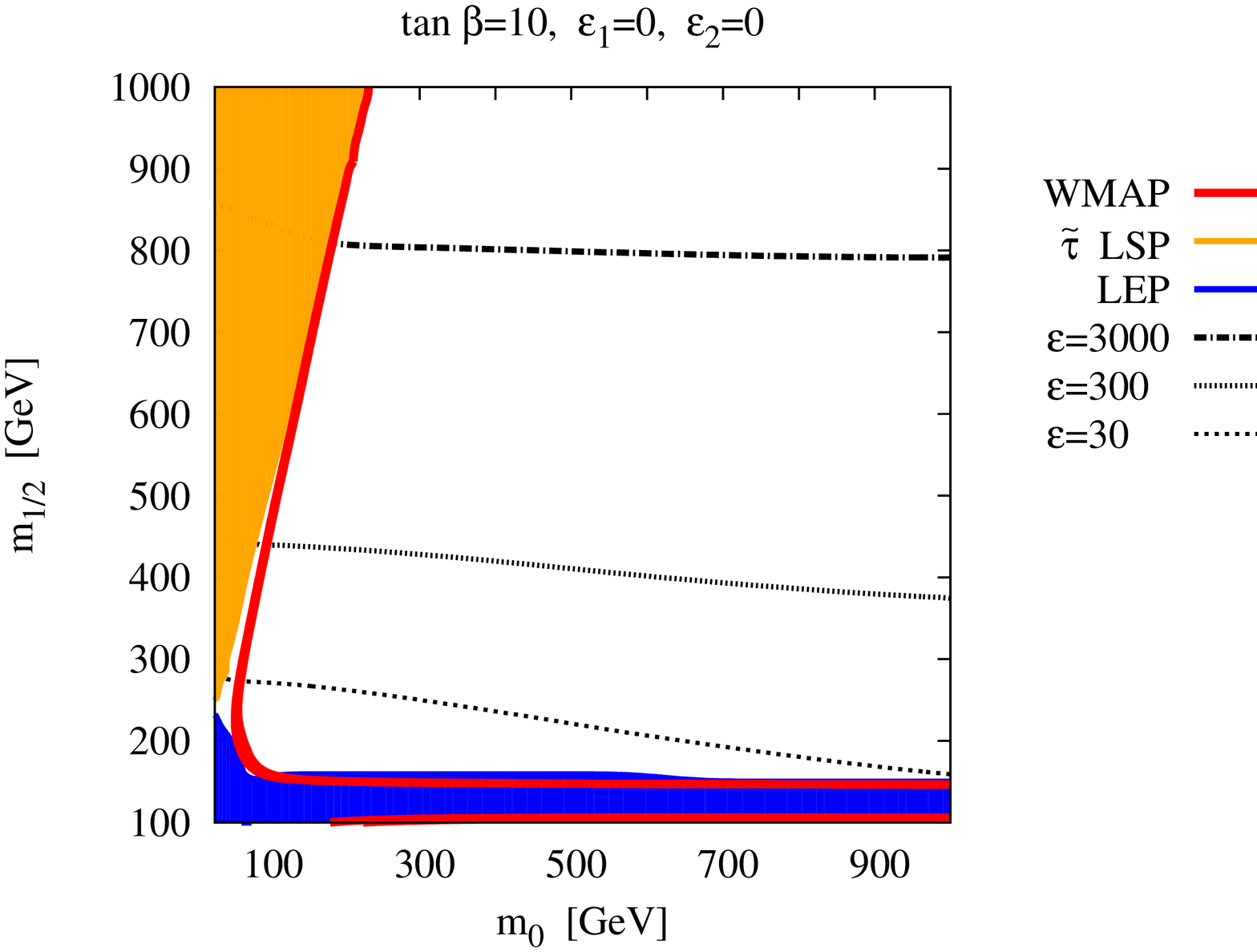}\\
\vspace{0.6cm}%\hspace{-2.5cm}
\includegraphics[width=5.5cm,angle=-90]{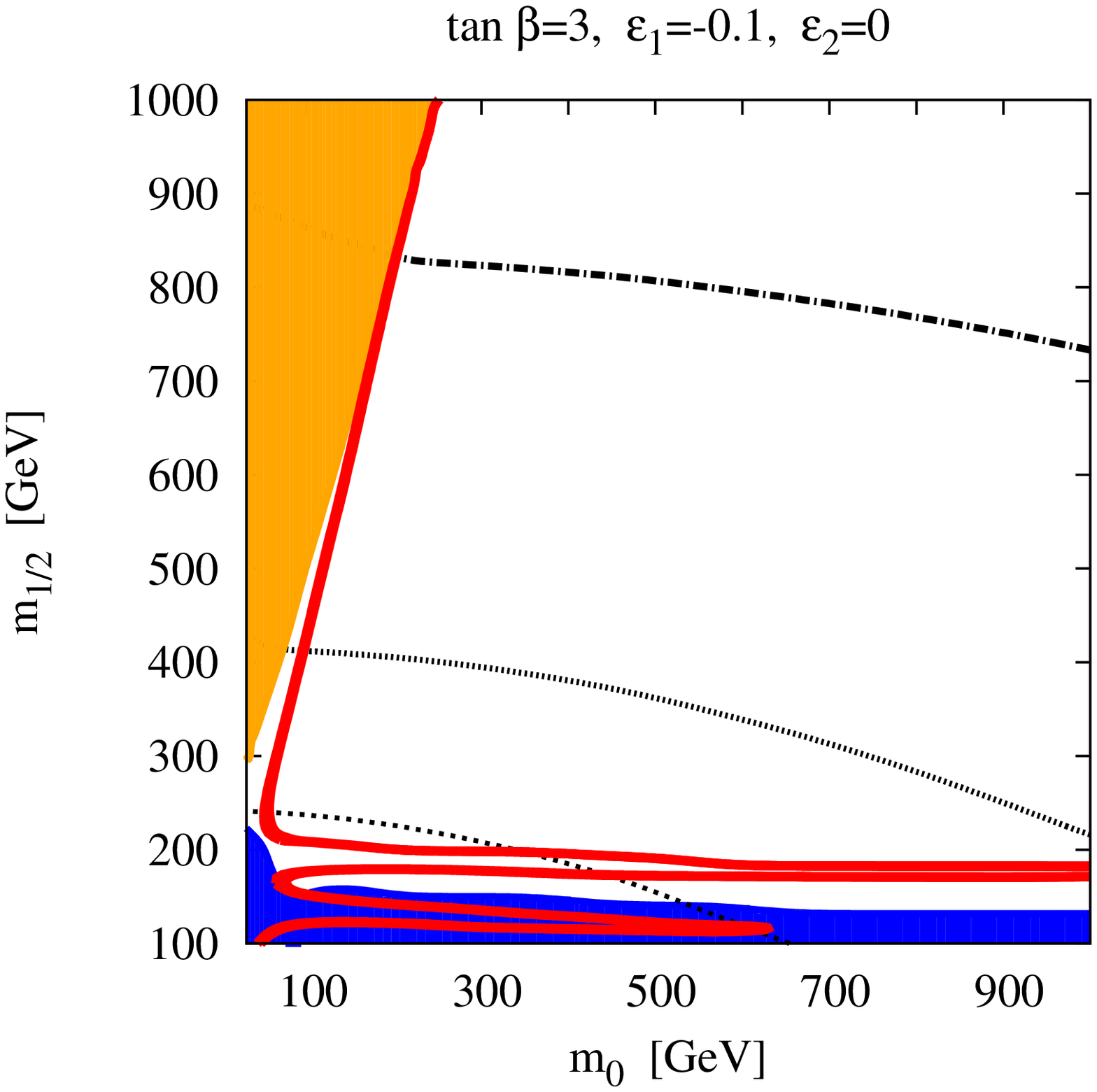}\hspace{0.2cm}
\includegraphics[width=5.5cm,angle=-90]{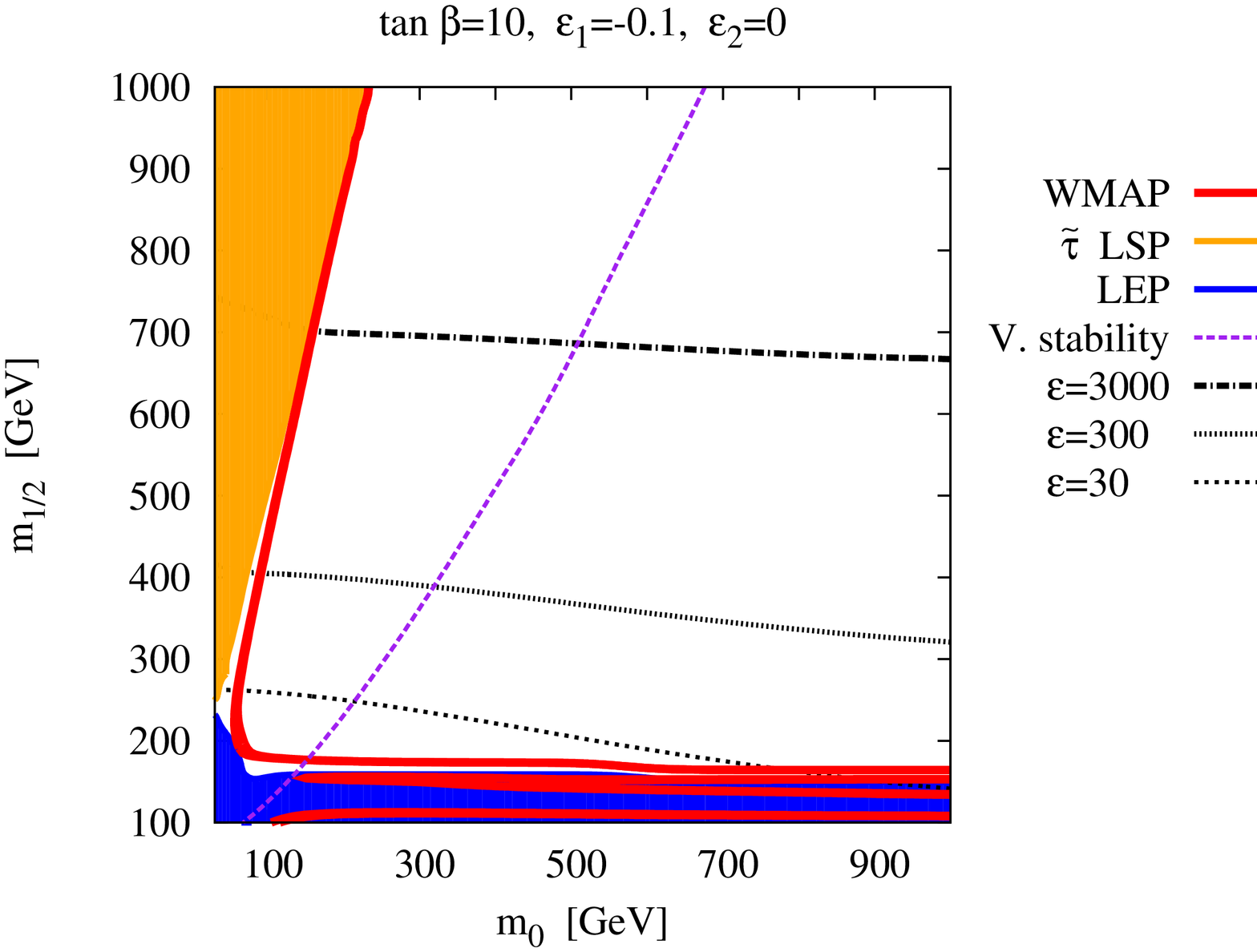}\\
\vspace{0.6cm}%\hspace{-2.5cm}
\includegraphics[width=5.5cm,angle=-90]{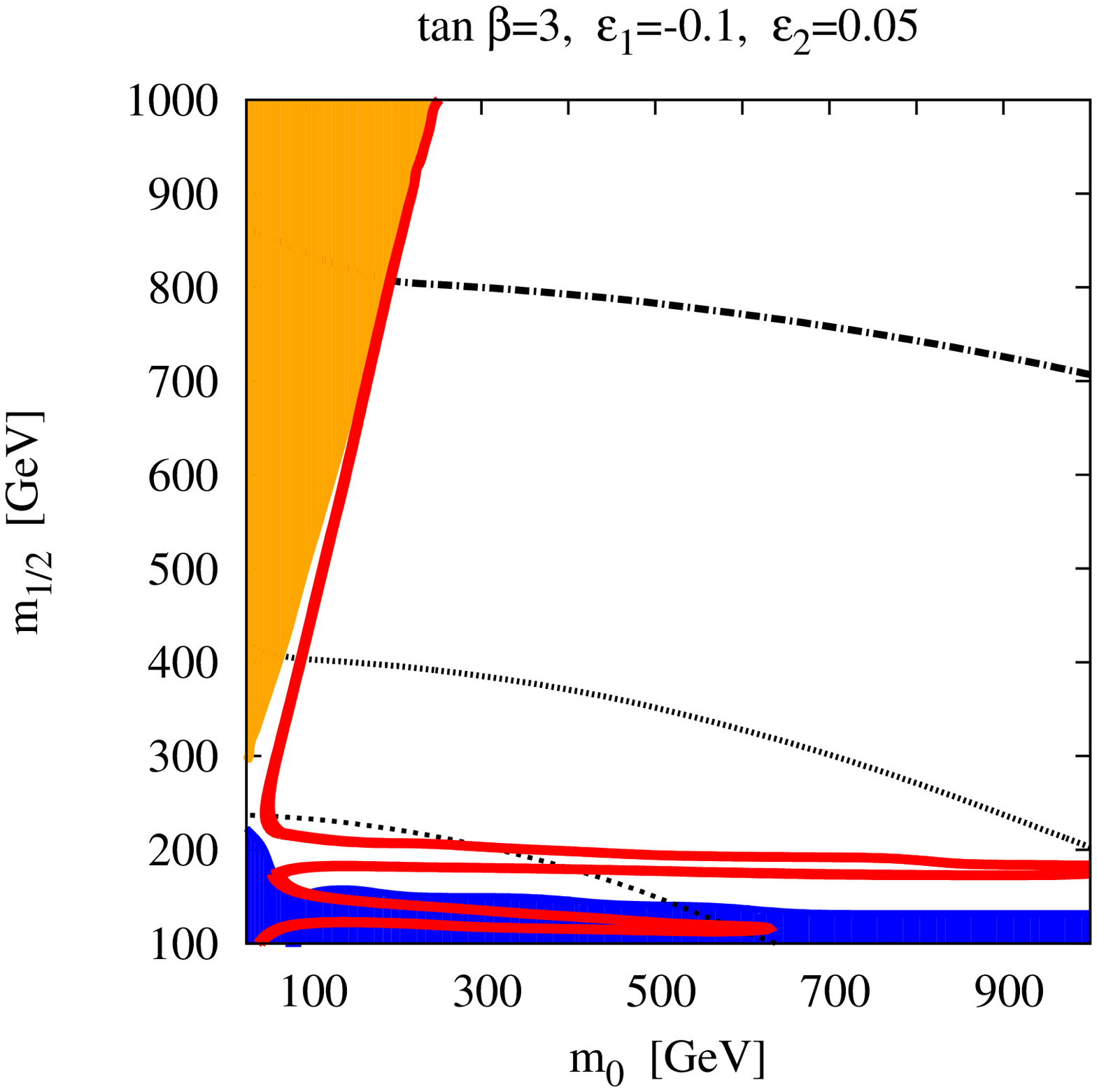}\hspace{0.2cm}
\includegraphics[width=5.5cm,angle=-90]{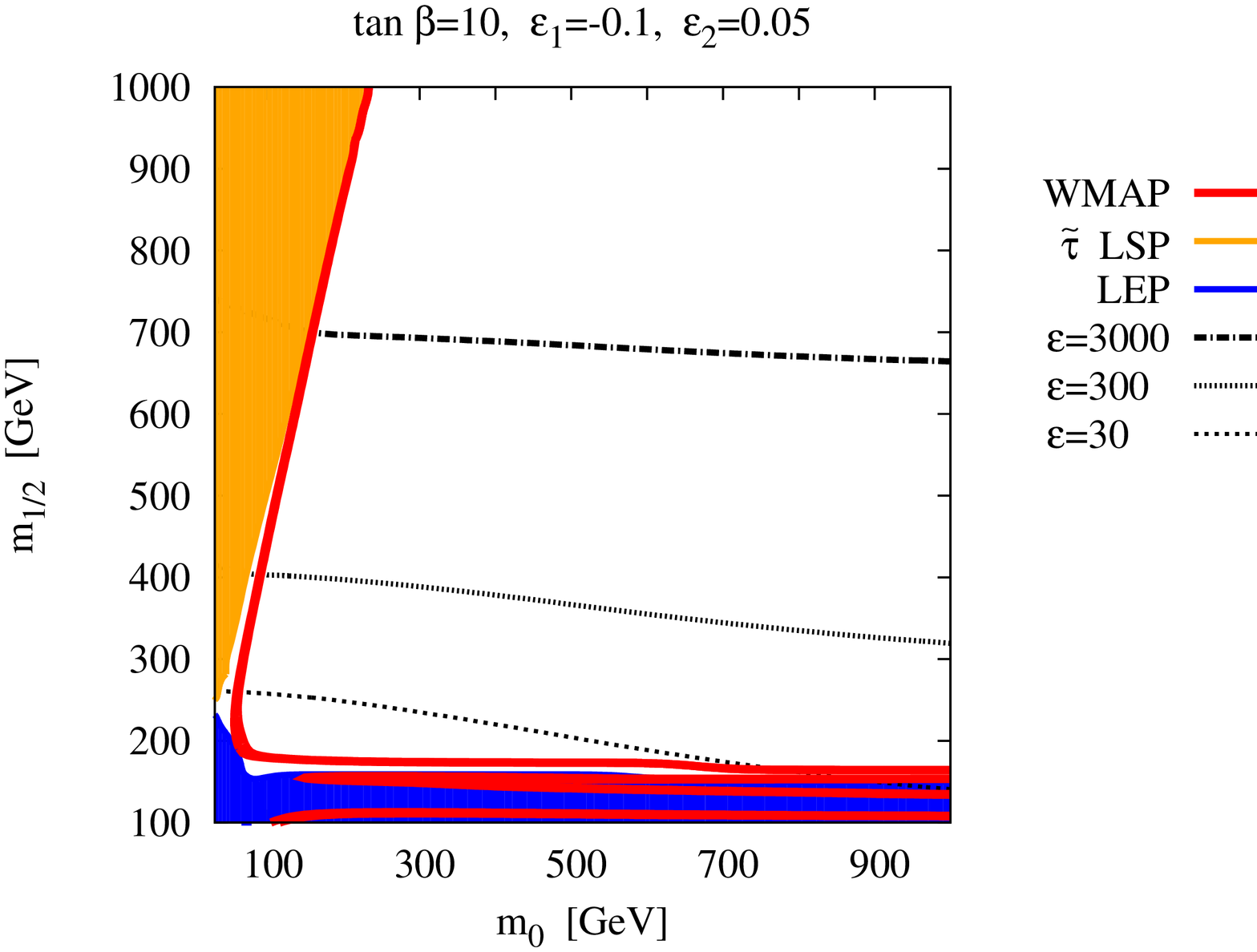}
\end{center}
%\vspace{-0.9cm}
\caption{{\footnotesize
Regions in the $[m_0,\,m_{1/2}]$ plane that can be detected by XENON using exposures
$\varepsilon=30$, $300$ and $3000$ kg$\cdot$year, for our mSUGRA-like scenario. 
The black lines depict the detectability regions: the area below the lines can be probed.
Whenever a line is absent, this means that the whole parameter space can be tested
by the experiment. The blue and orange regions depict the areas that are excluded by
direct LEP chargino searches and the requirement for a 
neutralino LSP respectively. The area above the violet line is excluded by the
metastable vacuum constraint.}}
\label{dir1}
\end{figure}

Figure \ref{dir1} shows the sensitivity lines (black lines) for 
exposures $\varepsilon=30$, $300$ and $3000$ kg$\cdot$year, 
on the $[m_0,\,m_{1/2}]$ parameter space, for all other parameters as defined in paragraph \ref{CorrSSmasses}.
We repeat for the sake of convenience that
the first-row plots correspond to plain CMSSM scenarios whereas the second and third to the
`mSUGRA-like' benchmark, with the $\epsilon_1$ and $(\epsilon_1, \epsilon_2)$ parameters turned on 
respectively.
The plots on the left correspond to a choice $\tan\beta = 3$ whereas the right-hand side ones
to $\tan\beta = 10$.
\footnote{At this point, we should notice once more something that we remarked previously:
passing from the second-row plots to the third-row ones (i.e. turning on $\epsilon_2$) has just
a small impact on dark matter phenomenology, due to the fact that the Higgs-Higgs-Higgsino-Higgsino
interaction lagrangian does not depend on $\epsilon_2$. The scalar potential, on the other hand, 
\textit{does} depend on it, hence the significant changes in the vacuum stability constraints.}
These curves reflect the XENON sensitivity and represent its ability to test and 
exclude different regions of the mSUGRA and BMSSM relevant model at $95\%$ CL: 
all points lying below the black lines are detectable.
When some line is absent, this means that the whole parameter space can be 
probed for the corresponding exposure.

Some further information is included in the plot (as well as the ones to follow):
The red regions depict the parameter space points yielding relic densities compatible
with the WMAP limits.
The regions in orange (light gray) or blue (dark gray) are excluded
due to the fact that the LSP is the stau or because of the null searches for charginos
at LEP.
For large $\tan\beta$, an important fraction of the $[m_0,\,m_{1/2}]$ plane,
corresponding to the region above the violet line, generates an unstable vacuum
and is then excluded.
An interesting remark is that the introduction of $\epsilon_2$ alleviates the vacuum 
stability constraint \cite{Blum:2009na}, and slightly increases the Higgs mass.

As a general rule, we can see from all plots that 
the detection prospects are maximized for low values of the $m_0$
and $m_{1/2}$ parameters. For higher $m_0$ values, the masses of the
squarks in the internal propagators increase, causing the scattering cross-section to decrease.
In the same way, the increase of $m_{1/2}$ augments the WIMP mass and
leads to a deterioration of the detection perspectives.
On the other hand, the region of low $m_{1/2}$ is also preferred because in that case the
lightest neutralino is a mixed bino-Higgsino state, favoring the $\chi_1^0-\chi_1^0-h$
and $\chi_1^0-\chi_1^0-H$ couplings, and thus the scattering cross-section. Let us recall
that a pure Higgsino or a pure gaugino state does not couple to the Higgs bosons, as can be seen
in Appendix \ref{NeutralinoCouplings}. We note that whereas the couplings to both $CP$-even
Higgses are enhanced, it is mostly the coupling to the light one that dominates.

On the other hand, by comparing the left- and right-hand figures, we can see that
the detection prospects are also maximized for low values of $\tan\beta$.
For large values, besides the increase of the lightest Higgs boson mass, the coupling
of the latter to a $\chi_1^0$ pair decreases significantly because it is proportional 
to $\sin 2\beta$, for $|\mu|\gg M_1$.

The introduction of the NR operators gives rise to an important deterioration of the detection prospects.
The main effect enters via the important increase in the lightest CP-even Higgs mass.
This behavior is attenuated for larger values of $\tan\beta$, since 
the corrections to the Higgs masses are suppressed by $1/\tan\beta$ (see e.g. reference \cite{Dine:2007xi}).
On the other hand, the neutralino couplings are not really influenced by the NR operators
in this regime, since $\chi_1^0$ is mostly bino-like. So, the impact
on its couplings with Higgs bosons is marginal. 

It is important to note at this point that the deterioration in the detection perspectives, while
existing, is nevertheless relative: we must not forget that the plain MSSM scenarios presented
here are already excluded because of the light Higgs mass.

Concerning the plots in figure \ref{dir1}, an overall remark that can be made is that, 
even for low exposures, a sizable amount of the parameter space can be probed.
The XENON experiment will be particularly sensible to low values of $m_{1/2}$.
However, larger exposures could be able to explore almost the whole parameter space taken
into account.

\subsubsection{Light stops, heavy sleptons}
\begin{figure}[tbp!]
\begin{center}
\vspace{-0.2cm}%\hspace{-2.5cm}
\includegraphics[width=5.5cm,angle=-90]{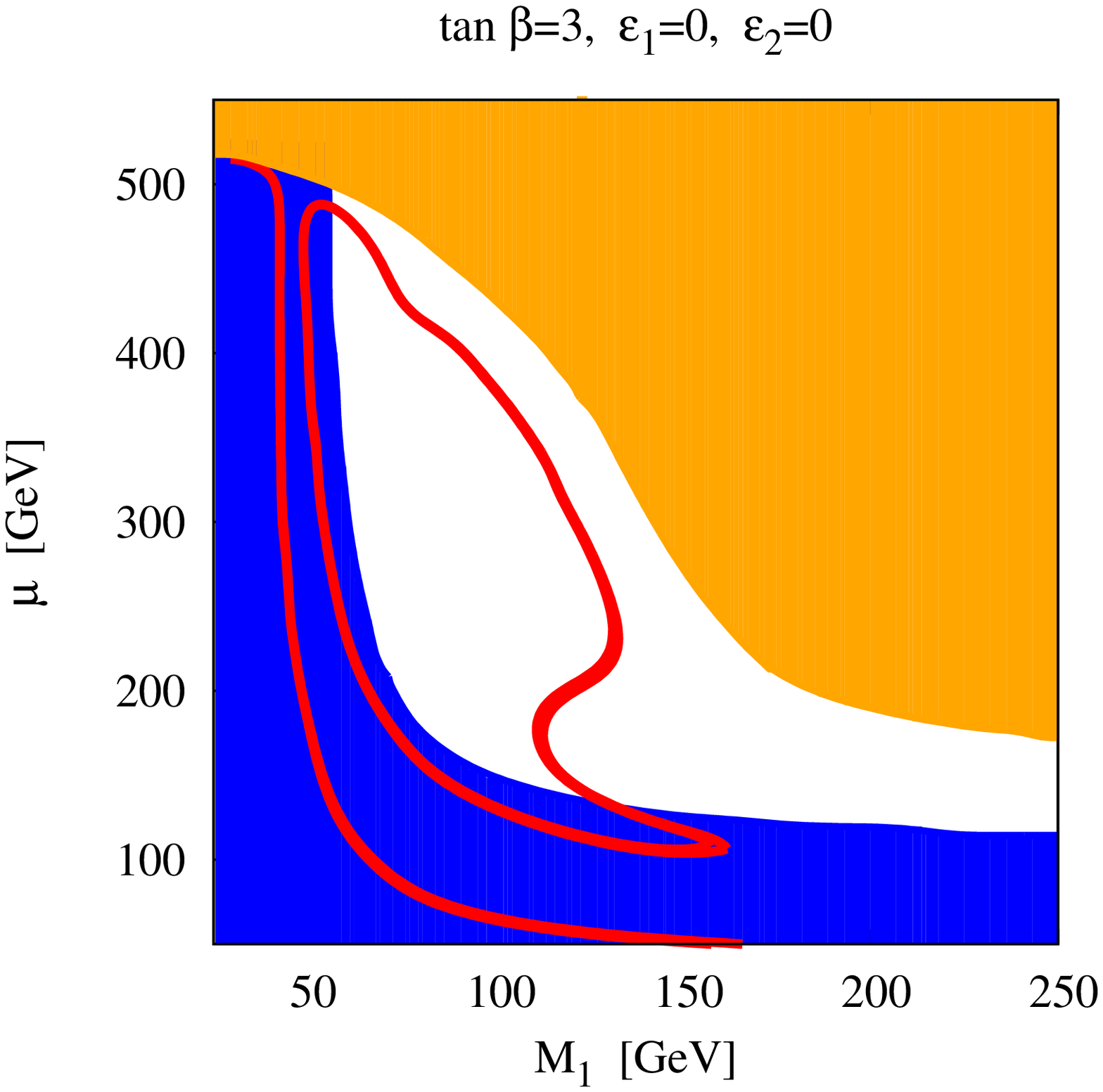}%\hspace{-2.3cm}
\includegraphics[width=5.5cm,angle=-90]{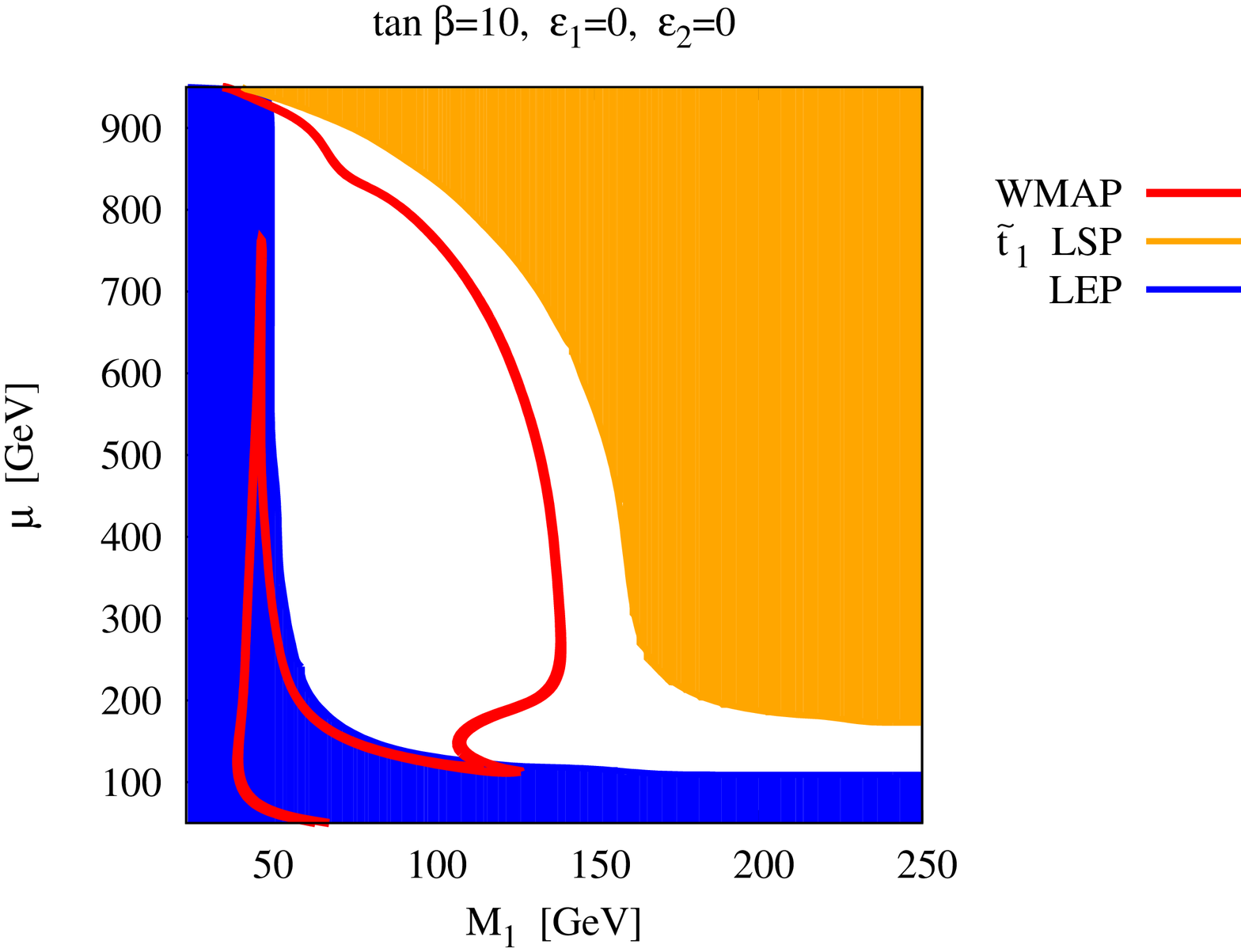}\\
\vspace{0.6cm}%\hspace{-2.5cm}
\includegraphics[width=5.5cm,angle=-90]{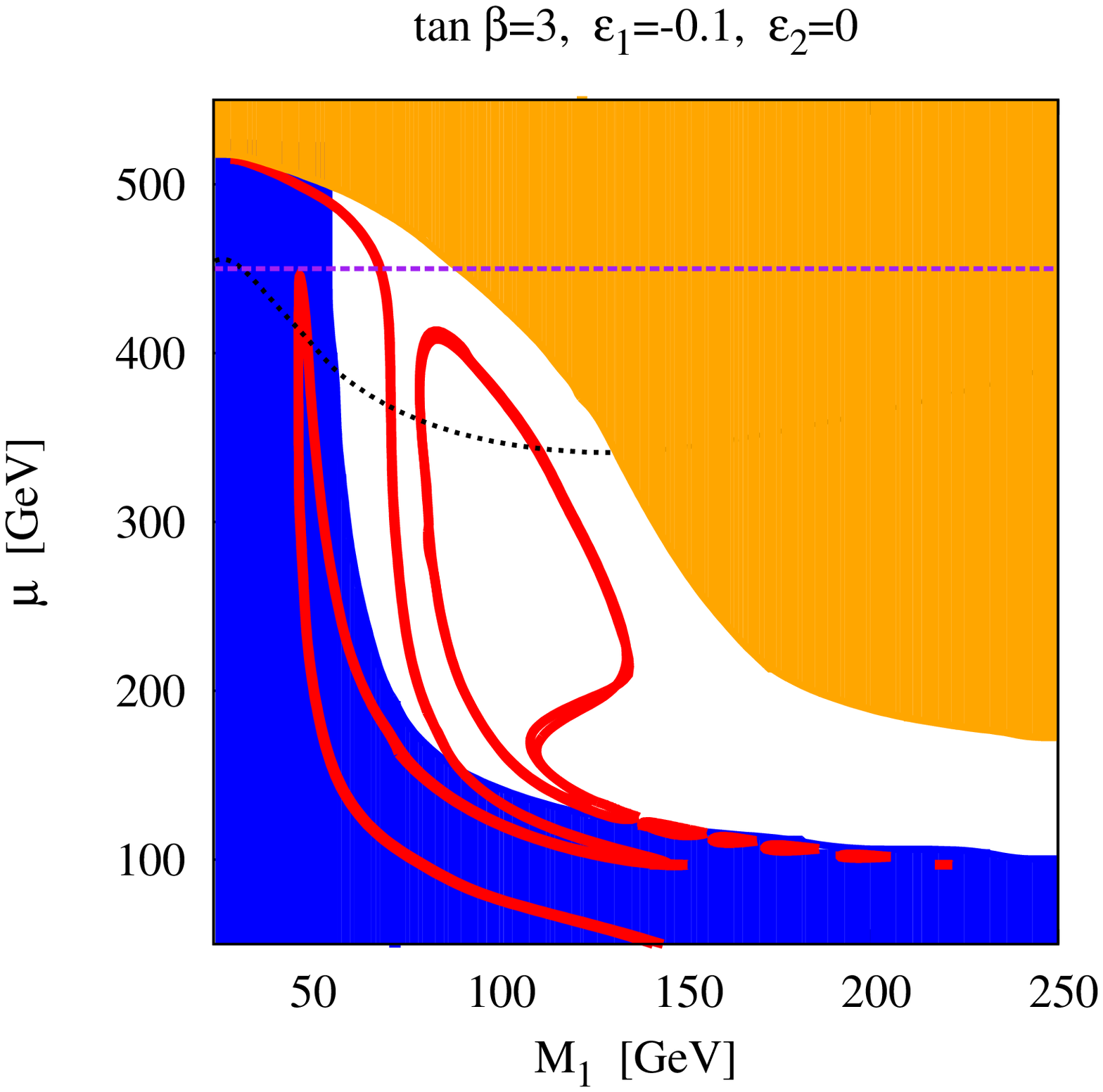}%\hspace{-2.3cm}
\includegraphics[width=5.5cm,angle=-90]{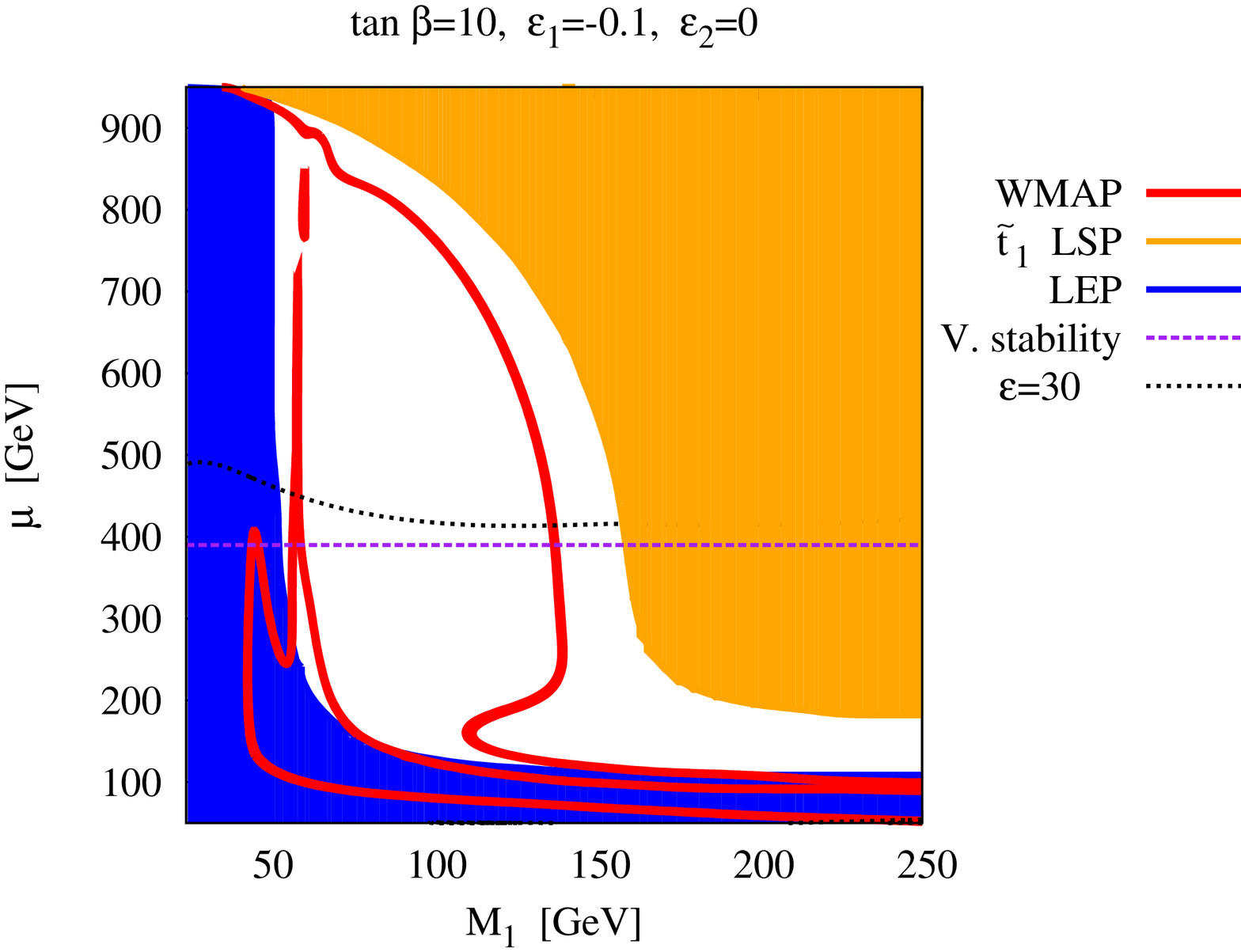}\\
\vspace{0.6cm}%\hspace{-2.5cm}
\includegraphics[width=5.5cm,angle=-90]{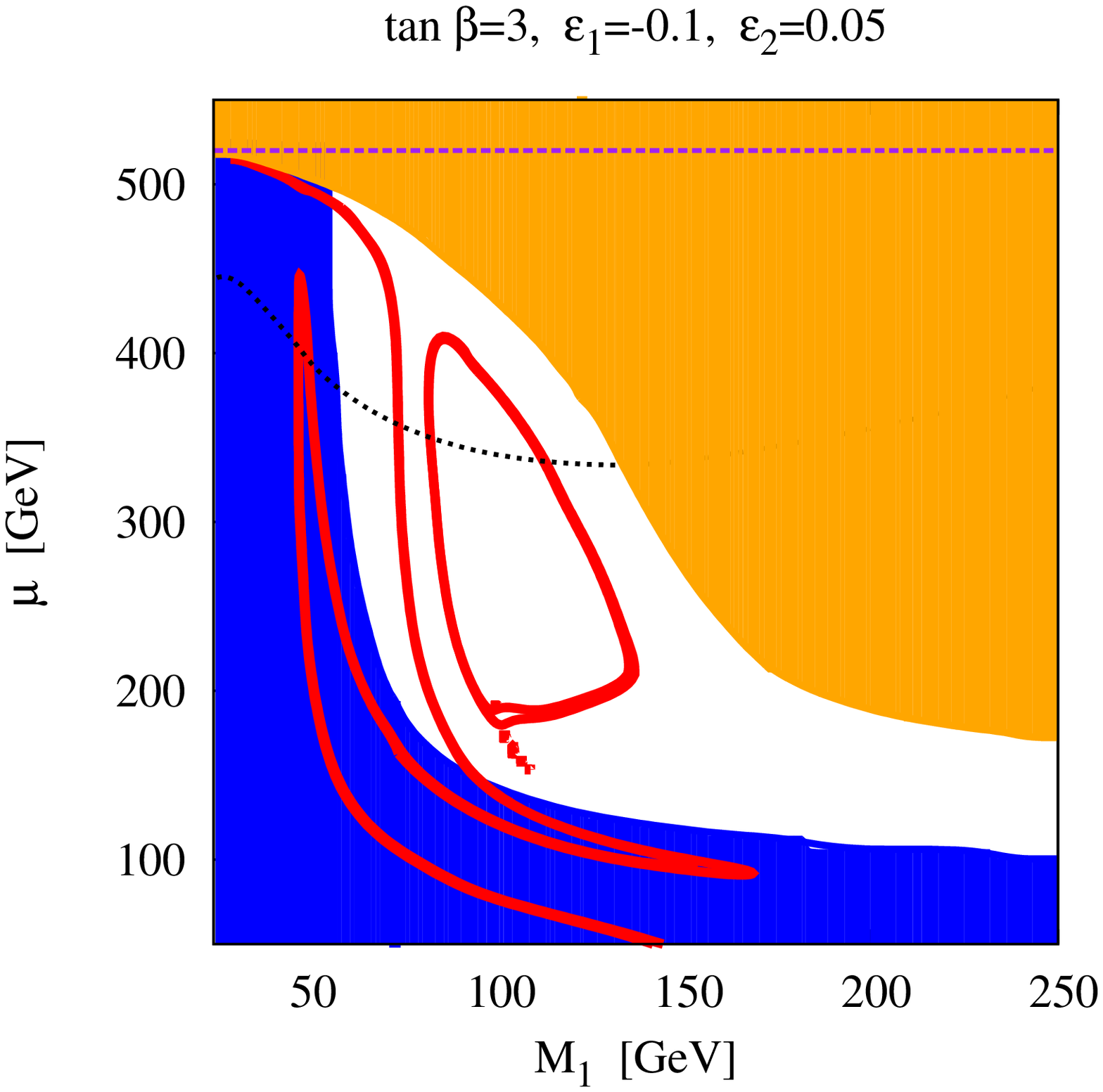}%\hspace{-2.3cm}
\includegraphics[width=5.5cm,angle=-90]{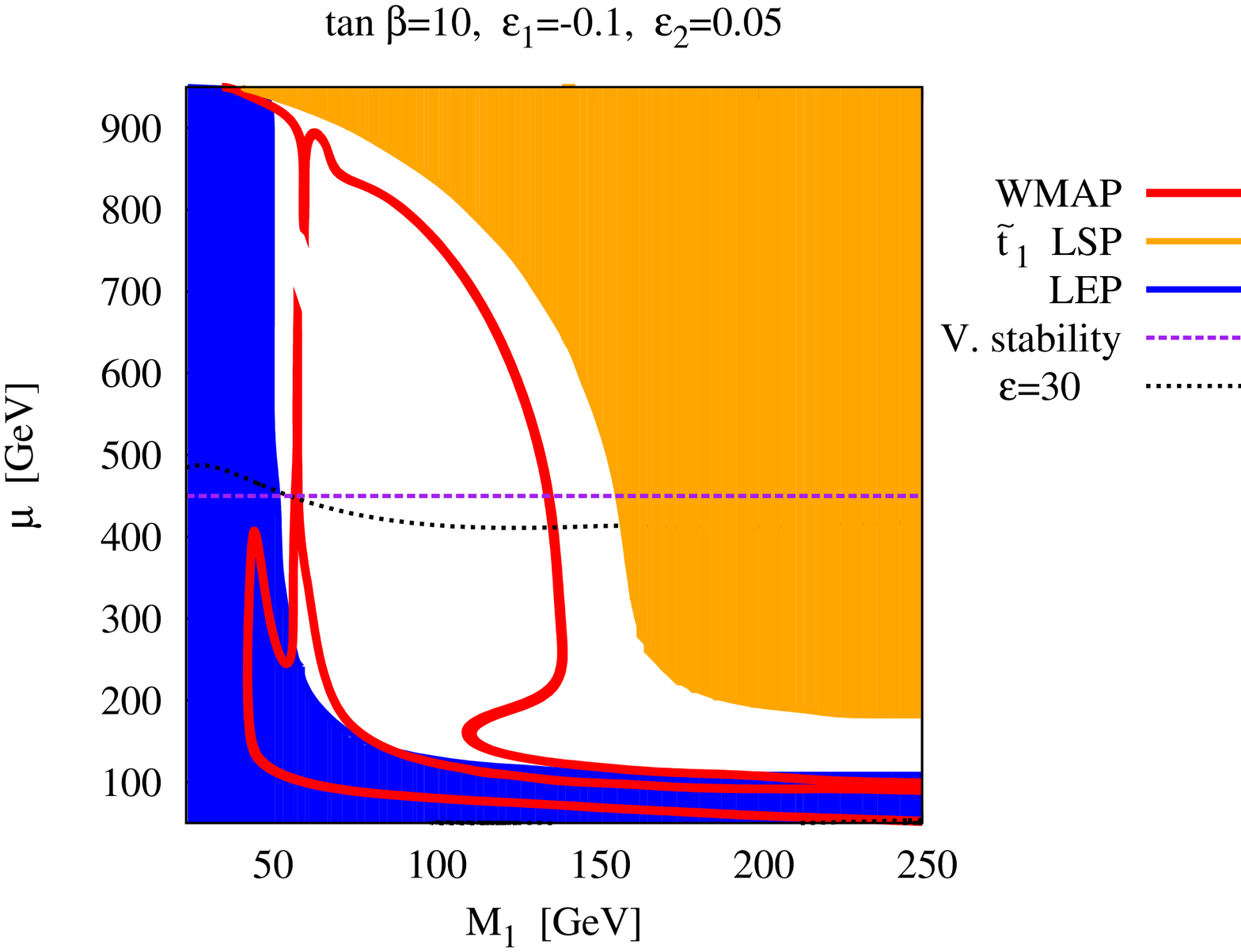}\\
\end{center}
%\vspace{-0.9cm}
\caption{{\footnotesize
Regions in the $[M_1,\,\mu]$ plane that can be detected by
XENON for the scenario with light stops and heavy sleptons. 
The black lines depict the detectability regions for the corresponding XENON
detector with exposures
$\varepsilon=30$, $300$ and $3000$ kg$\cdot$year: the area below the lines can be probed.
Whenever a line is absent, this means that the whole parameter space can be tested
by the experiment. The blue and orange regions depict the areas that are excluded by
direct LEP chargino searches and the requirement for a 
neutralino LSP respectively. The areas above the violet lines are excluded by the
metastable vacuum constraints.
}}
\label{dir2}
\end{figure}

Figure \ref{dir2} shows the exclusion lines for XENON with exposures
$\varepsilon=30$, $300$ and $3000$ kg$\cdot$year,
on the $[M_1,\,\mu]$ parameter space for our LSHS scenario, 
with the other parameters as defined in section \ref{LSHS} for
$\tan\beta=3$ (left panel) and $10$ (right panel).
Here again, the experiment will be sensitive to the regions below the black contours.
It can be seen that in general, the detection prospects are maximized for low values of the $M_1$
and/or the $\mu$ parameters. These regions correspond to a light $\chi_1^0$.
Although this might not be obvious in the figure, we have further seen that 
the scattering cross-section is enhanced near the region
$M_1\sim\mu$. In this case, the lightest neutralino is a mixed bino-Higgsino state,
favoring the $\chi_1^0-\chi_1^0-h$ and $\chi_1^0-\chi_1^0-H$ couplings.
Again, the detection prospects are also improved for low values of $\tan\beta$.
This is due to the fact that for $|\mu|\gg M_1$
the coupling between the LSP and the Higgs bosons is suppressed by
a factor $\sin 2\beta$.
We further checked that 
the first line of the figure (corresponding to the case without
the NR operators), besides being excluded by the Higgs mass, is partially ruled out
by the recent XENON10 \cite{Angle:2007uj} and CDMS \cite{Ahmed:2009zw} searches.

When introducing the dimension $5$ operators the detection prospects deteriorate, in a
similar way as in the last subsection. The main effect is again the rise of the lightest Higgs mass.
Furthermore, the $\chi_1^0-\chi_1^0-h$ coupling is suppressed.
The latter effect is very accentuated in the region where the LSP is Higgsino-like.
It is interesting however to notice that, for the case of large $\tan\beta$,
almost the whole area that falls outside the reach of XENON for $\varepsilon=30$ kg$\cdot$year
is already excluded by the vacuum stability constraint 
(i.e. the region above the violet line).
It should be noted that the BMSSM scenarios evade the aforementioned constraints from XENON10
and CDMS.

This scenario seems to offer exceptionally good detection perspectives.
Even with middle exposures, XENON will be able to detect dark matter in the whole viable region
for all three benchmarks and in the two models. 
%%%%%%%%%%%%%%%%%%%%%%%%%%%%%%%%%%%%%%%%%%%%%%%%%%%%%%%%%%%%%%%%%%%%%%%%%%%%%%%%%%%%%%%%%%%%%%%%%%%%%%%%%%%%%

\subsection{Gamma - rays from the Galactic Center}
Next, we examine the capacity of the Fermi mission to detect gamma-rays from neutralino 
annihilations coming from the galactic center region. We calculate the corresponding fluxes
and extract detectability limits considering gamma-rays within a cone of 
$\Delta\Omega \approx 3\cdot 10^{-5}$ around the galactic center. Our results are computed for
three halo profile cases already discussed in the first chapter: the Navarro, Frenk and White one, 
the Einasto profile as well as 
a NFW - like profile including adiabatic compression effects. The relevant values for the 
$\bar{J}$ quantity are shown in table \ref{tabProfilesBMSSM}.

\begin{table}
\begin{center}
\begin{tabular}{|c|ccccc|}
\hline
 & $a$ [kpc] & $\alpha$ & $\beta$ & $\gamma$ & $\bar{J}(3\cdot10^{-5}$ sr$)$\\
\hline
Einasto &  -   &   -   &       &   -    & $6.07\cdot10^3$\\
NFW     & $20$ & $1.0$ & $3.0$ & $1.0$  & $8.29\cdot10^3$\\
NFW$_c$ & $20$ & $0.8$ & $2.7$ & $1.45$ & $5.73\cdot10^6$\\
\hline
\end{tabular}
\caption{{\footnotesize Einasto, NFW and NFW$_c$
density profiles with the corresponding parameters,
and values of $\bar{J}(\Delta\Omega)$.
The latter has been computed by means of a VEGAS Monte-Carlo integration
algorithm, imposing a constant density for $r\le 10^{-7}$kpc so as to 
avoid divergences appearing in the NFW-like profiles.}}
\label{tabProfilesBMSSM}
\end{center}
\end{table}

The energy range we examine is $[1, 300]$ GeV which we divide into $20$ logarithmically evenly 
spaced energy bins and calculate the chisquare quantities as previously. The assumed
data acquisition period is $5$ years.

Concerning the background, we take into account two sources already described in the second chapter:
The bright source nearly coincident with the Galactic Center as detected by the HESS mission, as
well as the HESS measurements of the diffuse gamma-rays in the area surrounding the GC.

\subsubsection{Correlated stop-slepton masses}

\begin{figure}[tbp!]
\begin{center}
\vspace{-0.2cm}%\hspace{-2.5cm}
\includegraphics[width=5.5cm,angle=-90]{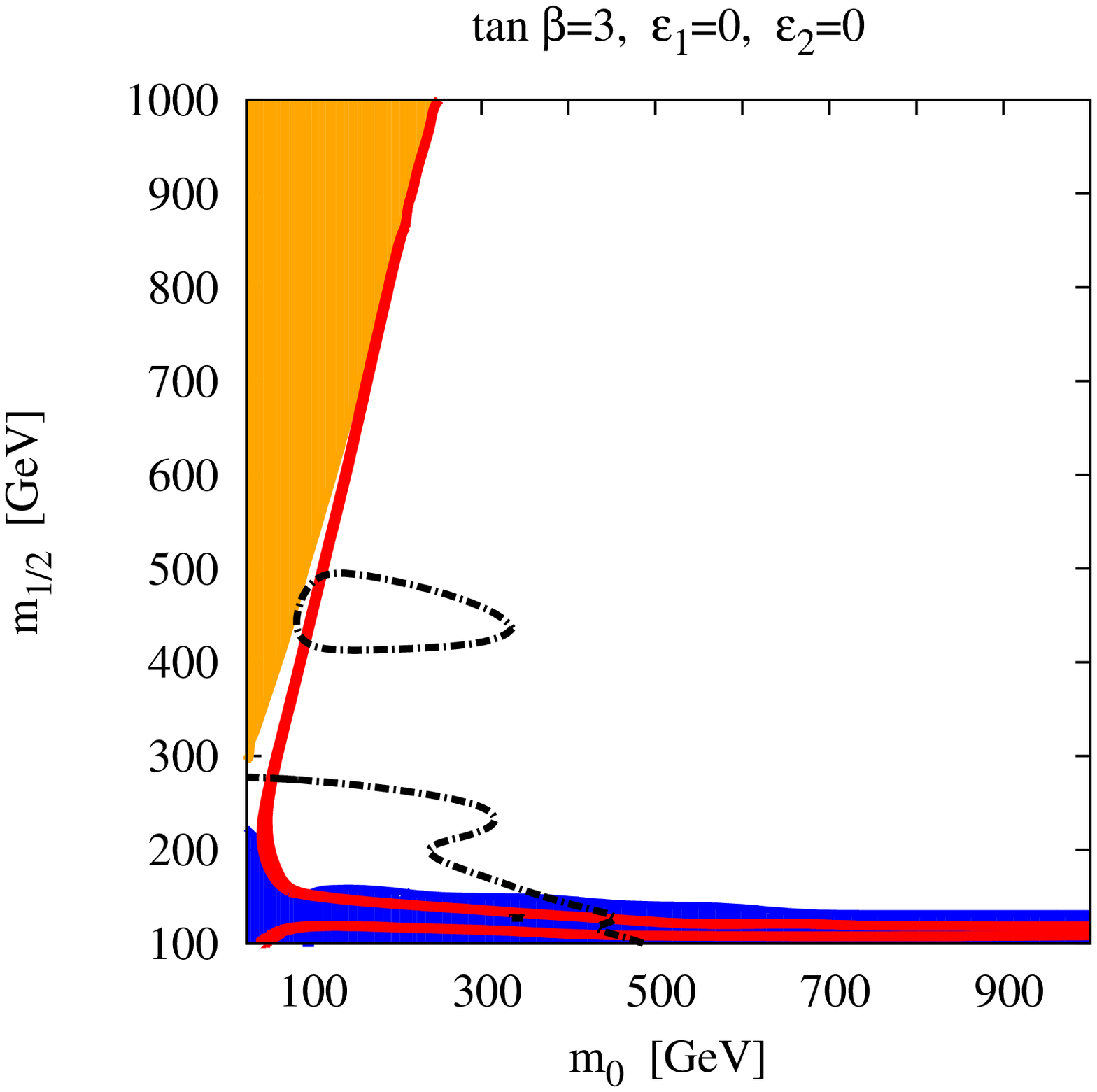}\hspace{0.2cm}
\includegraphics[width=5.5cm,angle=-90]{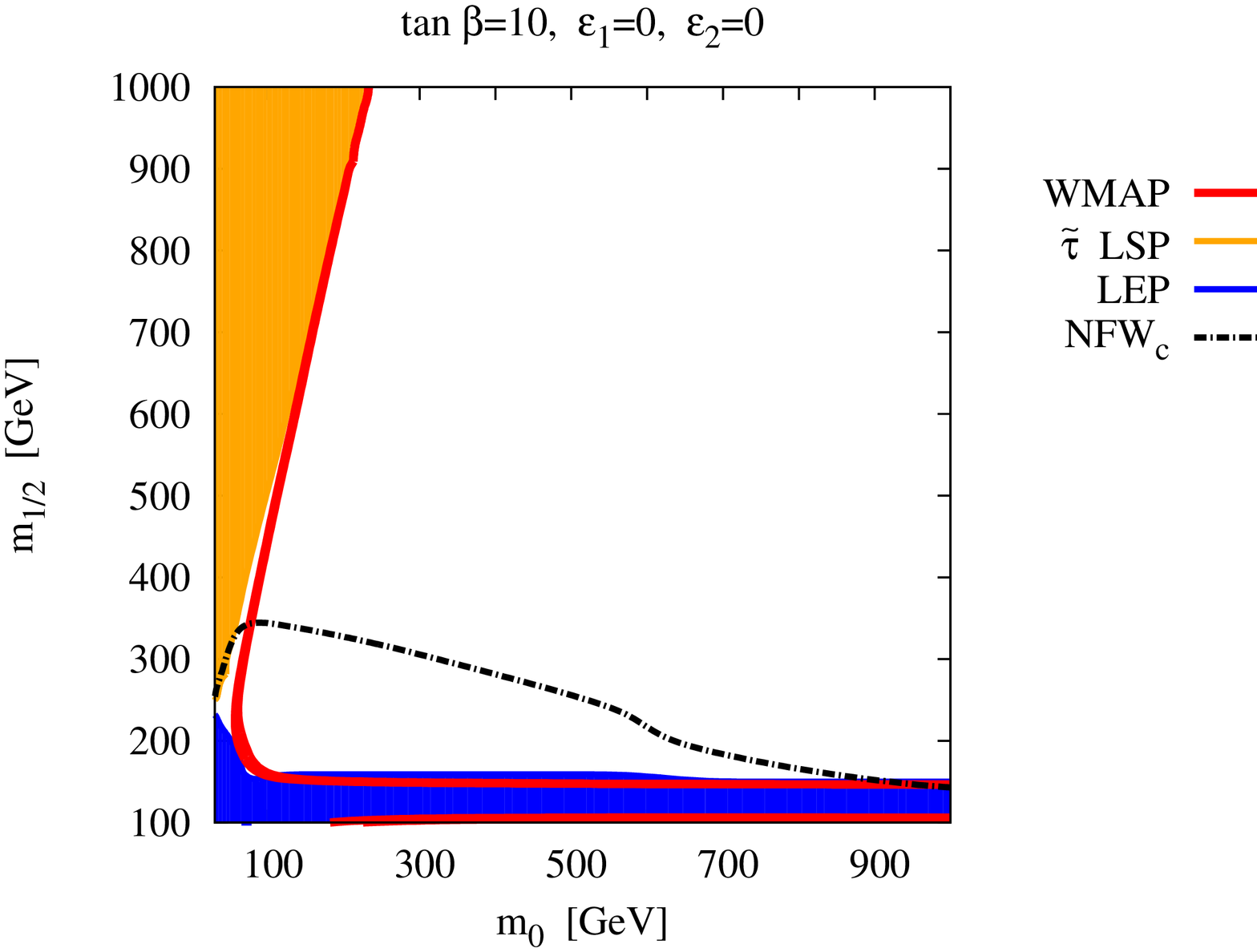}\\
\vspace{0.6cm}%\hspace{-2.5cm}
\includegraphics[width=5.5cm,angle=-90]{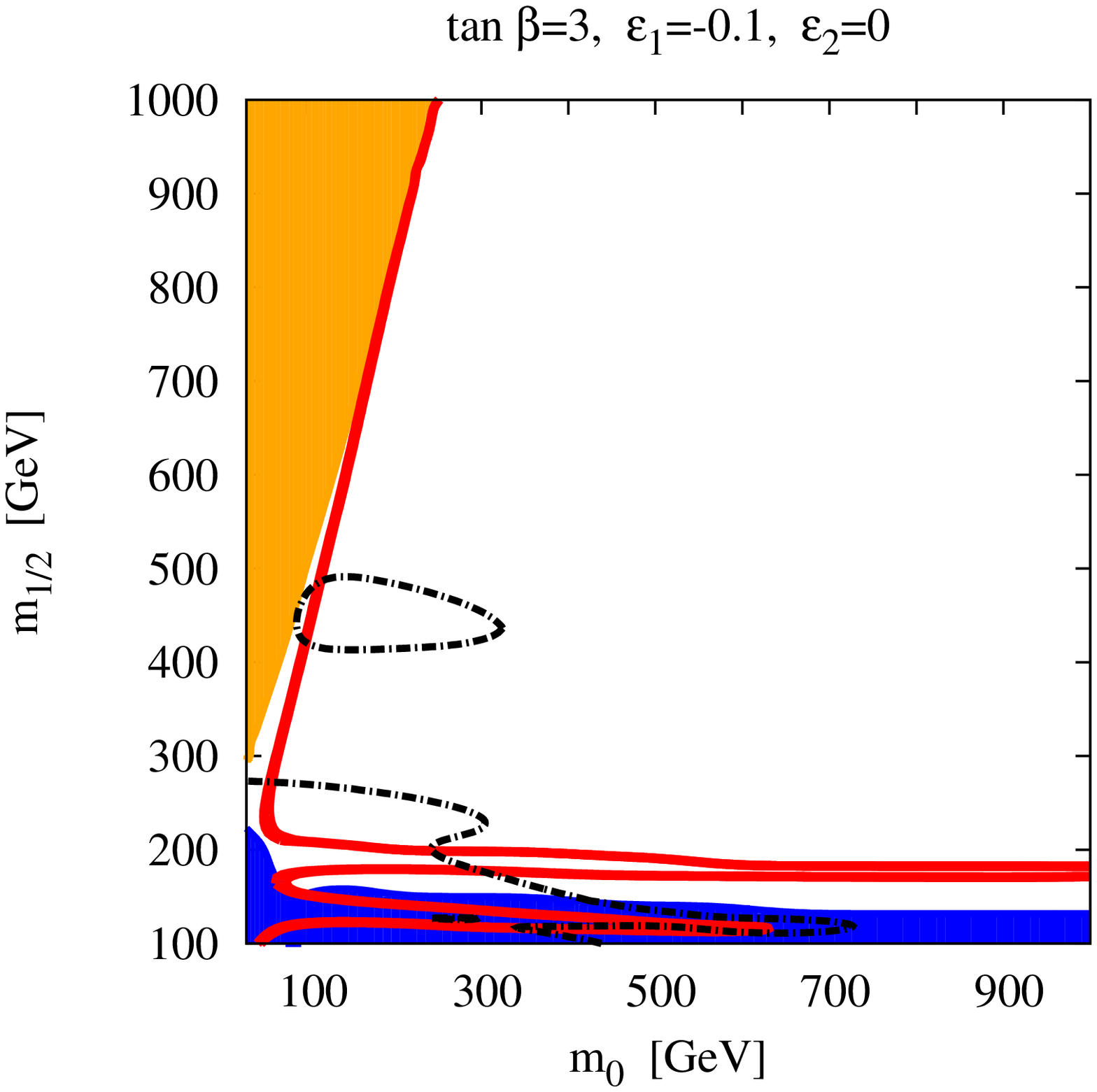}\hspace{0.2cm}
\includegraphics[width=5.5cm,angle=-90]{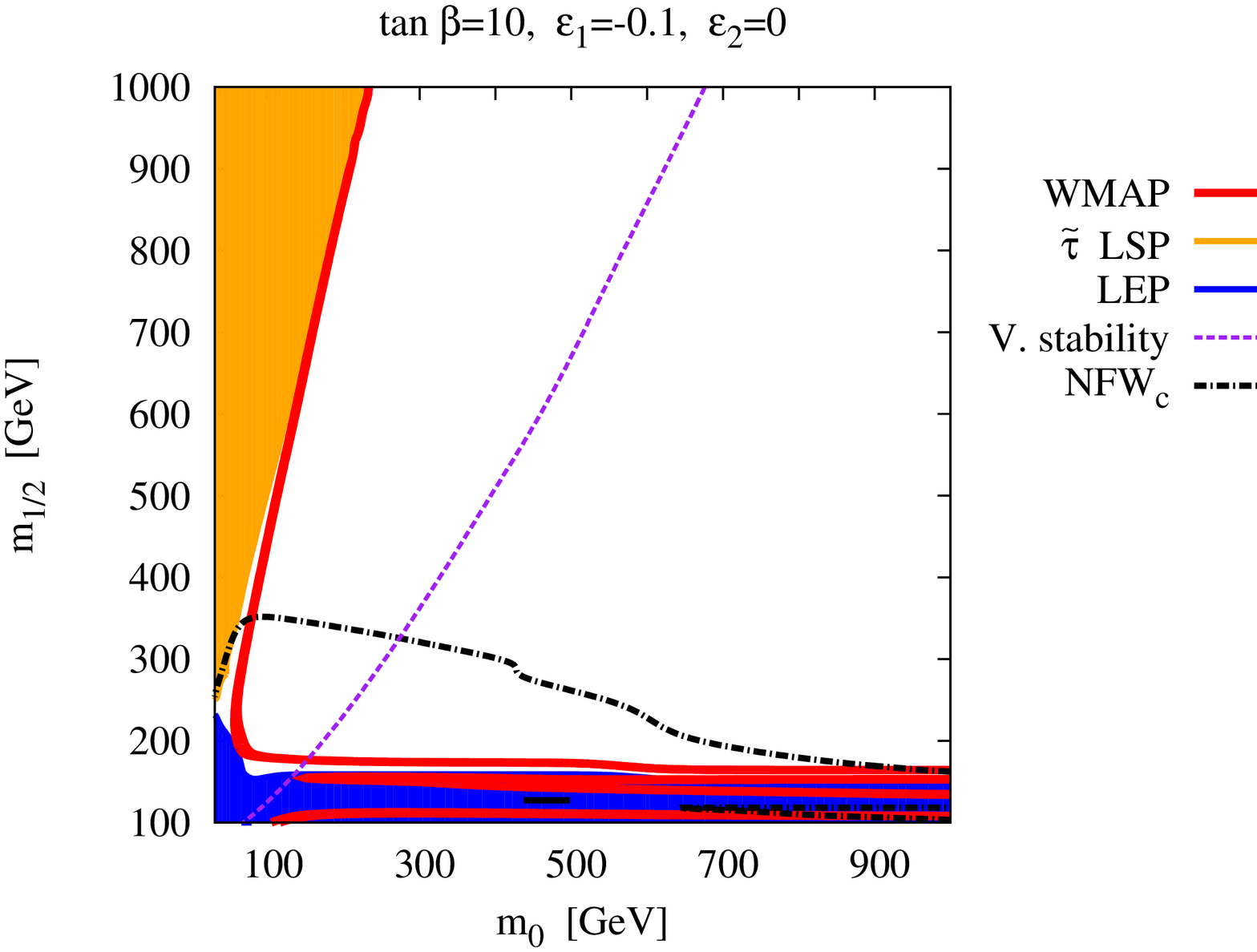}\\
\vspace{0.6cm}%\hspace{-2.5cm}
\includegraphics[width=5.5cm,angle=-90]{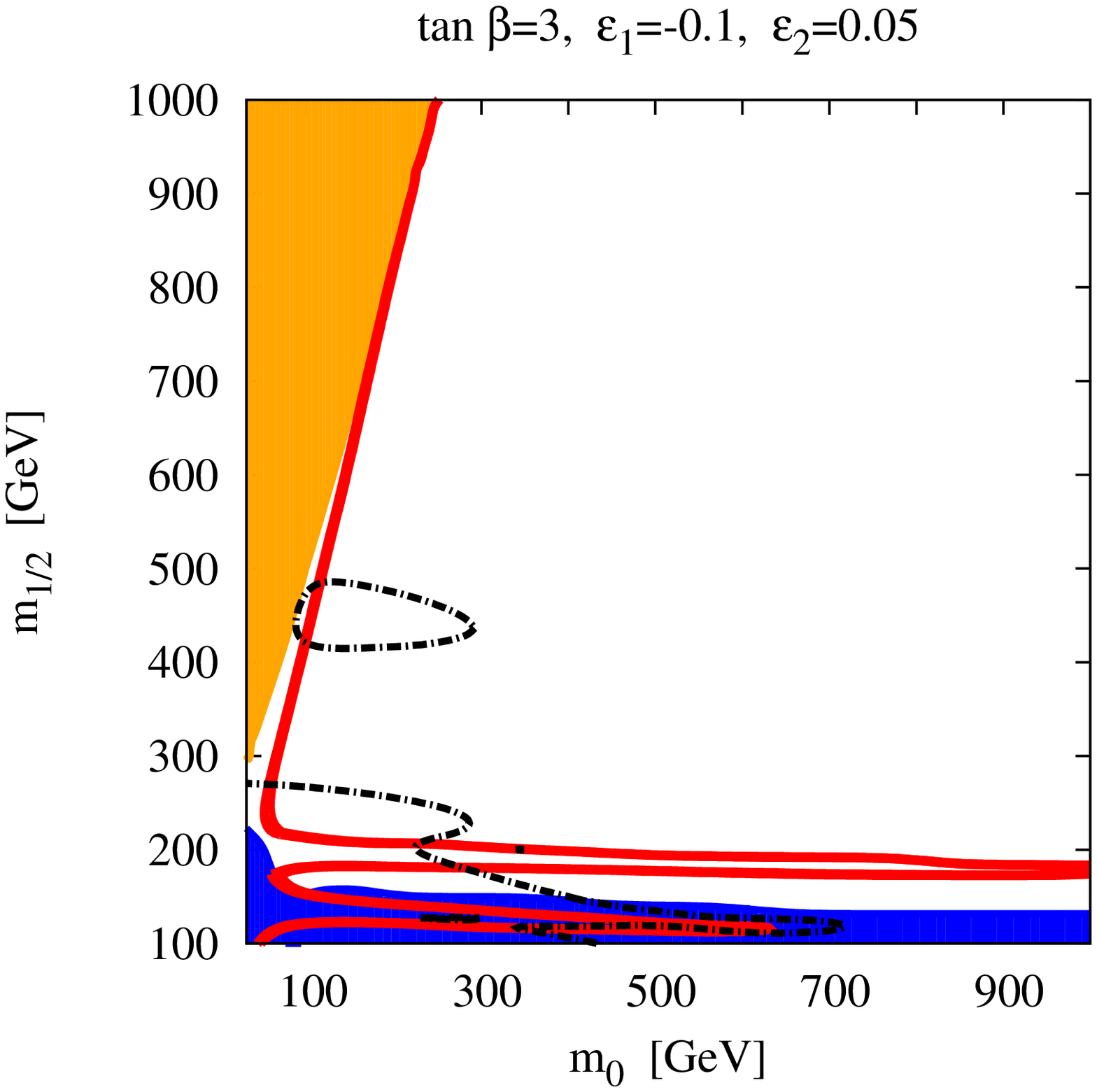}\hspace{0.2cm}
\includegraphics[width=5.5cm,angle=-90]{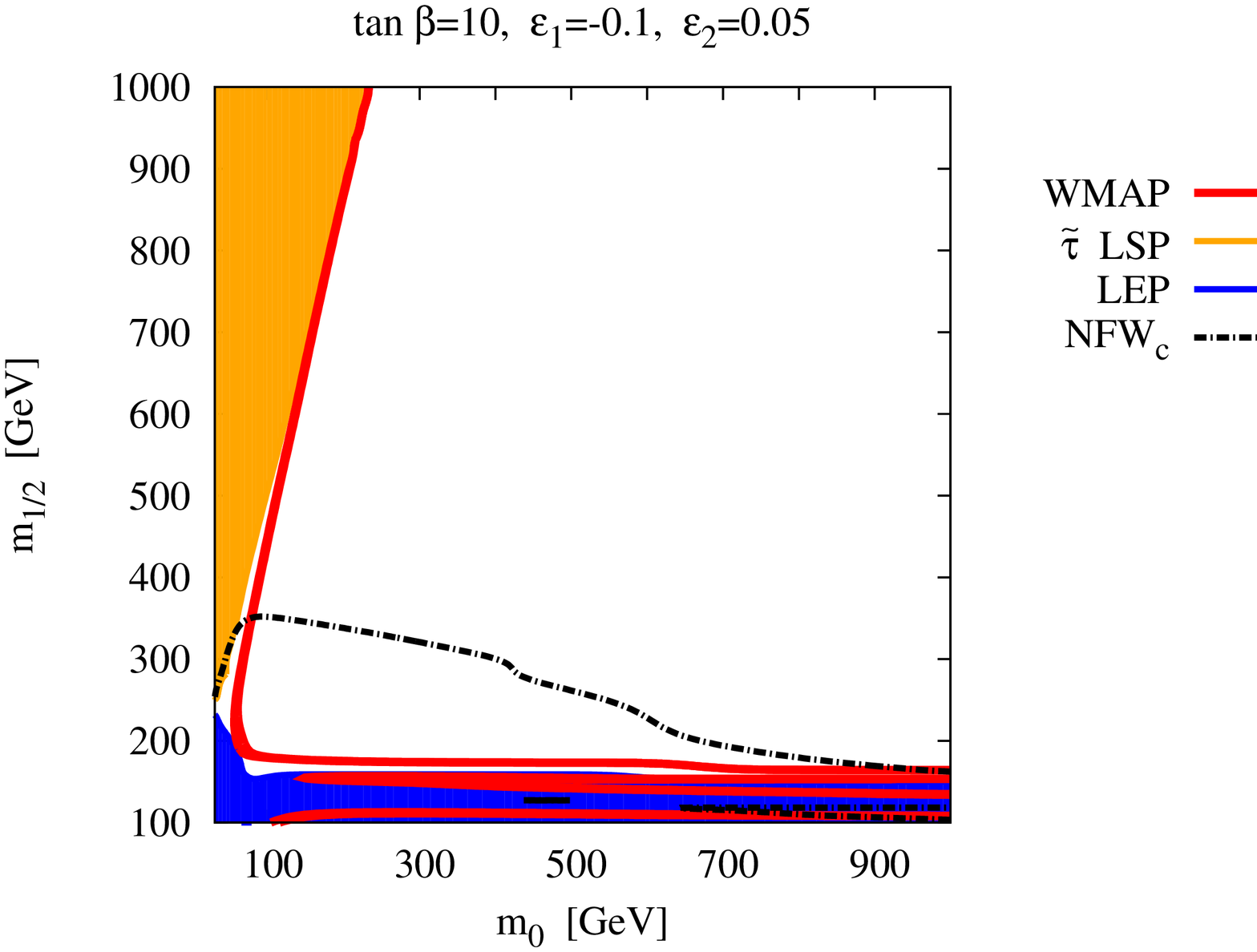}
\end{center}
% \vspace{-0.4cm}
\caption{\footnotesize{
Regions in the $[m_0,\,m_{1/2}]$ plane that can be detected by the
Fermi satellite mission for our mSUGRA-like scenario. 
The black lines depict the detectability regions for the corresponding halo profile assumptions 
and $5$ years of data acquisition: the area below and on the left of the lines can be probed.
The same applies to the top-resonance blob at $m_{1/2}\sim 450$ GeV appearing on the left-hand side
plots. For NFW and Einasto profiles, the model could not be tested.}}
\label{gam1}
\end{figure}

In figure \ref{gam1} we present our results concerning the detection perspectives at
the Fermi mission for the three halo profiles.
Fermi will be sensitive to the regions below the contours and, for 
$\tan\beta=3$, to the area inside the blob.

It can be seen that the detection prospects are maximized for low values of the $m_0$
and $m_{1/2}$ parameters. This is due to the fact that for higher $m_0$ values, the masses of the
squarks increase, causing the annihilation cross-section to decrease.
However, the growth of $m_{1/2}$ gives rise to resonnances or to 
the opening of some relevant production channels,
after passing some thresholds, increasing significantly $\langle\sigma v\rangle$.
These thresholds appear as features, especially in the left-hand side
plots, where the detectability lines follow a less smooth behavior.

The first feature corresponds to a light neutralino, with mass $m_\chi\sim m_Z/2$
($m_{1/2}\sim 130$ GeV). In that case the annihilation is done via the $s$-channel exchange
of a real $Z$ boson, decaying in hadrons ($\sim 70\%$), neutrinos ($\sim 20\%$) and charged
leptons ($\sim 10\%$). We can see that in this region, although it is excluded by the LEP constraints, 
the detection prospects are good.

Secondly, a threshold appears for $m_\chi\sim m_W$ ($m_{1/2}\sim 220$ GeV). The annihilation cross-section
is enhanced by the opening of the production channel of two real $W^\pm$ bosons in the final state.
This process takes place solely through chargino exchange, since both $Z$ and Higgs bosons
exchange are suppressed by taking the limit $v\to 0$. 
In fact, this feature is even more interesting:
right below the opening of the gauge boson final states, the detectability lines seem to ``avoid'' the
$h$-pole region, since the process $\chi_1^0 \chi_1^0 \rightarrow h \rightarrow f \bar{f}$ that
dominated at freeze-out and augmented the self-annihilation cross-section is an inefficient 
mechanism at present times, as discussed in Appendix \ref{FeynmanGraphs}. As the neutralino mass
increases a little, the chargino exchange process starts entering the game as the gauge boson
final state becomes accessible. And, contrary to annihilation into a $h$ propagator, the cross-section
for this process does not decease so dramatically as $v\to 0$.

The last threshold corresponds to the opening of the channel $\chi_1^0\chi_1^0\to t\bar{t}$
($m_{1/2}\sim 400$ GeV). The diagrams involved in such a process contain contributions from $t$-
and $u$-channel exchange of stops, and from $s$-channel exchange of $Z$'s and pseudoscalar Higgs bosons.
The aforementioned threshold appears as a particular feature on the left-hand side plots:
An isolated detectable region for $m_{1/2}\sim 400$-$500$ GeV and $m_0\lesssim 300$ GeV
corresponding to the annihilation into a pair of real top quarks.

Larger values for the annihilation cross-section can be reached for higher values of $\tan\beta$.
In that case, the production process of a pair of down-type quarks (in particular $b\bar b$ pairs) and
charged leptons, dominates the total cross-section. In fact, the diagrams containing exchanges
of a pseudoscalar Higgs boson or a sfermion are enhanced by factors $\tan\beta$ and $1/\cos\beta$
respectively.
On the other hand, for high values of $\tan\beta$, the channels corresponding to
the annihilations into $W^+W^-$ and $t\bar{t}$ vanish. The first because of the reduction of the
coupling $\chi_1^0-\chi_i^\pm-W^\mp$; the second because of important destructive interference
between diagrams containing the exchange of a $Z$ boson and stops.

For the present scenario, the introduction of the NR operators gives rise to a very mild signature.
Actually, as in almost the whole parameter space the lightest neutralino is bino-like, its couplings
do not vary drastically. Moreover, the increment in the Higgs masses has a small impact on the
$\langle\sigma v\rangle$ factor. For indirect detection prospects, the main effect corresponds to a
slight increase in the LSP mass. Let us emphasize on the fact that, however, the detectable regions
are in the BMSSM case more cosmologically relevant than in the corresponding plain MSSM one.

Concerning figure \ref{gam1}, let us note that the
only astrophysical setup in which some useful information can be extracted
is the NFW$_c$ one. This means that in this scenario, in order to have some positive detection
in the $\gamma$-ray channel, there should exist some important enhancement of the signal by some
astrophysical mechanism (as the adiabatic contraction mechanism invoked in this
case). We note that, and this will be different from the case of antimatter signals, there is
however no important constraint on astrophysical boosts from the Galactic Center.
Gamma-ray detection does not rely, as is the case for positrons that we shall examine 
followingly, that much on local phenomena. 
In this respect, the NFW$_c$ results can be characterized as optimistic (it has been pointed
out that even by changing the gravitational collapse conditions, the results can get even more
pessimistic in the case, e.g., of a binary black hole formation in the GC), but not excluded.

\subsubsection{Light stops, heavy sleptons}
\begin{figure}[tbp!]
\begin{center}
\vspace{-0.2cm}%\hspace{-2.5cm}
\includegraphics[width=5.5cm,angle=-90]{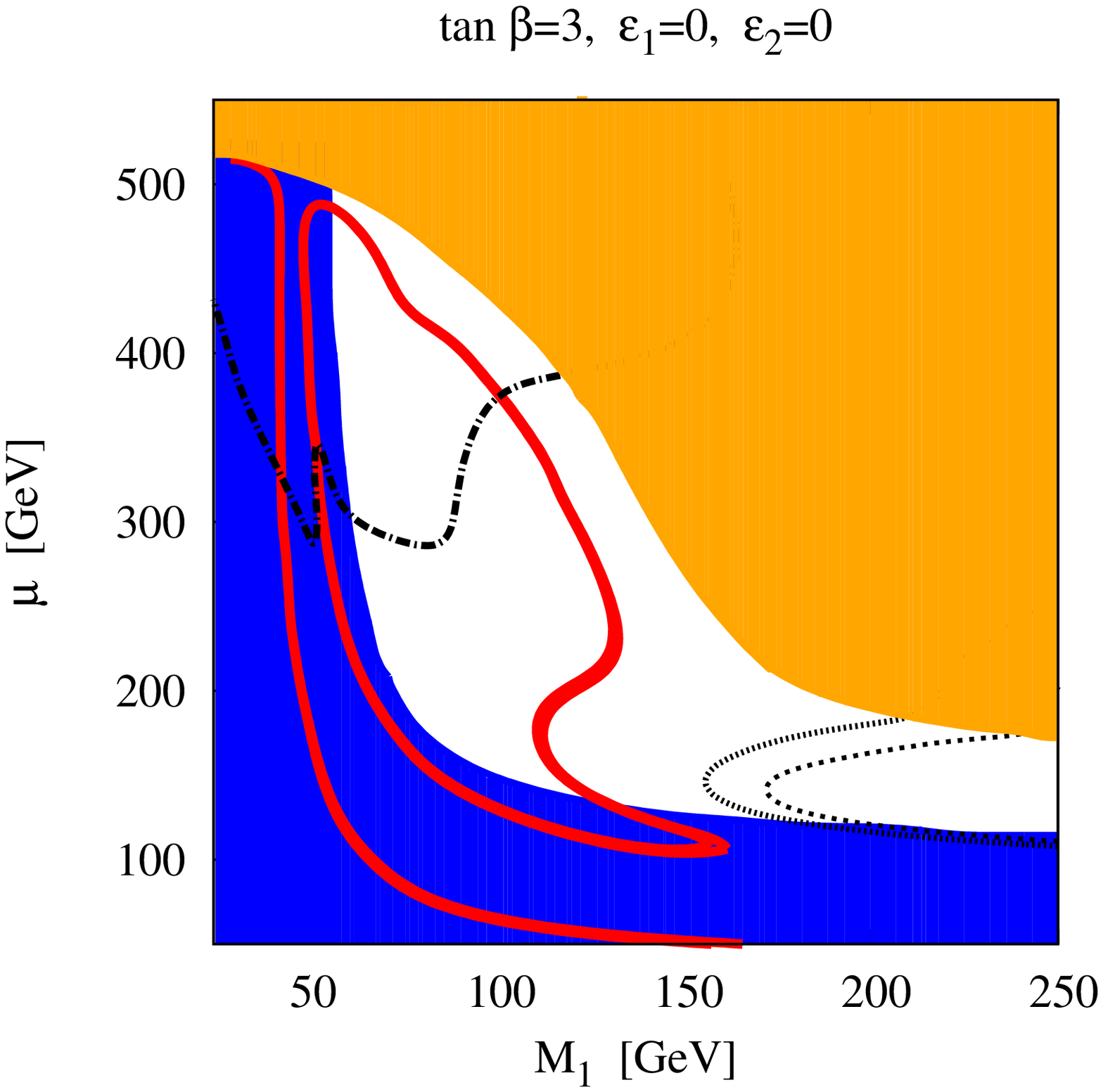}%\hspace{-2.3cm}
\includegraphics[width=5.5cm,angle=-90]{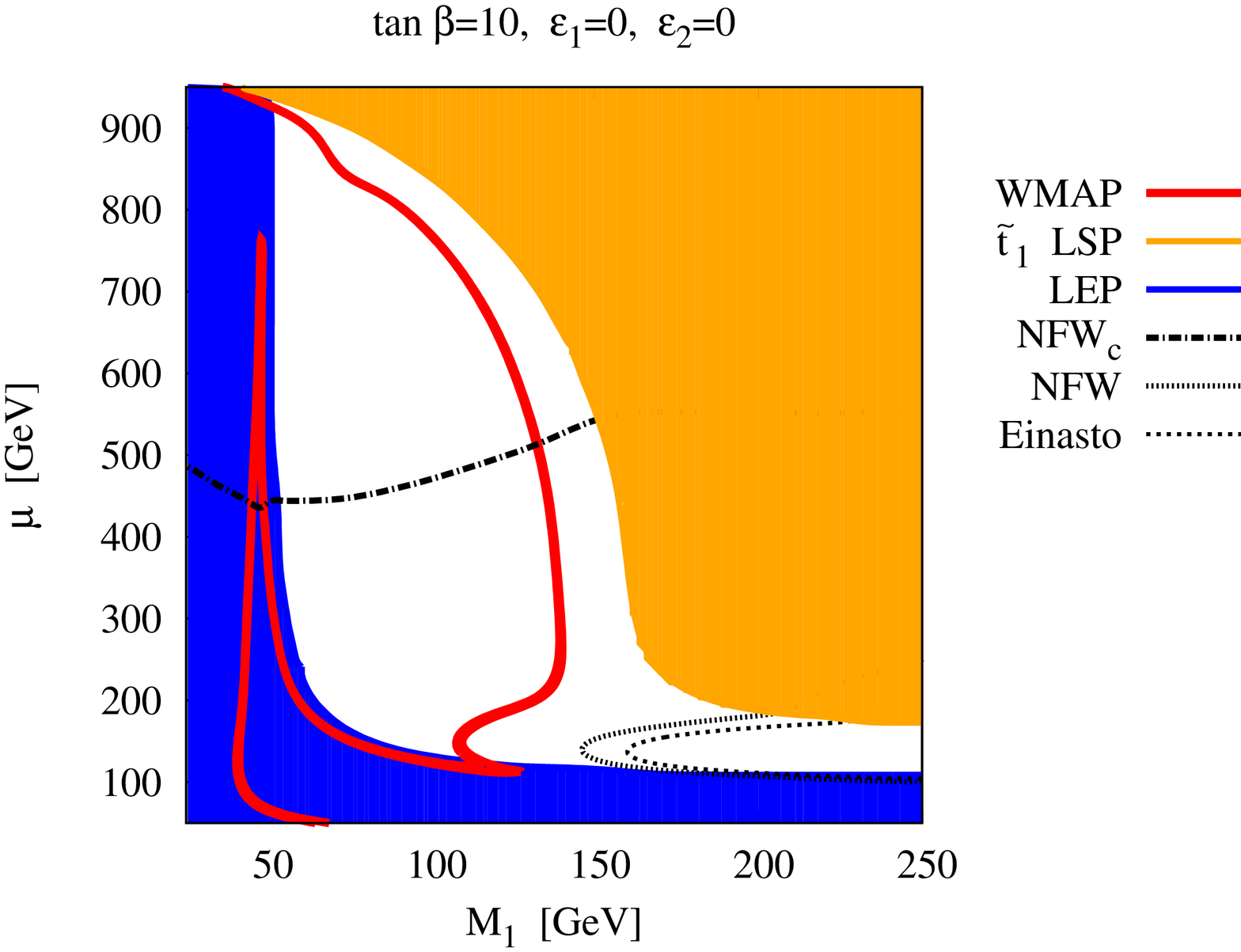}\\
 \vspace{0.6cm}%\hspace{-2.5cm}
\includegraphics[width=5.5cm,angle=-90]{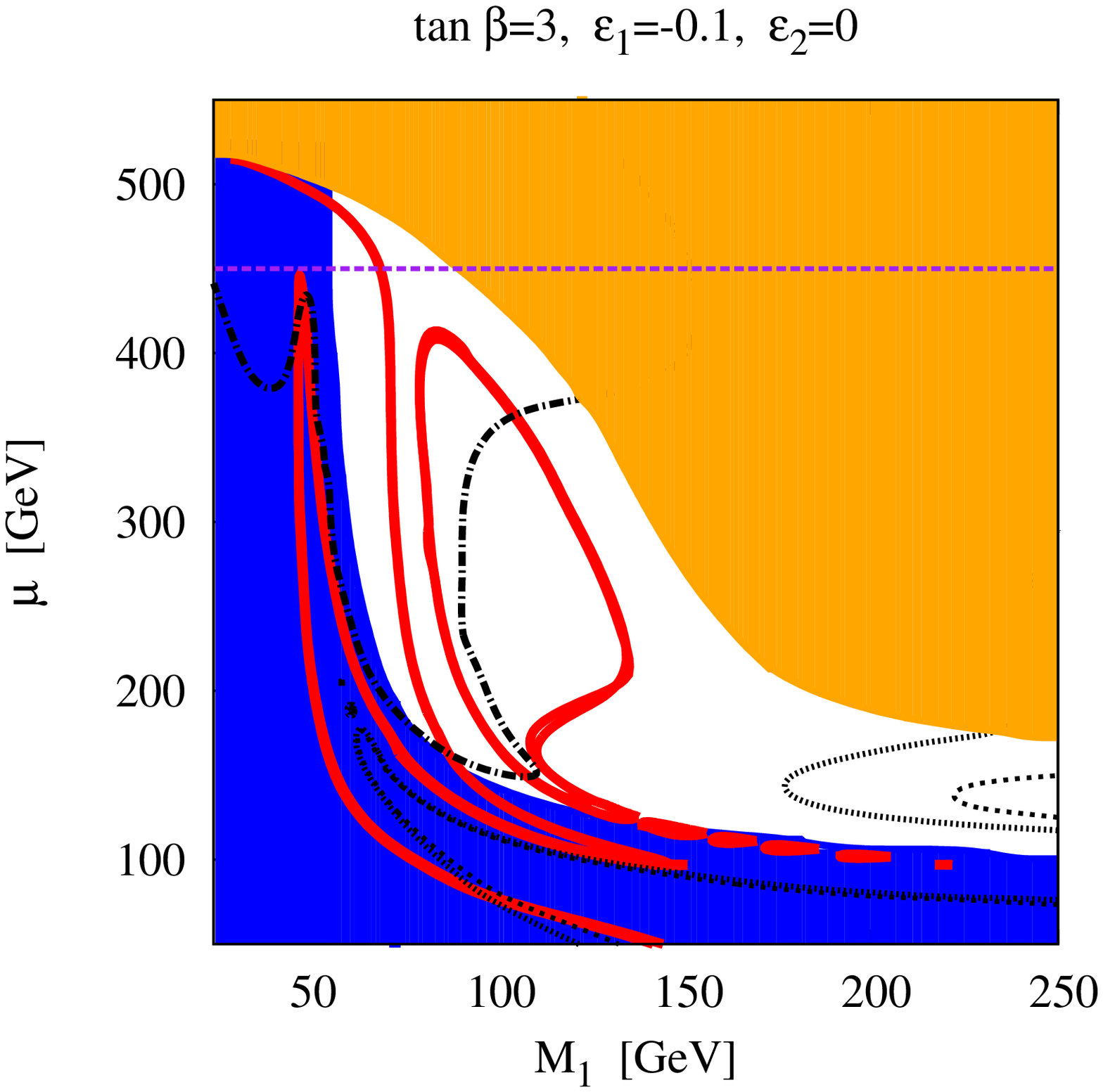}%\hspace{-2.3cm}
\includegraphics[width=5.5cm,angle=-90]{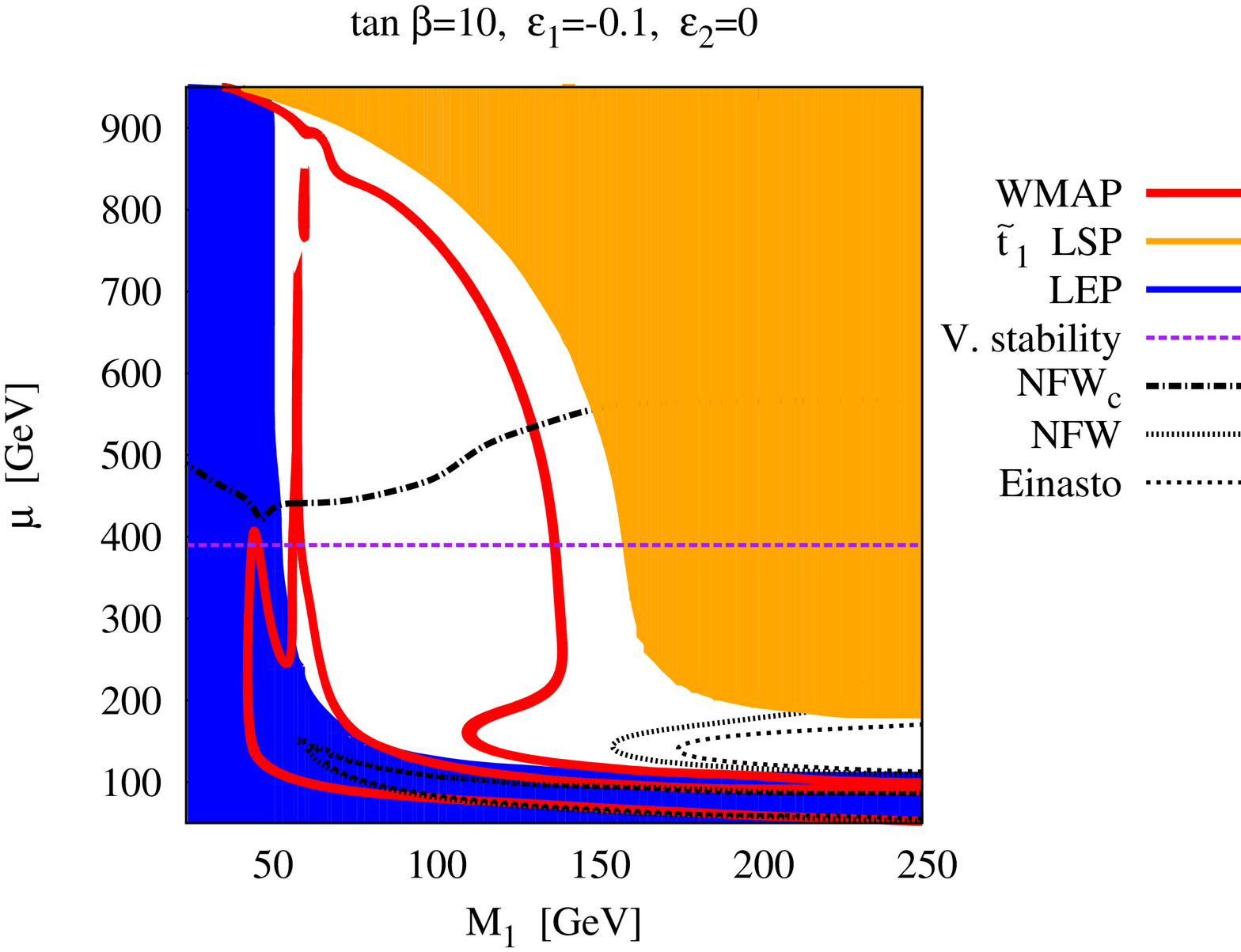}\\
 \vspace{0.6cm}%\hspace{-2.5cm}
\includegraphics[width=5.5cm,angle=-90]{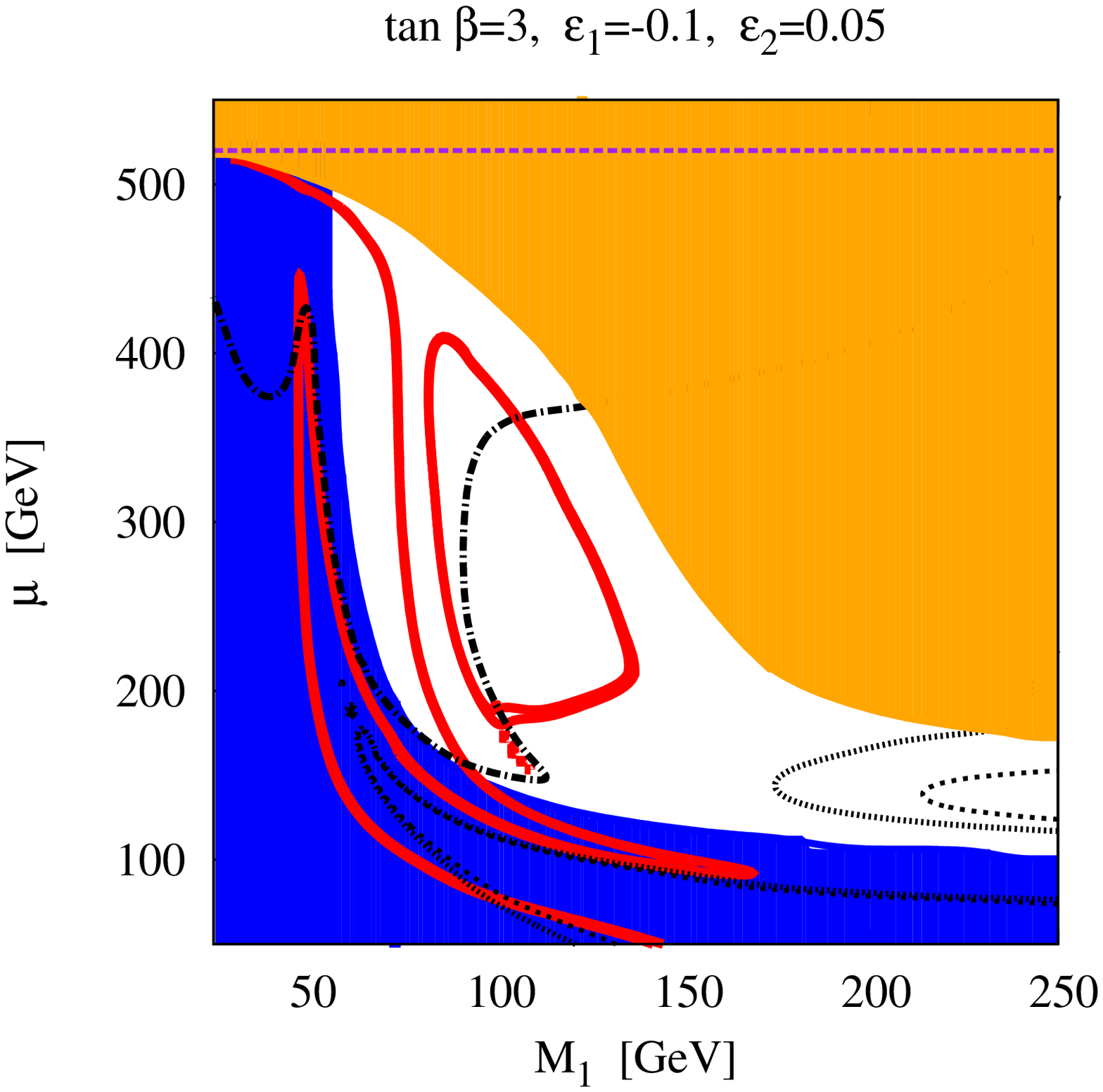}%\hspace{-2.3cm}
\includegraphics[width=5.5cm,angle=-90]{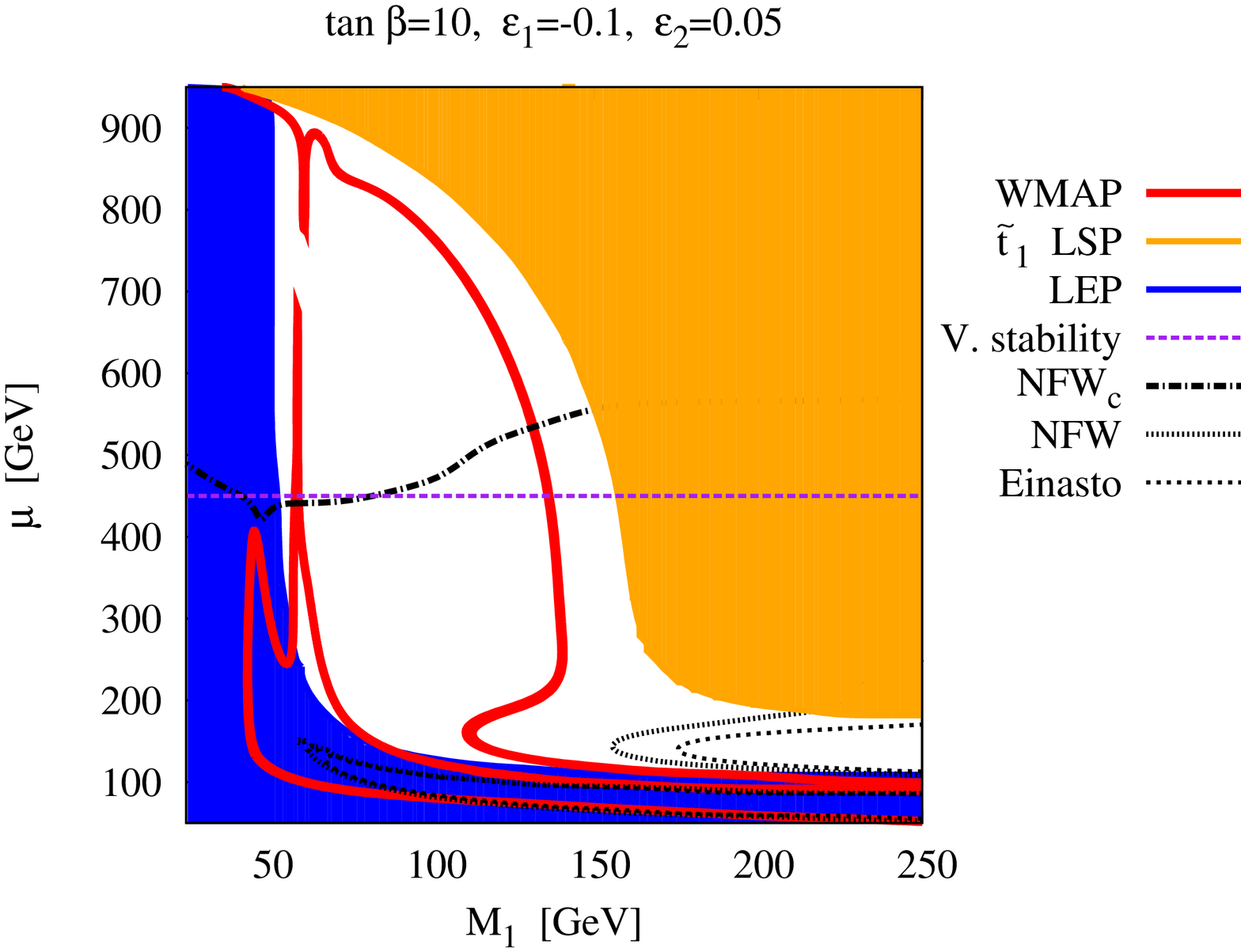}
\end{center}
% \vspace{-0.9cm}
\caption{{\footnotesize
Regions in the $[M_1,\,\mu]$ plane that can be detected by the
Fermi satellite mission for our scenario with light stops and heavy sleptons. 
The black lines depict the detectability regions for the corresponding halo profile assumptions 
and $5$ years of data acquisition: the area below the lines can be probed. The same applies
to lines forming closed regions with respect to the axes, as is the case for the NFW and NFW$_c$
profiles: the parameter space points lying in the interior of these regions yield signals that
are detectable.}}
\label{gam2}
\end{figure}

Figure \ref{gam2} presents the results for the second scenario with light stops and heavy sleptons.
The experiment will be sensitive to the regions below/on the right of the contours.
Again, the detection prospects are maximized for low values of the $M_1$
and $\mu$ parameters, corresponding to light WIMPs.
However, the growth of any of the latter parameters gives rise to the opening of some production
channels or to some resonnances, enhancing significantly the neutralino self-annihilation cross-section.
The first feature appears for $m_\chi\sim m_Z/2$ and corresponds to the $s$-channel exchange
of a real $Z$ boson. 

The second one concerns the production channel of two real $W$ bosons. Once again, 
we notice the important difference in the detectability lines \textit{at} and right \textit{after} the
$h$ - pole region.
Let us note that in this scenario the neutralino LSP can be as heavy as
$\sim 110$ GeV, implying that the annihilation into a pair of top quarks is never kinematically allowed.

Finally, the region where $M_1\gg\mu$ is highly favored for indirect detection due to
the fact that the LSP is Higgsino-like, maximizing its coupling to the $Z$ boson. Let us recall
that the $Z$ boson does not couple to a pure gaugino-like neutralino. This feature appears as an
isolated region detectable by all three halo profiles in the right area of all plots. Unfortunately,
these regions are cosmologically disfavored, as they yield too low a relic density.

Large values for the annihilation cross-section can be reached for high values of $\tan\beta$,
mainly because of the enhanced production of $b\bar b$ pairs.
On the other hand, for high values of $\tan\beta$, the threshold corresponding to the opening
of the annihilation into $W^+W^-$ is suppressed or enhanced for $\mu\gg M_1$ or $\mu\ll M_1$ respectively, 
due to the dependence of the $\chi_1^0-\chi_i^\pm-W^\mp$ coupling on the texture of the LSP.

For the present scenario, the introduction of the NR operators gives rise to an important
increase of the $\chi_1^0-\chi_1^0-A$ coupling when $\mu> M_1$, and therefore to a boost
in the annihilation into fermion pairs. On the other hand, as the Higgs boson $h$
becomes heavier, the processes giving rise to the final state $h\,Z$ get kinematically closed.

In the case presented in figure \ref{gam2}, there is a positive detection for all three halo profiles;
however, the regions that can be probed for either the NFW or the Einasto cases are cosmologically
irrelevant.

In fact, they could give rise to a positive detection near the $Z$-funnel and in the region where
the LSP is a Higgsino state ($M_1\gtrsim 150$ GeV); nevertheless the first is already excluded by LEP
(at least for minimal scenarios) and the second generates too small a dark matter
relic density, below the WMAP limits.
On the other hand, the profile NFW$_c$ could test a large amount of the parameter space we examine,
particularly for high values of $\tan\beta$. Only the Higgs peak and the regions with a heavy LSP
escape from detection.

%%%%%%%%%%%%%%%%%%%%%%%%%%%%%%%%%%%%%%%%%%%%%%%%%%%%%%%%%%%%%%%%%%%%%%%%%%%%%%%%%%%%%%%%%%%%%%%%%%%%%%%

\subsection{Positron detection}

\subsubsection{Correlated stop-slepton masses}
The results concerning the detectability perspectives for
the CMSSM-like scenario in the positron detection channel 
are quite pessimistic.
We already mentioned in our discussion of the singlet scalar model that
since the PAMELA and Fermi measurements, 
and according to our conservative treatment of considering the whole combined measurements as
the background for our study, the main issue in the positron channel is an
extreme domination of all measurements by a large background severely obscuring the signal.

One could invoke large boost factors of an astrophysical nature
as was the case in the first efforts to explain
the PAMELA excess through dark matter annihilations, a case in which a larger portion of the 
parameter space would be visible. However, it has been pointed out  
that it is highly unlikely to expect large boost factors due, e.g., to substructures
in the halo \cite{Lavalle:1900wn}. In this respect, if we assume a maximal clump-due signal enhancement 
by a factor $\sim 10$,  the only hope for positive detection of a non-LEP excluded area
might come for the bulk region, as it is the only one lying at the limits of detectability.
For the sake of brevity, we omit the relevant plots for the mSUGRA-like benchmark, since
no point of the parameter space can be tested.

We note that the considered background and AMS02 setup is exactly the same as in the singlet scalar
treatment presented in the previous chapter, with the data acquisition period being fixed at $3$ years.

\subsubsection{Light stops, heavy sleptons}
\begin{figure}[tbp!]
\begin{center}
\vspace{-0.2cm}%\hspace{-2.5cm}
\includegraphics[width=5.5cm,angle=-90]{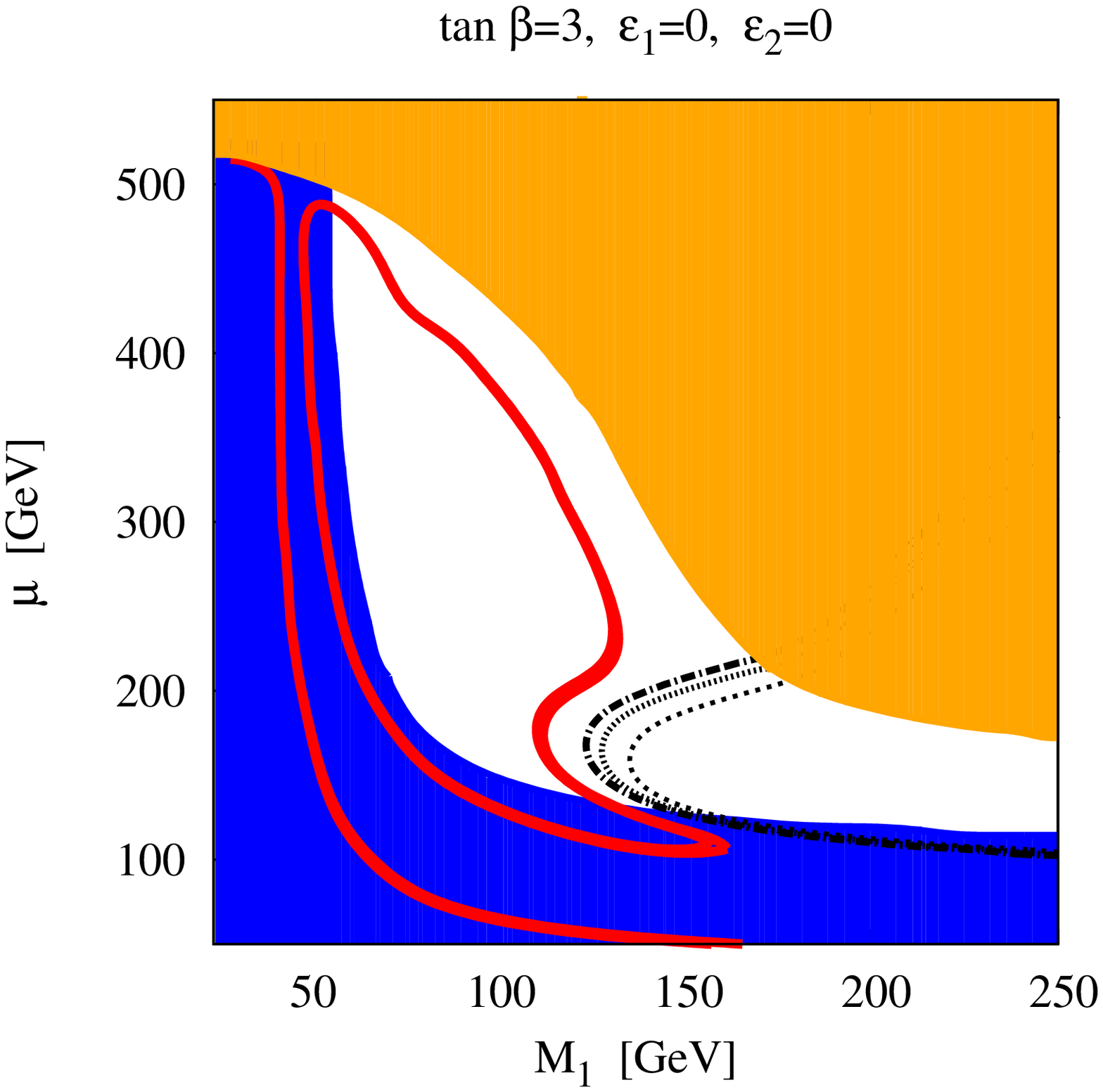}%\hspace{-2.3cm}
\includegraphics[width=5.5cm,angle=-90]{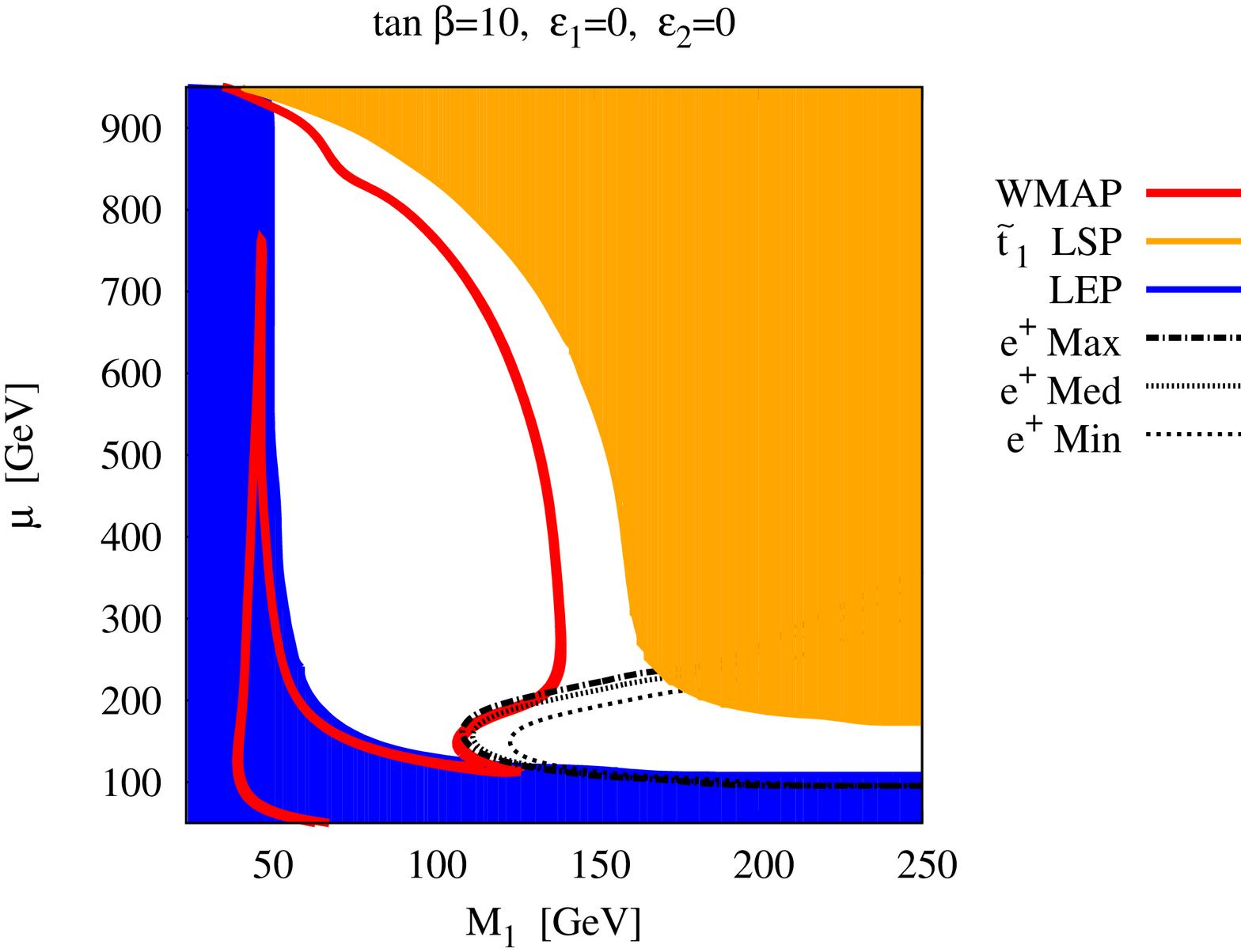}\\
\vspace{0.6cm}%\hspace{-2.5cm}
\includegraphics[width=5.5cm,angle=-90]{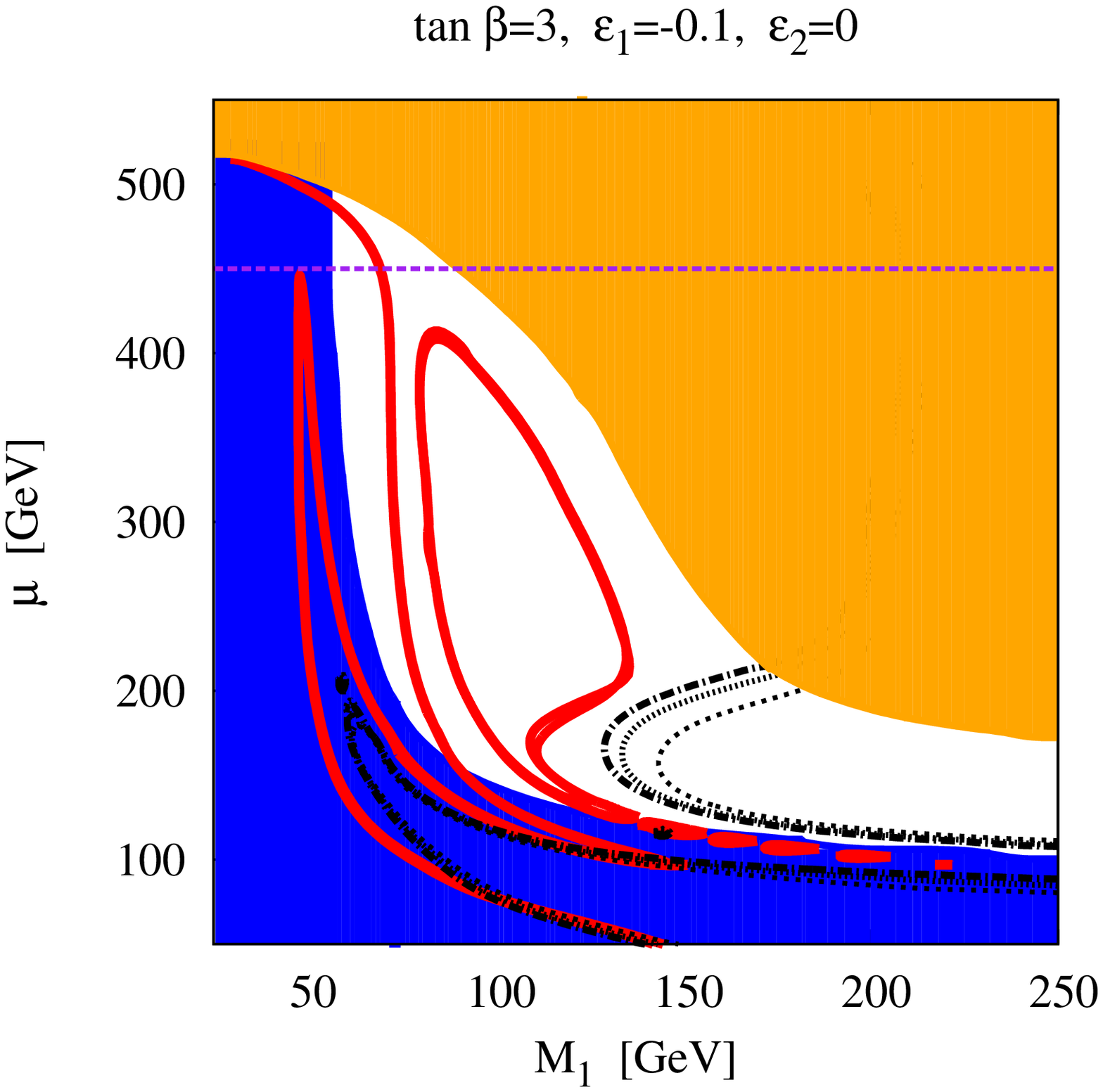}%\hspace{-2.3cm}
\includegraphics[width=5.5cm,angle=-90]{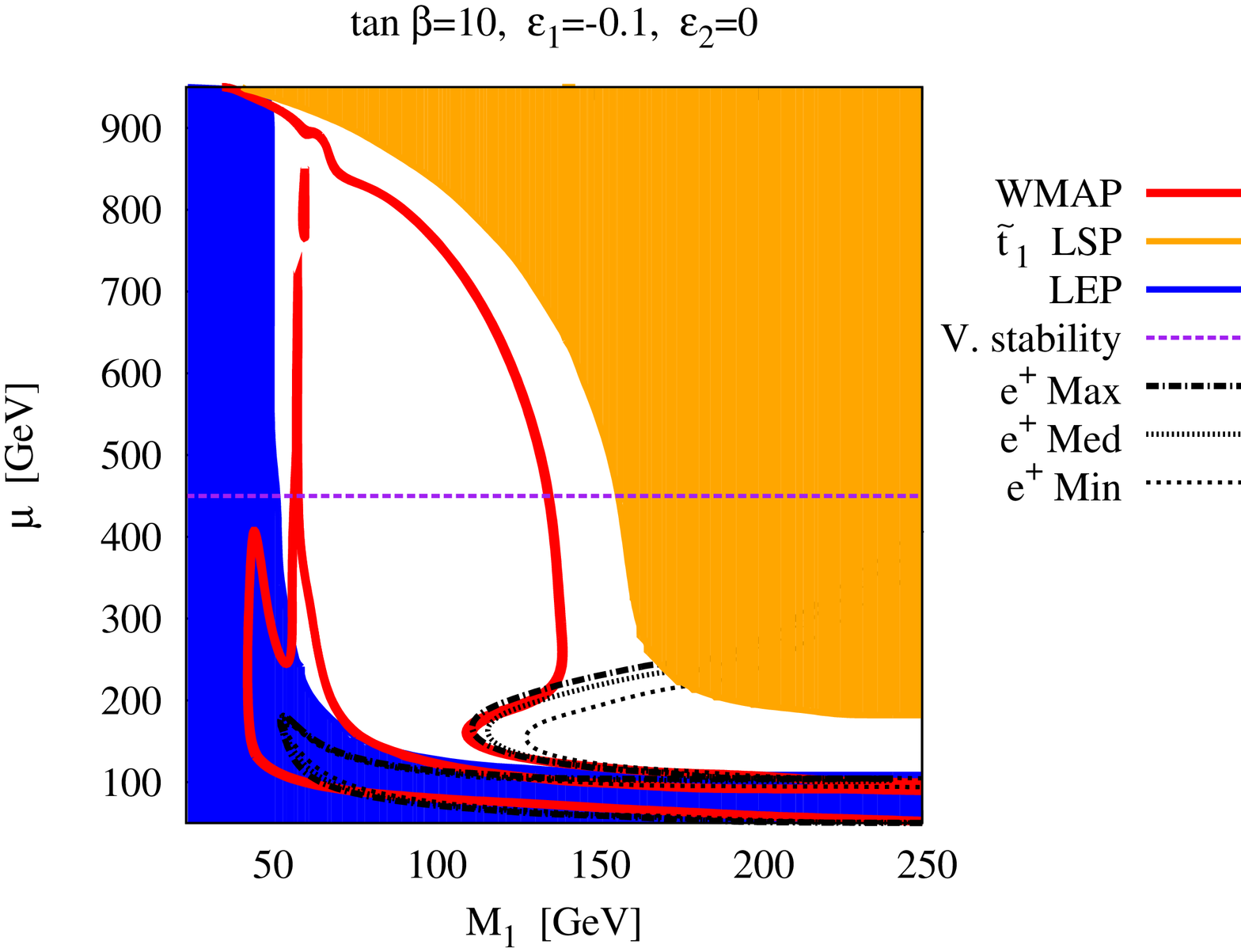}\\
\vspace{0.6cm}%\hspace{-2.5cm}
\includegraphics[width=5.5cm,angle=-90]{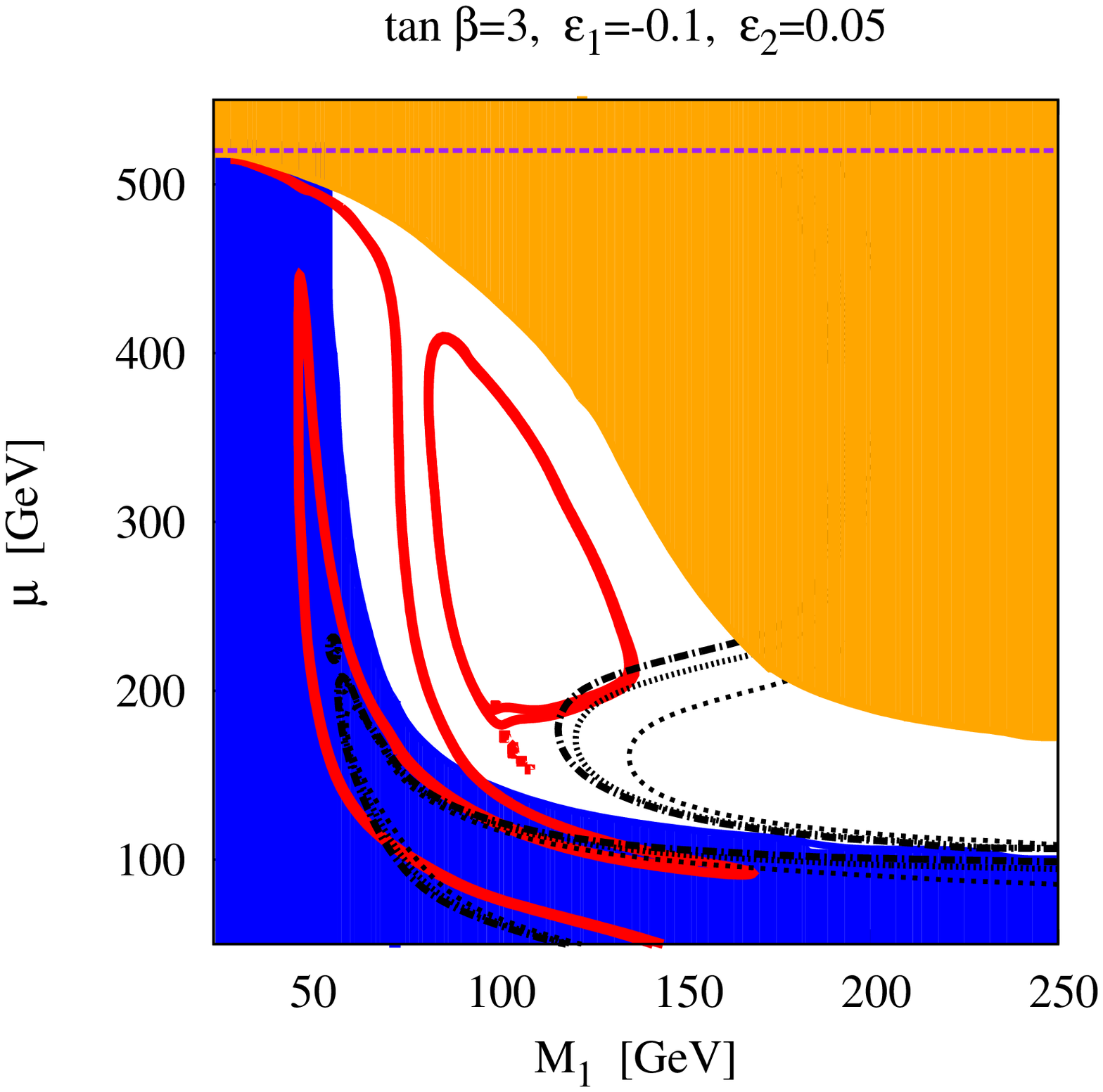}%\hspace{-2.3cm}
\includegraphics[width=5.5cm,angle=-90]{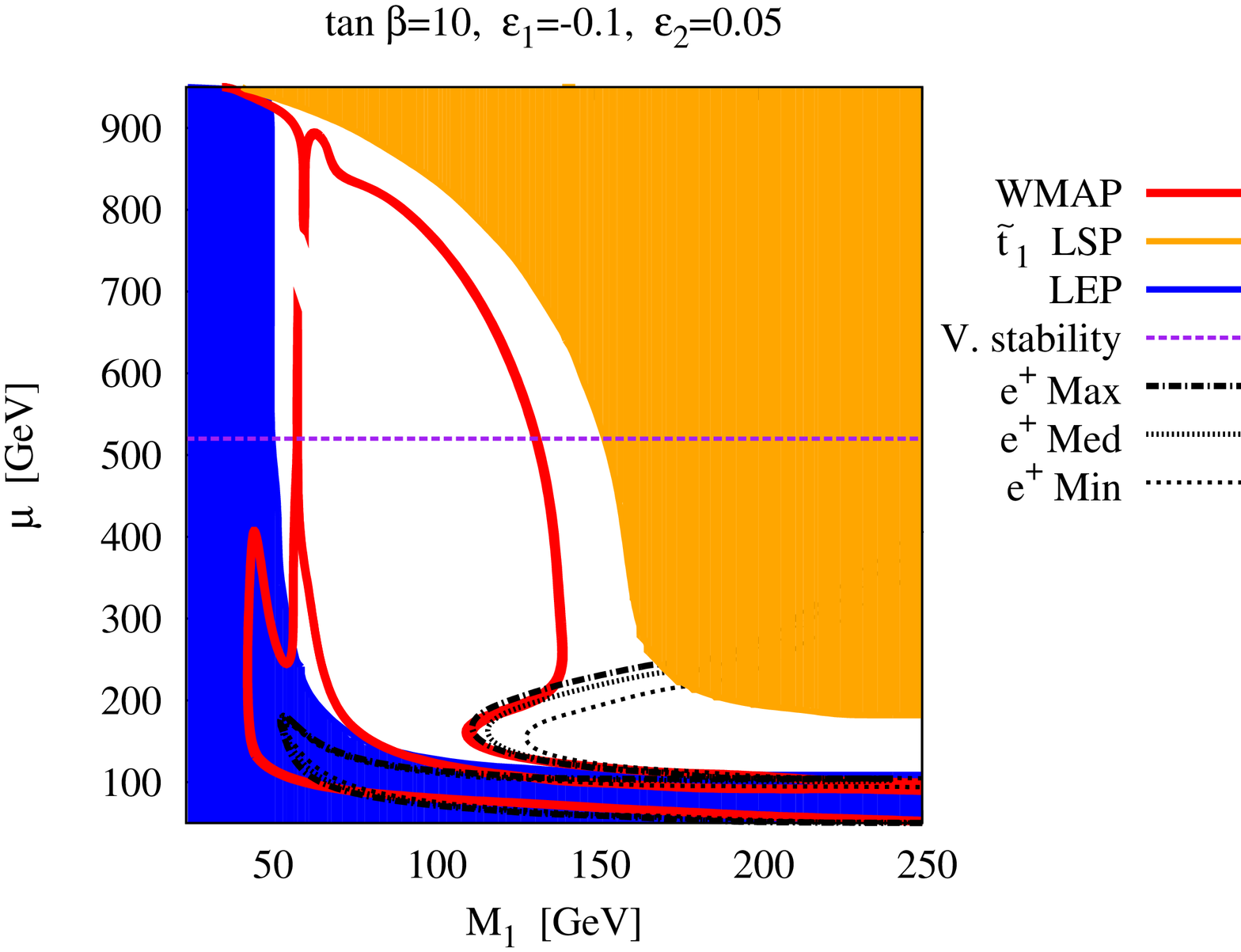}
\end{center}
% \vspace{-0.9cm}
\caption{{\footnotesize
Regions in the $[M_1,\,\mu]$ plane that can be detected by a $3$-year run of the 
AMS-02 satellite mission for the scenario with light stops and heavy sleptons, in the positron channel. 
The black lines depict the detectability regions for the $3$ considered propagation
models, MIN, MED and MAX: 
the parameter space points lying within the regions delimited by the black lines can be
probed, assuming the corresponding propagation models. 
Part of the mixed bino-Higgsino region, as well as (marginally) some part of the $Z$ funnel
region can be probed.}}
\label{po2}
\end{figure}

In figure  \ref{po2} we present the detection perspectives in the positron channel for our scenario
with light stops and heavy sleptons.
The detectable parameter space regions lie within the zones delimited by the black lines
for the three propagation models: the oval-shaped blobs as well as the banana-shaped ones.
Once again, we notice the general features already present in the $\gamma$-ray channel.
The regions giving rise to a positive detection lie within the zone where the
LSP is a Higgsino-like state, with mass $m_\chi>m_W$, in order to have the final state
$W^+W^-$ kinematically available. This region in general does not fulfill the WMAP limit.
However, and this is a novel feature of the BMSSM, with both $\epsilon_1$ and $\epsilon_2$
couplings turned on, a small region of the mixed Higgsino-bino regime can be detected for
the MAX (and even the MED) propagation model. As we pointed out before, 
in this regime the total annihilation cross-section can be quite significantly enhanced, leading
to better detection perspectives.

%%%%%%%%%%%%%%%%%%%%%%%%%%%%%%%%%%%%%%%%%%%%%%%%%%%%%%%%%%%%%%%%%%%%%%%%%%%%%%%%%%%%%%%%%%%%%%%%%%%%%%%
\subsection{Antiproton detection}
The last step in our analysis is to examine the antiproton channel predictions for the AMS02
experiment. We already mentioned the relevant experimental and background parameter values in
the singlet scalar model case. Once again, we stick to antiprotons with kinetic energies larger 
than $10$ GeV, whereas we consider that the AMS02 mission will collect data during three
years.

\subsubsection{Correlated stop-slepton masses}
\begin{figure}[tbp!]
\begin{center}
%\vspace{-0.2cm}\hspace{-2.5cm}
\includegraphics[width=5.5cm,angle=-90]{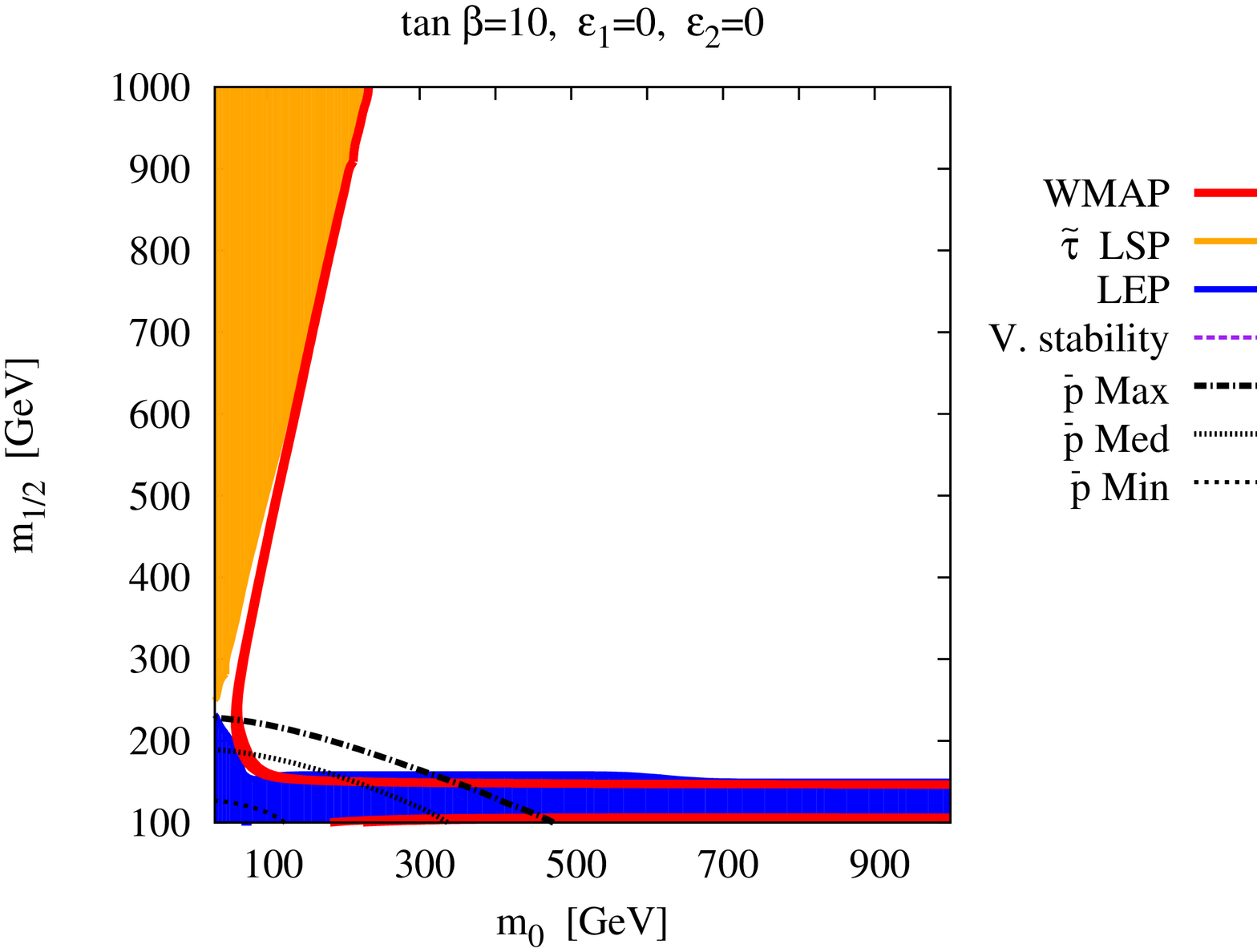}\\
\vspace{0.6cm}%\hspace{-2.5cm}
\includegraphics[width=5.5cm,angle=-90]{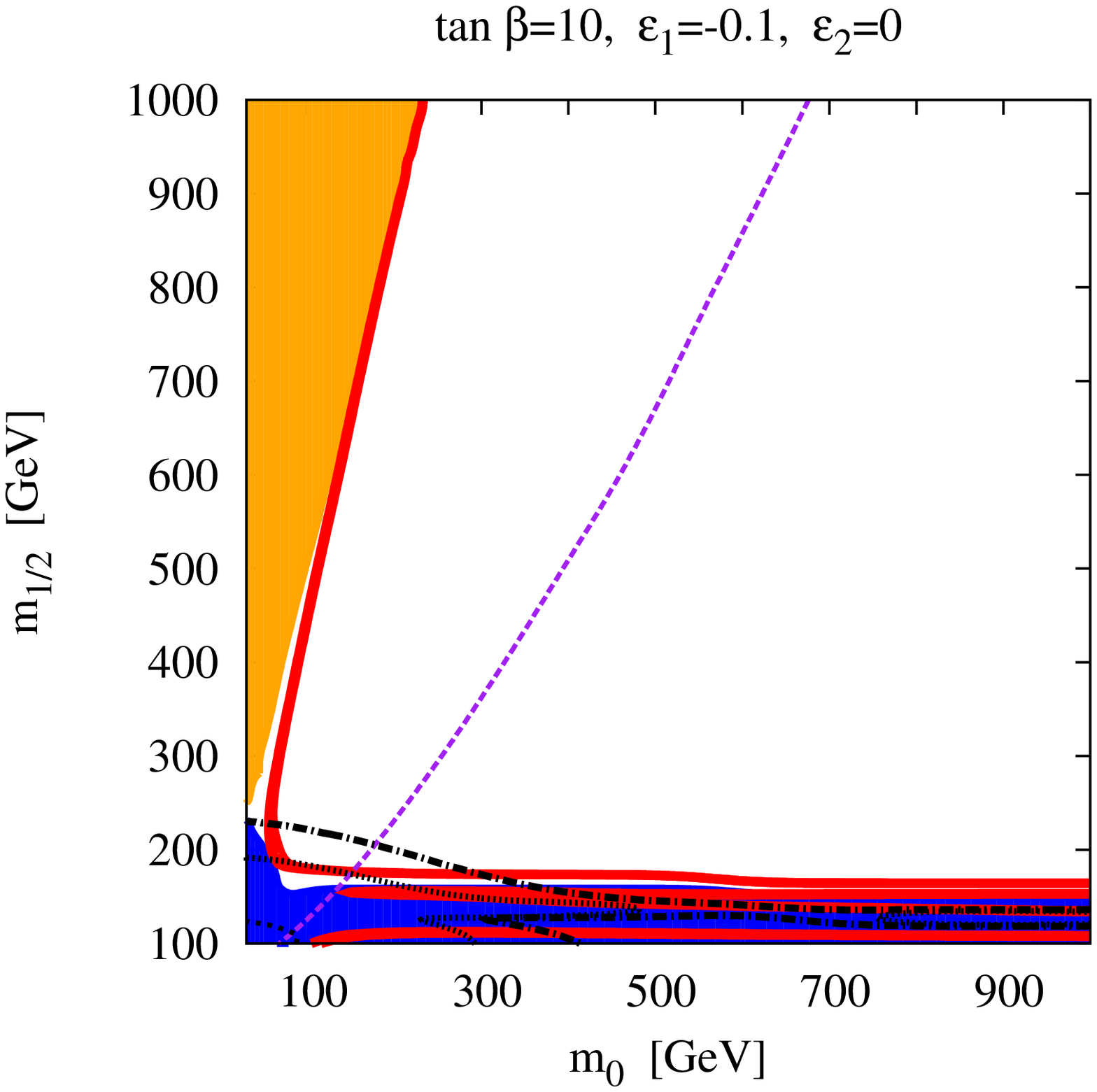}\hspace{0.2cm}
\includegraphics[width=5.5cm,angle=-90]{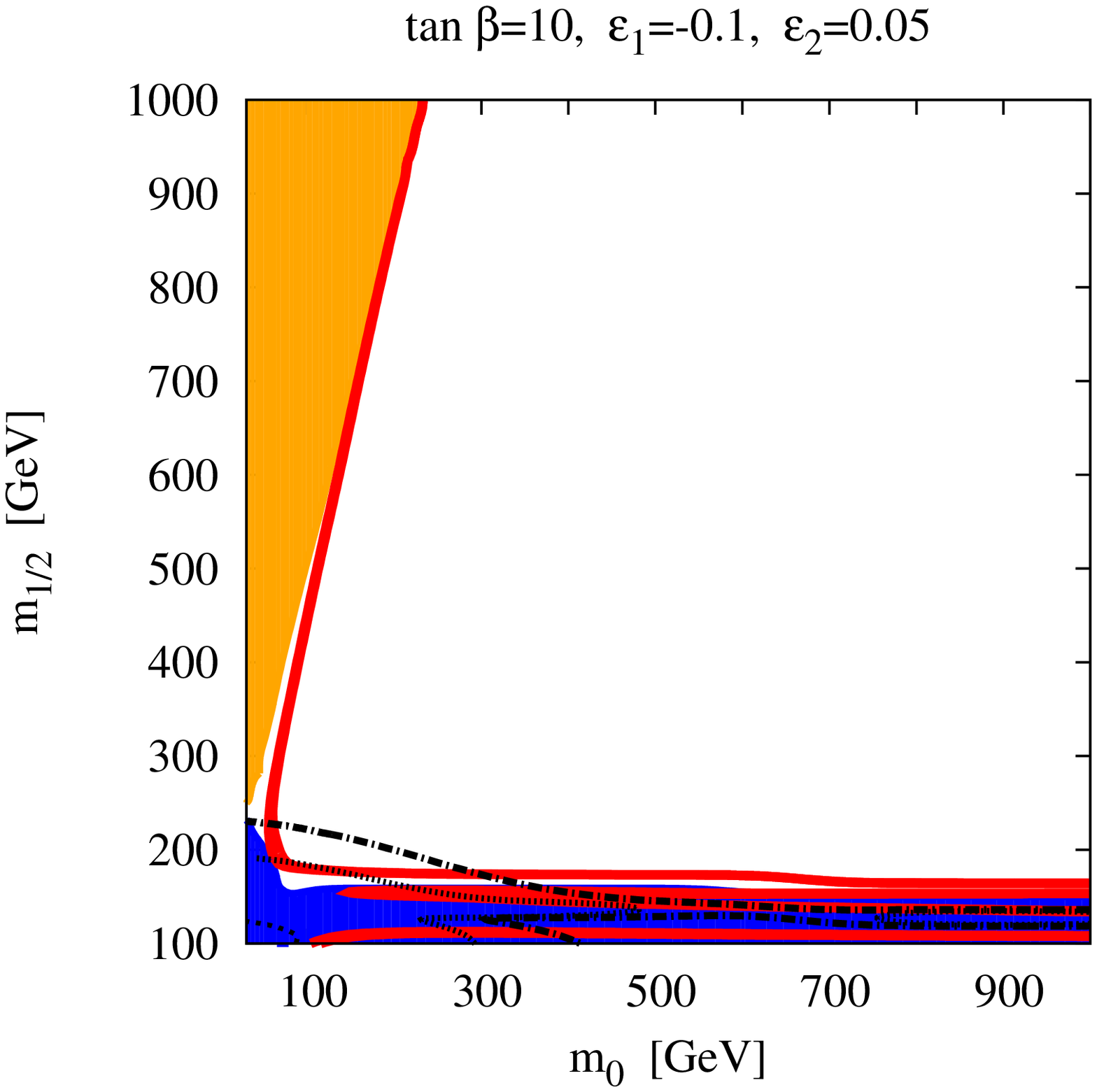}
\end{center}
%\vspace{-0.9cm}
\caption{{\footnotesize
Regions in the $[m_0,\,m_{1/2}]$ plane that can be detected by a $3$-year run of the 
AMS-02 satellite mission for our mSUGRA-like scenario in the antiproton channel. 
The black lines depict the detectability regions for the $3$ considered propagation models: 
the area delimited by the axes and the black lines can be probed for the corresponding propagation
model (i.e. the region towards the lower left corner in each plot).}}
\label{an1}
\end{figure}

In figure \ref{an1} we present our results for the detectability  of the BMSSM in comparison to
the CMSSM by the AMS-02 experiment for the antiproton channel. The detectable regions lie below the
black lines.
In the case $\tan\beta=3$, the experiment is not sensitive to any point in the parameter space satisfying also the
collider constraints and, hence, the corresponding results are once again omitted. 
A first remark here should concern the fact that the perspectives for antiproton detection are
significantly ameliorated with respect to the corresponding positron ones, at least for large values of $\tan\beta$.
We saw that this was also the case in the singlet scalar model of dark matter and attributed it to 
the important difference in background levels among the two channels.

Important areas of the viable parameter space are at the limits of detectability: the bulk region, but also, 
for some cases, part of the Higgs funnel where sfermion exchange continues being efficient.
Now, as we stressed out before, the possible enhancements due for example to substructures are quite constrained.
Given however that some regions are marginally out of reach, it would not be impossible to state that 
even small boosts could render important (in a qualitative sense, due to their cosmological relevance) 
regions of the parameter space detectable by AMS-02.

\subsubsection{Light stops, heavy sleptons}
\begin{figure}[tbp!]
\begin{center}
\vspace{-0.2cm}%\hspace{-2.5cm}
\includegraphics[width=5.5cm,angle=-90]{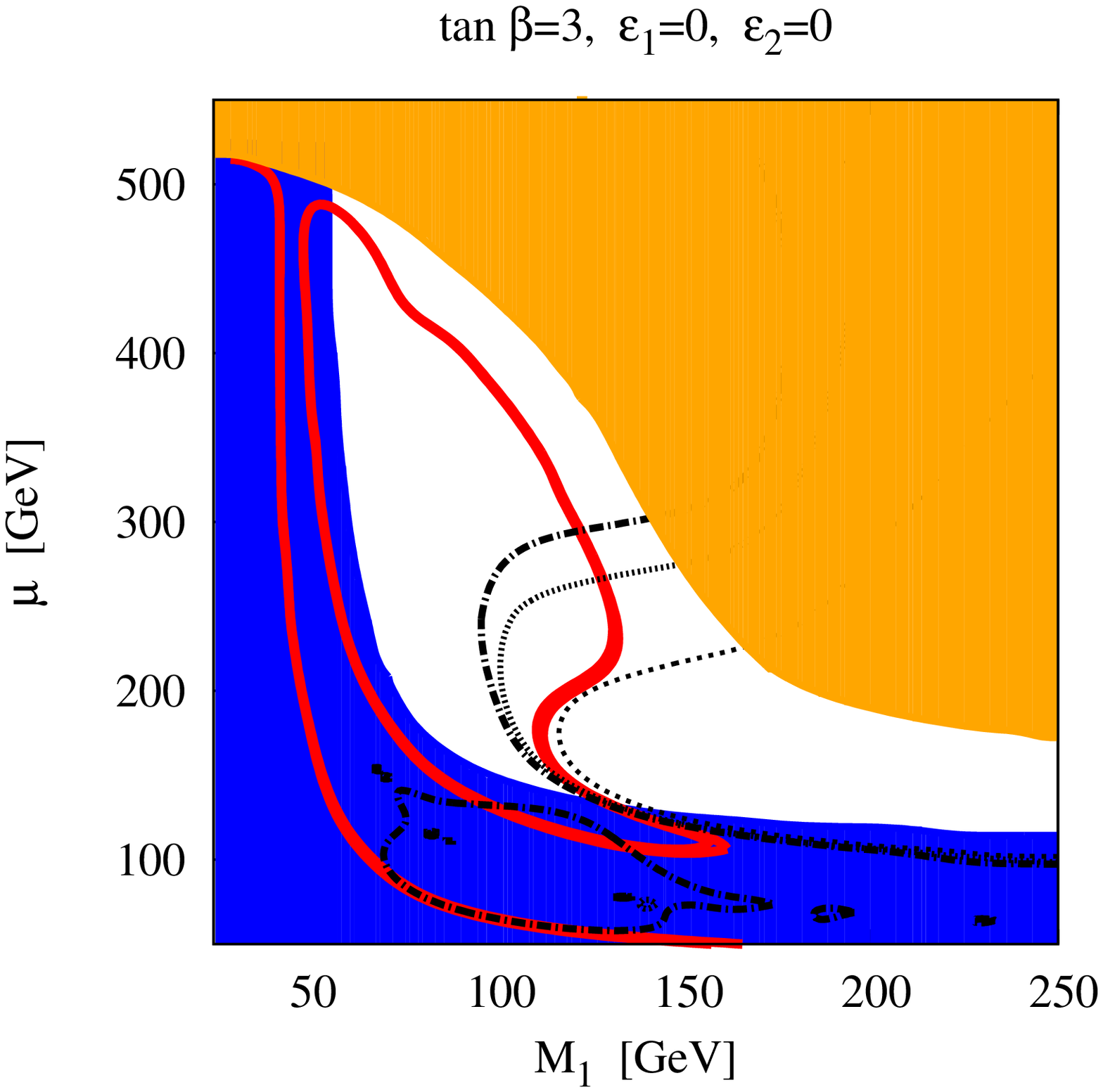}%\hspace{-2.3cm}
\includegraphics[width=5.5cm,angle=-90]{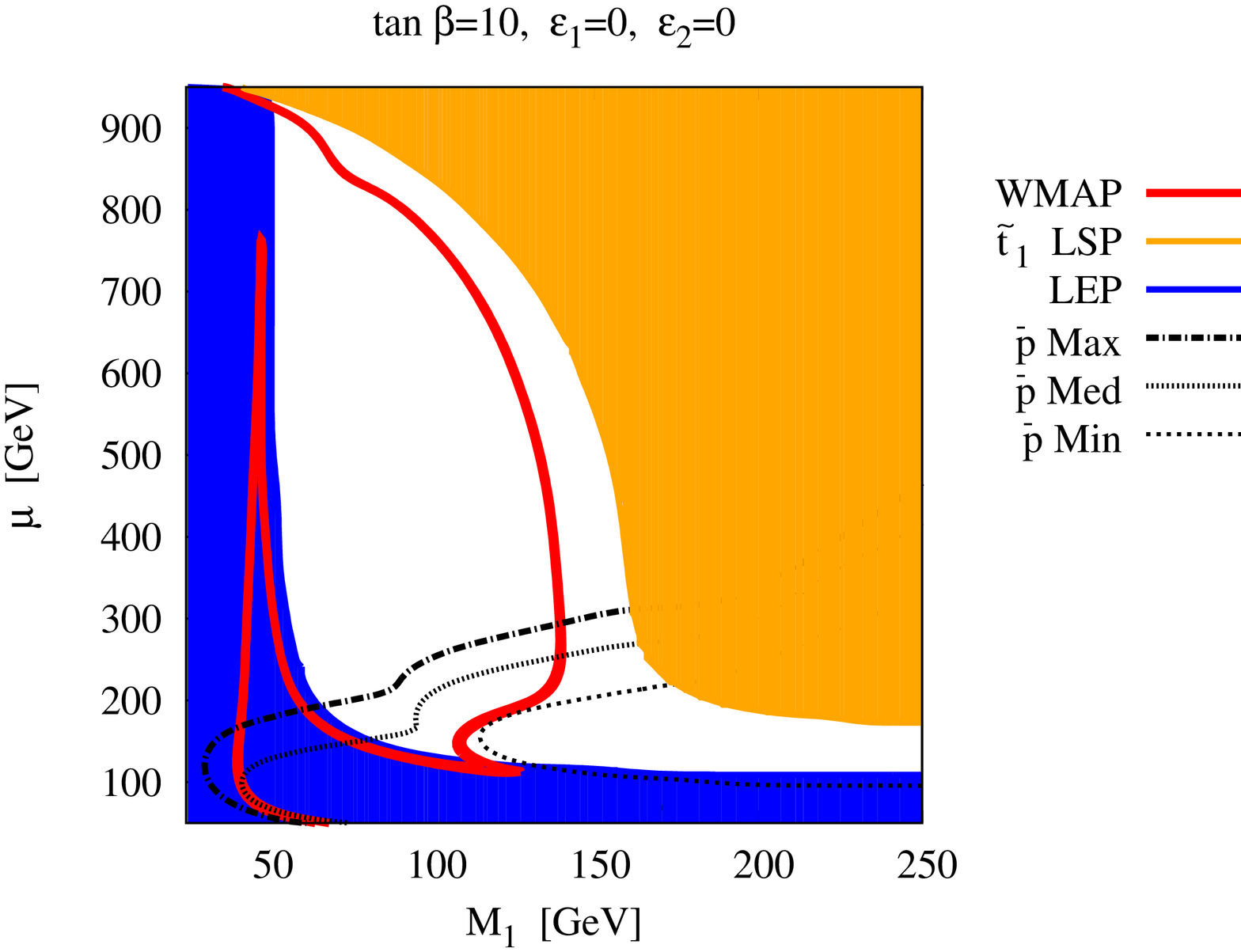}\\
\vspace{0.6cm}%\hspace{-2.5cm}
\includegraphics[width=5.5cm,angle=-90]{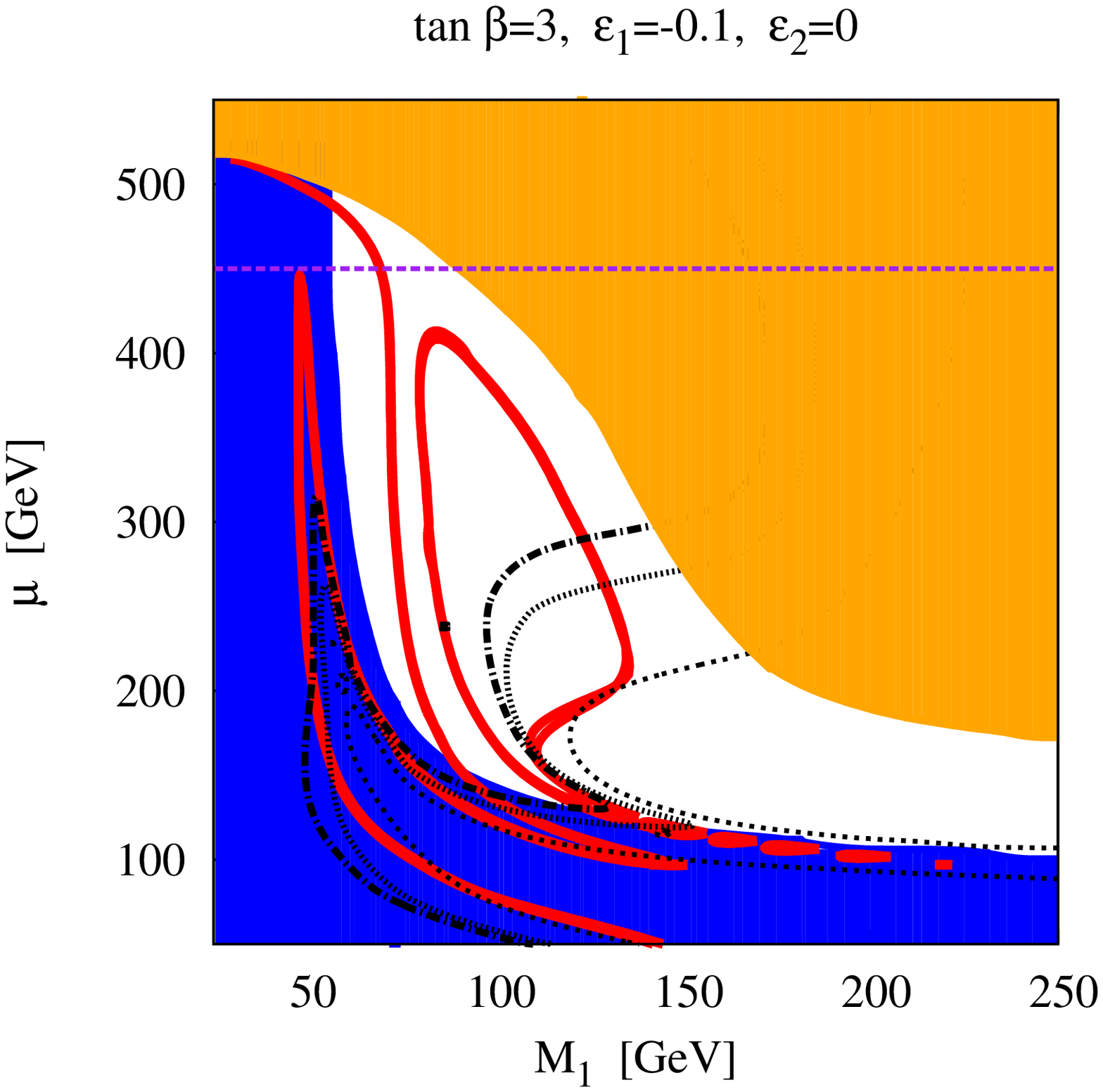}%\hspace{-2.3cm}
\includegraphics[width=5.5cm,angle=-90]{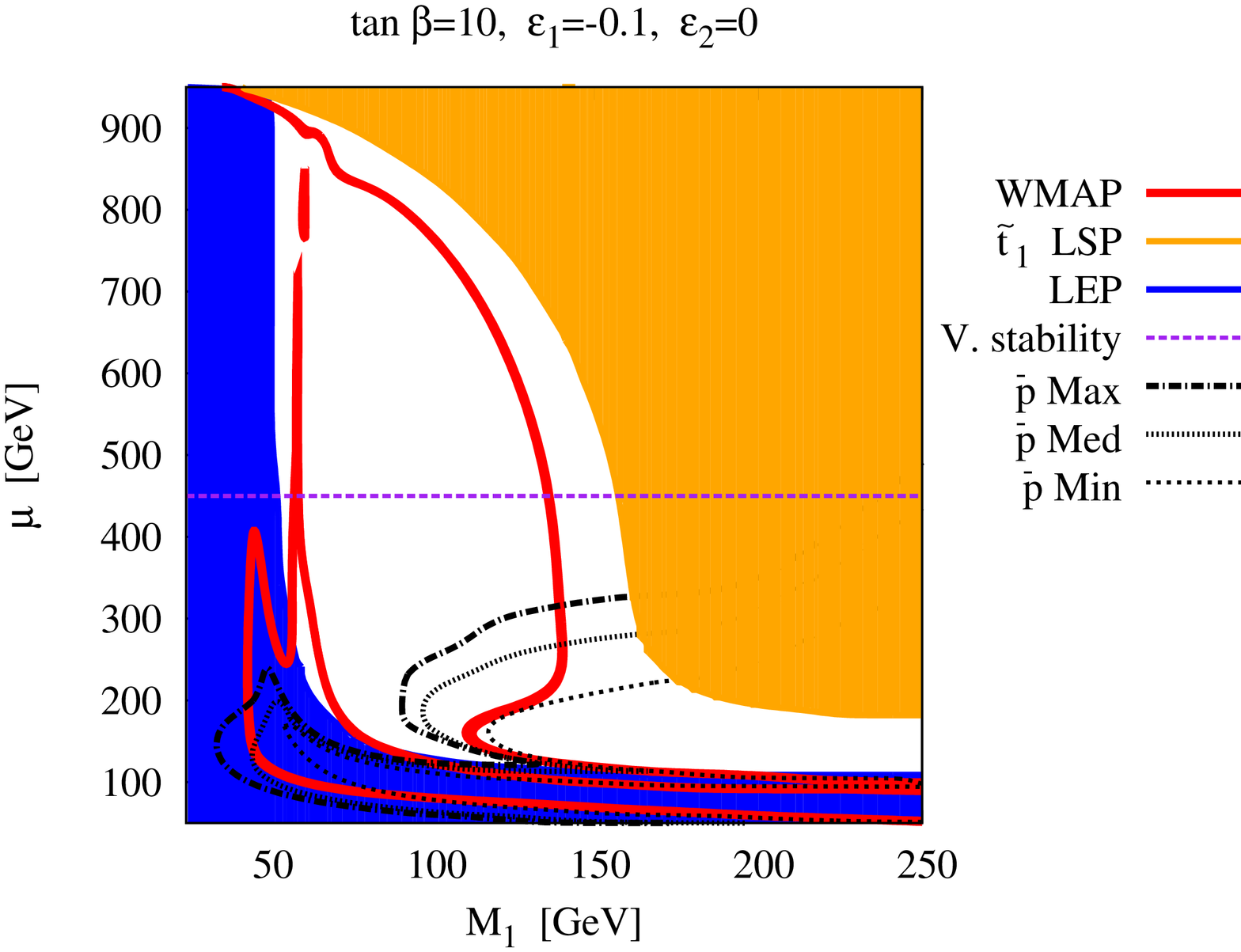}\\
\vspace{0.6cm}%\hspace{-2.5cm}
\includegraphics[width=5.5cm,angle=-90]{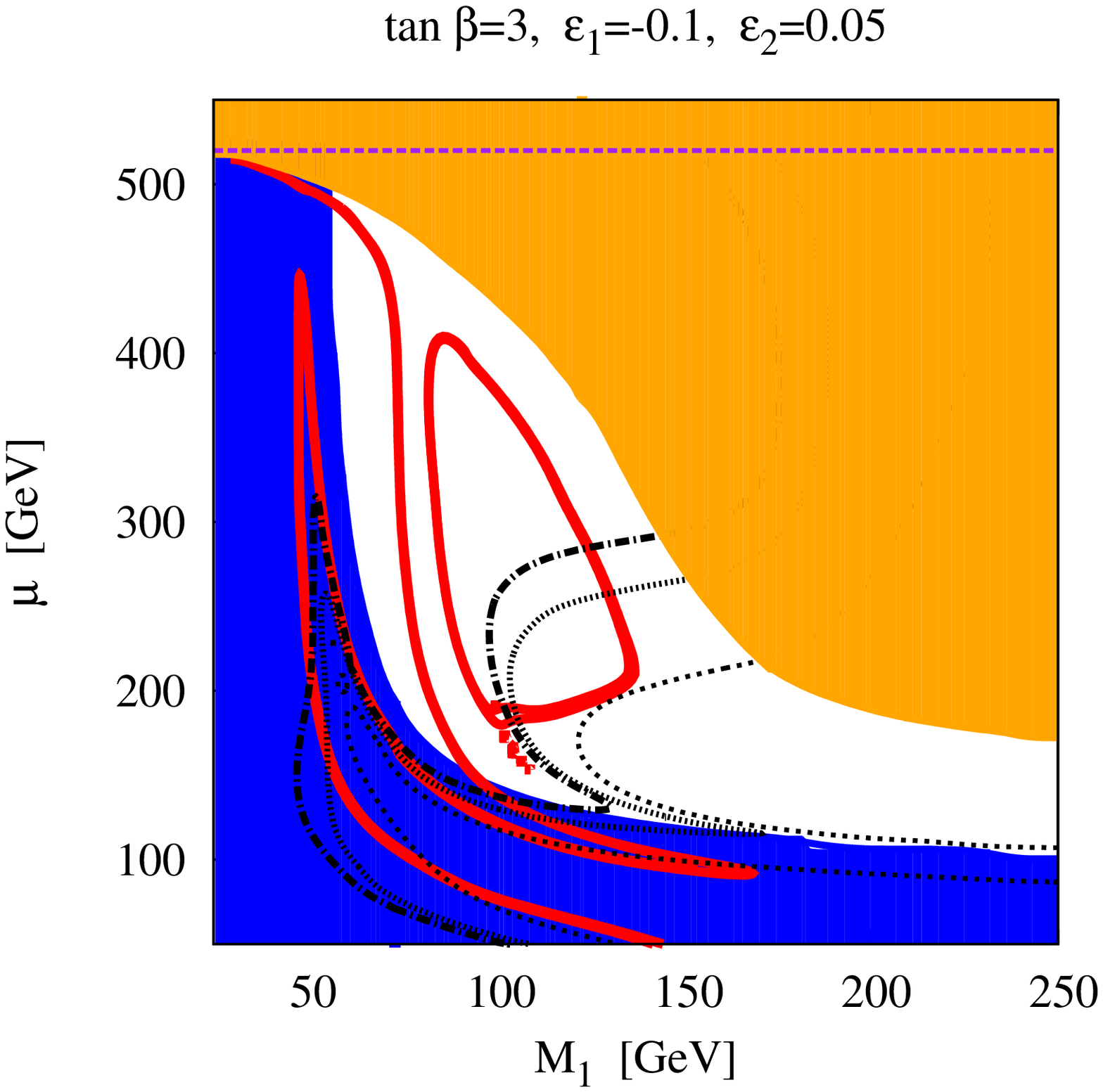}%\hspace{-2.3cm}
\includegraphics[width=5.5cm,angle=-90]{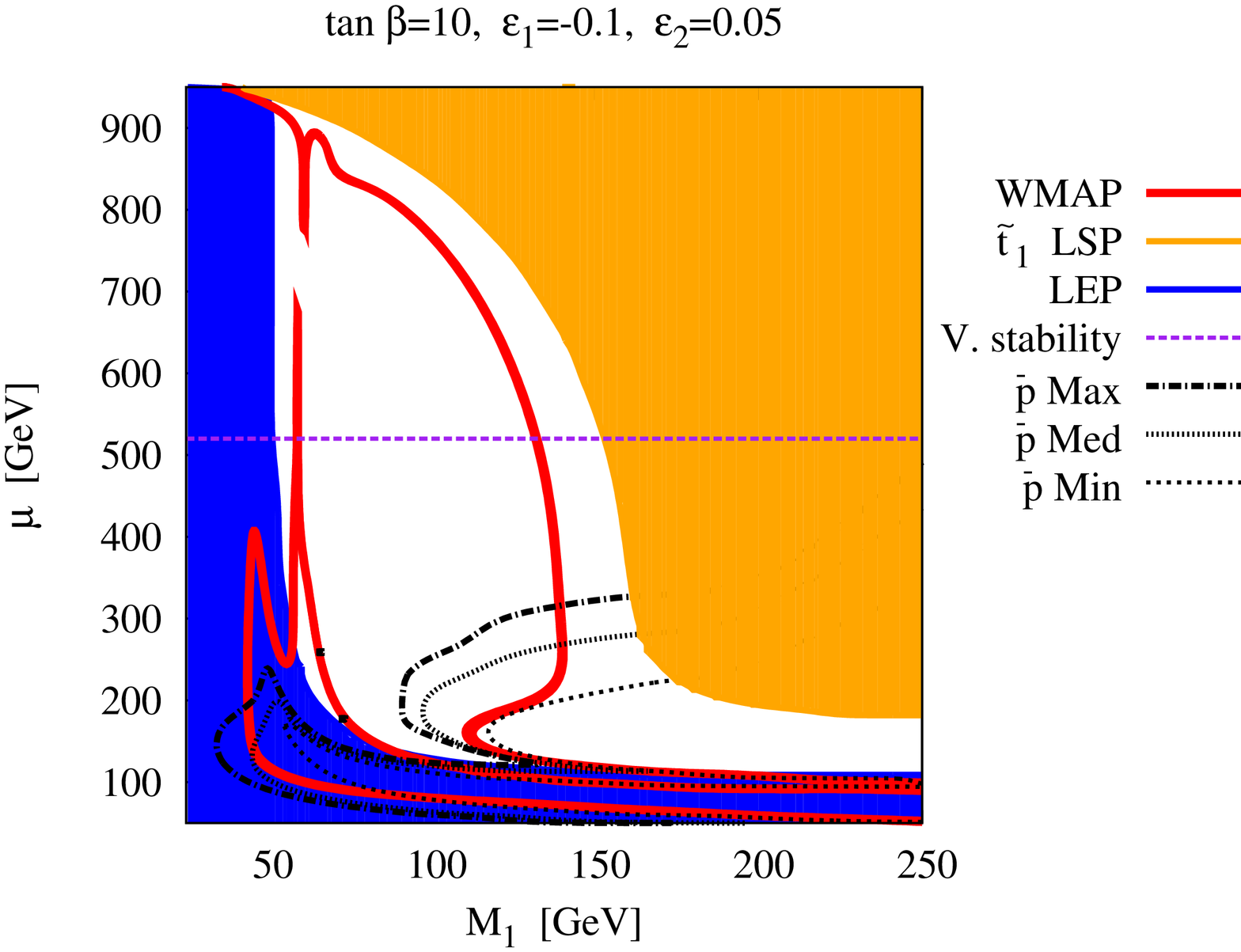}
\end{center}
%\vspace{-0.5cm}
\caption{{\footnotesize
Regions in the $[M_1,\,\mu]$ plane that can be detected by a $3$-year run of the 
AMS-02 satellite mission for the scenario with light stops and heavy sleptons, in the antiproton channel. 
The black lines depict the detectability regions for the $3$ considered propagation models: 
the areas delimited by the axes and the black lines can be probed for the corresponding propagation
model (i.e. the regions towards the lower right side in each plot), as well as the areas delimited by closed
lines.}}
\label{an2}
\end{figure}

Figure \ref{an2} presents the results for antiprotons and for the second scenario under consideration.
AMS-02 will be able to probe the regions lying within the oval-like blobs and the banana-shaped regions
delimited by the black contours and the axes.

Once again, the BMSSM turns out to be more favorable for DM detection than the ordinary case
of light stops and heavy sleptons without NR operators.
Detectable regions fall either into the case of the Higgsino-like neutralino regime, or in the
low-mass $Z$ funnel region. We point out that an important part of the area where the dark matter
relic density is fulfilled via coannihilation with the lightest stop could also be tested.
This last point might appear strange, since the coannihilation mechanism is inefficient at
present times. However, it should be noted that the detectable regions are mostly near the 
end of this region, where the correct relic density is in fact obtained through a 
combination of coannihilation and self-annihilation enhancement due to the mixed
gaugino - Higgsino nature of the lightest neutralino.

An interesting remark concerns the different behavior in the detectability lines among gamma-rays and
antiprotons just above the $h$ funnel. Whereas the opening of the gauge boson final state channels
lead to an important increase in the self-annihilation cross-section as well as the photon yield
at present times, this seems to be less the case for antiprotons. At this point, we should recall 
fig.\ref{antiprotonyieldmchi120GeV} that demonstrates that gauge boson final states are not the
most favorable ones for antiproton detection, since the corresponding yield is much lower 
compared to the hadronic one.

\subsection{Summarizing}
We saw that taking into account higher-dimensional contributions that could come from beyond the
MSSM physics can contribute significantly in resolving the little hierarchy problem of the plain
MSSM. The lightest
Higgs mass increases without demanding for large radiative corrections, whereas the model 
has been shown to be testable at the LHC. A very interesting
interplay appears with respect to dark matter phenomenology, since new regions yielding the
correct relic density appear and manage to evade collider constraints.

Dark matter detection gets quite challenged in this scenario, mainly due to the increase in mass
of practically the entire sparticle spectrum compared to the CMSSM case. The most promising detection
mode appears to be the direct detection one, although interesting information could also come 
mainly from the gamma-ray and the antiproton channel. 

Especially concerning the last one, we see that important regions of the parameter space can
be probed. As an encouraging comment, we should again point out that the assumptions made in
this particular channel are quite conservative.

Overall, WIMP detection in this framework is best for small masses and when 
the neutralino obtains a significant Higgsino component. On the other hand, resonant
annihilation into a $h$ boson and then in fermion pairs is a mechanism that can be efficient 
in yielding large enough cross-sections and hence the correct relic density at early times
but is an inefficient mechanism at present times, since the relevant cross-section tends to
zero as $v \to 0$. This remark has already been made, for example, in \cite{Jungman:1995df} and is 
further explained in Appendix \ref{FeynmanGraphs}.
%%%%%%%%%%%%%%%%%%%%%%%%%%%%%%%%%%%%%%%%%%%%%%%%%%%%%%%%%%%%%%%%%%%%%%%%%%%%%%%%%%%%%%%%%%%%%%%%%%%%%%%%%%%%%
%%%%%%%%%%%%%%%%%%%%%%%%%%%%%%%%%%%%%%%%%%%%%%%%%%%%%%%%%%%%%%%%%%%%%%%%%%%%%%%%%%%%%%%%%%%%%%%%%%%%%%%%%%%%%
%%%%%%%%%%%%%%%%%%%%%%%%%%%%%%%%%%%%%%%%%%%%%%%%%%%%%%%%%%%%%%%%%%%%%%%%%%%%%%%%%%%%%%%%%%%%%%%%%%%%%%%%%%%%%
%%%%%%%%%%%%%%%%%%%%%%%%%%%%%%%%%%%%%%%%%%%%%%%%%%%%%%%%%%%%%%%%%%%%%%%%%%%%%%%%%%%%%%%%%%%%%%%%%%%%%%%%%%%%%
%%%%%%%%%%%%%%%%%%%%%%%%%%%%%%%%%%%%%%%%%%%%%%%%%%%%%%%%%%%%%%%%%%%%%%%%%%%%%%%%%%%%%%%%%%%%%%%%%%%%%%%%%%%%%
%%%%%%%%%%%%%%%%%%%%%%%%%%%%%%%%%%%%%%%%%%%%%%%%%%%%%%%%%%%%%%%%%%%%%%%%%%%%%%%%%%%%%%%%%%%%%%%%%%%%%%%%%%%%%
\section{The Light Higgs Scenario}
The BMSSM framework discussed in the previous section tries to alleviate the little
hierarchy problem by invoking new physics inducing a positive contribution to the
Higgs mass. However, in section \ref{LittleHierarchyProblem} we discussed another
possible way out.

As we saw in Eq.\eqref{HiggsProdLEP}, the Higgs production cross-sections for the channels favored at LEP2 
depend not only on its mass, but also on its couplings. In particular, we saw that the $hZZ$
coupling receives a $\sin(\beta - \alpha)$ suppression with respect to the Standard Model 
one. If  $\sin(\beta - \alpha) < 1$, then the LEP bound
could potentially be evaded (or, more precisely, modified towards lower masses).
The problem in CMSSM models arose from the fact that typically $\sin(\beta - \alpha) \sim 1$, 
a regime which is usually called the \textit{decoupling regime}, since the pseudoscalar Higgs is much
heavier than the lightest $CP$ - even Higgs boson.

Departing from the mSUGRA/CMSSM framework allows to open a parameter space where the $114.4$ GeV
limit on the Higgs mass no longer holds, due to the decrease in the $hZZ$ coupling. In this regime,
all Higgs bosons can have comparable masses, hence it is often called the \textit{non-decoupling zone}.
The masses of the neutral Higgses can reach down to the value of the $Z$ boson \cite{Djouadi:2005gj}
without conflicting the LEP2 constraints. Interestingly, such a LHS with $m_h \sim 98$ GeV can 
apparently also accommodate the $2.3 \sigma$ LEP2 excess \cite{Barate:2003sz}. The little hierarchy
problem is concretely evaded due to the fact that since the Higgs mass can now be much lower than
the LEP2 conventional limit, radiative corrections need not be large and hence the assumptions on
the stop sector can be rendered less restrictive.

Several approaches towards the possibility for a light Higgs have been examined in the
literature 
\cite{Kane:2004tk, Kim:2006mb, Belyaev:2006rf, Chattopadhyay:2008hk, Bhattacharya:2009ij, 
Drees:2005jg, Asano:2007gv, Asano:2009kj, Cao:2010ph, Boos:2002ze, Boos:2003jt, Boos:2004cd, Djouadi:2006zu}.
One such framework are non-universal Higgs mass (NUHM) models 
\cite{Chattopadhyay:2008hk, Bhattacharya:2009ij, Polonsky:1994sr, Matalliotakis:1994ft, Polonsky:1994rz, Nath:1997qm, 
Ellis:2002iu, Ellis:2004bx, Baer:2005bu, Baer:2004fu, Ellis:2009ai, Ellis:2008eu, Calibbi:2007bk, Dudas:2008qf, Everett:2008qy,
Holmes:2009mx, Endo:2009uj}
where the Higgs masses are attributed different values than the other scalars at the GUT scale.
Assuming such a condition does by no means necessarily abandon Grand Unification. On the 
contrary, the Higgs bosons often live in different representations of GUTs and can therefore
acquire different masses. Furthermore, it is known that specific supersymmetry breaking
mediation schemes, such as gauge mediation, generate non-universal scalar soft breaking
terms at the GUT scale. This offers a further motivation to study such scenarios.

Describing the full phenomenology of such a framework goes beyond the scope of the present work.
We shall focus on the analysis performed in \cite{Das:2010kb}, where we examined a particular variant of 
a NUHM model and its dark-matter related phenomenology.

\subsection{The model and its constraints}
The Light Higgs Scenario can be realized by slightly relaxing the constraints of the CMSSM. In particular,
if one assumes non-universal Higgs masses at the GUT scale, then it turns out possible to obtain 
viable models which yield electroweak scale Higgs masses lower than $114.4$ GeV, 
without violating the LEP2 bounds, through the
reduction of the $\sin(\beta - \alpha)$ coefficient.
The model can be described by seven parameters
\begin{equation}
 m_{1/2}, \ A_0,  \ \mbox{sign}(\mu), \ \tan\beta, \ m_0, \ m_{H_u}^2(M_{GUT}), \  m_{H_d}^2(M_{GUT})
\end{equation}
where $m_{1/2}$ is a common mass for gauginos, $A_0$ is a universal trilinear coupling, $\mbox{sign}(\mu)$ is
the sign of the Higgs mass parameter, $\tan\beta$ is the ration of the two Higgs doublet vacuum expectation
values and the GUT-scale common scalar mass $m_0$ concerns all scalars but the two Higgs bosons. The
GUT-scale masses of the latter are denoted as $m_{H_u}^2(M_{GUT})$ and $m_{H_d}^2(M_{GUT})$.

Starting from this set of parameters it is possible to retrieve the low-energy quantities  
as well as a viable parameter space passing all electroweak constraints. $\mu$ and $m_A$
can be computed using the Renormalization Groups Equations as well as the requirement
for radiative EWSB, through the following equations
\begin{eqnarray}
 \mu^2 & = &
-\frac{1}{2} M^2_Z +\frac {m_{H_d}^2-m_{H_u}^2 \tan^2\beta} {\tan^2\beta -1}
+ \frac {\Sigma_1 -\Sigma_2 \tan^2\beta} {\tan^2\beta -1}\\
\sin2\beta & = & 2B\mu/(m_{H_d}^2+m_{H_u}^2+2\mu^2+\Sigma_1+\Sigma_2)\\
m_A^2 & = & m_{H_d}^2+m_{H_u}^2 +2\mu^2 \sim m_{H_d}^2-m_{H_u}^2
\label{mumA}
\end{eqnarray}
where $\Sigma_i$'s represent the one-loop radiative corrections 
\cite{Arnowitt:1992qp, Gamberini:1989jw, Barger:1993gh}
and the last of the three equations holds at tree-level given that one can approximate 
$\mu^2 \sim -m_{H_u}^2$ for $\tan\beta \ge 5$. Here $m_{H_u}$ and $m_{H_d}$ are of course
defined at the electroweak scale
\footnote{To fix the Higgs mass notation for the following, we adopt the convention that whenever
$m_{H_u}$ or $m_{H_d}$ refer to GUT-scale parameter, we shall include the scale of definition in parenthesis
whereas whenever we refer to EW-scale quantities the scale shall be ommited.}.
All the EW scale parameters and masses
are computed using the SuSpect package \cite{Djouadi:2002ze}.

Then, a set of electroweak observable constraints can be imposed in order to produce realistic
models:
\begin{itemize}
\item {{\bf Higgs boson mass limit:}} 
In the non-decoupling region $m_A$ becomes very light so that one has $m_A \sim m_H \sim m_h \sim M_Z$.
Then, the lower limit of $m_h$ goes down to 93 GeV or even lower\cite{Barate:2003sz}. 
We define the \textit{light Higgs boson scenario} by demanding that 
$\sin(\beta-\alpha)^2~<~0.3$ (or $\sin(\beta-\alpha)~<~0.55$) and at the same
time $93<m_h<114$. In practice, we slightly relax the limit on $\sin(\beta - \alpha)$
and demand $\sin(\beta-\alpha)~<~0.6$. An interesting feature is that in this case, the
LEP2 limit of roughly $114$ GeV now starts to apply for the heavier Higgs boson, since its
coupling for the Higgstrahlung process is $g_{zzH} \propto \cos(\beta-\alpha)$, which
now becomes dominant. So, in order to obtain an acceptable
SUSY spectrum with $93<m_h<114$, in addition to the desired 
value for $\sin(\beta-\alpha)$, one also requires $m_H>114~$GeV. 

On the other hand, in the decoupling region 
($\sin(\beta-\alpha) \sim 1$ ), the $m_h \gtrsim 114$ GeV limit 
needs to be respected. 
However we note that there is an uncertainty of about 3~GeV
in computing the mass of the light Higgs boson
\cite{Heinemeyer:2004gx, Heinemeyer:2004kg, Degrassi:2002fi, Allanach:2004rh}. This
theoretical uncertainty primarily originates from momentum-independent
as well as momentum-dependent two-loop corrections and higher loop corrections
from the top-stop sector. Consequently, a lower limit of $111$~GeV is often 
accepted for the SUSY light Higgs boson mass $m_h$.

\item {\bf $Br(b\rightarrow s\gamma)$ constraint:} 
In models like mSUGRA, 
the most significant contributions to $b\rightarrow s\gamma$ 
originate from charged Higgs and chargino exchange diagrams.
The charged Higgs 
($H^- - t$ loop) contribution has the same sign and comparable strength 
with respect to the $W^- - t$ loop contribution of the SM, which already 
saturates the experimental result. Hence, in scenarios where the charged Higgs mass
can be very small, satisfying the $b \rightarrow s \gamma$ constraint requires 
a cancellation between the two diagrams.  
This in turn requires a particular sign of $\mu $ or 
more precisely, that $\mu$ and $A_t$ are of opposite sign. The effect
of this constraint on the Light Higgs boson zone has been discussed
in \cite{Kim:2006mb}.  We have 
used the following
 3$\sigma$ level constraint from $b \rightarrow s \gamma $ with 
the following limits\cite{Koppenburg:2004fz, Aubert:2002pd, Barberio:2006bi}. 
\begin{equation}
2.77 \times 10^{-4} < Br (b \rightarrow s \gamma) < 4.33 \times 10^{-4}.
\label{bsgammalimits}
\end{equation}
\item {\bf $Br(B_s\rightarrow \mu^+ \mu^-)$ constraint:} 
Similarly, the flavor physics observable 
$B_s\rightarrow \mu^+ \mu^-$ may become very significant in 
this particular parameter space. The current experimental limit 
for $Br(B_s \to \mu^+ \mu^-)$ 
coming from CDF\cite{Aaltonen:2007kv} can be written as (at ${\rm 95\,\%\,C.L.}$) 
\begin{eqnarray}
{\rm Br} ( B_s \to \mu^+ \mu^-) < 5.8 \times 10^{-8}. 
\label{Bsmumu}
\end{eqnarray}
The estimate of 
$ B_s \to \mu^+ \mu^- $ in the MSSM depends strongly
on the mass of the A-boson and on the value of $\tan \beta$. In particular, the  
neutral Higgs boson contribution scales as $ m_A^{-4}$ whereas 
there is an additional dependence on $(\tan \beta)^6$. However, in the
present analysis, we choose $\tan\beta = 10$ which makes this constraint
less restrictive for most of the parameter space.
\item {\bf WMAP constraint :}
As far as the relic density constraint is concerned, we consider 
the following 3$\sigma$ limit of the WMAP data\cite{Dunkley:2008ie}, 
\begin{equation}
0.091 < \Omega_{CDM}h^2 < 0.128.
\label{relicdensity}
\end{equation}
Here $\Omega_{CDM}h^2$ is the dark matter 
relic density in units of the critical
density and $h=0.71\pm0.026$ is the Hubble constant in units of
$100 \ \rm km \ \rm s^{-1}\ \rm Mpc^{-1}$. We have used the 
code micrOMEGAs \cite{Belanger:2008sj} to 
compute the neutralino relic density.

\end{itemize}

%%%%%%%%%%%%%%%%%%%%%%%%%%%%%%%%%%%%%%%%%%%%%%%%%%%%%%%%%%%%%%%%%%%%%%%%%%%%%%%%%%%%%%%%%%%%%%%%%%%%%%%%%%%%%
%%%%%%%%%%%%%%%%%%%%%%%%%%%%%%%%%%%%%%%%%%%%%%%%%%%%%%%%%%%%%%%%%%%%%%%%%%%%%%%%%%%%%%%%%%%%%%%%%%%%%%%%%%%%%
%%%%%%%%%%%%%%%%%%%%%%%%%%%%%%%%%%%%%%%%%%%%%%%%%%%%%%%%%%%%%%%%%%%%%%%%%%%%%%%%%%%%%%%%%%%%%%%%%%%%%%%%%%%%%
\section{Dark matter in a LHS with Non - Universal Higgs Masses}

\subsection{Relic density and electroweak observables}
\begin{figure}[ht!]
\begin{center}
\includegraphics[width=12cm,height = 10cm]{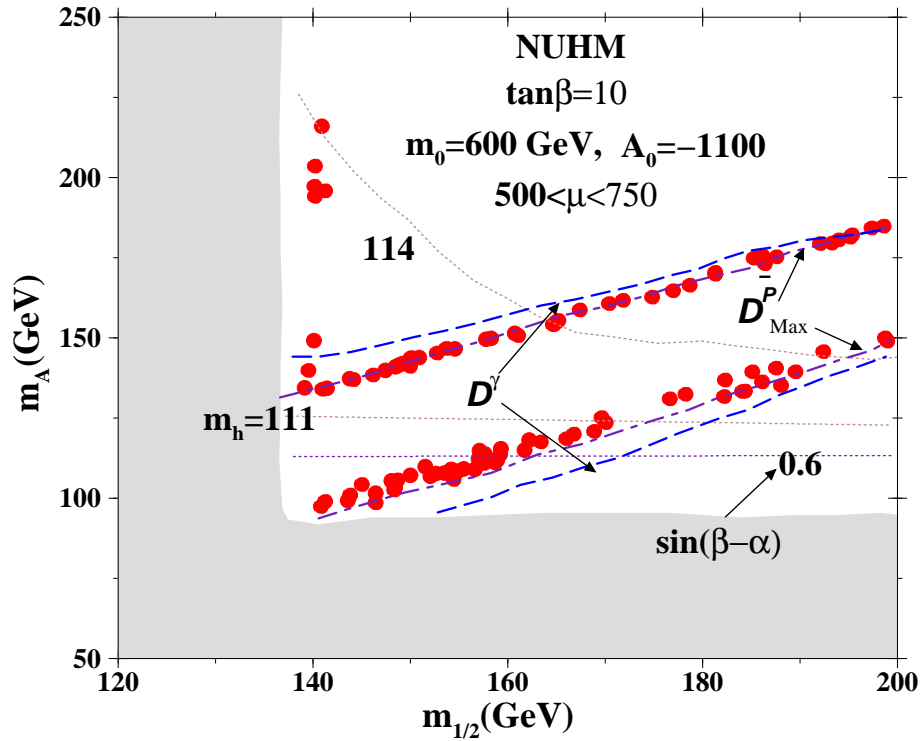}\\
\end{center}
\caption{{\footnotesize
WMAP - compliant parameter space points (red dots) in the $m_{1/2}-m_A$ plane. 
The correct relic density is obtained via $s-$channel $h$ 
or $A,H$ exchange annihilations. Neutralino masses with $m_\chi \sim 55-65$ GeV 
correspond to the light Higgs boson region. 
The two brown-dotted contours correspond to $m_h = 114$ GeV (upper) and $m_h = 111$ GeV
(lower), whereas the violet-dotted contour corresponds to $\sin(\beta-\alpha) ) 0.6$.
The violet dotted-dashed lines delimit detectable regions in the antiproton channel
and the blue-dashed ones in the gamma-ray one.
The complete $A$ pole annihilation region is within the reach of the Fermi and 
upcoming AMS-02 experiments.
}}
\label{a01100}
\end{figure}

\begin{figure}[ht!]
\begin{center}
\includegraphics[width=12cm,height = 10cm]{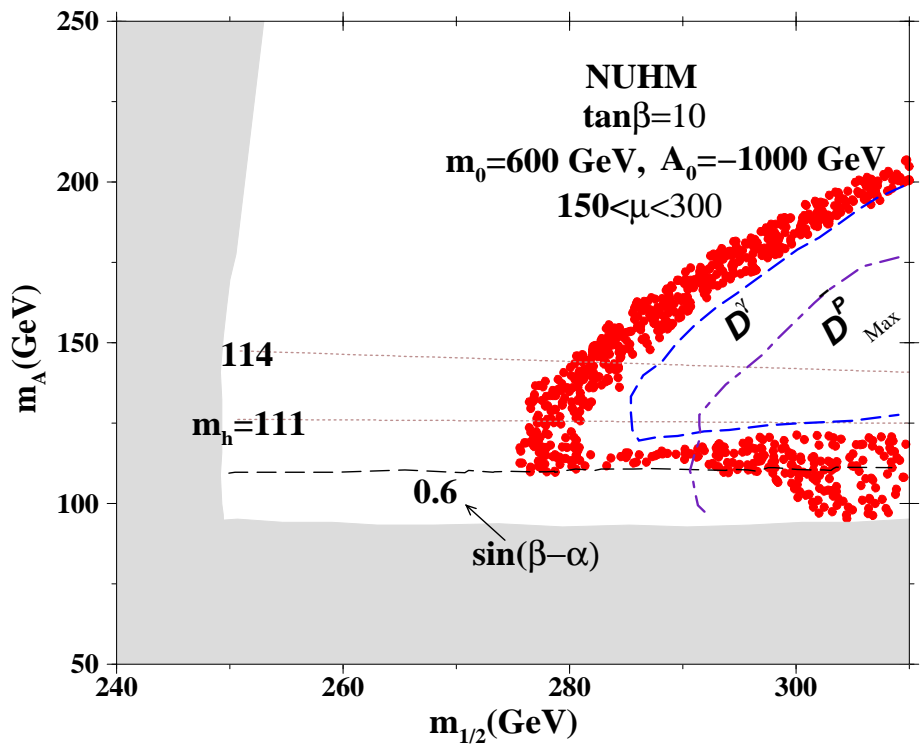}\\
\end{center}
\caption{{\footnotesize
Same as Fig.\ref{a01100}, except the light Higgs boson zone
is shifted to larger gaugino mass values. 
}}
\label{a01000}
\end{figure}

The first step in our analysis is to impose the various constraints as explained before
and compute the regions where the WMAP bounds are satisfied.
Our main results are presented in figs.\ref{a01100} and \ref{a01000}. In both cases,
we depict the parameter space points consistent with the WMAP data in the 
$m_{1/2}-m_A$ plane using red dots. We assume a moderate value for $\tan\beta=10$. 

Let us first concentrate on figure \ref{a01100}. 
The other parameters are set at $m_0=600~$ GeV, $A_0=-1100$ GeV
\footnote{The top quark mass is taken to be $173.1$ GeV, whereas the $\mu$ parameter
is taken to be positive throughout our analysis.}. 
We have 
varied the mass parameters $m_{H_u}^2(M_{GUT})$ and $m_{H_d}^2(M_{GUT})$ in the regions 
$\left[0, m_0^2\right]$ and $\left[-1.5 m_0^2, -0.5 m_0^2\right]$ respectively so as
to obtain light neutralino dark matter consistent with light Higgs masses
($m_{H,A} < 250$ GeV).
% $\mu$ and $m_A$ parameter values in the desired range. 
%within $0<\mhu<m_0^2$ and $-1.5 m_0^2<\mhd<-0.5 m_0^2$. 
% This provides us with
% reasonably large values of the $\mu$ parameter for the valid parameter points.
The lightest neutralino is mostly $\tilde B$ - like 
(although with a non negligible $\tilde H$ component).

In the same plot, we also show contours for the lightest Higgs mass ($m_h$, brown-dotted
in the plot) and
$\sin(\beta-\alpha)$ (violet-dotted). All the parameter space points with $\sin(\beta-\alpha)< 0.6$
are characterised, according to our conventions, as belonging to the  Light Higgs boson scenario,  
where the Higgs mass can evade the LEP2 limit due to
the reduced coupling with the $Z$ boson. On the other hand, admitting 
a 3 GeV uncertainty in the Higgs mass calculation, as mentioned, we delineate
the regions corresponding to $111 < m_h < 114$ GeV. Satisfying WMAP for smaller $m_{1/2}$ values, 
imposes lighter squarks as well as lighter gluinos.
\\
There are two distinct regions in the parameter space satisfying the 
relic abundance constraint:

\begin{itemize}
 \item The light Higgs pole annihilation region (funnel region) where
 neutralino annihilation 
produces an acceptable relic density via
the $s$ - channel exchange of a light Higgs. This
region extends in the direction parallel to the $m_A$ axis with gaugino mass 
value $\sim 140$ GeV.
In plain mSUGRA models, this zone is highly bound due to 
flavor physics constraints \cite{Djouadi:2005dz,Chattopadhyay:2010vp}. 
The spin independent cross-sections\cite{Chattopadhyay:2010vp},
on the other hand, could reach the CDMS-II\cite{Ahmed:2009zw} limits.

 \item The region (also a funnel region) where 
annihilations are principally due to $s$ - channel exchange of $A$ and $H$ bosons, 
since now $2m_{\chi_1^0} \simeq m_A, m_H$. 
Similar to the case of mSUGRA, this WMAP - satisfying
region in the NUHM model is principally characterized by the pseudoscalar 
Higgs boson -  mediated resonant annihilation.  The exact or near-exact 
resonance regions have very large annihilation cross-sections resulting in  
under-abundance of dark matter. In fact, an acceptable relic density
can be produced when the $A$-width is 
quite large and $2m_{\chi_1^0}$ can be appreciably away from the exact 
resonance zone. This is precisely the reason for the two branches of red circles 
that extend along the direction of $m_{1/2}$ in figure \ref{a01100}. The same
effect was also present in the $h$ - pole region in our BMSSM analysis.
\end{itemize}

Since we wish to focus on the LHS scenario, of particular interest
is the region where $\sin(\beta-\alpha)< 0.6$. 
In this regime, the lightest neutralino is characterized by a quite small
mass, $55<m_{\chi^0_1}<65$ GeV, whereas the $A$ boson plays a dominant role 
in the annihilation process. Now, apart from the mass of the $A$ bosons, 
neutralino pair annihilation also depends on the coupling 
${\cal{C}}_{\chi_1^0 \chi_1^0 A} \sim Z_{11} Z_{13}$. 
We recall that the Higgsino component $Z_{13}$ of the LSP
is essentially determined by the $\mu$ parameter. For a relatively large $\mu$
parameter ($500 \mbox{GeV} < \mu < 750 \mbox{GeV}$), 
the Higgsino components are relatively small. One hence needs
quite small neutralino masses ($m_{\chi_1^0} \sim 55-65$ GeV) in order to satisfy
the relic density constraint in the non-decoupling limit, where $m_A$ is of the order
of $100$ GeV.
In other words, the neutralino 
in this case cannot be too far from the exact resonance condition.
In this respect, as the neutralino mass increases, the WMAP-compliant
regions extend in the direction of larger $m_A$.
However then the Higgs bosons fall into the other category, i.e. in the decoupling
region. Interestingly, even this WMAP - satisfying zone does not require
large values for gaugino masses. We shall followingly see that this whole region
can lead to large gamma-ray as well as antiproton signals in present or oncoming
experiments.
\\ \\
Before presenting our results on indirect detection, we should
also discuss the flavor physics observables $b \rightarrow
s \gamma$ and $B_s \rightarrow \mu^+ \mu^-$. Since we choose a rather 
moderate value for $\tan\beta$($=10$), the $B_s \rightarrow \mu^+ \mu^-$ constraint is 
not very stringent for the parameter space as shown in Fig.\ref{a01100}.
On the other hand $Br(b \rightarrow s \gamma)$ constitutes a strong
constraint particularly in the non-decoupling region 
where charged Higgs bosons are very light. However, 
for large negative $A_0$ values, hence negative values of $A_t$ at the EW scale, 
one of the stop eigenstates
becomes lighter due to large mixing. This in turn provides a cancellation
between charged Higgs and chargino - induced diagrams. Choosing $A_0=-1100~$GeV 
at $M_{GUT}$, almost all parameter space points and more importantly the whole WMAP
allowed region in the $m_{1/2}-m_A$ plane can satisfy the constraint. 
\\ \\
Figure \ref{a01100} also contains some gray zones, which correspond to some further
constraints:\\
 (i) For $m_{1/2} \ge 135$ GeV, parameter space points with
$m_A$ smaller than roughly $100$ GeV 
are not compatible with the Higgs mass limit in the non-decoupling zone i.e,
here one has $m_h<93$ GeV.\\ 
 (ii) For $m_{1/2} \le 135$ GeV, the gluino becomes 
lighter (we demand $m_{\tilde g}> 390$ GeV for a parameter space point
to be valid \cite{Amsler:2008zzb}) and
then very soon the chargino becomes too light with $m_{\chi_1^\pm}<103.5$ GeV. We should
note here that a light Higgs  with mass $m_h\le 93$ GeV may be allowed,
but then $\sin(\beta-\alpha)$ needs to be further suppressed. However this
region is then further constrained and we do not consider it in our analysis. 

For this first scenario, the parameter value choice 
of fig.\ref{a01100} yields relatively light neutralinos, of masses up to $80$ GeV.
\\ \\  
Next, we probe a parameter space region where neutralinos become heavier, remaining
nevertheless in the {\it Light Higgs boson} zone. Once again, the relic density constraint 
is mainly satisfied by $s$ - channel quasi-resonant $A$ exchange. It should nonetheless
be noted that in this second scenario, $s$-channel $Z$ and $t$-channel neutralino
exchange are also potentially significant contributions.
For illustration we fix $m_0$ at a very similar value as previously, 
i.e $m_0 = 600$ GeV while the trilinear coupling $A_0$ is fixed at
$A_0 = -1000$ GeV in order to make $b \rightarrow s \gamma$ less restrictive.  
Then, 
$m_{H_u}^2(M_{GUT})$ is set to $2.4 m_0^2$ and $m_{H_d}^2(M_{GUT})$ is varied in the region
$\left[-0.3 m_0^2 , 0.1\right]$ GeV in search for points satisfying WMAP. 
These choices yield small values for the $\mu$
parameter ($150<\mu<300$ GeV), hence the LSP can have a significant Higgsino component.

The corresponding results can be seen in fig.\ref{a01000}.
It can be seen that 
$m_{1/2} \sim 300$ GeV corresponds to the non-decoupling zone for the Higgs boson.
The lightest neutralino (with $m_{\chi_1^0} \sim 120$ GeV) 
is quite far from the resonant annihilation
condition $2m_{\chi_1^0} \simeq m_A$. 
This is due to the fact that the neutralino self-annihilation cross-section is
in this case further enhanced by the presence of a significant Higgsino component
in the LSP, augmenting its couplings to the Higgs bosons.
In comparison 
to the previous case, where the $b \bar{b}$ final state has the maximum branching
ratio, here several other final states involving the Higgs as well as gauge bosons ($Zh, 
ZH, W^\pm H^\pm, hA, HA, hh$) open up. We should note that
$m_{\chi_1^0}$ cannot be too large as then the neutralino would be 
even further away from the resonance condition, which in turn makes the pair
annihilation via $A$-boson exchange less efficient. As in the previous scenario, 
the $b \rightarrow s \gamma$ constraint does not influence the WMAP - compliant 
zones. The gray areas correspond to the same constraints as before.

\label{results}

\subsection{Indirect detection in the NUHM model}
The next step is to compute constraints from  and perspectives for dark matter detection experiments.
This shall be done for gamma-rays coming from DM annihilation at intermediate galactic latitudes and
for the Fermi mission, as well as antiprotons at the PAMELA and AMS02 experiments.

We omit the relevant study for positrons, since the previous two model examples in this work 
reveal rather pessimistic perspectives for this channel.

\subsubsection{Exclusion and detectability}
Let us first present the criteria we shall adopt in this analysis in order to characterize
a parameter space point as being detectable. As we have seen in all previous treatments, 
in order to assess whether a parameter space point is excluded by current data
from Fermi or PAMELA, we should have some estimate of the background spectra 
for both gamma-rays and antiprotons. In the previous analyses of the singlet scalar DM
model and the BMSSM, we just considered that the current measurements of Fermi or PAMELA
essentially consist of background events, and that the measured spectra could be used in order to 
estimate the background. In the treatment of \cite{Das:2010kb} we adopt a slightly different 
philosophy, which begins from the fact that up to the present day there does not exist
some precise background functional form enjoying global acceptance in the community. The point where
agreement seems to exist, however, is the general form that this background should follow,
that of a power-law: $\Phi_{bkg} = a E^{b}$. This is expected to be the case in
both detection channels.

In reality, every flux measurement contains both background as well as
(hopefully) signal events. For every parameter space point of our NUHM, we can compute the
gamma-ray or antiproton fluxes as described previously. Then, each point shall
be considered as excluded if there is no $(a,b)$ combination 
(i.e. the generic power-law background) for which 
the sum of the signal and the background can provide a good fit to the current data.
Hence, we vary $(a,b)$, compute the corresponding backgrounds and then subsequently 
add the signal contribution
to check if there exists some background form for which
this sum provides a sufficiently good fit to the data. If no such $(a,b)$ can be found, 
the corresponding parameter space point can be considered as excluded. In practice, the criterion 
we demand is that there should exist at least one $(a,b)$ combination for which the sum of the 
signal and the background fallq within the $95\%$ CL error
bars as given by the Fermi or PAMELA collaborations.

The method we follow in order to characterize a parameter space point as being detectable is, in
some sense, the inverse one:
First, we need some estimate of what future data could look like. We expect that this data
should also comprise of both some power-law background and the eventual signal events.
In order to minimize the signal's significance,
we choose this background at the higher $68\%$ CL limits of the Fermi or PAMELA experimental points.
Then, for each parameter space point, we add the signal to this background, 
creating a set of pseudo-data that could appear in the future.
As pointed out, the exact form of the background is in general unknown, but its
general form is expected to be a power-law. So, we could look for deviations of the pseudo-data
from such a behaviour.
If it is impossible to find a power law form that fits this pseudo-data well enough, 
then the corresponding parameter space point is characterized as detectable. If such an $(a,b)$ combination can
be found, then the signal shall be indistinguishable from the background (unless some other measurements
allow to constrain the viable $(a,b)$ combination, a possibility that we do not consider here).
The goodness-of-fit criterion we choose is based on the $\chi^2$ quantity, which is defined as
\begin{equation}
 \chi^2 = \sum_{i = 1}^{\mbox{\begin{tiny}nbins\end{tiny}}} 
\frac{(N_{\mbox{\begin{tiny}bkg\end{tiny}}} - N_{\mbox{\begin{tiny}exp\end{tiny}}})^2}
{N_{\mbox{\begin{tiny}bkg\end{tiny}}}}
\end{equation}
where nbins is the number of bins, taken to be $20$ in both cases, 
$N_{\mbox{\begin{tiny}exp\end{tiny}}}$ is the pseudo-data, whereas $N_{\mbox{\begin{tiny}bkg\end{tiny}}}$ is 
the background-only number of events that we try to fit to the pseudo-data.

If the best fitting power-law has a $\chi^2$ larger than $28.87$ (our problem
has $20 - 2 = 18$ degrees of freedom, since we are trying to fit $2$ variables $(a,b)$), 
this means that there is no (background-only) power-law that can fit the pseudo-data. Hence, if
$\chi^2 > 28.87$ the corresponding parameter space point is detectable, since 
the signal it generates is distinguishable from the background.

If we wish to sum up our method in ``hypothesis testing'' terms, we could say that in the case
of exclusion we are testing a null hypothesis according to which existing data can be well-described 
by dark matter annihilations plus some background form. In the detectability case, the null hypothesis
is that the pseudo-data (containing both signal and background) can be well fitted by a 
background only function.

Obviously, the former results shall be subject to changes according to
deviations from Gaussian statistics, further experimental errors, systematics etc.
More detailed analyses are certainly performed within the experimentalist community.
We note that our calculations concerning the detection perspectives in gamma-rays are done 
assuming a $3$-year data acquisition period, with the region of interest being within the field-of-view
$100\%$ of the time. The same data acquisition period is used for antiprotons.

\subsubsection{Gamma-rays from intermediate galactic latitudes}
We have seen that the most common region of the sky that is examined in the literature as a source of $\gamma$-rays
is the galactic center, since it is the region where $N$-body simulations predict a maximization
of the dark matter density distribution and, hence, the corresponding gamma-ray flux.
However, the galactic center is quite poorly
understood as a region: there are large uncertainties in the background modelizations as well
as the density profile itself. 

It has been proposed (see, for example, ref.\cite{Stoehr:2003hf}) that one could maximize the
signal/ background ratio by actually excluding the region around the galactic center. Following
this reference, we perform our computations in an annular region extending from $20^\circ$ up to 
$35^\circ$ from the galactic center, excluding at the same time the regions within $10^\circ$ from the
galactic plane. It has actually been shown that within the framework of such an analysis, 
one can enhance the signal/background ratio by up to roughly an order of magnitude. This consitutes,
in some sense, a ``change in strategy'' with respect to the previous analyses we have presented in this
work: instead of looking for regions where the signal becomes maximal (i.e. the GC), we look for
a region where the signal's relative significance with respect to the background increases. A
model-independent analysis of the Fermi discovery potential - among others - at intermediate latitudes has
been performed in \cite{Baltz:2008wd}.

In the meantime, we present in table \ref{JbarsLight} the values obtained for the $\bar{J}$ quantity
defined in Eq.(\ref{Jbar}), for the three different halo profiles also discussed in the BMSSM analysis:
the Navarro, Frenck and White one, the Einasto profile and a NFW-like profile that has tried to take into account
the effects of baryons in the inner galactic regions.

\begin{table}
\begin{center}
\begin{tabular}{|c|ccccc|}
\hline
 & $a$ [kpc] & $\alpha$ & $\beta$ & $\gamma$ & $\bar{J}$\\
\hline
Einasto &  -   &   -   &       &   -    & $ 10.486$\\
NFW     & $20$ & $1.0$ & $3.0$ & $1.0$  & $8.638$\\
NFW$_c$ & $20$ & $0.8$ & $2.7$ & $1.45$ & $ 12.880$\\
\hline
\end{tabular}
\caption{{\footnotesize Einasto, NFW and NFW$_c$
density profiles with the corresponding parameters,
and values of $\bar{J}(\Delta\Omega)$ for the galactic
region under consideration.}}
\label{JbarsLight}
\end{center}
\end{table}
The $\bar{J}$ values obtained in the table demonstrate another virtue of searching
for dark matter at intermediate latitudes, namely the fact that the results become
quite robust with respect to the various dark matter density distribution
modelizations. We saw in all previous analyses that in the galactic center case, there can be differences of
orders of magnitude in this factor, whereas in this case the differences are of
$O(1)$. In the following, we shall be presenting our results for an Einasto
profile, since it yields results somewhere in the middle among the other two
scenarios.

The Fermi collaboration has published its 1-year observation results outside the GC
\cite{Abdo:2010nz, Porter:2009sg, Abdo:2009mr}. In this paper, the collaboration presents its
observations for a period of $19$Msec and for various galactic latitudes. 
In the companion paper, the results for latitudes $20^\circ<b<60^\circ$ are
also presented, which lie actually in our region of interest. The data between $b = 10^\circ$ and
$20^\circ$ are included in the same paper, presenting an enhancement by a factor of roughly 
$1.5 - 2$  with respect to the higher latitude data. In this analysis, 
for the sake of simplicity, we shall
be focusing on the data from higher latitudes ($20^\circ<b<60^\circ$), integrating them over the whole
region of interest. This is  justified, since we have excluded 
from our analysis the region within $20^\circ$ from the galactic center, which should provide
one of the major contributions to this spectrum.

In the paper, the authors could fit the data quite well using a Diffuse Galactic
Emission model based on the GALPROP code. We find this model to be well reproduced, in our 
region of interest, by a simple power-law
\begin{equation}
 \Phi_{\mbox{\begin{tiny}bkg\end{tiny}}}^{\mbox{\begin{tiny}Th\end{tiny}}} = 2.757\cdot 10^{-6} E^{-2.49}
\end{equation}
in units of GeV$^{-1}$ sec$^{-1}$ cm$^{-2}$ sr$^{-1}$.

In the same analysis, the collaboration presents the detector effective area values 
that should be used in order to compare predictions with observations, as a function of
the gamma-ray energy. In the following, we shall be using these values rather than the usual
nominal effective area of $10000$ cm$^2$.
We consider a $3$ - year data acquisition period under the previous assumptions.

\subsubsection{Antiproton detection}
The PAMELA collaboration recently published its updated antiproton measurements
in the kinetic energy range from $60$ MeV up to $180$ GeV \cite{Adriani:2010rc}. The data acquisition period
was $850$ days and the results seem to be in quite good agreement with several theoretical
predictions for secondary production. Model-independent
constraints from this data have been discussed, for example, in \cite{Cholis:2010xb}.

Above $10$ GeV, which is the region of interest in our case, the data can be well described by a simple
power law
\begin{equation}
 \Phi_{\mbox{\begin{tiny} bkg \end{tiny}}} = 5.323\times 10^{-4} E^{-2.935}
\ \ \mbox{GeV}^{-1} \mbox{sec}^{-1} \mbox{sr}^{-1} \mbox{cm}^{-2}.
\end{equation}

When we examine the detection perspectives in the antiproton channel, we shall take the data acquisition
period as $3$ years. We have performed our calculations for the three propagation models MIN, MED
and MAX but we shall only be showing our results for the MAX model  
and comment upon the results for the other two propagation models.

\subsubsection{Results for indirect detection}
According to our findings, no point of our parameter space is presently excluded
by the existing Fermi or PAMELA data while passing all constraints analyzed previously
(electroweak or relic density). The only points which are actually excluded are those that
possess such large self-annihilation cross-sections that they lead to under-abundance
of dark matter at present times. From now on, we shall therefore only stick to predictions
concerning the detection perspectives of our model.
\\ \\
In fig.\ref{a01100} we plot the contours where the $\chi^2$ between the background-only fit and 
the pseudo-data becomes equal to $28.87$, a case in which the background-only hypothesis can be rejected at
$95\%$ CL and the corresponding parameter space points are thus detectable for the case $A_0 = -1100~$GeV. 
The parameter space region lying between the contours $D^\gamma$ or $D^{\bar P}_{Max}$ has 
$\chi^2 \geq 28.87$ and is, thus, detectable. The contours for
gamma-rays are blue-dashed whereas for antiprotons violet-dotted-dashed. 
In this plot, we assume the Einasto profile and the MAX propagation model for gamma-rays 
and antiprotons respectively. 

We see that the $A$-pole region falls within the detectability limits in both channels, whereas
the $h$-pole region is completely invisible both for Fermi and AMS-02. We shall comment
on these points in the following. In the case of gamma-rays, switching to another profile does 
not significantly alter the results.
In the case of antiprotons however, the results \textit{do} change. For the sake of clarity we omit plotting
these results, but we have calculated that both for the MIN and the MED propagation model the entire
parameter space evades detection.
\\ \\
Before explaining these results, we make the introductory remark that as can be seen for example 
in \cite{Baltz:2008wd}, the Fermi satellite should in principle be able to exclude WIMPs with 
thermal cross-sections lying in our neutralino mass range and for $b \bar{b}$ final states. 
\\ \\
Passing on to our results, it is interesting that practically all of the points offer quite 
good detection perspectives. This is related to 
two facts: firstly, the present values of the annihilation cross-sections 
for the parameter space points satisfying the WMAP constraints
are quite high, namely of the same order as during the time of decoupling (i.e., in the thermal region). This is
mainly due to the mechanism through which the correct relic density is actually obtained. 

We already pointed out that in this scenario the mechanism that drives neutralino annihilation is resonnant 
$s$-channel pseudoscalar Higgs boson exhange, apart from the small region at low $m_{1/2}$ and relatively large 
$m_A$, where the dominant mechanism is CP-even light Higgs exchange. 
In the case of annihilation through an $A$ propagator, the cross-section is practically insensitive to
velocity changes as pointed out for example in \cite{Jungman:1995df} and demonstrated in Appendix \ref{FeynmanGraphs}. 
This leads to the conclusion that the self-annihilation cross-section stays quite high even at present times. 
It is really instructive to compare this regime with the corresponding points where the acceptable 
relic density is produced via neutralino pair annihilation into $h$. In this case, $<\sigma v>$ tends 
to zero as the LSP velocity does so (i.e., at present times, which is relevant for 
indirect detection experiments). 

This is indeed an interesting effect, which renders the $h$-pole points practically invisible to 
indirect detection experiments. If one observes the indirect detection results for the BMSSM, 
one can deduce that detectability limits (although defined
differently) seem to systematically ``avoid'' the $h$-pole region. Furthermore, we should note that
the decay modes in the present scenario are dominated by the $b \bar{b}$ final state. 
This is due first of all to kinematics but also
to the fact that annihilation into $A$ and then into down-type fermions 
is proportional to the quark mass, through the dependence of the 
relevant amplitude on the Yukawa couplings. We have seen that this is a final state
yielding relatively rich photon spectra if compared, for example, to the leptonic case. 
This is less the case for the antiproton yield, where the decays of light quarks have the tendency of
yielding more antiprotons than $b \bar{b}$ pairs. This could be an explanation of the relatively better 
detection perspectives at Fermi than at AMS-02 - although such a comparison could be misleading, since
what matters is not only the absolute magnitude of the signal, but rather its relative magnitude with 
respect to the background, their precise relative form and so on.
\\ \\
Let us proceed to our second scenario, i.e., fig.\ref{a01000}. Once again, 
the blue-dashed line depicts region where $\chi^2 = 28.87$ for gamma-rays whereas the 
violet-dotted-dashed line represents 
the same condition for antiprotons. Astrophysical assumptions are the same as in the previous case. Points lying 
above, below or on the left of the contours are detectable. If we plotted the gamma-ray results for the
other two profiles we examined, results would be practically unchanged. In the case of the two other
propagation models for antiprotons, AMS-02 will be blind to the relic density satisfying points.

We can see that in this scenario, the perspectives are also quite good. We should note that we are still lying 
in the $A$-pole region (with significant contribution from $s-$channel $Z$ and $t-$channel 
neutralino exchange): neutralino annihilation
is driven by the $s$-channel pseudoscalar exchange. Once again, $<\sigma v>$ lies roughly in the 
typical thermal region. But in this case, the lightest neutralino has a higher mass than previously.
This is the reason why for relatively large values of $m_{1/2}$, we have a certain
deterioration in the detection perspectives, particularly in the lower branch of the WMAP
compliant parameter space. This is mostly visible in the antiproton channel, 
where we see that practically all LHS points are invisible at AMS-02. We have checked that if we
consider a more stringent gamma-ray detectability criterion, the same tendency would be visible for the
corresponding contour as well. This behaviour could be connected to the fact that
in this particular parameter space region of small $m_A$ and large $m_{1/2}$, the final state
is comprised, to a large extent, by Higgs and gauge bosons. This is not the case
for the upper branch, where the dominant channel is $b \bar b$. The main effect
of a final state including Higgses is to shift the energy spectrum towards lower energies
(since we consider the Higgs bosons to decay predominantly into $b\bar{b}$ pairs), where the
background is larger. It is thus more difficult to disentagle the non-power law component of the
spectrum (i.e. the signal) from the background.

In the two scenarios we examine two regimes, one with a quite light neutralino 
($50 \lesssim m_{\chi_1^0} \lesssim 80$ GeV) and one with a relatively heavier one
($100 \lesssim m_{\chi_1^0} \lesssim 130$ GeV), finding that a good part of our parameter
space should be visible at Fermi and AMS-02. There is, 
however, one question that could arise, namely what happens in the intermediate mass regime, 
particularly in the context of the light Higgs boson zone.  
The answer could be given once more by considering that it is well-known that
increasing the WIMP mass tends to aggravate detection perspectives, if the same final states and
self-annihilation cross-section are assumed. Both the $Br_i$'s and $\left\langle \sigma v \right\rangle$
remain quite stable in value from lighter to higher masses in our model: the final state is mostly $b \bar{b}$ 
(in the second case Higgs final states are also significant which subsequently decay mostly 
into $b \bar b$)
and the cross-section is of the typical thermal value, both during decoupling \textit{and} at present times.
Thus, it is easy to infer that the intermediate mass regime would also be able
to produce rich $\gamma-$ray or anti-proton signals. Overall, this $A$-pole scenario that we have examined
can be considered as quite promising for indirect detection. Later on, we shall further comment
on some more general conclusions that could be drawn concerning this particular region.

%%%%%%%%%%%%%%%%%%%%%%%%%%%%%%%%%%%%%%%%%%%%%%%%%%%%%%%%%%%%%%%%%%%%%%%%%%%%%%%%
%%%%%%%%%%%%%%%%%%%%%%%%%%%%%%%%%%%%%%%%%%%%%%%%%%%%%%%%%%%%%%%%%%%%%%%%%%%%%%%%%%%%%%%%%%%%%%%%%%%%%%%
%%%%%%%%%%%%%%%%%%%%%%%%%%%%%%%%%%%%%%%%%%%%%%%%%%%%%%%%%%%%%%%%%%%%%%%%%%%%%%%%%%%%%%%%%%%%%%%%%%%%%%%
%%%%%%%%%%%%%%%%%%%%%%%%%%%%%%%%%%%%%%%%%%%%%%%%%%%%%%%%%%%%%%%%%%%%%%%%%%%%%%%%%%%%%%%%%%%%%%%%%%%%%%%
\subsection{Direct detection in the NUHM model}
\begin{figure}[tb!]
\begin{center}
\includegraphics[width=0.40\textwidth,angle=-90]{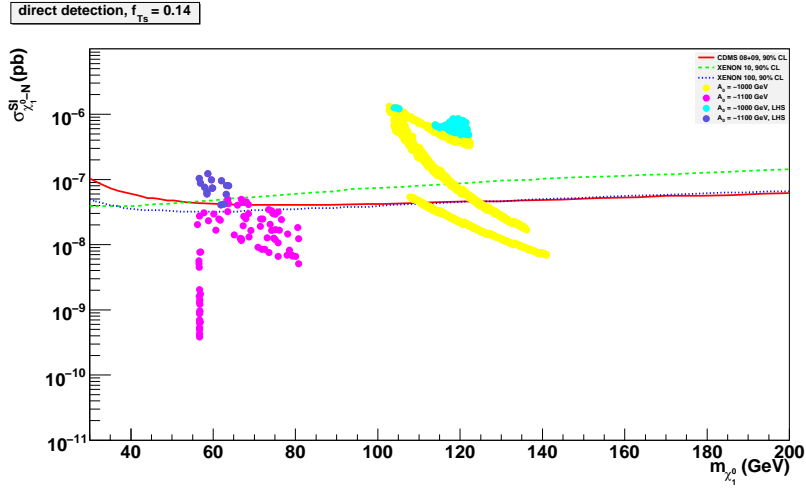}
\end{center}
\caption{{\footnotesize
$(m_{\chi^0_1}, \sigma_{\chi^0_1-N}^{SI})$ combinations 
along with the relevant exclusion limits from direct detection
experiments. The points lying above the lines are in principle excluded according to the published
limits. Yellow points correspond to a trilinear coupling value of 
$A_0 = -1000~$GeV whereas pink ones to
$A_0 = -1100~$GeV. The light blue and  the dark blue points represent the
{\it light Higgs boson} regime for the two scenarios respectively. 
$f_{Ts}^{(p, (n))}$ is taken at the default DarkSusy value, namely $0.14$.}}
\label{DirectLimits}
\end{figure}

In fig.\ref{DirectLimits} we scatter the WMAP-compliant parameter
space points on the $(m_\chi, \sigma_{\chi-N}^{SI})$ (neutralino mass - neutralino-nucleon spin-independent
scattering cross-section) plane and compare them against the three strongest bounds available in the 
literature: The combined 2008 and 2009 CDMS-II results, the constraints from the XENON10 experiment as
well as the latest bounds from XENON100. We take the two former ones from ref.\cite{Kopp:2009qt} and the
latter from \cite{Aprile:2010um}. We further highlight the points falling into the LHS
with different colors. The neutralino-nucleon spin-independent 
scattering cross-section is computed by means of the public code DarkSUSY 
\cite{Gondolo:2000ee, Gondolo:2002tz, Gondolo:2004sc, Gondolo:2005we}.

We see that many of our points fall largely within the region that is supposed to be excluded
from the existing data. Whereas the $A_0 = -1100~$GeV scenario (pink points) 
is more or less satisfying the constraints, the scenario with $A_0 = -1000~$GeV 
(yellow points) is in most cases largely above the limits, exceeding 
by more than an order of magnitude the CDMS-II and XENON100 allowed cross-sections. 
The large values for $\sigma_{\chi^0_1-N}^{SI}$ in the second scenario can be 
attributed 
to the large Higgsino components of the neutralino which enhance the coupling
${\cal{C}}_{\chi_1^0 \chi_1^0 h(H)}$. 
In both cases, the points falling in the LHS lie within the excluded zones.

However, in chapter \ref{Chapter2} it was argued that there can be significant uncertainties
that complicate the assertion on whether a particular model is excluded or not. More specifically:
\begin{itemize}
 \item Uncertainties can arise in the calculation of the neutralino-nucleon elastic scattering cross-section,
which can be due to a number of factors. For example, as described in
detail in \cite{Ellis:2008hf}, significant uncertainties can arise in the passage from the parton-level cross-section
to the hadronic level one. 
 \item Some uncertainties might be present in the passage from the hadronic to the nuclear level. Indeed, 
at the end of the day the primarily constrained quantity is the WIMP-\textit{nucleus} elastic scattering cross-section
and not the WIMP-\textit{nucleon} one. 
 \item The local dark matter density is by no means a perfectly well-known quantity and is in fact a normalization
factor in the overall procedure of computing the WIMP-nucleus scattering rate. 
 \item Little is known on the true velocity distribution of WIMPs in the detector rest frame as well as on the
escape velocity at which the integral over the velocity distribution should be truncated.
\end{itemize}
The first point concerns our own calculation of the spin-independent
neutralino-nucleon elastic scattering cross-section. 
In other words, and refering to fig.\ref{DirectLimits}, we 
expect that a certain variation in the position of parameter space points on the 
$(m_{\chi^0_1}, \sigma_{\chi^0_1-N}^{SI})$ plane should be allowed. 
The other remarks apply to the experimental limits published
from the various collaborations, i.e., they can amount to a change in the position of the exclusion lines. 

In ref.\cite{Ellis:2008hf}, a systematic study of the hadronic uncertainties entering the neutralino-nucleon
scattering cross-section is performed. It turns out that the most striking and influential uncertainty
comes from the pion-nucleon $\sigma$ term related to the strange quark content of the nucleon
which is poorly known but an essential ingredient for a precise calculation of the relevant cross-section. 
This source of uncertainty alone can give rise to a variation in the spin-independent cross-section 
of more than an order of magnitude\cite{Cao:2010ph,Ellis:2008hf}. This means
that the relevant neutralino-nucleon scattering cross-sections that we 
have calculated can in fact vary by a factor of more than $10$. 

In order to better comprehend this argument, we should digress for a moment and present some 
formalism on the passage from the parton to the nucleon - level cross-section.
The effective Lagrangian that describes neutralino elastic scattering at
small velocities is given by
\begin{equation}
{\cal L} = \alpha'_{qi}\bar{\chi_1^0} \gamma^\mu \gamma^5 \chi_1^0 \bar{q_{i}}
\gamma_{\mu} \gamma^{5} q_{i} +
\alpha_{qi} \bar{\chi_1^0} \chi_1^0 \bar{q_{i}} q_{i}~.
\label{lagxsection}
\end{equation}
The first term represents spin - dependent scattering while the second term refers to spin - independent scattering.
Eq.\eqref{lagxsection} assumes summing over both the quark generations $q$ while the subscript $i$ runs for up ($i=1$) 
and down type ($i=2$) quarks respectively. The neutralino-quark coupling coefficients
$\alpha_{q}$ and $\alpha'_{q}$ contain all SUSY model-dependent 
information.
The spin-independent scattering cross-section of a neutralino with a target
nucleus of proton number (atomic number) 
$Z$ and neutron number $A-Z$ ($A$ being the
mass number) is given by
\begin{equation}
\sigma^{SI} = \frac{4 m_{r}^{2}}{\pi} \left[ Z f_{p} + (A-Z) f_{n}
\right]^{2}~,
\label{sitotal}
\end{equation}
where $m_r$ is the reduced mass
defined by $m_r=\frac{m_{\chi_1^0} m_N}{(m_{\chi_1^0}+ m_N)}$ and
$m_N$ refers to the mass of the nucleus.
The quantities $f_p$ and $f_n$
contain all the information of short-distance physics and nuclear
partonic strengths. These are given by
\begin{equation}
\frac{f_{p, (n)}}{m_{p, (n)}} = \sum_{q=u, d, s} f_{Tq}^{(p, (n))} 
\frac{\alpha_{q}}{m_{q}} +
\frac{2}{27} f_{TG}^{(p, (n))} \sum_{c, b, t} \frac{\alpha_{q}}{m_q}~, 
\label{fpn}
\end{equation}
where $f_{Tq}^{(p, (n))}$ defined as
\begin{equation}
m_{p, (n)} f_{Tq}^{(p, (n))} = \langle p, (n) | m_{q} \bar{q} q | p, (n) 
\rangle \equiv 
m_q B_q~. 
\label{defbq}
\end{equation}
The quantities $f_{Tq}^{(p, (n))}$ can be evaluated using hadronic data \cite{Ellis:2000ds}.
The gluon - related part namely $f_{TG}^{(p, (n))}$ is given by
\begin{equation}
 f_{TG}^{(p, (n))} = 1 - \sum_{q=u, d, s} f_{Tq}^{(p, (n))}~. 
\end{equation}
The numerical values of $f_{Tq}^{(p, (n))}$ may be seen in \cite{Ellis:2000ds,Hooper:2009zm}. 

Here the parameter $f_{Ts}^{(p, (n))}$ requires the information of the strange quark content 
of the nucleon $y$ which, on the other hand, depends on the pion-nucleon sigma term 
$\sigma_{\pi N}$ and the size of the $SU(3)$ symmetry breaking-$\sigma_0$ through
$y=1-\frac{\sigma_0}{\sigma_{\pi N}}$. More specifically,
 $f_{Ts}^{(p, (n))} \propto \sigma_{\pi N}y \propto (\sigma_{\pi N}-\sigma_0)$,
so that $\sigma^{SI} \sim (\sigma_{\pi N}-\sigma_0)^2$. 

In DarkSUSY  the above coefficient is chosen as $f_{Ts}^{(p, (n))} \equiv 0.14$.   
Recent lattice results however, hint towards much smaller values of $y$
($y <0.05$) which leads to $f_{Ts}^{(p, (n))}\sim 0.02$ \cite {Cao:2010ph,Ohki:2008ff}, 
a value much smaller than previous estimates. Considering even larger
uncertainty in $\sigma_{\pi N}$ and thus in $y$ one may assume 
$\sigma_{\pi N}=\sigma_0$ 
which leads to $y=0$ or $f_{Ts}^{(p, (n))} = 0$\cite{Ellis:2008hf}.
This could provide a significant change in the results of the $\sigma^{SI}$. 
In fact, in \cite{Cao:2010ph,Ellis:2008hf}, the
variation in the spin-independent cross-section due to this reduced $f_{Ts}^{(p, (n))}$
has been estimated. 

\begin{figure}[tb!]
\begin{center}
\includegraphics[width=0.40\textwidth,clip=true,angle=-90]{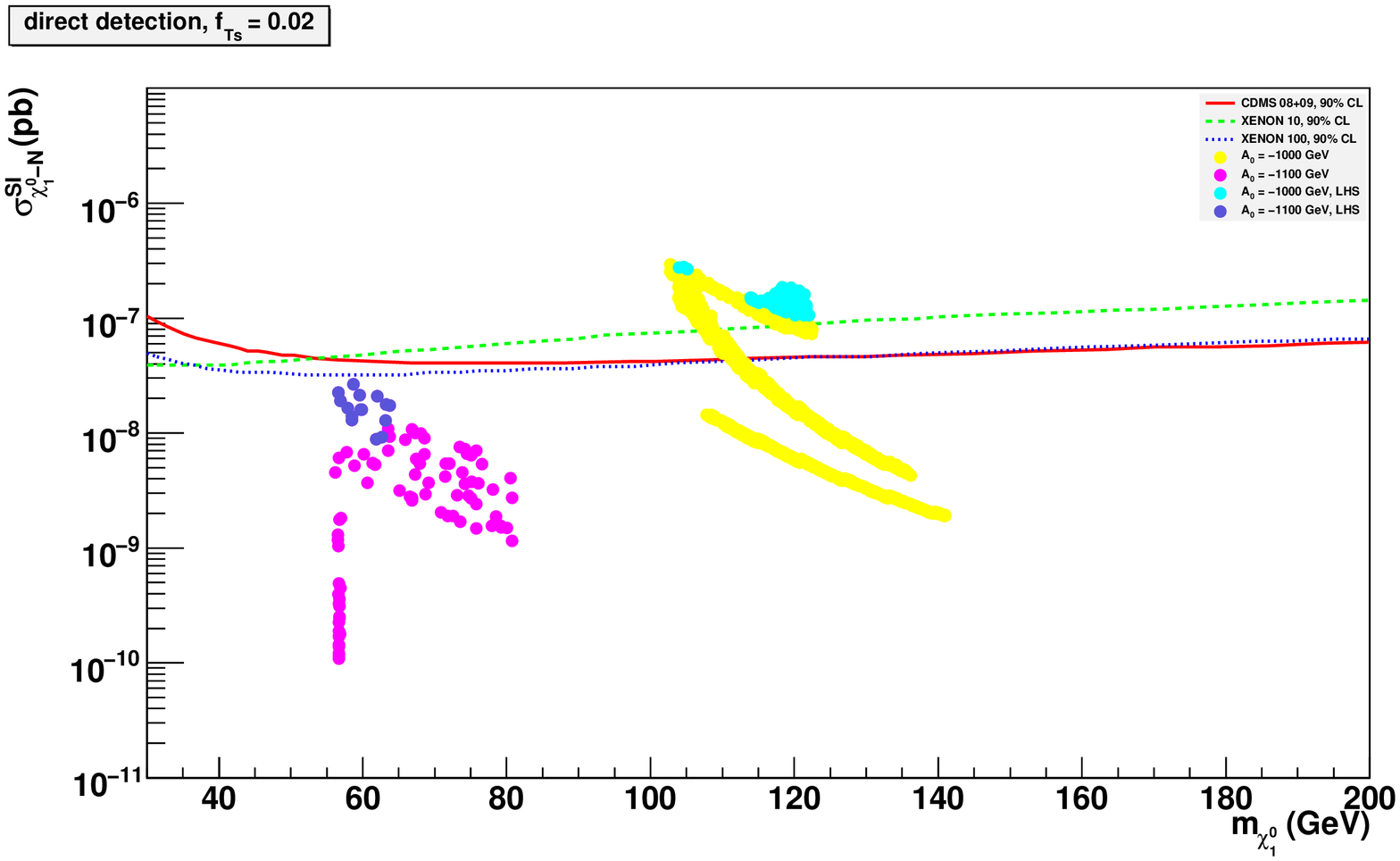}\\
\includegraphics[width=0.40\textwidth,clip=true,angle=-90]{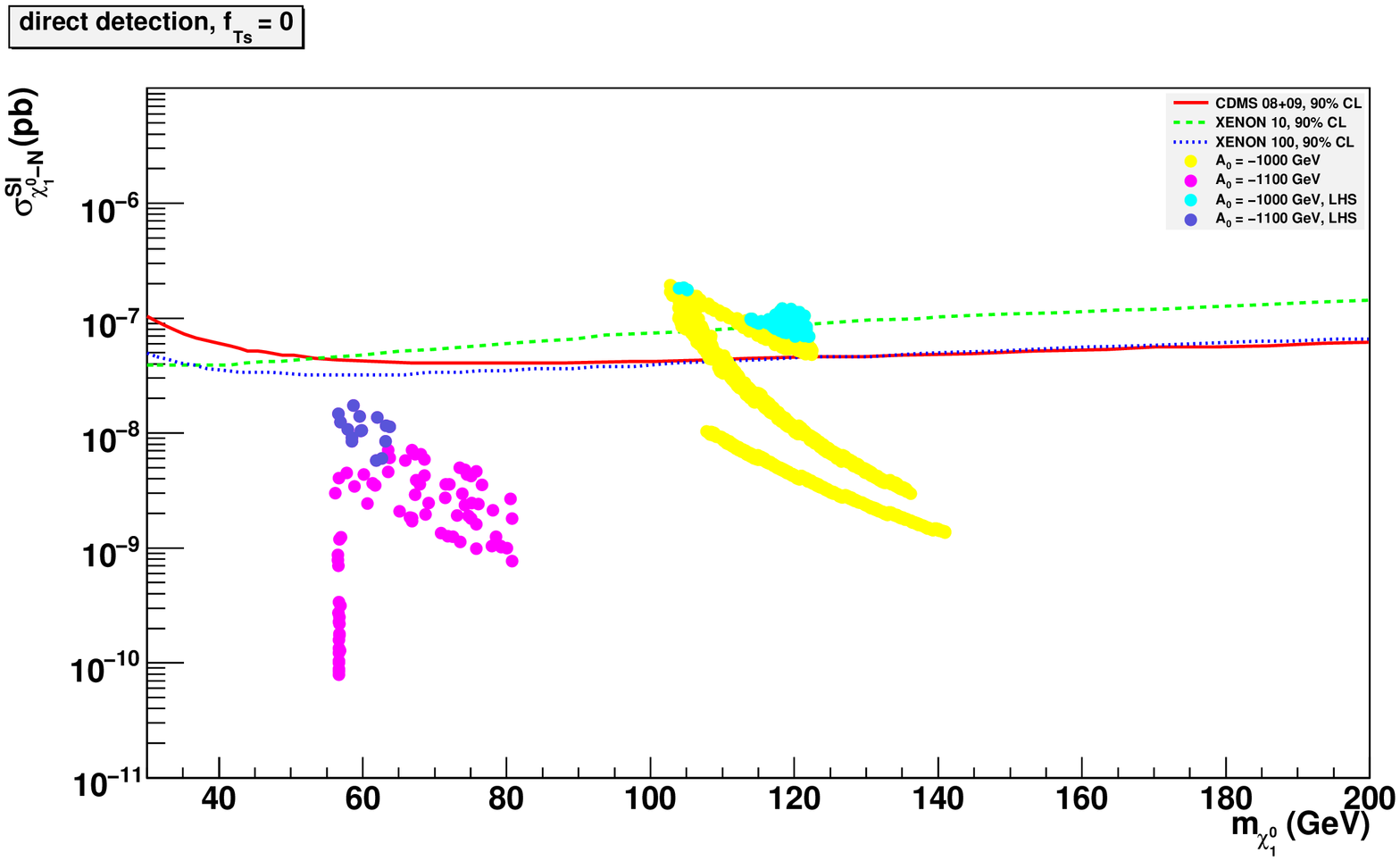}
\end{center}
\caption{{\footnotesize
As in fig.\ref{DirectLimits} but for $f_{Ts}^{(p, (n))} = 0.02$ (top) and $0$ (bottom).}}
\label{DirectLimitsfTs}
\end{figure}

In order to quantify the effect of the strange quark content uncertainties, we consider two 
representative values for $f_{Ts}^{(p, (n))}$ namely $0.02$ and $0$. We present our results in 
fig.\ref{DirectLimitsfTs}. Indeed, we can
see that the corresponding cross-sections decrease by significant factors, reaching up to an order of magnitude (particularly for $f_{Ts}^{(p, (n))}=0$). 
This clearly starts raising questions on whether a good portion of our parameter
space is excluded (as one would naively expect from Fig.\ref{DirectLimits}) 
or not.

We see, however, that - especially in the heavier neutralino scenario - there are still some
parameter points lying above the exclusion lines (roughly a factor of 
$2 - 3$). This is particularly true for the LHS scenario. However, this conclusion may become weaker
if one considers the other uncertainties that we mentioned.

Passing to the nuclear level requires modelling of the nucleon density within the nucleus. The most commonly
used parametrization is the one presented by Engel in \cite{Engel:1991wq}. An analysis of potential 
deviation
that might arise from different form factor parametrizations has been performed, for example, in
\cite{Duda:2006uk}, where the authors find that the exclusion lines can shift vertically by roughly a
factor of $1.5$.

Apart from these points, we have already seen in chapter \ref{Chapter2} that the uncertainties 
related to the local dark matter density in the solar neighborhood as well as the ones related 
to the velocity distribution in the detector rest frame and the escape velocity can induce
further uncertainties which in our case could shift the exclusion lines towards larger values
of cross-sections by roughly a factor of $2$.

We thus see that overall, and despite the apparent exclusion of a large portion of our parameter space, there is still
quite some margin for changing the relation among the predicted $(m_{\chi^0_1}, \sigma_{\chi^0_1-N}^{SI})$ as derived
from the model and the exclusion limits as presented by the corresponding collaborations. We feel it is
reasonable to say that we cannot assess that easily whether the parameter space points lying above the exclusion
lines in Fig.\ref{DirectLimits} are actually excluded or not.

We should clarify at this point that the previous remarks have by no means the purpose of demeaning the
remarkable works that are done both by theorists and experimentalists in order to develop tools for calculations
and extract reliable bounds. Our goal was just to illustrate that it might still be
meaningful to examine models which at first sight appear to be excluded. This becomes particularly apparent
from our calculation of the spin-independent cross-section for different values of $f_{Ts}^{(p, (n))}$.

%%%%%%%%%%%%%%%%%%%%%%%%%%%%%%%%%%%%%%%%%%%%%%%%%%%%%%%%%%%%%%%%%%%%%%%%%%%%%%%%%%%%%%%%%%%%%%%%%%%%%%%%%%%%%
%%%%%%%%%%%%%%%%%%%%%%%%%%%%%%%%%%%%%%%%%%%%%%%%%%%%%%%%%%%%%%%%%%%%%%%%%%%%%%%%%%%%%%%%%%%%%%%%%%%%%%%%%%%%%
%%%%%%%%%%%%%%%%%%%%%%%%%%%%%%%%%%%%%%%%%%%%%%%%%%%%%%%%%%%%%%%%%%%%%%%%%%%%%%%%%%%%%%%%%%%%%%%%%%%%%%%%%%%%%

\newpage
\chapter{Conclusions and outlook}

\section{Summarizing}
The existence of a significant quantity of non-baryonic, non-luminous, 
Cold Dark Matter in the universe is today considered to be quite well-established.
Up to the writing of this work, the CDM approach seems to be the only 
coherent way of explaining a whole series of cosmological observations.

At the same time, the Standard Model of particle physics does not include
a neutral stable particle that could play the role of CDM. In fact, the
existence of dark matter is today one of the few experimental indications
that there should be physics beyond the SM: the interface between particle
physics and cosmology turns into a driving force for the search for new 
physics.

Among the several classes of dark matter candidates that could be envisaged, a particularly
interesting one is that of Weakly Interacting Massive Particles. Among their
most attractive features appart from the fact that their production
mechanism is rather simple in its principle, is that they can arise naturally in various
extensions of the SM and that there are strong chances that they can be detected
in present or future experiments: Direct detection, indirect detection and TeV - scale
colliders.

In this work we examined WIMP dark matter mostly from the point of view of its
detection perspectives in such experiments. At first, we discussed dark matter 
in a model-independent framework, examining what could be the prospects for the 
determination of some of its characteristics, notably the WIMP mass. We saw that
for moderate or low values of the WIMP mass, there could be hope not only to
detect such particles but also to constrain their properties.

The next step was to study dark matter in a minimal extension of the 
Standard Model by a real singlet scalar field. We presented some existing
results concerning various constraints of the model. Then, we tested potential
constraints coming from the PAMELA experiment in the positron and the antiproton
channel, to find that for the time being all of the viable parameter space
survives the test. Interestingly though, some part of it is excluded if one
assumes somehow ``optimistic'' astrophysical configurations including clumps. 
We also checked the detection perspectives for the model in these two channels
and in the oncoming AMS02 experiment. The conclusion is that significant regions
of the parameter space shall be probed in this experiment. If clumpiness is invoked, 
practically the whole model is testable.

Finally, we examined neutralino dark matter in two supersymmetric models trying to resolve
the so called ``little hierarchy problem'' of the MSSM, related to the lightest
higgs mass and the LEP2 bounds on it. The first approach introduces non-renormalizable
contributions coming from physics beyond the MSSM in order to augment the lightest
higgs mass. The second model tries instead to evade the LEP constraints by reducing
the higgs boson coupling to the $Z$ boson. We saw that for important regions of the parameter
space of the corresponding models, we can not only obtain the correct relic density
in accordance to the WMAP measurements, but also seriously hope that these candidates
could be detected in the years to come. Furthermore, we discussed some particular 
features of the viable supersymmetric parameter space and the interface between
the dominant mechanism for obtaining the correct relic density and the detection
prospects at present times.

Generically speaking, dark matter detection is more favourable for relatively light
candidates maintaining large self-annihilation cross-sections from decoupling up to
present times. We saw as a counter - example the $h$ - pole region, where the
cross-section is high enough at earlier times but tends to zero with the neutralino
velocity. This particular (x)MSSM region gives the correct relic density but is
only visible in direct detection experiments. On the opposite side there is the
$A$ - pole region as well as parameter space domains where the LSP has a significant
higgsino component, since the cross-section does not depend that strongly on the
neutralino kinetic energy.

\section{Perspectives}
As the present work progressed, we have had the chance to witness repeated excitement
due to unexpected signals that could be interpreted as coming from dark matter:
The PAMELA/Fermi positron excess, the CDMS-II events, the DAMA/CoGeNT/CRESST signals.
Whether these signals have indeed anything to do with dark matter is still an
open discussion. It is clear today that the number of uncertainties entering
dark matter detection is so large that it is hard to produce immediate conclusions.

Today most of the excitement seems to concern the signals coming from direct detection
experiments. If the triple DAMA/CoGeNT/CRESST excess is indeed due to dark matter, this
challenges several of the dominant dark matter models since they point to a very low
mass WIMP, of the order of $10$ GeV, and interacting at a relatively high rate, 
which is not that easy to obtain without violating current bounds from other sources.

In order to identify the nature of potential excesses, a huge amount of work is required
in pinpointing the various factors and uncertainties entering not only the determination
of viable models, but also the identification of the signals themselves and whether or
not they can be explained through standard physics. This is one of the main difficulties
in dark matter detection: the observation object is simply our natural environement
as a whole and it is hard to identify what is the ``background'' or the ``signal''
physics, since both are quite poorly known. 
In this respect, it is of crucial importance to try and quantify these uncertainties.
The combination of different experimental
sources can certainly contribute significantly in this direction.
At this point it should not be omitted that calculations often bare uncertainties
also from the quantum field theoretical side: loop corrections, multi - particle 
final states etc need often be taken into account in order to obtain reliable
results.

Especially in the advent of the Large Hadron Collider, there is hope that we could 
detect signals with a significant amount of missing transverse energy that could be
due to the existence of some neutral stable particle. If such a prediction is confirmed,
the next step should be to study whether this particle could be the essential ingredient
of dark matter. This is not at all a trivial task, as often it includes reconstructing
entire models in order to check if indeed the particle can reproduce the correct
relic abundance.

As final comment, it should be said that dark matter phenomenology 
englobes a large amount of information coming from 
very different fields: particle physics, classical astrophysics, cosmology, nuclear
physics. This is at the same time a major difficulty and a major challenge. And, if we are
lucky enough so that dark matter is not so dark after all, this could open a whole
new era in high energy physics.

\appendixpage
\appendix
%\addappheadtotoc

\chapter{Propagation of Cosmic Rays}
\label{cosmicrayprop}

In this appendix we give some details on the solution of Eq.\ref{masterProp}
for positrons and antiprotons.

\section{Positrons}
We largely follow the formalism utilized in \cite{Lavalle:2006vb}.
We saw that the master equation \ref{masterProp} gets simplified in the form \ref{masterPos}
\begin{equation}
K_0\,\epsilon^\alpha \nabla^2 \psi  + 
\frac{\partial}{\partial \epsilon}\left( \frac{\epsilon^2}{\tau_E} \psi \right) + q = 0\,,
\end{equation}
with $\epsilon\equiv E/E_0$.
\\ \\
The first step to the solution of the last equation consists of transcribing equation 
(\ref{masterPos}) with respect to a pseudo-time variable \cite{Baltz:1998xv}
\begin{equation}
 \tilde{t}(E) = \tau_E \left( \frac{\epsilon^{\alpha - 1}}{1 - \alpha} \right)
\label{pseudotimePos}
\end{equation}
which gives the following equation to solve
\begin{equation}
 \frac{\partial \tilde{\psi}}{\partial \tilde{t}} - K_0 \nabla^2 \tilde{\psi} = \tilde{q}(\vec{x}, t) \ .
\label{PosPseudoEq}
\end{equation}
\\
Suppose now we deposit a unit point-like charge in the coordinate origin and at pseudo-time equal zero.
Then, the previous equation obtains the form
\begin{equation}
 \frac{\partial \tilde{\psi}}{\partial \tilde{t}} - K_0 \nabla^2 \tilde{\psi} = \delta^3(\vec{x_s})\delta(t_s) \ .
\end{equation}
If we ignore boundary conditions, this equation is analytically solvable and actually defines the
Green's function for our problem in pseudo-spacetime, which can be found to be
\begin{equation}
 \tilde{G}(\vec{x}, \tilde{t};\vec{0}, 0) = 
\theta(\tilde{t}) \left( 4\pi K_0 \tilde{t}\right)^{-3/2} 
\exp \left[ -\frac{r^2}{4 K_0 \tilde{t}}  \right] \ .
\label{ModGreenPos}
\end{equation}
Then, we can actually express the general integral of equation \eqref{PosPseudoEq} with the help of the Green's 
function as
\begin{equation}
 \tilde{\psi}(\vec{x}, \tilde{t}) = 
\int_0^{\tilde{t}} d\tilde{t}_s
\int d^3\vec{x}_s \tilde{G}(\vec{x}, \tilde{t};\vec{x_s}, \tilde{t}_s)
\tilde{q}(\vec{x}_s, \tilde{t}_s)
\end{equation}
which, passing back into real spacetime gives us the flux
\begin{equation}
 \Phi (\vec{x}, E) = 
\int_{E}^{\infty} dE_s 
\int d^3 \vec{x}_s G_{e^+} (\vec{x}, E;\vec{x}_s, E_s) 
q(\vec{x}_s, E_s)
\end{equation}
the propagator can be computed via the modified Green's function \eqref{ModGreenPos} through
\begin{equation}
 G_{e^+} (\vec{x}, E;\vec{x}_s, E_s) = 
 \frac{\tau_E}{E_0 \epsilon^2} \tilde{G}(\vec{x}, \tilde{t};\vec{x_s}, \tilde{t}_s)
\end{equation}
Now, since positrons can clearly escape our cylindrical diffusive zone, boundary conditions
should be imposed. A first simplification that can be done quite safely is to ignore completely
the radial boundary conditions. This is justified by the fact that since positrons loose energy 
during their propagation, they cannot originate from very far away. So, if the diffusive zone 
coincides more or less with the milky way visible part, positrons received on the earth can originate 
from roughly a couple of kpc distance. It is much more probable that a positron could escape through the
vertical limits of the diffusive slab rather than the radial ones. 
The vertical boundary conditions are much more important, since
the vertical height of the diffusive zone is much smaller than the radial one.
In any case, ignoring radial boundary conditions allows us to separate the radial and vertical part
of the diffusion equation, with the modified propagator obtaining the general form
\begin{equation}
 \tilde{G} (\vec{x}, \tilde{t};\vec{x_s}, \tilde{t}_s) = 
\frac{\theta(\tilde{\tau})}{4 \pi K_0 \tilde{\tau}}
 \exp \left[ -\frac{R^2}{4 K_0 \tilde{\tau}} \right]
\tilde{V} (z, \tilde{t};z_s, \tilde{t}_s) 
\end{equation}
where $\tilde{\tau} = \tilde{t} - \tilde{t}_s$.\\ \\
Regarding the form of $\tilde{V}$, it turns out that we can identify two regimes in which the
diffusion equation can be solved in a different manner. The parameter determining the regime in
which we are situated is
\begin{equation}
 \zeta = \frac{L^2}{4 K_0 \tilde{\tau}}
\end{equation}
For large $\zeta$, which could be interpreted at a first level as the regime where the diffusion
time is small, then the positrons originate from small distances, much smaller than the $L$ limits
of the diffusive slab, and the vertical boundary conditions can be ignored. In this case, 
the diffusion equation is actually a 
1D Schrödinger equation and $\tilde{V}$ can be written in the form
\begin{equation}
 \tilde{V} (z, \tilde{t};z_s, \tilde{t}_s) = 
\frac{\theta(\tilde{\tau})}{4 \pi K_0 \tilde{\tau}}
 \exp \left[ -\frac{(z - z_s)^2}{4 K_0 \tilde{\tau}} \right]
\end{equation}
In the opposite regime of small $\zeta$ values, the vertical boundary conditions can no longer be ignored but
$\tilde{V}$ can be expanded as a series
\begin{equation}
\tilde{V} = \sum_{n = 1}^{\infty} 
\frac{1}{L} \left[ e^{-\lambda_n \tilde{\tau}} \phi_n(z_s)  \phi_n(z_\odot) + 
 e^{-\lambda_n' \tilde{\tau}} \phi_n'(z_s)  \phi_n'(z_\odot)   \right] 
\end{equation}
where
\begin{equation}
\phi_n(z) = \sin[k_n(L - |z|)] \ \ , \ \ \phi_n'(z) = \sin[k_n'(L - z)]\,,
\end{equation}
and
\begin{eqnarray}
 k_n  = & \left( n - \frac{1}{2}\right) \frac{\pi}{L} \ \ , \ \ \  &k_n'  = n \frac{\pi}{L} \,,\\
 \lambda_n  = & K_0\,k_n^2 \ \ , \ \ \ &\lambda_n' = K_0\,(k'_n)^2
\end{eqnarray}

\begin{figure}[ht]
\begin{center}
\includegraphics[width = 6cm, angle=270]{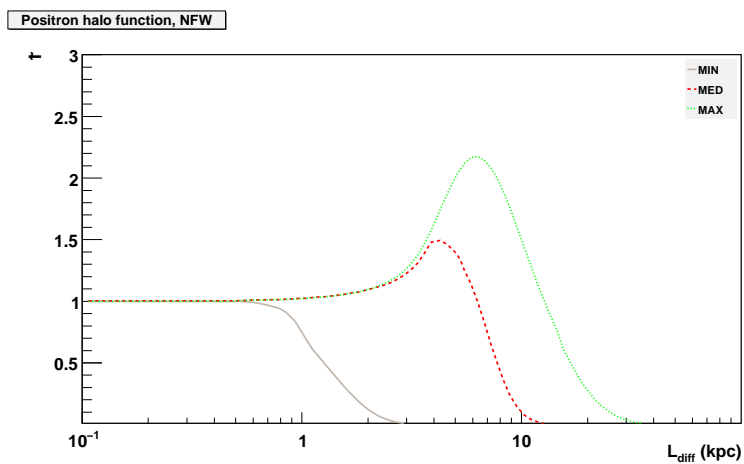}
\caption{{\footnotesize
Halo function for positrons for the three propagation models MIN, MED and MAX as a function
of the diffusion length. The assumed profile is NFW.}}
\label{fighalofuncpos}
\end{center}
\end{figure}

In order to compute the halo function with respect to the diffusion length, we developed a FORTRAN
code that calculates the relevant integral. The results were then fitted and used throughout the
calculations presented in this work.
In figure \ref{fighalofuncpos} we plot the halo function, defined in Eq.\eqref{halofuncpos} as a function
of the diffusion length and for the three propagation models defined in Table \ref{PropParametersPos}, assuming
a Navarro, Frenk and White halo profile.

As we mentioned, this halo function completely encodes the astrophysics of positron propagation
in the diffusion model we adopt.

\begin{figure}[ht]
\begin{center}
\includegraphics[width = 6cm, angle=270]{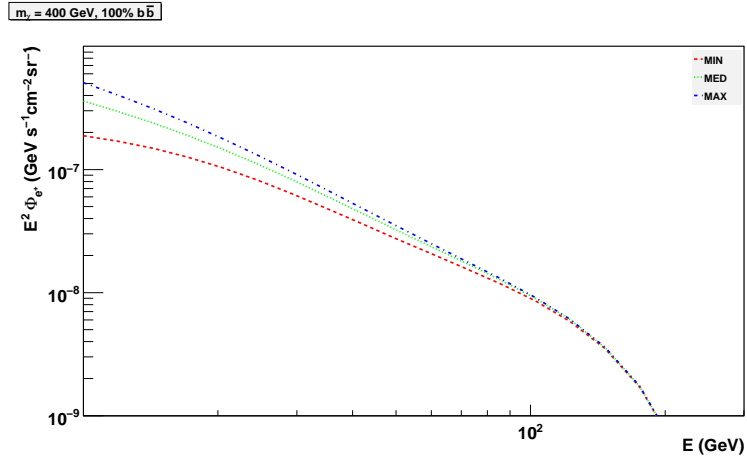}
\caption{{\footnotesize
Expected differential positron flux times squared energy for a $400$ GeV Majorana particle 
annihilating into $b \bar{b}$ pairs and for the three propagation models discussed in the
text.
}}
\label{PosFluxNoClumps}
\end{center}
\end{figure}
In order to illustrate the impact of different propagation models as explained in the main body of
the work, we demonstrate in fig.\ref{PosFluxNoClumps} the flux received on the earth (times the
squared energy) in the case of a Majorana particle with $m_\chi = 400$ GeV and 
$\left\langle \sigma v \right\rangle = 3\cdot 10^{-26}$ cm$^3$ sec$^{-1}$ annihilating into a
pure $b \bar{b}$ final state.

%%%%%%%%%%%%%%%%%%%%%%%%%%%%%%%%%%%%%%%%%%%%%%%%%%%%%%%%%%%%%%%%%%%%%%%%%%%%%%%%%%%%%%%%%%%%%%%%%%%%%
%%%%%%%%%%%%%%%%%%%%%%%%%%%%%%%%%%%%%%%%%%%%%%%%%%%%%%%%%%%%%%%%%%%%%%%%%%%%%%%%%%%%%%%%%%%%%%%%%%%%%
%%%%%%%%%%%%%%%%%%%%%%%%%%%%%%%%%%%%%%%%%%%%%%%%%%%%%%%%%%%%%%%%%%%%%%%%%%%%%%%%%%%%%%%%%%%%%%%%%%%%%
\section{Antiprotons}
\begin{figure}[ht]
\begin{center}
\includegraphics[width = 6cm, angle=270]{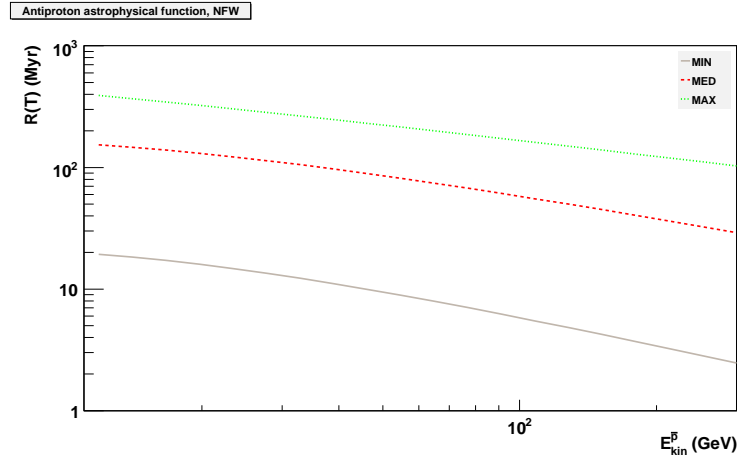}
\caption{{\footnotesize
Astrophysical function $R(T)$ for the three propagation models MIN, MED and MAX as a function
of the antiproton kinetic energy and for our energy region of interest. The assumed profile is NFW.}}
\label{fighalofuncantip}
\end{center}
\end{figure}

In the case of antiproton propagation, the equation that needs to be solved is
\begin{equation}
\left[ -K\,\nabla + V_c\,\frac{\partial}{\partial z}
+2\,h\,\Gamma^{\mbox{ann}}_{\bar p}\,\delta(z) \right] G = 
q(r, t)
\end{equation}
or, if we again assume a unit point source at the origin of our spacetime coordinates, 
\begin{equation}
\left[ -K\,\nabla + V_c\,\frac{\partial}{\partial z}
+2\,h\,\Gamma^{\mbox{ann}}_{\bar p}\,\delta(z) \right] G = 
\delta\left(\vec{r} - \vec{r'}\right)\,,
\end{equation}
with $h = 100$ pc being the half-thickness of the galactic disc.
The antiproton propagator at the solar position can then be written (in cylindrical coordinates) as
\begin{equation}
G^{\odot}_{\overline{p}}(r,z) = 
\frac{e^{-k_v\,z}}{2 \pi\,K\,L}\,
\sum_{n=0}^{\infty} c_n^{-1}\,K_0\left(r\sqrt{k_n^2 + k_v^2}\right)
\sin(k_n\,L)\,\sin(k_n\,(L-z))\,,
\label{GreenPbars}
\end{equation}
where
$K_0$ is a modified Bessel function of the second kind and
\begin{eqnarray}
c_n & = & 1 - \frac{\sin(k_n L) \cos(k_n L)}{k_n L}\,,\\
k_v & = & V_c/(2K)\,,\\
k_d & = & 2\,h\,\Gamma_{\overline{p}}^{\mbox{\tiny{ann}}}/K + 2\,k_v\,.
\end{eqnarray}
$k_n$ is obtained as the solution of the equation
\begin{equation}
n\,\pi - k_{n}\,L - \arctan(2\,k_n/k_d) = 0, \ \ n\in\mathbb{N}\,.
\end{equation}
Then, in order to compute the flux expected on  earth, we should
convolute the Green function \eqref{GreenPbars} with the source 
distribution $q(\vec{x}, E)$. For dark matter annihilations in the 
galactic halo, the source term is given by
\begin{equation}
q(\vec{x}, E) = \frac{1}{2}
\left( \frac{\rho(\vec{x})}{m_\chi} \right)^2
\sum_i 
\left( 
\langle\sigma v\rangle \frac{dN_{\bar{p}}^i}{dE_{\bar{p}}}
\right)\,,
\label{eq:q}
\end{equation}
where the index $i$ runs
over all possible annihilation final states. 
Regarding the distribution of dark matter in the Galaxy, $\rho(\vec{x})$, 
we assume a NFW profile. The final expression for 
the antiproton flux on the Earth takes the form
\begin{equation}
\Phi_{\odot}^{\bar{p}} (E_{\mbox{\tiny{kin}}}) = 
\frac{c\,\beta }{4\pi}
\frac{\langle\sigma v\rangle}{2}
\left(   \frac{\rho(\vec{x}_{\odot})}{m_\chi} \right)^2
\frac{dN}{dE}(E_{\mbox{\tiny{kin}}})
\int_{DZ} \left(\frac{\rho(\vec{x_s})}{\rho(\vec{x}_{\odot})} \right)^2
G^{\odot}_{\overline{p}}(\vec{x}_s)\,d^3x\,,
\label{PbarFluxApp}
\end{equation}
where none of the integrated quantities depends on the antiproton energy. 
\\ \\
In figure \ref{fighalofuncantip} we plot the quantity
\begin{equation}
 R(T) = \int_{DZ} 
\left(\frac{\rho(\vec{x_s})}{\rho(\vec{x}_{\odot})} \right)^2
G^{\odot}_{\overline{p}}(\vec{x}_s)\,d^3x\,
\label{halofuncantip}
\end{equation}
entering Eq.\ref{PbarFluxApp}.
This function $R(T)$ encodes the entire astrophysics of antiproton propagation and was calculated
with the help of a dedicated FORTRAN code that was developed in the framework of the present
work.

\begin{figure}[htb!]
\begin{center}
\includegraphics[width = 6cm, angle=270]{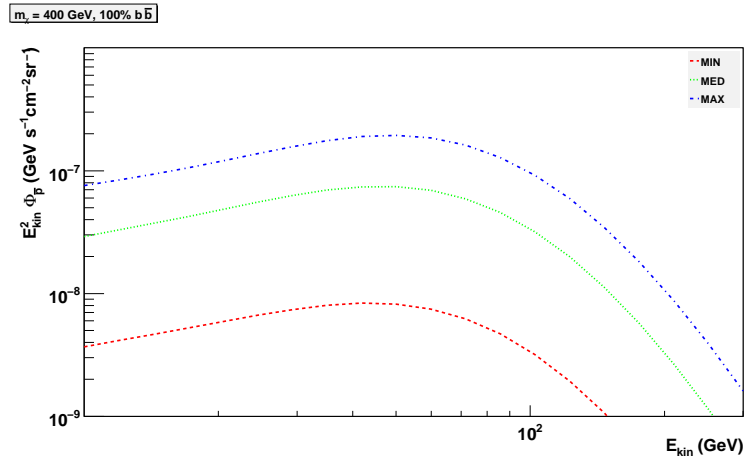}
\caption{{\footnotesize
Expected differential antiproton flux times squared kinetic energy for a $400$ GeV Majorana particle 
annihilating into $b \bar{b}$ pairs and for the three propagation models discussed in the
text.
}}
\label{AntipNoClumps}
\end{center}
\end{figure}
Once more, we demonstrate the effects of changing the propagation parameters in fig.\ref{AntipNoClumps}
where we plot the expected differential antiproton flux on the earth according to the three propagation
models MIN, MED and MAX. The flux is computed, once again, for a $400$ GeV Majorana particle annihilating into
$b \bar{b}$ with a cross-section $\left\langle \sigma v \right\rangle = 3\cdot 10^{-26}$ cm$^3$ sec$^{-1}$.

It is interesting to notice the difference among the MIN and MAX models with respect to the relevant
differences in the positron case. The change in the flux is much larger in the antiproton case. 
This large difference also manifests itself in the
results presented in this work: In all of the performed studies, it is clear that changes in the
antiproton propagation model bring along drastic modifications in exclusion/detection limits, 
whereas this is less the case for positrons.

%%%%%%%%%%%%%%%%%%%%%%%%%%%%%%%%%%%%%%%%%%%%%%%%%%%%%%%%%%%%%%%%%%%%%%%%%%%%%%%%%%%%%%%%%%%%%%%%%%%%%
%%%%%%%%%%%%%%%%%%%%%%%%%%%%%%%%%%%%%%%%%%%%%%%%%%%%%%%%%%%%%%%%%%%%%%%%%%%%%%%%%%%%%%%%%%%%%%%%%%%%%
%%%%%%%%%%%%%%%%%%%%%%%%%%%%%%%%%%%%%%%%%%%%%%%%%%%%%%%%%%%%%%%%%%%%%%%%%%%%%%%%%%%%%%%%%%%%%%%%%%%%%
\section{Substructure in the galactic halo}
Once again, we closely follow the approach outlined in \cite{Lavalle:2006vb}. 
Since the distribution of dark matter clumps in the Galaxy  is unknown, the enhancement of
the positron or antiproton flux due to substructures cannot be computed 
from first principles; it can only be studied from a statistical point of view.
What is of relevance in the present work is not the contribution of individual clumps but rather
the average contribution of the halo's clumpy component, which gives rise to the so-called
\textit{effective boost factor}.

In \cite{Lavalle:2006vb}, it is argued that assuming a particular - somehow intuitive - statistical
distribution of the subhalos $p(\vec{x}) = \rho_s(\vec{x})/M_H$, where $\vec{x}$ is the clumps
position and $\rho_s, M_H$ are the smooth component's distribution and the mass of the 
protohalos respectively, does not affect the results, at least qualitatively.

Assume that a fraction $f$ of the dark matter component of the Milky Way
is bound in clumps and that the boost factor of each clump is a constant number $B_c$, which
can differ from one clump to another. Since the positrons received on the earth cannot
originate from very large distances, it is quite reasonable to consider that $B_c$ is
constant throughout the clumps that could contribute (i.e. in the vicinity of the
solar system). Assuming further that the masses of clumps are comparable (in practice, the same)
and that they are point-like, the authors of the paper demonstrate that the effect of clumpiness
can be encoded in an \textit{energy dependent} function (the effective boost) $B_{eff}$ that can be written
as
\begin{equation}
B_{\mbox{\tiny{eff}}} \equiv \frac{\langle\phi\rangle}{\phi_{\mbox{\tiny{sm}}}} = 
(1-f)^2 + f B_c \frac{{\cal{I}}_1}{{\cal{I}}_2}\,,
\label{EffBoost}
\end{equation}
where $\langle\phi\rangle$ is the average flux coming from the clumpy DM distribution, and $\phi_{\mbox{\tiny{sm}}}$ is the 
flux that we would expect if the whole halo were smooth.
\\
The functions ${\cal{I}}_{n=1,2}$ are given by 
\begin{equation}
{\cal{I}}_n = 
\int_{\mbox{\tiny{DM halo}}} 
G(\vec{x},E) \left( \frac{\rho_{\mbox{\tiny{sm}}}(\vec{x})}{\rho_0} \right)^n d^3\vec{x}\,.
\end{equation}
The effective boost factor, then, depends on $f$ and $B_c$. When invoking  clumpiness, 
we follow \cite{Lavalle:2006vb} and use $f=0.2$ as a representative value 
(see e.g. \cite{Bertone:2005xz,Diemand:2005vz}). Regarding the constant boost factor, $B_c$, 
it could vary from just a few up to two orders of magnitude 
\cite{Lavalle:1900wn,Berezinsky:2003vn,Diemand:2005vz}. 
In this work we use  $B_c=3, 10, 100$, which give rise to effective boost factors in the approximate 
ranges ($1,2$), ($3,5$) and ($10,40$) respectively. This last range roughly coincides with the 
upper limit for the boost factor found in \cite{Lavalle:1900wn}, for the case of a NFW smooth halo 
and clumps with a Moore et al internal profile. 
Once again, the effective boost factors are calculated by developing a dedicated FORTRAN code.
\begin{figure}[ht]
\begin{center}
\includegraphics[width = 6cm, angle=270]{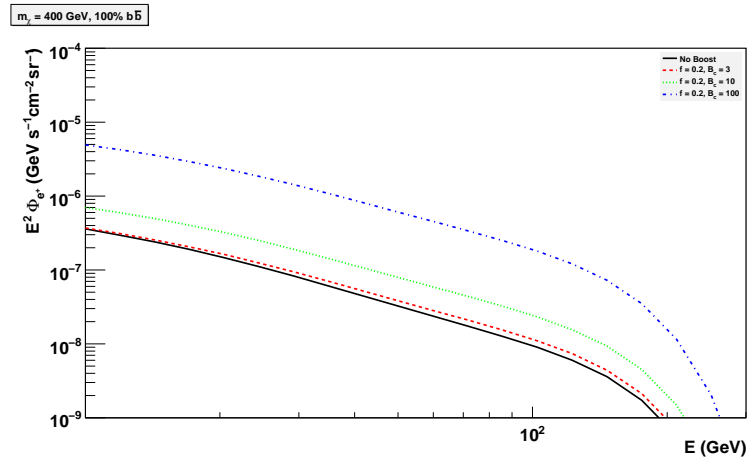}
\caption{{\footnotesize
Expected differential positron flux times squared energy for a $400$ GeV Majorana particle 
annihilating into $b \bar{b}$ pairs and for four different halo setups as discussed in the text.
The MED propagation model is assumed.
}}
\label{PosFluxClumps}
\end{center}
\end{figure}
In order to illustrate the impact of substructures on the expected flux, we plot in fig.\ref{PosFluxClumps}
the positron flux expected on the earth in the case of a Majorana particle as described in the case
with no clumps, assuming three different astrophysical setups: We choose the MED propagation model and
impose that $20\%$ of the halo's matter is in clumpy form. Then, we choose three different values for
the constant boost factor for every clump (as discussed in the text) and compute the corresponding 
fluxes. For comparison, we repeat the MED case with no clumps already present in fig.\ref{PosFluxNoClumps}.

As a last comment, it must be kept in mind
that in some exceptional cases --for example when there is a large dark matter clump very close 
to the Earth-- $B_{eff}$ can deviate significantly from $B$. Since  the probability of such an 
event is quite small \cite{Hooper:2003ad}, we do not consider such a possibility in this work.

\chapter{Some useful simple amplitudes}
\label{FeynmanGraphs}
In this Appendix we give the expression for two simple Feynman diagrams playing
an important role for neutralino annihilation through the $h$ and the $A$
pole. The notation followed is very condensed, since the main interest here is
the general form of the vertices and not the exact expressions. The exact form of
the vertices can be found in Appendix \ref{NeutralinoCouplings}.

\section{$\chi \chi \longrightarrow h \longrightarrow f \bar{f}$}
\begin{figure}[ht]
\begin{center}
\includegraphics[width = 8cm]{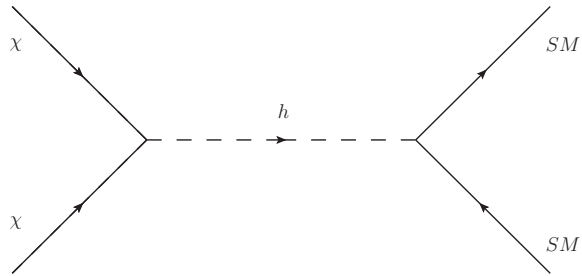}
\caption{{\footnotesize
$s$ - channel neutralino self-annihilation into standard model particles through
a light Higgs propagator.
}}
\label{chichihSMSM}
\end{center}
\end{figure}

In fig.\ref{chichihSMSM} we show the leading contribution to the neutralino
self-annihilation cross-section in case of resonant $s$ - channel annihilation 
to fermions through a light Higgs boson.
Other contributions are of course present, that must be computed and added. But this is
the leading one in the case of interest.

For the sake of brevity, we neglect the exact form of the vertices of $h$ denoting, for example, by $A$ the
$h\chi\chi$ vertex and $B$ the $h f \bar{f}$ one. The point of this simple calculation
does not depend on the exact expression of this coupling, only on that it is a simple number.
The amplitude for such a process of fig.\ref{chichihSMSM} can be written as
\begin{equation}
 i {\cal{M}} = 
u(p,s) A \bar{v}(p,s) 
\frac{i}{p^2 - m_h^2 + i m_h \Gamma}
\bar{u}(k,s) B v(k,s) 
\end{equation}
Squaring the matrix element, then averaging over initial state spins and summing
over final state ones, we get the simple expression
\begin{equation}
 \overline{ \left| {\cal{M}} \right|^2} = \frac{A^2 B^2 (4 m_\chi^2 - s)(s - 4 m_f^2)}{m_h^4 + (\Gamma^2 - 2s)m_h^2 + s^2}
\end{equation}
where $s$ is the usual Mandelstam variable, $s = (p_1 + p_2)^2 = 4p^2$ if we work in the 
center-of-mass (CM) frame and $\Gamma$ is the $h$ width.
Now, in the zero velocity limit, $s \rightarrow 4 m_\chi^2$ and hence the amplitude obviously
vanishes.

\section{$\chi \chi \longrightarrow A \longrightarrow f \bar{f}$}
\begin{figure}[ht]
\begin{center}
\includegraphics[width = 8cm]{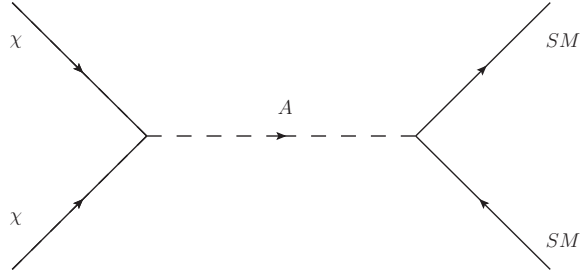}
\caption{{\footnotesize
$s$ - channel neutralino self-annihilation into standard model particles through
a light Higgs propagator.
}}
\label{chichiASMSM}
\end{center}
\end{figure}

In fig.\ref{chichiASMSM} we show the leading contribution to the neutralino
self-annihilation cross-section in case of resonant $s$ - channel annihilation 
into fermions through a pseudoscalar Higgs boson.
Once again, we point out that this is just the leading contribution at tree-level at the
$A$-resonance.

Working in the same way as before, we collectively denote by $A$ and $B$ the couplings
of the pseudoscalar to the neutralinos and the fermions. Only that in this case, the
couplings are not actually numbers, they include a $\gamma_5$ factor. The relevant
amplitude should thus be written as
\begin{equation}
 i {\cal{M}} = 
u(p,s) A \gamma_5 \bar{v}(p,s) 
\frac{i}{p^2 - m_h^2 + i m_h \Gamma}
\bar{u}(k,s) B \gamma_5 v(k,s) 
\end{equation}
This expression gives
\begin{equation}
 \overline{ \left| {\cal{M}} \right|^2} = \frac{A^2 B^2 s^2}{m_h^4 + (\Gamma^2 - 2s)m_h^2 + s^2}
\end{equation}
which by no means vanishes in the zero velocity limit.

We thus see a point repeatedly pointed out in the main body of this work, namely
that the zero velocity limit of resonant annihilation into an $f\bar{f}$ pair 
exhibits a completely different behavior depending on the scalar or pseudoscalar
nature of the propagator.

\chapter{Masses and couplings}
\label{NeutralinoCouplings}

In this Appendix, we provide some useful formulae concerning the masses and couplings
of neutralinos to various MSSM particles in order to illustrate some points
in the main body of this work.

Furthermore, we provide the relevant modifications brought about by the
introduction of NR operators in the BMSSM case. In this case, we also include 
the corrections to the Higgs mass matrices.

For convenience, we repeat several formulae already found in the main body
of this work, notably the (mass) matrix expressions for the various physical
states. The aim is by no means to give a full listing of all the mass matrices,
rotation matrices or couplings in the MSSM and the BMSSM, but just to 
provide formulae which are useful for a better understanding of some 
of the arguments in the relevant chapters.

\section{MSSM}

\subsection{Physical spectrum and neutralino mass matrix}
In the Higgs sector, we have the following transformation matrices
that relate the gauge eigenstates to the physical ones:
\\ \\
1) Neutral goldstone boson and the $CP$ - odd Higgs\\
\begin{equation}
\left( \begin{array}{c}   G^0 \\ A \end{array} \right) 
= \left( \begin{array}{cc} \cos \beta & \sin \beta \\
- \sin \beta & \cos \beta \end{array} \right) \ 
\left( \begin{array}{c}   P_1^0 \\ P_2^0 \end{array} \right)
\end{equation}
2) Charged Goldstone bosons and charged Higgses
\begin{equation}
\left( \begin{array}{c}   G^\pm \\ H^\pm \end{array} \right) 
= \left( \begin{array}{cc} \cos \beta & \sin \beta \\
- \sin \beta & \cos \beta \end{array} \right) \ 
\left( \begin{array}{c}   H_1^\pm \\ H_2^\pm \end{array} \right) 
\end{equation}
3) $CP$ - even Higgs bosons
\begin{equation}
\left( \begin{array}{c}   H \\ h \end{array} \right) 
= \left( \begin{array}{cc} \cos \alpha & \sin \alpha \\
- \sin \alpha & \cos \alpha \end{array} \right) \ 
\left( \begin{array}{c}   H_1^0 \\ H_2^0 \end{array} \right) 
\end{equation}
\\ \\
The neutralino mass matrix can be written as
\begin{equation}
{\cal M}_0 =
 \begin{pmatrix}
    M_1 & 0 & -M_Z s_W c_\beta & M_Z s_W s_\beta \\
    0 & M_2 & M_Z c_W c_\beta & -M_Z c_W s_\beta \\
    -M_Z s_W c_\beta & M_Z c_W c_\beta & 0 & -\mu \\
    M_Z s_W s_\beta & -M_Z c_W s_\beta & -\mu & 0
  \end{pmatrix}.
\label{mneutraApp}
\end{equation}\\
where $s_\beta = \sin\beta$, $c_\beta = \cos\beta$, $s_W = \sin\theta_W$,  $c_W = \cos\theta_W$,
$\theta_W$ is the Weinberg angle and $M_Z$ is the $Z$ boson mass.
\\
This matrix can be diagonalized by a unitary matrix $Z_0$ as
\begin{equation}
 {\cal M}_0 = Z_0 D_0 Z_0^\dag
\end{equation}
resulting in four eigenstates which are mixtures of the bino, the wino and the two Higgsinos.
Analytical diagonalization has been performed in \cite{ElKheishen:1992yv}.
\\
We denote the lightest of these eigenstates by 
\begin{equation}
 \chi_1^0 = Z_{11} \tilde{B} + Z_{12} \tilde{W}^3 + Z_{13} \tilde{H}_1^0 + Z_{14} \tilde{H}_2^0
\label{NeutralinoGenericApp}
\end{equation}
where we assume that we have rearranged the neutralino matrix in order to have the lightest one
at the top row and we have dropped the subscript $0$ in the $Z$ matrix. 
The gaugino and Higgsino fraction are defined by
\begin{eqnarray}
 f_G & = & Z_{11}^2 + Z_{12}^2\\ \nonumber
 f_H & = & Z_{13}^2 + Z_{14}^2
\end{eqnarray}
%%%%%%%%%%%%%%%%%%%%%%%%%%%%%%%%%%%%%%%%%%%%%%%%%%%%%%%%%%%%%%%%%%%%%%%%%%%%%%%%%%%%%%%%%%%%%%%%%%%%%%%%%%%%%
\subsection{Couplings}
\textbf{1) Couplings to neutral Higgs bosons:}\\
We write these couplings in a generic manner as $g^{L}_{ijk} P_L + g^{R}_{ijk} P_R$
where $g^{L,R}_{\chi^0_i \chi^0_j H^0_k} = g^{L,R}_{ijk}$. The indices correspond to
$h = 1, H = 2, 1 = 3$. $P_L$ and $P_R$ are left
and right handed projection operators, defined as $P_{L,R} = 1/2(1 \mp \gamma^5)$ 
and the relevant $g_L, g_R$ are:
\\
\begin{tabular}{ll}
\begin{picture}(150,70)(0,-10)
%left horizontal line
\DashLine(60,0)(10,0){5}
\Text(0,0)[c]{$H^0_k$}
%right horizontal line
\ArrowLine(60,0)(110,0)
\Text(120,0)[c]{$\chi^0_i$}
%upper vertical line
\ArrowLine(60,50)(60,0)
\Text(65,45)[l]{$\chi^0_j$}
%blob
\Vertex(60,0){2}
\end{picture}
&
\raisebox{40\unitlength}{
\begin{minipage}{5cm}
\begin{eqnarray} \nonumber
g^L_{ijk} & = & \frac{1}{2 s_W} \left( Z_{j2}- \tan\theta_W Z_{j1} \right) 
\left(e_k Z_{i3} + d_kZ_{i4} \right) \ + \ i \leftrightarrow j \\ \nonumber
g^R_{ijk} & = & \frac{1}{2 s_W}  \left( Z_{j2}- \tan\theta_W Z_{j1} 
\right) \left(e_k Z_{i3} + d_kZ_{i4} \right) \epsilon_k \ + \ i 
\leftrightarrow j \nonumber
\end{eqnarray}

\end{minipage}
}
\end{tabular}\\ 
In these equations $Z$ is the $4\times 4 $ matrix that 
diagonalizes the neutralino matrices and $\epsilon_{1,2}=- 
\epsilon_3 =1$. The coefficients $e_k$ and $d_k$ are
\begin{eqnarray}
e_1=+ \cos \alpha \ , \
e_2=- \sin \alpha \ ,  \
e_3=- \sin\beta \nonumber \\ 
d_1=  -\sin\alpha \ , \
d_2= -\cos\alpha \ ,  \
d_3= +  \cos\beta
\label{ed-coefficients}
\end{eqnarray}
We note that a pure gaugino ($f_H = 0$) or a pure Higgsino ($f_G = 0$) does not
couple to the Higgs bosons.
\\ \\
\textbf{2) Couplings to the $Z$ boson:}\\
We follow the same notation writing the couplings under the generic form
$G^{L}_{ijZ} P_L + G^{R}_{ijZ} P_R$ where
$G^{L,R}_{\chi^0_i \chi^0_j Z} = G^{L,R}_{ijZ}$ and the relevant coefficients
are given by:\\
% NNZ
\begin{tabular}{ll}
\begin{picture}(150,70)(0,-10)
%left horizontal line
\Photon(10,0)(60,0){3}{4}
\Text(0,0)[c]{$Z^0_{\mu}$}
%right horizontal line
\ArrowLine(60,0)(110,0)
\Text(120,0)[c]{$\chi^0_i$}
%upper vertical line
\ArrowLine(60,50)(60,0)
\Text(65,45)[l]{$\chi^0_j$}
%blob
\Vertex(60,0){2}
\end{picture}
&
\raisebox{40\unitlength}{
\begin{minipage}{5cm}
\begin{eqnarray}\nonumber
G^L_{ijZ} & = & - \frac{1}{2s_Wc_W} [Z_{i3} Z_{j3} - Z_{i4} Z_{j4}]  \\
G^R_{ijZ} & = & + \frac{1}{2s_Wc_W} [Z_{i3} Z_{j3} - Z_{i4} Z_{j4} ] \nonumber
\end{eqnarray} 
\end{minipage}
}
\end{tabular}\\
In this case, we should note that a pure gaugino neutralino does
not couple to the $Z$ boson, a point which is already commented upon in the
main body of this work.

A complete set of Feynman rules, along with the (numerous) ones which are omitted here
can be found, for example, in \cite{Rosiek:1995kg}, although the notations vary with
respect to the ones used here. The notation followed in the present work draws largely
from \cite{Djouadi:2001fa, Djouadi:2005gj}.

\section{BMSSM}
The introduction of NR operators brings along modifications in a series of quantities:
The Higgs, neutralino and chargino mass matrices, the Higgs trilinear and quartic self-couplings, 
the Higgs sector couplings to neutralinos and charginos. We do not include here a full set of 
Feynman rules for the BMSSM. We only give some couplings that help illustrating some arguments
in the main body of the text.
\\ \\
\textbf{1) Modification in the lightest Higgs mass:}\\
Considering $M_Z$, $m_A$ and $\tan\beta$ as input parameters, we obtain the
correction to the lightest Higgs mass:
\begin{equation}
 \delta_\epsilon m_h^2 = 
2v^2\left(\epsilon_{2}-2\epsilon_{1}s_{2\beta}
-\frac{2\epsilon_{1}(m_A^2+M_Z^2)s_{2\beta}
+\epsilon_{2}(m_A^2-M_Z^2)c^2_{2\beta}}
{\sqrt{(m_A^2-M_Z^2)^2+4m_A^2M_Z^2s^2_{2\beta}}}\right)
\end{equation}
In the same time, the Higgs mixing angle is shifted from its MSSM value:
 \begin{eqnarray}
s_{2\alpha} & = & \frac{-(m_A^2+M_Z^2)s_{2\beta}+4v^2\epsilon_{1}}
{(m_H^2-m_h^2)s_{2\beta}}\\
& = & -\frac{(m_A^2+M_Z^2)s_{2\beta}}
{(m_A^4-2m_A^2M_Z^2c_{4\beta}+M_Z^4)^{1/2}}
-4v^2c_{2\beta}^2\frac{2\epsilon_{1}(m_A^2-M_Z^2)^2
-\epsilon_{2}s_{2\beta}(m_A^4-M_Z^4)}
{(m_A^4-2m_A^2M_Z^2c_{4\beta}+M_Z^4)^{3/2}} \nonumber
\end{eqnarray}
where we should note that generically, these equations should contain only the real parts
of $\epsilon_1$ and $\epsilon_2$. In this work, we neglect the possibility for $CP$
violating imaginary parts in the two parameters.
\\ \\
\textbf{2) Modification of the neutralino mass matrix:}\\
The neutralino mass matrix receives contributions like
\begin{equation}
M_{0}=\begin{pmatrix}
M_1 & 0 & -M_Zs_Wc_\beta & M_Zs_Ws_\beta \\
0 & M_2 & M_Zc_Wc_\beta & -M_Zc_Ws_\beta \\
-M_Zs_Wc_\beta & M_Zc_Wc_\beta & 0 & -\mu \\
M_Zs_Ws_\beta & -M_Zc_Ws_\beta & -\mu & 0 \end{pmatrix}
+\frac{4\epsilon_1 m_W^2}{\mu^* g^2}
\begin{pmatrix} 0 & 0 & 0 & 0 \\ 0 & 0 & 0 & 0 \\
0 & 0 & s^2_\beta & s_{2\beta} \\ 0 & 0 & s_{2\beta} & c^2_\beta
\end{pmatrix}.
\end{equation}
\\ \\
\textbf{3) Modification of the neutralino couplings to Higgs bosons:}\\
The neutralino couplings to neutral Higgs bosons are also modified,
according to the following formulae:
% \begin{eqnarray}
% \delta_{\epsilon h} & = & \frac{2\sqrt{2}\lambda_1^* v}{gM}\times
% \frac{-c_\beta s_\alpha Z_{14}^2 +s_\beta c_\alpha
%   Z_{13}^2+2c_{(\alpha+\beta)}Z_{13}Z_{14}}
% {(Z_{12}-\tan\theta_W Z_{11})(s_\alpha Z_{13}+c_\alpha Z_{14})}\\
% \delta_{\epsilon H} & = & \frac{2\sqrt{2}\lambda_1^* v}{gM}\times
% \frac{c_\beta c_\alpha Z_{14}^2 +s_\beta s_\alpha
%   Z_{13}^2+2s_{(\alpha+\beta)}Z_{13}Z_{14}}
% {(Z_{12}-\tan\theta_W Z_{11})(-c_\alpha Z_{13}+s_\alpha Z_{14})}\\
% \delta_{\epsilon A} & = & \frac{\sqrt{2}\lambda_1^* v}{gM}\times
% \frac{s_{2\beta} (Z_{14}^2+Z_{13}^2)+4Z_{13}Z_{14}}
% {(Z_{12}-\tan\theta_W Z_{11})(s_\beta Z_{13}-c_\beta Z_{14})}
% \end{eqnarray}

\begin{eqnarray}
 g^L_{h \chi^0_i \chi^0_j} & = & 2 i \left( \frac{\epsilon_1}{\mu^*} \right) \Big(- v\sqrt{2} \cos\beta \cos\alpha N^*_{i4} N^*_{j4} - v \sqrt{2} \sin\beta \sin\alpha N^*_{i3} N^*_{j3} \nonumber \\
& - & 2\sqrt{2} v \sin(\alpha+\beta) \frac{1}{2}(N^*_{i4} N^*_{j3} + N^*_{j4} N^*_{i3})\Big) \\
 g^R_{h \chi^0_i \chi^0_j} & = & i \left( g^L_{H^0_1 \chi^0_i \chi^0_j} \right)^* \\
 g^L_{H \chi^0_i \chi^0_j} & = & 2 i \left( \frac{\epsilon_1}{\mu^*} \right) \Big( v\sqrt{2} \cos\beta \sin\alpha N^*_{i4} N^*_{j4} - v \sqrt{2} \sin\beta \cos\alpha N^*_{i3} N^*_{j3} \nonumber \\
& - & 2\sqrt{2} v \cos(\alpha+\beta) \frac{1}{2}(N^*_{i4} N^*_{j3} + N^*_{j4} N^*_{i3}) \Big) \\
 g^R_{H \chi^0_i \chi^0_j} & = & i \left( g^L_{H^0_2 \chi^0_i \chi^0_j} \right)^* \\
 g^L_{A \chi^0_i \chi^0_j} & = & 2 i \left( \frac{\epsilon_1}{\mu^*} \right) \Big( -i v\frac{1}{\sqrt{2}} \sin{2\beta} N^*_{i4} N^*_{j4} - i v \frac{1}{\sqrt{2}} \sin{2\beta} N^*_{i3} N^*_{j3} \nonumber \\
& - & i 2 \sqrt{2} v \frac{1}{2}(N^*_{i4} N^*_{j3} + N^*_{j4} N^*_{i3}) \Big) \\
 g^R_{A \chi^0_i \chi^0_j} & = & i \left( g^L_{H^0_3 \chi^0_i \chi^0_j} \right)^* 
\end{eqnarray}
where once again the couplings are decomposed as $g_{ijk}^L P_L + g_{ijk}^R P_R$.

\bibliographystyle{utphys}
\addcontentsline{toc}{chapter}{References}
\bibliography{biblio}

%...Switch again to empty pagestyle
%  \pagestyle{empty}
% \newpage
% \include{remerciements/remerciements}

\end{document}